\documentclass{aastex62}
\usepackage[T1]{fontenc}
\usepackage[utf8]{inputenc}
\usepackage[english]{babel}
\usepackage{esint}
\usepackage{booktabs}
\usepackage{url}
\usepackage{graphicx}
\usepackage{longfigure}
\usepackage{lineno}
\usepackage{soul}
\usepackage{fixmath}
\usepackage{longtable}
\usepackage{footmisc}

\usepackage[caption=false]{subfig}
\usepackage{verbatim}

\usepackage{xcolor}
\usepackage{natbib}
\newcommand\fermilat{{\it Fermi}-LAT }

\graphicspath{{./}{figures/}}

\shorttitle{Detection of new $\gamma$-ray transients}
\shortauthors{\fermilat}

\begin{document}

\title{Catalog of Long-Term Transient Sources in the First 10 Years of \fermilat Data}

\correspondingauthor{Isabella Mereu, Sara Cutini, Gino Tosti, Elisabetta Cavazzuti}
\email{isabella.mereu@pg.infn.it, sara.cutini@pg.infn.it, gino.tosti@unipg.it, elisabetta.cavazzuti@asi.it}

\collaboration{The \fermilat collaboration}

%\begin{linenumbers}

\author{L.~Baldini}
\affiliation{Universit\`a di Pisa and Istituto Nazionale di Fisica Nucleare, Sezione di Pisa I-56127 Pisa, Italy}
\author{J.~Ballet}
\affiliation{AIM, CEA, CNRS, Universit\'e Paris-Saclay, Universit\'e de Paris, F-91191 Gif-sur-Yvette, France}
\author{D.~Bastieri}
\affiliation{Istituto Nazionale di Fisica Nucleare, Sezione di Padova, I-35131 Padova, Italy}
\affiliation{Dipartimento di Fisica e Astronomia ``G. Galilei'', Universit\`a di Padova, I-35131 Padova, Italy}
\author{J.~Becerra~Gonzalez}
\affiliation{Instituto de Astrof\'isica de Canarias, Observatorio del Teide, C/Via Lactea, s/n, E-38205 La Laguna, Tenerife, Spain}
\author{R.~Bellazzini}
\affiliation{Istituto Nazionale di Fisica Nucleare, Sezione di Pisa, I-56127 Pisa, Italy}
\author{A.~Berretta}
\affiliation{Dipartimento di Fisica, Universit\`a degli Studi di Perugia, I-06123 Perugia, Italy}
\affiliation{Istituto Nazionale di Fisica Nucleare, Sezione di Perugia, I-06123 Perugia, Italy}
\author{E.~Bissaldi}
\affiliation{Dipartimento di Fisica ``M. Merlin" dell'Universit\`a e del Politecnico di Bari, via Amendola 173, I-70126 Bari, Italy}
\affiliation{Istituto Nazionale di Fisica Nucleare, Sezione di Bari, I-70126 Bari, Italy}
\author{R.~D.~Blandford}
\affiliation{W. W. Hansen Experimental Physics Laboratory, Kavli Institute for Particle Astrophysics and Cosmology, Department of Physics and SLAC National Accelerator Laboratory, Stanford University, Stanford, CA 94305, USA}
\author{E.~D.~Bloom}
\affiliation{W. W. Hansen Experimental Physics Laboratory, Kavli Institute for Particle Astrophysics and Cosmology, Department of Physics and SLAC National Accelerator Laboratory, Stanford University, Stanford, CA 94305, USA}
\author{R.~Bonino}
\affiliation{Istituto Nazionale di Fisica Nucleare, Sezione di Torino, I-10125 Torino, Italy}
\affiliation{Dipartimento di Fisica, Universit\`a degli Studi di Torino, I-10125 Torino, Italy}
% ...

\begin{abstract}

We present the first \textit{Fermi} Large Area Telescope (LAT) catalog of long-term $\gamma$-ray transient sources (1FLT). This comprises sources that were detected on monthly time intervals during the first decade of \textit{Fermi}-LAT operations.
The monthly time scale allows us to identify transient and variable sources that were not yet reported in other \fermilat catalogs. 
The monthly datasets were analyzed using a wavelet-based source detection algorithm that provided the candidate new transient sources. The search was limited to the extragalactic regions of the sky to avoid the dominance of the Galactic diffuse emission at low Galactic latitudes.
The transient candidates were then analyzed using the standard \fermilat Maximum Likelihood analysis method. All sources detected with a statistical significance above $4\sigma$ in at least one monthly bin were listed in the final catalog.

The 1FLT catalog contains 142 transient $\gamma$-ray sources that are not included in the 4FGL-DR2 catalog. 
Many of these sources (102) have been confidently associated with Active Galactic Nuclei (AGN): 24 are associated with Flat Spectrum Radio Quasars; 1 with a BL Lac object; 70 with Blazars of Uncertain Type; 3 with Radio Galaxies; 1 with a Compact Steep Spectrum radio source; 1 with a Steep Spectrum Radio Quasar; 2 with AGN of other types. The remaining 40 sources have no candidate counterparts at other wavelengths. 
The median $\gamma$-ray spectral index of the 1FLT-AGN sources is softer than that reported in the latest \fermilat AGN general catalog. This result is consistent with the hypothesis that detection of the softest $\gamma$-ray emitters is less efficient when the data are integrated over year-long intervals.

\end{abstract}


\section{Introduction}\label{sec:intro}
The Large Area Telescope (LAT) on board the {\it Fermi} Gamma-ray Space Telescope is an imaging pair-conversion detector operating in the energy range from 20 MeV to >300 GeV. It has a field of view of 2.7 sr at 1 GeV and can provide an image of the entire sky approximately every three hours \citep{atwood2009large}. This, together with its large collecting area, makes the LAT well suited to the investigation of $\gamma$-ray variable and transient sources. 
The general (xFGL) catalogs (\citealt{abdo2010fermi1FGL}, \citealt{nolan2012fermi2FGL}, \citealt{acero2015fermi3FGL}, \citealt{20204FGL}) were produced by integrating over years of LAT observations. Studies of the counterparts of these LAT catalogued sources are reported in dedicated papers, e.g. the Active Galactic Nuclei (AGN) catalogs named xLAC (\citealt{abdo2010firstLAC}, \citealt{ackermann2011secondLAC}, \citealt{ackermann20153LAC}, \citealt{20204LAC}). In the \fermilat Third Source Catalog (aka 3FGL, 4 years of integration;  \citealt{acero2015fermi3FGL}), the last catalog that included monthly light curves for the detected objects, only $\sim 22\%$ of the sources were found to be variable. The majority of the latter were associated with AGN, with only a few classes of Galactic sources being found to show variability, i.e. Crab Pulsar Wind Nebula \citep{2011Sci...331..736T}, High-mass X-ray binaries \citep{fermi2009modulated} and Galactic novae \citep{2014Sci...345..554A}.
In the \fermilat Fourth Source Catalog Data Release 2 (hereafter 4FGL-DR2; \citealt{ballet2020fermi}), based on 10 years of data, the fraction of sources associated with flat-spectrum radio quasars (FSRQs) is less than that associated with BL Lacs. This difference in populations was also found in the previous xFGL catalogs, with the fraction of new sources slightly higher for hard-spectrum sources than for soft-spectrum ones relative to the previous catalog.
This could be related to the different variability properties for FSRQs and BL Lacs  as well as to the typically soft $\gamma$-ray spectra of FSRQs, while BL Lacs generally show harder spectra \citep{abdo2010gamma}.
In fact the sensitivity of the  \textit{Fermi-}LAT\footnote{\label{fn:slac}\url{https://www.slac.stanford.edu/exp/glast/groups/canda/lat_Performance.htm}}, improves at a faster rate as a function of exposure for harder sources
\footnote{This is because the LAT point-spread function (PSF) at low energies is broader and, therefore, the soft-spectrum sources reach the confusion limit. On the other hand, at higher energies the PSF is relatively narrow. Hard-spectrum sources emit many more high-energy photons than soft sources, leading to more precise localizations and, in turn, more reliable associations.}.

Based on the luminosity function of radio-loud AGN however, we would have expected to find more soft $\gamma$-ray AGN whose emission peaks at high frequencies in the Inverse Compton regime \citep{radiogamma} than those that were detected in the xFGL catalogs (\citealt{abdo2010firstLAC}, \citealt{ackermann2011secondLAC} and \citealt{ackermann20153LAC}).

Following these premises, soft-spectrum objects that have low duty cycle would not be detectable in the years-long integration of xFGL catalogs but could be detectable during brief periods of enhanced $\gamma$-ray activity. 
Searches for variable sources  have already been carried out for this reason by the \fermilat collaboration using different techniques. The Monitored Source List\footnote{\url{https://fermi.gsfc.nasa.gov/ssc/data/access/lat/msl_lc/}} provides daily and weekly light curves of the brightest sources and transients found during LAT observations. Variability on time scales of 6 hours and 1 day is monitored by the {\it Fermi} Flare Advocate program \citep{2012AIPC.1505..697Ciprini2012} using the quick-look science data products of the Automated Science Processing pipeline \citep{2012amld.book...41C}. Finally, the $Fermi$ All-sky Variability Analysis (\citealt{ackermann2013fermiFAVA} and \citealt{abdollahi2017second2FAV}) uses a photometric technique to  blindly search the data for transients over the entire sky on weekly time intervals. This latter analysis is independent of any model for the diffuse and isotropic $\gamma$-ray emission, and it produced a list of 215 and 518 flaring $\gamma$-ray sources (including those already listed in the xFGL catalogs) over 3.9 years and 7.4 years of integration, respectively.

In this work we present the  First {\it Fermi} Gamma-ray LAT Transient Catalog (1FLT), a census of sources located at $|b|>10^\circ$ and detected by scanning a decade of \fermilat data over monthly time intervals. 
In Section \ref{sec:analysis} we describe the analysis procedure used to search for new source candidates and the Maximum Likelihood (ML) analysis used to estimate their significance. In Section \ref{sec:assoc} we describe the methods used to perform a positional association of the new sources with those listed in several radio, infrared optical and X-ray catalogs. In Section \ref{sec:properties} we discuss the spectral properties of the 1FLT catalog, and the results of the comparison with the sources reported in the fourth \fermilat catalog of active galactic nuclei (4LAC, \citealt{20204LAC}) in order to investigate the differences between the $\gamma$-ray sources detected in short and long integration times. In Section \ref{sec:discussion} we discuss the comparison with the second catalog of flaring $\gamma$-ray sources (2FAV, \citealt{abdollahi2017second2FAV}) and we describe some particular sources. Finally in Section \ref{sec:conc} we report our results and conclusions of this new strategy for the detection of variable $\gamma$-ray sources.

\section{\fermilat data analysis}\label{sec:analysis}

The data used in this work encompass a 10-year period from 2008 August 4 15:43:36 UTC (Mission Elapsed Time, MET 239557417) to 2018 August 05 10:23:32 UTC (MET 555157417). They are analyzed in monthly time bins, where one time bin (TBIN) is defined as 2630000 seconds ($\sim$ 30 days).
The 1FLT is constructed in the energy range from 100 MeV to 300 GeV using the \texttt{Pass8} SOURCE class events, in combination with  \texttt{P8R3$\_$SOURCE$\_$V2} Instrument Response Functions (IRFs)\footref{fn:slac}.  
The exposure of the \fermilat is fairly uniform across the sky, but the brightness of the interstellar diffuse $\gamma$-ray background, and hence the sensitivity for source detection,  depends strongly on the Galactic latitude. For this reason we apply a further cut, considering only data with Galactic latitude $|b|>10^\circ$.

We applied a zenith angle cut of $<90^\circ$ in order to reduce the contamination from the Earth limb  and the standard data quality cuts (DATA\_QUAL > 0 \&\& LAT\_CONFIG == 1) for the extraction of good time intervals. We selected only ``SOURCE'' event class (LAT EVENT\_CLASS = 128) and event type ``FRONT+BACK'' (LAT EVENT\_TYPE = 3).
We repeated the monthly binned analysis selecting data with a 15-day shift, that is, with the first one-month time bin starting on 2008 August 19 13:39:59 UTC (MET 240846000). 
This improves the sensitivity for sources that may display variable activity at the edges of the time bins.
Adopting this procedure we have analyzed  a total of two sets of 120 monthly data sets, which are not independent since they partially overlap. In the following we refer to these two sets as follows: ``nominal'' for the first set starting on 2008 August 4, and ``shifted'' for the second set starting on 2008 August 19.

\subsection{Source detection and localization}\label{subsec:souSel}
We have partitioned each of the 240 monthly TBINs into 192 circular regions of interest (ROIs) centered on points defined by means of a HEALPix\footnote{\url{http://healpix.sourceforge.net}} sky pixelization \citep{gorski2005healpix} with $N_{side} = 4$ and adopting celestial coordinates. The ROIs include events in cones of $15^\circ$ radius about the center of the pixel. The ROIs constructed in this way are not independent as they partially overlap. 

To identify the candidate sources (seeds) we used a wavelet transform analysis \citep{damiani1997method} using the \textit{\textit{PGWave}} tool \citep{ciprini20071d}. \textit{PGWave} works on the square images inscribed in ROIs of $15^\circ$ radius; therefore, the side of the square images is $15^\circ\sqrt{2} \sim 21.2$ degrees. We adopted a pixel size of 0.25$^\circ$ and  the stereoscopic projection for the images. We performed the analysis independently on each ROI and TBIN deleting the seeds located at the border of the ROI (distance from the ROI center $>10^\circ$) in order to avoid edge effects. In this way we obtained 129,802 seeds in the nominal months and 133,673 seeds in the shifted ones.

After the seed extraction we performed a ML analysis on all seeds that had Galactic latitude $|b|>10^\circ$ and that had an angular distance greater than 50 arcmin from any 4FGL-DR2 source (with this selection we excluded $\sim$ 10\% of the extragalactic sky).
The conservative 50 arcmin cross correlation radius corresponds to the average plus 1$\sigma$ value of the distribution of the 1FLT catalog’s semimajor axis of the error ellipses at 95\% confidence level (Conf\_95\_SemiMajor in Table \ref{table:catalogdescription}).
The ML analysis provides the evaluation of the Test Statistic, TS, defined as TS $= 2\mathrm{\Delta}$log$\mathcal{L}$ \citep{mattox1996likelihood} which quantifies how significantly a source emerges from the background comparing the likelihood function $\mathcal{L}$ with and without a source.
We performed the final localization only for seeds with TS $>$ 10 (4872 for the nominal months and 4737 for the shifted ones, respectively). 

We evaluated the significance of the detection and the spectral parameters for the newly found sources by performing a binned ML with eight bins per decade in energy and 0.1$\degr$ binning in Galactic coordinates. For each point source we assumed a power-law spectrum defined as
\begin{math}
dN/dE = K*(E/E_0)^{-\Gamma} 
\end{math},
where $K$ is the flux scale factor, $\Gamma$ the spectral index, and $E_0$ is the reference energy ($E_0=$ 1 GeV). 
The ROI model used in the likelihood analysis contained all \textit{PGWave} seeds found (including also the ones associated with 4FGL-DR2 sources) in the selected month in a ROI of 10$^\circ$ x 10$^\circ$ square, centered at the position of the target, along with the Galactic and isotropic diffuse backgrounds\footnote{\url{https://fermi.gsfc.nasa.gov/ssc/data/access/lat/BackgroundModels.html}}. To perform the analysis we used the \textit{Fermitools} 1.2.1 package available from the Fermi Science Support Center (FSSC)\footnote{\url{https://fermi.gsfc.nasa.gov/ssc/data/analysis/}} and the \textit{Fermipy} 0.18.0 software package \citep{wood2017fermipy}. 
We use TS = 25 as the detection threshold, corresponding to a significance of $\sim 4\sigma$ assuming a $\chi^2$ distribution with 4 degrees of freedom (position and spectral parameters of a power-law source, see \citealt{mattox1996likelihood}).

\subsection{False positive estimation and source detection efficiency}\label{subsec:eff}

The evaluation of the false positive rate and efficiency are fundamental to assess the reliability of the adopted method in the detection of transient sources.
In order to calculate the false positive rate we simulated 120 month-long data sets, including in the model only the Galactic and the extragalactic diffuse emissions. Over month-long time scales the false positives are dominated by statistical fluctuations of the background rather than systematic effects. For the simulations we adopted the analysis of the 1FLT. 
We identified the seeds as described in Sect.\ref{subsec:souSel} and performed the binned ML analysis assuming a power-law spectrum for each seed.
The distributions of the statistical significance of the candidate sources found in the fake and true skies are shown in Figure \ref{figure:sim}. 
In the 120 simulated skies we found 12 spurious detections with TS $>$ 25 over $\sim$ 24000 seeds, and 2 and 1 detections with TS > 28 and TS $>$ 30, respectively. 
To be conservative, in the estimation of the spurious detections in the real data these numbers should be multiplied by a factor of two to account also for the shifted months. Therefore we may expect 24 fake detections with TS $>25$. 

\begin{figure}[hbt!]
 \plotone{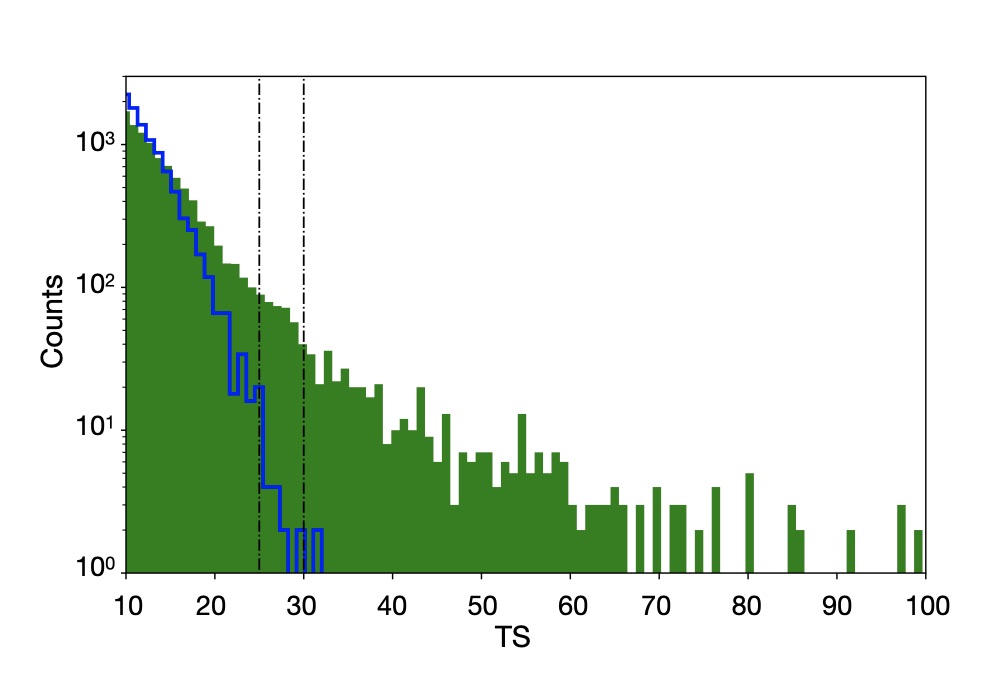}
 \caption{Distributions of TS evaluated for the simulated skies (blue curve) and the actual skies (shaded green). The results obtained in the simulation on 120 months are represented multiplied by two to account also for the shifted months. The two vertical lines show the TS range of low-confidence sources ($25<TS<30$).}
\end{figure}\label{figure:sim}

Since our adopted threshold of TS $>$ 25 still includes a non-negligible number of spurious sources, in  the 1FLT catalog TS $>$ 25 candidate sources are flagged as low-confidence when they are detected in only one TBIN and have a TS value between 25 and 30. 72 1FLT sources are flagged as low-confidence. Each individual low-confidence source has a probability of about 34\% of being spurious.

To assess the detection efficiency we used astrophysical $\gamma$-ray sources known to be steady in flux such as pulsars (PSRs) located at high latitude reported in the 4FGL catalog. The majority of sources taken in consideration for the estimation of efficiency are millisecond pulsars.

For each TBIN of data we applied the analysis procedures targeting the selected PSRs and we compared the number of detections obtained as a function of energy flux to the total number of pulsars (efficiency of 100$\%$ when we had a significant detection of the PSRs in all 240 months). Figure \ref{figure:eff} represents the rate of detection over 240 months for the selected PSRs as a function of energy flux. As suggested in \citet{principe2018first}, we modeled the detection efficiency with a hyperbolic tangent function tanh$\lambda(f - f_0)$,  where the two parameters $\lambda$ and $f_0$ are determined by fitting to the detection efficiency points. We obtained the following values: $\lambda =  4.0 \times 10^4 ~MeV^{-1}~cm^{2}$ s and $f_0=8.41 \times 10^{-7}$ MeV cm$^{-2}$ s$^{-1}$ with a $\chi^2$ of 0.93 with 9 degrees of freedom.

\begin{figure}[hbt!]
 \plotone{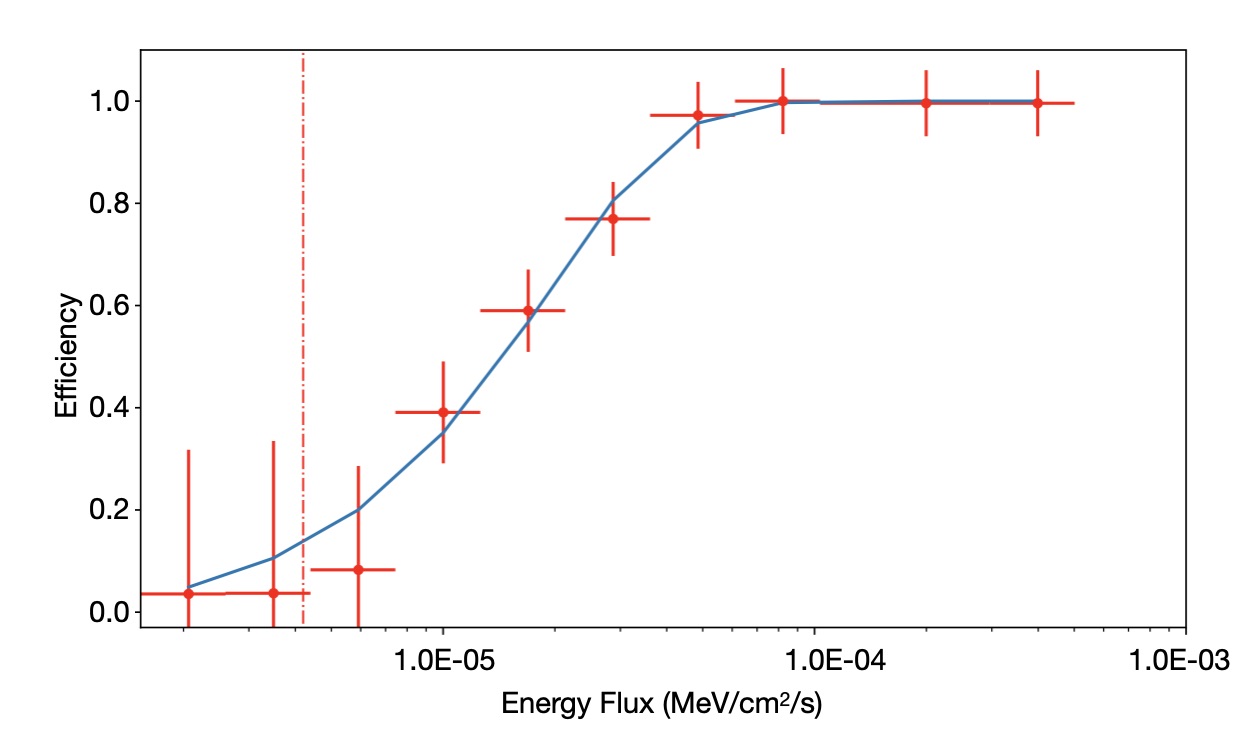}
 \caption{Detection rate as a function of the energy flux of high-latitude PSRs. Each point represents the ratio of detected sources over 240 overlapping months with respect to the number of true 4FGL PSRs with $|b|>10$ binned by flux. The error bars correspond to the statistical uncertainty estimated using the binomial statistic. The blue curve is the hyperbolic tangent function tanh$\lambda(f - f_0)$ and the vertical line is at sensitivity flux limit extrapolated for one month integration time.}
\end{figure}\label{figure:eff}

\subsection{The 1FLT Catalog construction}\label{subsec:catCon}
%%%--------------------------%%%%
The 1FLT catalog lists all the sources that satisfy the following criteria: a TS $>$ 25, a ratio between the flux error and the flux $\rm \Delta Flux/Flux < 0.5$ and a power-law index $\rm \Gamma$ $<4.5$. 
We associated multiple detections with the same source when they are  spatially coincident with 2D-sky cross-correlation using the error ellipses at 95\% confidence level. We manually checked each individual multiple association to verify the automatic procedure and in three cases we did not consider the sources as multiple detections since the centroids of the $\gamma$-ray emission were visually not co-spatial (1FLT J0240$-$4657 and 1FLT J0240$-$4739; 1FLT J0519$-$3709 and 1FLT J0527$-$3747; 1FLT J1322$-$4521 and 1FLT J1323$-$4439).
When the same source was detected on different but overlapping ROIs, we reported the results of the most significant detection. A reference of this repeated occurrence is reported in the catalog column ``repROI''. For instance see Appendix \ref{sec:catdesc} and Appendix \ref{sec:srclist}, as well as the catalog file.

The final 1FLT catalog is composed of 142 unique transient sources not associated with any 4FGL-DR2 emitters: 64 sources detected in the nominal 120 months and 78 in the shifted 120 months. Seventy-two 1FLT sources with 25 < TS < 30 are flagged as low-confidence (as reported in Sec. \ref{subsec:eff}).
The complete list of sources is available in electronic format (FITS format) as supplementary material. The columns are described in Appendix \ref{sec:catdesc} in Table \ref{table:catalogdescription}, Table \ref{table:SED} and Table \ref{table:EB}. In Appendix \ref{sec:srclist}, Table \ref{table:extract} has the 1FLT list. The source designation is 1FLT JHHMM$+$DDMM.

Out of these 142 1FLT sources, 108 1FLT sources were detected only once, i.e. in one single TBIN, while 34 other distinct 1FLT sources displayed significant gamma-ray emission in more then one TBIN.
These 34 sources listed in 1FLT correspond to their most significant detection in each relative cluster of multiple flaring episodes. 
Forty-four remaining flaring episodes are listed in the extension \texttt{FLARES} of the electronic release that includes other useful information specific to each flaring episode, such as time, localization, flux, significance and spectral parameters. Table \ref{table:extract} in the column ``Flares'' contains the number of flaring episodes of each 1FLT source. In case of multiple detections of positionally consistent candidates, the source name is based on the position of the candidate detection with the largest value of TS.

The 1FLT method allowed us to collect also 60 detections associated with the Sun and 27 associated with 14 of the brightest \fermilat $\gamma$-ray bursts (GRBs). In Appendix \ref{sec:srclist} Table \ref{table:sun} lists the Sun detections, Table \ref{table:grb} contains GRB detections.

The sky locations of the 1FLT sources together with Sun and GRB detections are shown in Figure \ref{figure:assoc}. 

\begin{figure}[hbt!]
 \epsscale{1.2}
 \plotone{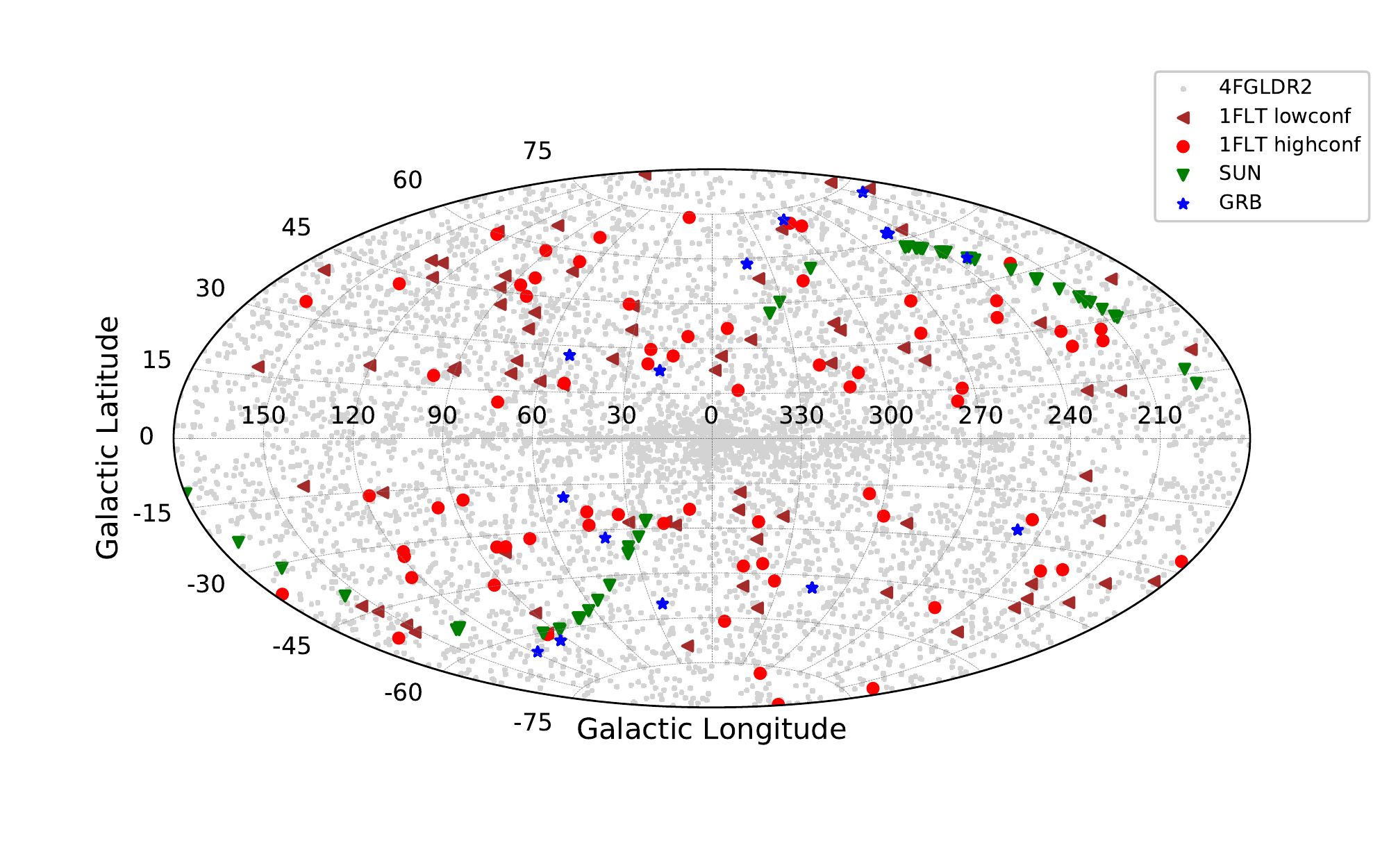}
 \caption{Aitoff projection of 1FLT sources represented in Galactic coordinates (high-confidence sources: red filled circles; low-confidence sources: brown filled triangles), Sun detections (green filled triangles) and GRB detections (blue filled stars) superimposed on 4FGL-DR2 sources represented in gray.} 
\end{figure}\label{figure:assoc}

To evaluate the possible contamination by the Moon of the 1FLT source catalog, we checked the times and trajectories of the passage of the Moon through the 1FLT fields. For the overlaps between the Moon's path and 1FLT source localization we extracted the 1FLT fluxes with a 1-day time scale in the referring month, to verify simultaneously the coincidence in time and space. 
We found no evidence for the Moon being confused with any 1FGL detection
because when we detected the higher 1-day time bin flux the Moon localization was more then 5$^{\degr}$ away from the 1FLT derived position.

\subsection{Gamma-ray Spectra and Light Curves}\label{sec:SEDLC} 

For each source in the 1FLT catalog we extracted the $\gamma$-ray spectrum and monthly binned light curve (LC). The spectral energy distributions are produced in three energy bands (i.e., 0.1--1 GeV, 1--10 GeV, 10--100 GeV) \footnote{We did not report in the catalog the last energy band (100--300 GeV) since it does not give any further information.} by modelling the 1FLT source with a power-law spectrum with free normalization and a spectral index fixed to 2. 
When the TS of the spectral bin was less than 4 or the value of spectral data point error was greater than 50\% of the measured value, we calculated the upper limit at 2$\sigma$-confidence, using the Bayesian computation if TS $<$ 1 \citep{helene1984errors}.
The $\gamma$-ray spectral data points ($\nu$F$_\nu$) are reported in the catalog available in electronic format as supplementary material.

The light curves were produced using the binned ML technique, over the energy range of 100 MeV to 300 GeV. The likelihood fit followed the same procedure described above (see Sect. \ref{sec:analysis}), fixing the photon index of the 1FLT source to the 1FLT catalog value. 
The  ROI model included all \textit{PGWave} seed detections of the TBIN located in a ROI of 10$^\circ$ centered at the 1FLT catalog position of the source of interest. When the target was detected with TS $<$ 4, or the number of its predicted photons was N$\mathrm{_{pred}<}$ 3, or the uncertainty on its flux estimate was large ($\mathrm{\Delta}$Flux/Flux > 0.5), upper limits were calculated based on the method of \citet{helene1984errors}.
For consistency with the 1FLT catalog procedure, the light curves start coherently with the nominal or shifted time selection. The statistical uncertainty on the fluxes is typically larger than the systematic uncertainty \citep{ackermann2012fermi} and only the former is considered in this paper. All light curves are shown in the Appendix.

\section{Source association} \label{sec:assoc}
%%%--------------------%%%%

To identify candidate counterparts of the 142 $\gamma$-ray sources reported in the 1FLT we used two approaches: the Bayesian method, extensively described in the xFGL catalogs (see, e.g. \citealt{acero2015fermi3FGL} and \citealt{20204FGL}), and the positional method, which relies solely on the location of the counterpart inside the error ellipse.
To estimate the probability that a specific counterpart association is likely to be real, the Bayesian approach trades the positional coincidence of possible counterparts with 1FLT sources against the expected number of chance coincidences. 
Similarly to what was done in the xFGL catalogs, in the 1FLT we retained associations if they had a posterior probability of at least 80\%. 
For the source-association analysis we first checked the same catalogs used to build the 4FGL (see \citealt{20204FGL}). 
We performed the Bayesian association procedure on each of the 240 skies for all the candidate seeds with a localization evaluation, each month ( i.e., TBIN) being an independent sky (see Sect. \ref{sec:analysis} for sky/TBIN correspondence definition).

For 1FLT sources without a Bayesian association, we searched the 95\% confidence error ellipse for the presence of promising, not yet known, $\gamma$-ray emitting sources. 
This search was carried out using the {\it SSDC Tools}\footnote{\url{https://tools.ssdc.asi.it/}} and {\it VOU-Blazar Tool}\footnote{\label{fn:open}\url{http://www.openuniverse.asi.it/}} \citep{chang2020open}.
In particular the {\it VOU-Blazar Tool} allowed us to retrieve all possible blazar candidates within a specified area of the sky. 
The complete list of catalogs used by the {\it VOU-Blazar Tool} can be found in \citet{brandt2018}. 
This method enables the association of the sources with radio surveys that are not used by the Bayesian association method. 
For 1FLT sources with large positional error ellipses more than one  candidate counterpart was found. In these cases, as well as when no candidate counterparts were found, we flagged the 1FLT source as unassociated.

\subsection{Class Designation}\label{sec:class} 
%%%--------------------%%%%
In order to classify candidate $\gamma$-ray counterparts we used either the optical classifications published in the BZCAT list (a compilation of sources  classified as blazars, \citealt{massaro20155th}) or the spectra available in the literature or from online databases. The main ones are: Sloan Digital Sky Survey (SDSS; \citealt{abolfathi2018fourteenth}, \citealt{albareti201713th}, \citealt{alam2015eleventh}) and the 6dF Galaxy Survey \citep{jones20096df}.
The classes have been assigned with the following criteria:
\begin{itemize}
    \item flat-spectrum radio quasar, BL Lac-type object, compact steep spectrum radio source (CSS), steep spectrum radio quasar (SSRQ) and radio galaxy (RG): sources with a well-established classification in the literature and/or through a good-quality optical spectrum;
    \item blazar candidate of uncertain type (BCU): an object in the listed BZCAT as blazar of uncertain/transitional type or a source with blazar multiwavelength characteristic - flat radio spectrum and double-humped, broad-band spectral energy distribution (SED);
    \item non-blazar active galaxy: for these candidate counterparts the existing data do not allow an unambiguous determination of the AGN type;
    \item unassociated: sources without an unequivocal counterpart or without any plausible candidate at other wavelengths
\end{itemize}
The class  designations for the 1FLT sources are listed in Table \ref{table:census}.

\subsection{Census}\label{sec:census}
As reported in Table \ref{table:census}, the 1FLT includes 142 sources, with 24 FSRQs, 1 BL Lac, 1 CSS, 1 SSRQ, 3 RGs, 70 BCUs and 2 other AGN.
This sample is composed exclusively of jetted AGN, which is also the largest population present in the xFGL catalogs. The sample composition however, is different when compared to that of the \fermilat general catalogs. There are 24 FSRQs which represents a higher fraction as compared to the xFGL catalogs.
Twenty-seven sources are associated by the Bayesian method, 75 by the positional method. Forty sources remain unassociated (28.2\%). This fraction is similar to that reported in the xFGL catalogs.
The low-confidence sample includes 72 sources among which associations are found with 29 BCUs, 1 BL Lac, 10 FSRQs, 2 RGs, 1 SSRQ and 2 other type AGN. 
These low-confidence sources were associated by the Bayesian method in 4 cases (1 BCU and 3 FSRQs).
The high-confidence sample includes 70 sources among which associations are found with 41 BCUs, 1 CSS, 14 FSRQs and 1 RG. In this sample, the Bayesian method finds 23 associations (1 CSS, 5 FSRQs and 17 BCUs).
1FLT includes 6 sources that are associated with a source from previous FGL catalogs (1FGL, 2FGL, 3FGL) but that do not have a counterpart in the 4FGL-DR2 list. These associations are reported in the catalog table in the column named {\it ASSOC\_FERMI}. In that column we also include the associations (14 1FLT sources) with the preliminary 8-year \fermilat source list (FL8Y)\footnote{\url{https://fermi.gsfc.nasa.gov/ssc/data/access/lat/fl8y/}}. In the catalog table column {\it ASSOC\_GAM} we report the associations with other $\gamma$-ray source lists: one association in the Second AGILE catalogue of gamma-ray sources \citep{2019A&A...627A..13B}, two associations in the Swift-BAT 105-Month Hard X-ray Survey \citep{2018ApJS..235....4O} and one association in The INTEGRAL/IBIS AGN catalogue \citep{2016MNRAS.460...19M}. 
\begin{table}
\centering
\caption{Census of 1FLT Sources}
\tablecaption{Census of 1FLT Sources}
\begin{tabular}{lll} 
%%\toprule
\hline
\hline
  %%\multicolumn{3}{c}{CENSUS} \\
%%\cmidrule(r){1-2}
%%\hline
  \multicolumn{1}{l}{Class}&
  \multicolumn{1}{l}{Class Description}&
  \multicolumn{1}{l}{Number}\\
%%\midrule
\hline
FSRQ & Flat Spectrum Radio Quasar & 24\\
BLL & BL Lacertae object &  1\\
CSS & Compact Steep Spectrum radio source & 1\\
SSRQ & Steep Spectrum Radio Quasar & 1\\
RG & Radio Galaxy & 3\\
BCU & Blazars of Uncertain Type & 70\\
AGN & Active Galactic Nuclei of other type & 2\\
UNASS & Unassociated & 40\\
%%\bottomrule
\hline
\end{tabular}

\end{table}\label{table:census}

\section{Properties of 1FLT sources}\label{sec:properties}

\subsection{Spectral properties}

Figure \ref{figure:histPL} shows the distribution of the power-law photon index $\Gamma$ for the 1FLT sources along with that for the 4LAC sources that are associated with FSRQs and BL Lacs.
The 1FLT distribution extends to softer $\Gamma$ values (median value of $\Gamma\sim$ 2.7) than that of the 4LAC sources (median value of $\Gamma\sim 2.2$ or $2.5$ if we consider only 4LAC FSRQs). 
As shown in Figure \ref{figure:histPL}, 26 1FLT sources exhibit softer spectral indices than the softest 4LAC FSRQ. We performed the Kolmogorov–Smirnov test (KS-test, \citealt{kol}) to evaluate whether the 1FLT spectral index distribution comes from the same parent population of 4LAC sub-class distributions (null hypothesis). We found that our null hypothesis is rejected in all cases at a confidence level of 99.9\% ($\alpha$=0.001).
The identification of these soft-spectrum sources shows that integrating over monthly timescales allowed us to investigate the soft part of the gamma-ray source distribution, which is usually suppressed (or confused with the low energy component of the $\gamma$-ray background) when analyzed over longer, multi-year timescales.

\begin{figure}[hbt!]
 \plotone{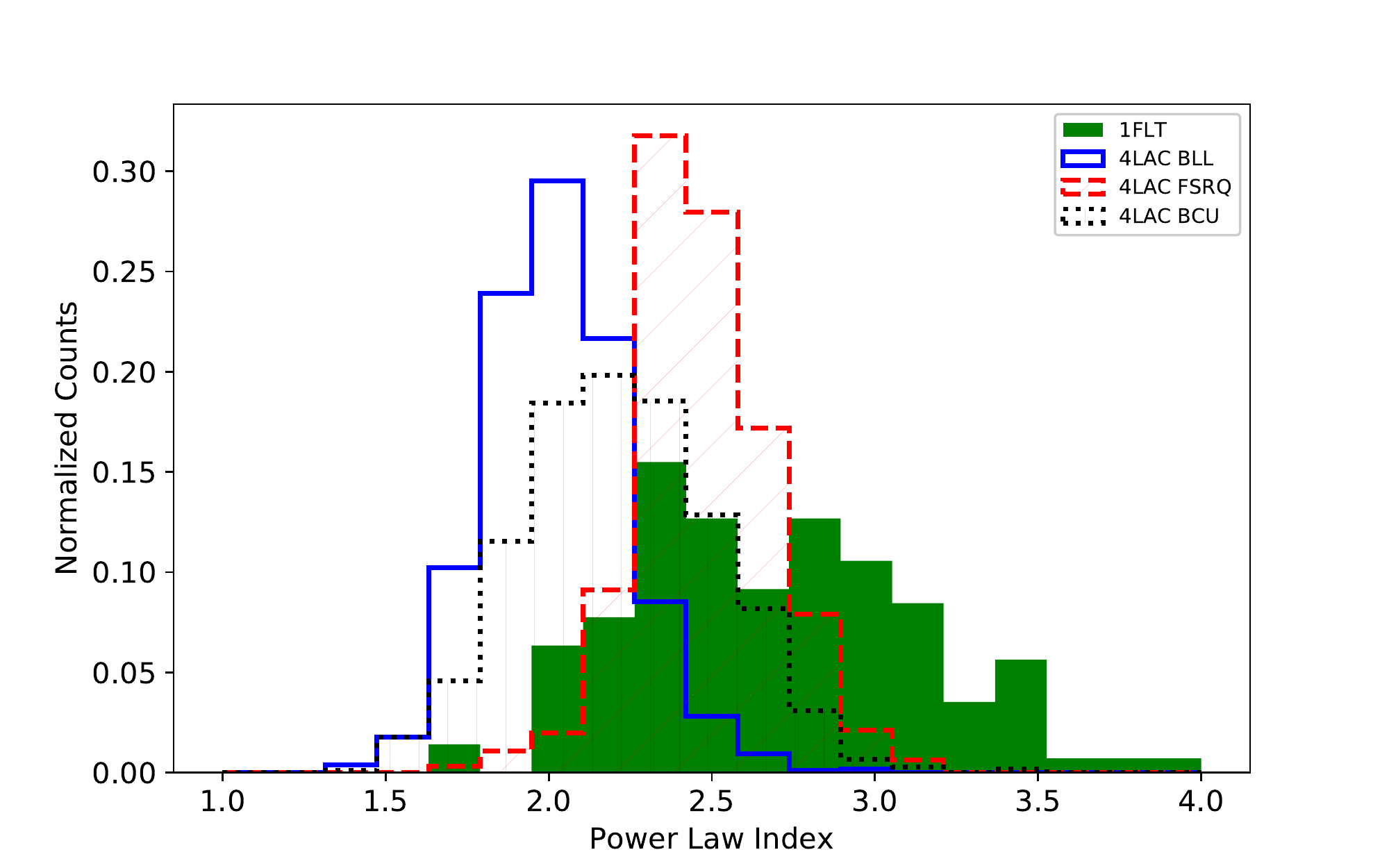}
 \caption{Normalized distribution of power-law index for 1FLT sources (green filled histogram) and 4LAC sources. The blue line represents the 4LAC BL Lac objects, the red dashed line 4LAC FSRQ and the black dotted line 4LAC BCU objects.}
\end{figure}\label{figure:histPL}

The population of soft-spectrum gamma-ray sources could include members of the so-called ``MeV-blazars'' (blazars that are exceptionally bright at MeV energies, \citealt{1995A&A...293L...1B}, \citealt{blom1995}).
These  objects have shown variability in their $\gamma$-ray flux when integrated over timescales of months. However their variability timescales are longer than those shown by harder-spectrum blazars \citep{sikora2002nature}. 
The MeV-peaked emission can be modeled with the external Comptonization of accretion disc radiation or of the near-IR emission of hot dust \citep{bednarek1996}. The transient nature of the MeV-blazar phenomenon is evidence for the presence of this population of soft FSRQs with a steep $\Gamma$(E >100 MeV) on an approximately 1-month time scale.
The month-scale variability and the pronounced increase of the emission in the keV – MeV regime is also reported by \citet{kreter2020search} in their search for high redshift blazars. The large distances of these sources and the subsequent redshifting of their emission means that the $\gamma$-ray emission is shifted towards longer wavelengths. 

\subsection{Spectral energy distribution classification}
For all of the 1FLT sources associated with blazars we collected the available multifrequency data in order to infer the synchrotron peak frequency $\nu_{\rm peak}^{\rm S}$ of the observed broadband SED which was in turn used to perform a SED-based classification. 
We classified blazars as low-synchrotron-peaked (LSP, $\nu_{\rm peak}^{\rm S}<10^{14}$ Hz), intermediate-synchrotron-peaked (ISP, $10^{14}$ Hz $<\nu_{\rm peak}^{\rm S}<10^{15}$ Hz ), or high-synchrotron-peaked (HSP, $\nu_{\rm peak}^{\rm S}>10^{15}$ Hz). 
In this procedure we used a compilation of broad-band archival data as described for the 2LAC \citep{ackermann2011secondLAC}. 
The estimation of $\nu_{\rm peak}^{\rm S}$ relies on a 3rd-degree polynomial fit of the low-energy hump of the SED which is performed on a source-by-source basis (the procedure was adopted in \citealt{ackermann20153LAC}). With this procedure we reconstructed the $\nu_{\rm peak}^{\rm S}$ for only 65 1FLT sources due to the limited multifrequency sampling. 
Figure \ref{figure:nusp} shows the power-law index versus the $\nu_{\rm peak}^{\rm S}$ for both the 1FLT and 4LAC sources. 
The strong correlation evident in the 4LAC sample appears to be diluted for the sources in the 1FLT catalog since we detected only LSP sources with soft spectra. We noted however that the number of 1FLT catalog sources with $\nu_{\rm peak}^{\rm S}$ estimation is just a few percent of those in the 4LAC catalog.

\begin{figure}[hbt!]
 \plotone{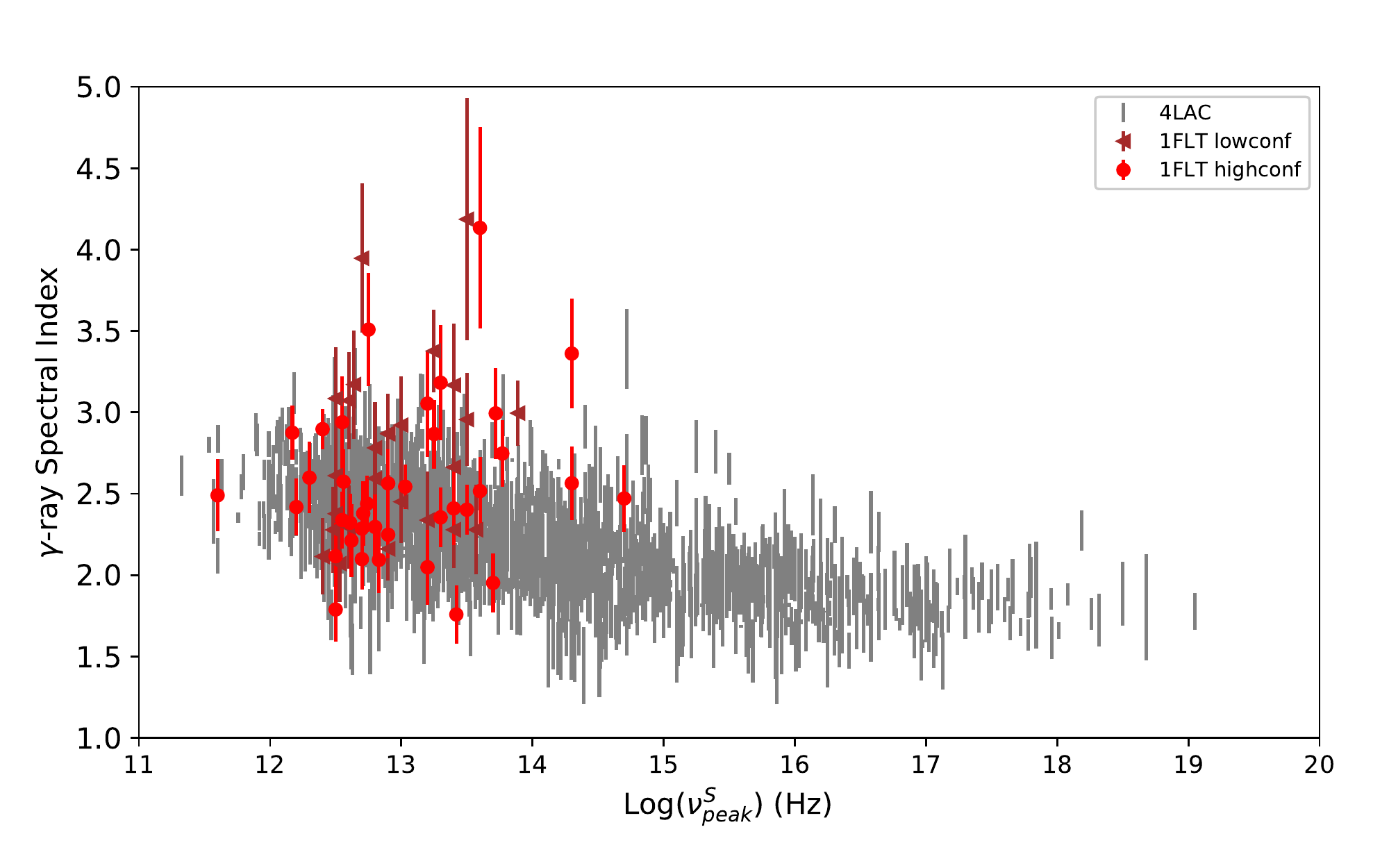}
 \caption{Power-law index of 1FLT sources plotted as a function of SED  $\nu_{\rm peak}^{\rm S}$ (high-confidence sources: red filled circles; low-confidence sources: brown filled triangles); in gray are shown 4LAC data points for comparison.}
\end{figure}\label{figure:nusp}

\subsection{Luminosity, radio flux and redshift}
For each 1FLT source that has a counterpart with a known redshift,  we computed the $\gamma$-ray and radio luminosity using the $\Lambda$CDM cosmological parameters obtained by \textit{Planck} \citep{2014A&A...571A..16P}; in  particular, we used $h$ = 0.67, $\Omega_m$ = 0.32, and $\Omega_{\Lambda}$ = 0.68, where the Hubble constant $H_0= 100h$ km s$^{-1}$Mpc$^{-1}$.
Redshifts are available for a total of 30 1FLT sources: all FSRQs, 1 BCU, 1 BLL, 3 RG and 1 non-blazar object.

\begin{figure}[hbt!]
  \centering
  \begin{tabular}{@{}c@{}}
    \includegraphics[width=0.85\textwidth]{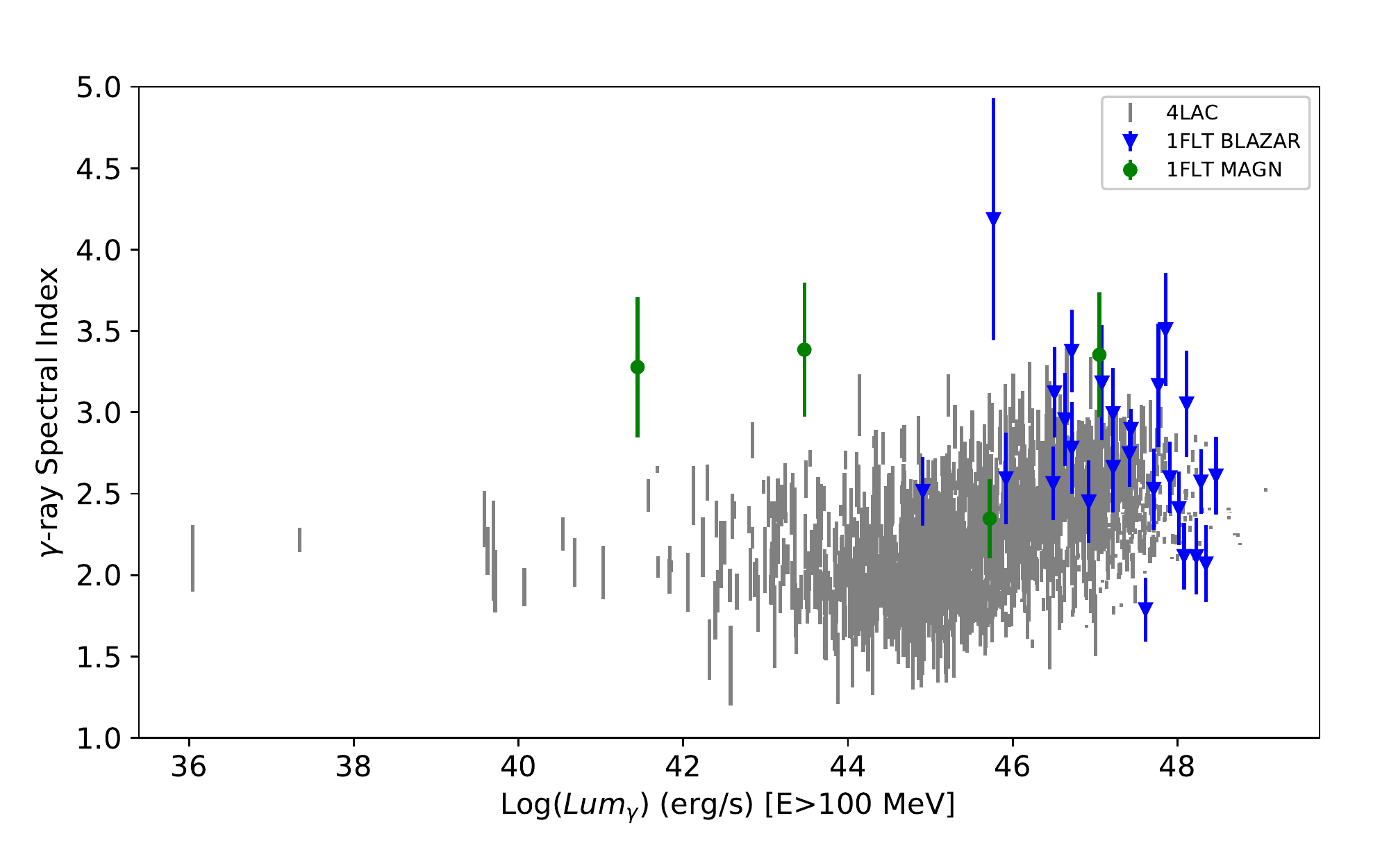} \\[\abovecaptionskip] %[width=0.7\textwidth]
    \small (a) 
  \end{tabular}

  %\vspace{\floatsep}

  \begin{tabular}{@{}c@{}}
    \includegraphics[width=0.85\textwidth]{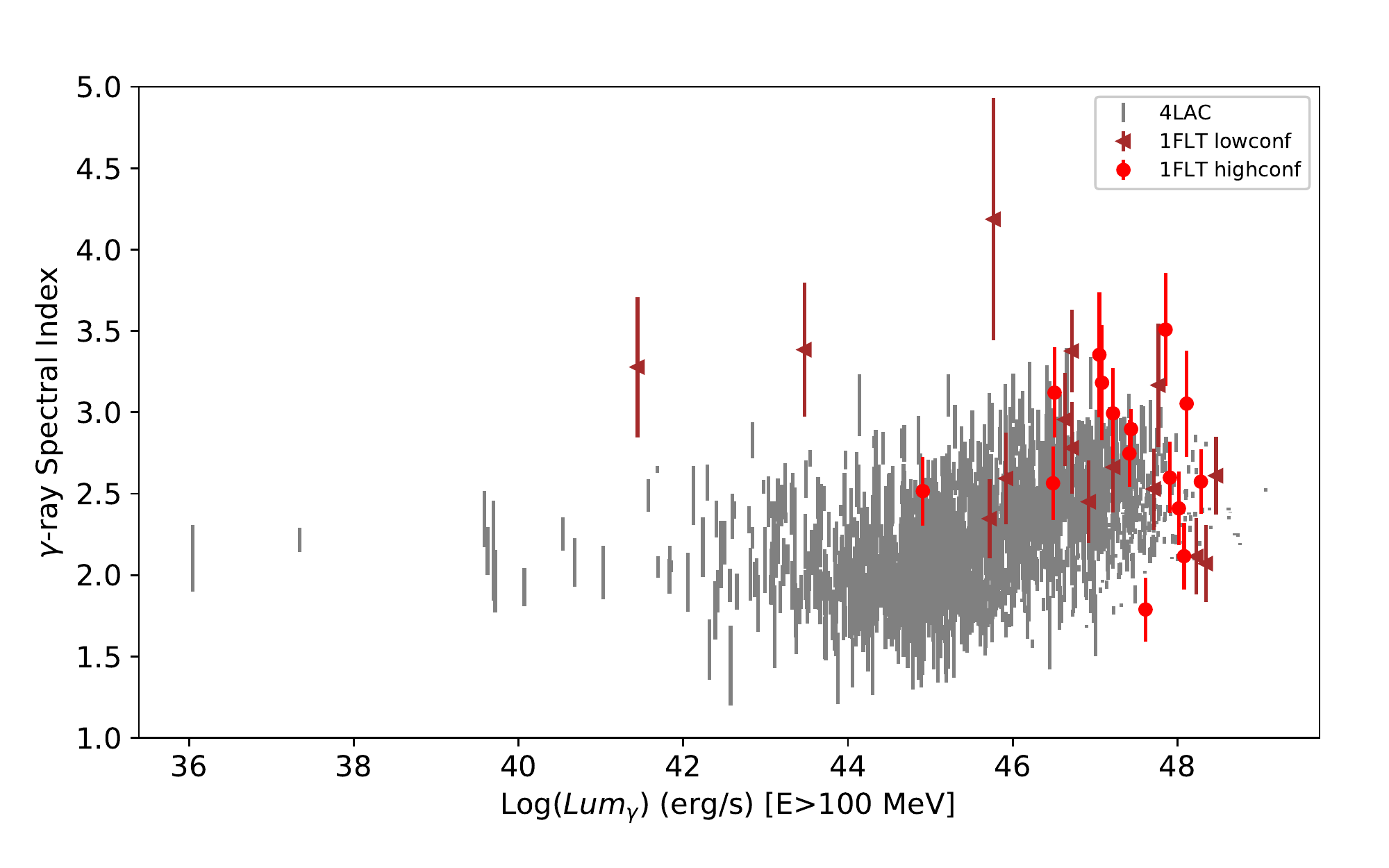} \\[\abovecaptionskip] %[width=0.7\textwidth]
    \small (b) 
  \end{tabular}

  \caption{Power-law index of all 1FLT sources, plotted as a function of the $\gamma$-ray luminosity (100 MeV -- 300 GeV). (a) Green filled circles represent 1FLT Misaligned AGN (MAGN); blue filled triangles represent 1FLT blazars. (b) Red filled circles represent 1FLT high-confidence sources; brown filled triangles represent 1FLT low-confidence sources. In gray are shown 4LAC data points for comparison.}\label{figure:lumsp}
\end{figure}

Analyzing the distribution of spectral index versus $\gamma$-ray luminosity for all the sources of our sample, along with the corresponding values for all sources of the 4LAC \citep{20204LAC}, plotted in Figure \ref{figure:lumsp}, we can see that the 1FLT sources and the 4LAC sources occupy different regions of the parameter space.
The 1FLT sources show greater $\gamma$-ray luminosities, possibly because they are detected during flaring states, and softer $\gamma$-ray spectra than the sources detected by integrating over years-long periods.
When integrating over short time intervals, sporadic flaring activity is detected from a sub-luminous population whose integrated emission would not have been detected over longer time intervals, due to their weak quiescent flux being merged into the integrated background gamma-ray emission.
It is noteworthy that more than half of the 1FLT sources lie in a region of the parameter space that is only sparsely populated by 4LAC sources.

Plotting the ratio of $\gamma$-ray luminosity to radio luminosity as a function of $\mathbold{\nu_{\rm peak}^{\rm S}}$ (Figure~\ref{figure:ratio}) we can see that the ratio is clearly greater for the 1FLT than for the 4LAC; this could be explained simply by the fact that the $\gamma$-ray flux detection and the radio flux measurement are not simultaneous and the gamma-ray measurements are by definition made during $\gamma$-ray flaring episodes. These sources are in fact very bright during a flare, but show a faint long-term integrated $\gamma$-ray flux, which limits their possibility to be detected significantly on the long-term periods typically considered for xFGL catalogs. 

\begin{figure}[hbt!]
 \plotone{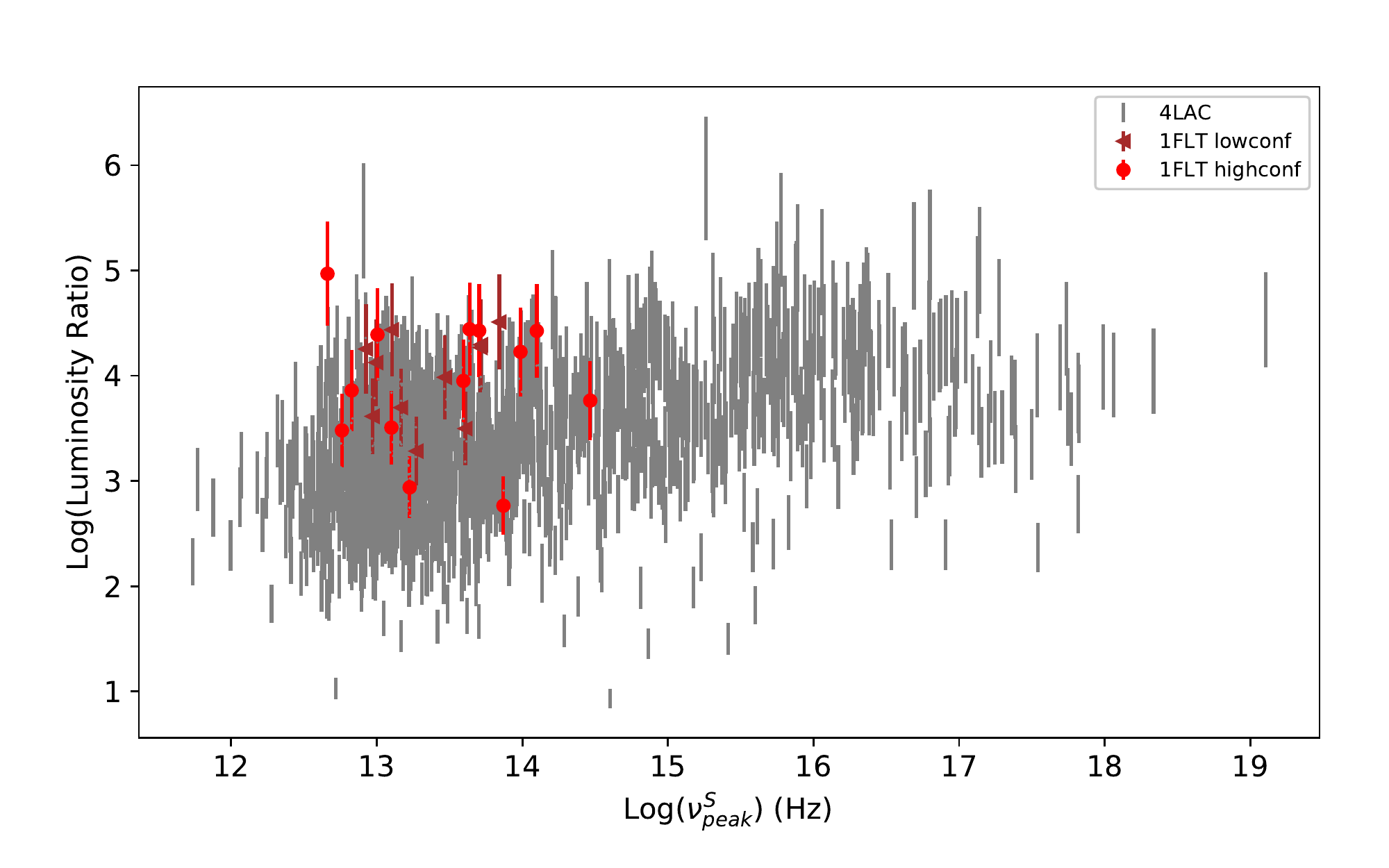}
 \caption{Ratio of $\gamma$ to radio luminosity plotted as a function of SED $\mathbold{\nu_{\rm peak}^{\rm S}}$. Red filled circles represent 1FLT high-confidence sources; brown filled triangles represent 1FLT low-confidence sources. In gray are shown 4LAC data points for comparison.}
\end{figure}\label{figure:ratio}

\begin{figure}[hbt!]
 \plotone{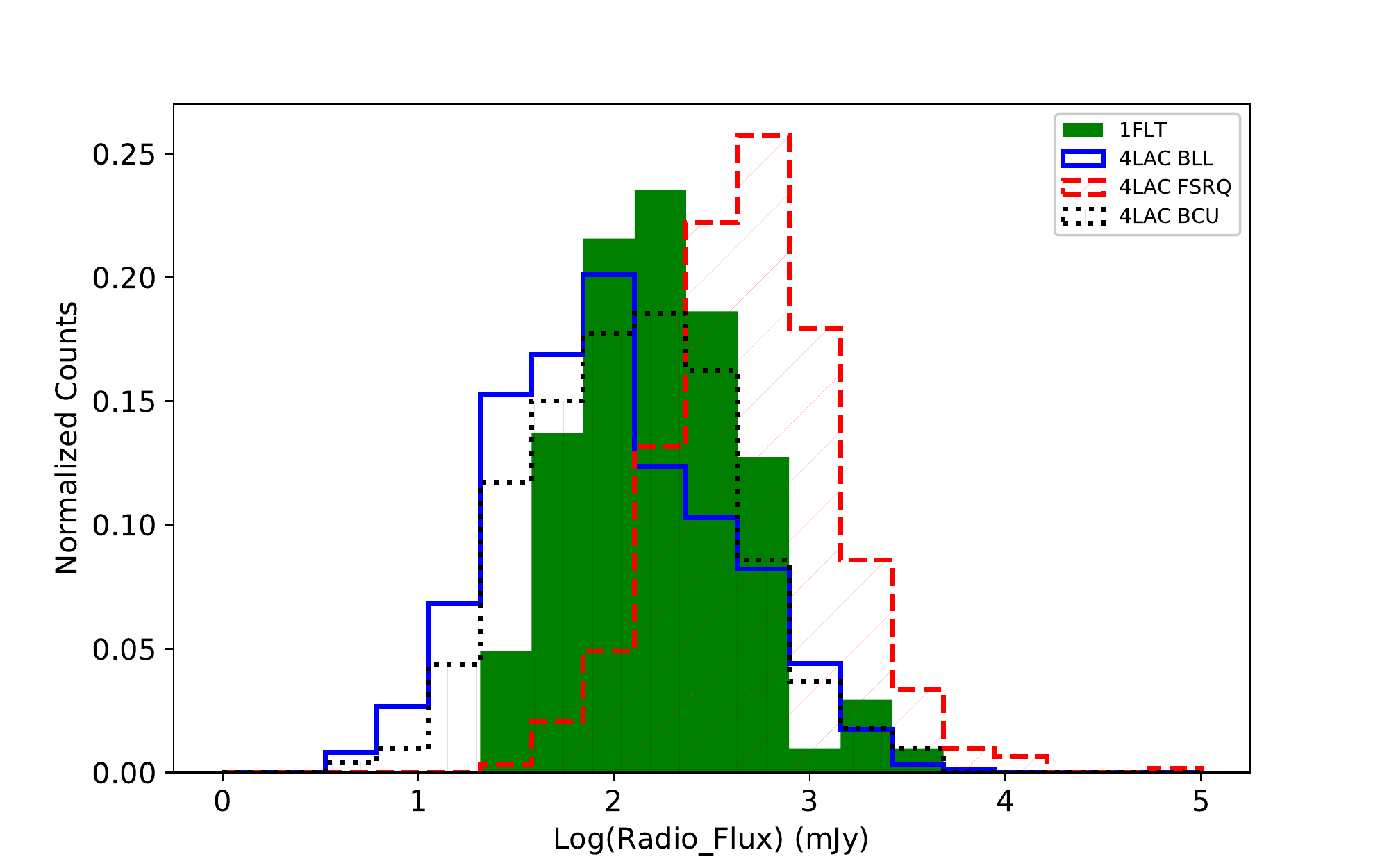}
 \caption{Histograms of radio flux density at 1.4 GHz for the 1FLT sources (green) and 4LAC sources. The blue continuous line represents  4LAC BL Lac objects, the red dashed line 4LAC FSRQs  and the black dotted line 4LAC BCU objects.}
\end{figure}\label{figure:histr}

In terms of the distribution of the radio fluxes for the 1FLT sources, our sources show fainter fluxes than those of the 4LAC FSRQs (using a KS-test we found that our  null hypothesis is rejected at a confidence level of 99.9\%)  but are comparable to those of the 4LAC BCUs as shown in Figure \ref{figure:histr} (using a KS-test we found that the two data sets are not very similar but do not differ by enough that we could reject the null hypothesis: p$=$2.14\% and distance$=$ 0.16).
This is because most of the 1FLT sources are BCUs.

\begin{figure}[hbt!]
 \plotone{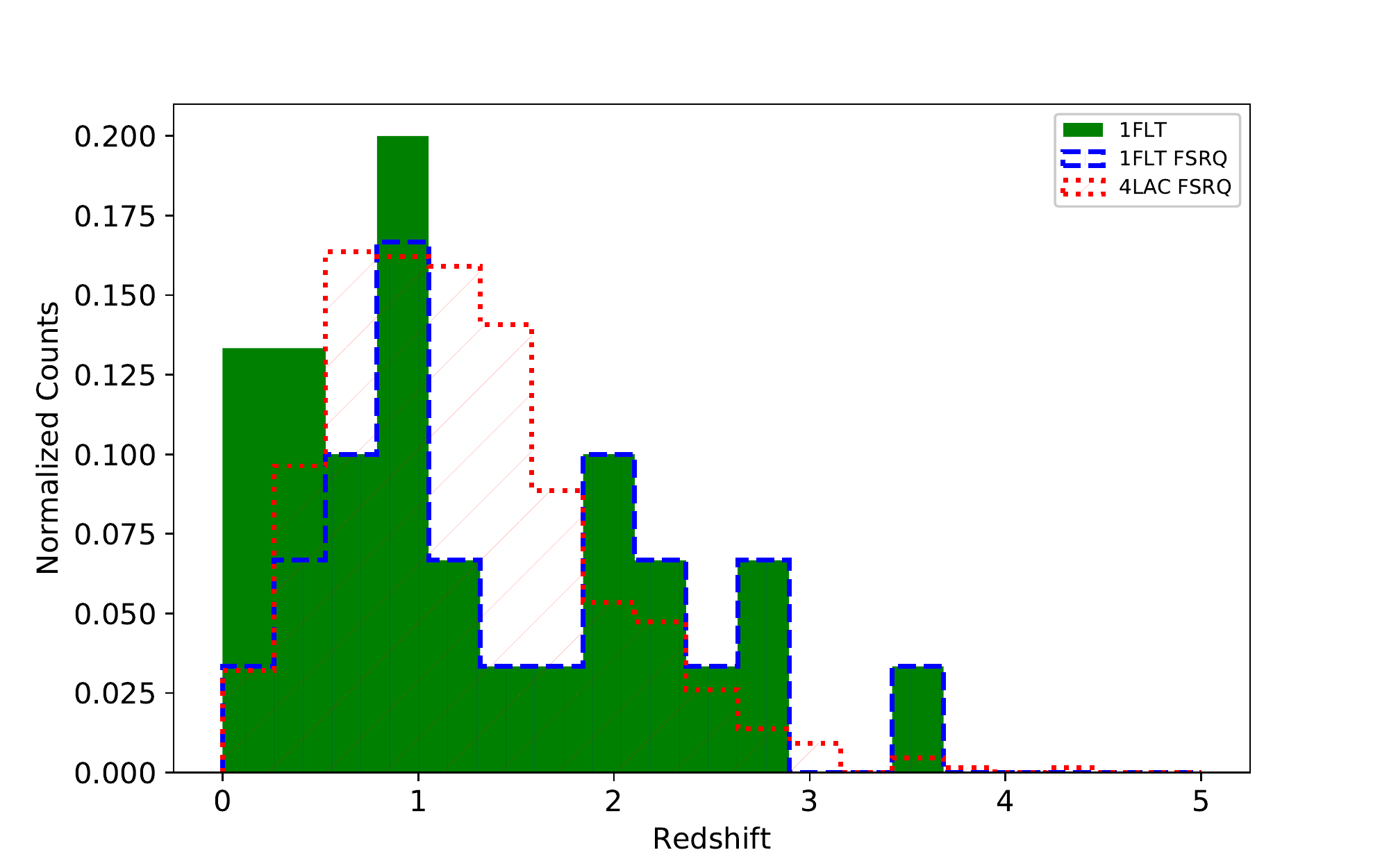} 
 \caption{Redshift distributions for 1FLT (green shaded region), 1FLT FSRQ (blue dashed line) and 4LAC FSRQ (red dotted line) sources.}
\end{figure}\label{figure:histz}

Figure \ref{figure:histz} shows the redshift distributions for 4LAC FSRQs and 1FLT sources. The redshift is available for all FSRQs, for 1 BCU, 1 BLL, 3 RG and 1 non-blazar object. The 1FLT redshift distribution shows a peak at low redshift ($<$1). This peak is due to the contribution of 1FLT misaligned AGN. These sources show a larger jet inclination angle, where de-boosted radiation makes the relativistic jet radiation weaker and more difficult to detect in $\gamma$ rays \citep{abdo2010magn}.  
However $\sim$ 50\% of FSRQs are detected at $z >1$, suggesting a different distribution.
Another effect, in fact, that can be seen in Figure \ref{figure:histz} is that there is an additional peak at z $>2$, unlike the 4LAC distribution which includes only a small fraction of objects at z $>2$.
This shows that the 1FLT method is well suited for detecting high-redshift blazars. 
These results are in agreement with the behavior of the population of high redshift blazars reported in \citet{kreter2020search}.

\section{Discussion}\label{sec:discussion}

\subsection{Comparison with the Second FAVA Catalog}\label{sec:FAVA}
The Fermi All-sky Variability Analysis (FAVA) searches for transient sources over the entire sky observed by the \fermilat on 1 week time intervals. The second catalog of flaring $\gamma$-ray sources (2FAV) spans the first 7.4 years of \fermilat observations and reports 518 variable $\gamma$-ray sources with significances of at least 6$\sigma$. Based on positional coincidence, candidate counterparts have been found for 441 sources, most of them blazars. The catalog also provides for each source the time, location, and spectrum of each flaring episode \citep{abdollahi2017second2FAV}.

Given the variable nature of 2FAV sources we decided to compare the 1FLT catalog to the 2FAV catalog. We have investigated positional (within the error ellipses) and time correspondence between 1FLT sources and 2FAV sources.\\
We found correspondence in position and time for a few sources: 1FLT J1732$+$1510 (TBIN 71.5) with 2FAV J1732$+$15.2 (reported in an Astronomer’s Telegram\footnote{\url{http://www.astronomerstelegram.org}}, ATel$\#$6395); 1FLT J1919$-$4543 (TBIN 21) with 2FAV J1919$-$45.7 (ATel$\#$2666); 1FLT J1936+5341 (TBIN 20.5) with 2FAV J1936+53.7; and 1FLT J2010$-$2523 (TBIN 72.5) with 2FAV J2009$-$25.4 (ATel$\#$6553). The associations with 2FAV catalog sources are reported in the column named {\it ASSOC\_FAVA} in the 1FLT catalog table. These 2FAV sources have the same counterparts as the corresponding 1FLT sources except in the case of 2FAV J1936$+$53.7, which is not associated, and 1FLT J1936$+$5341 which we associated with TXS 1935$+$536.

Looking at $\gamma$-ray candidate seeds in the 1FLT search with TS  $<$ 25 and not already in 4FGL-DR2, we found correspondence in position and time with two other 2FAV sources. 2FAV J0539$-$28.8 corresponds to a detection in TBIN 55 with TS = 9 and 2FAV J2056$-$11.5 to a detection in TBIN 60.5 with a TS = 14. We attribute this to the dilution over the month time scale of the $\gamma$-ray emission that occurred during a brief (about 1 week) flare.
This result confirms the potential of $Fermi$ to find new sources when the data are integrated over different time intervals. 
It also shows that choosing a monthly time interval allows us to detect a different and new population of sources than those found when the data are binned on weekly time scales.

\subsection{Duty cycle evaluation}\label{DC}
%%%--------------
The duty cycle for an astronomical object of a given class was defined by \citet{romero1999optical} as:
\begin{equation}
    DC = 100\frac{\sum_{i=0}^n N_i 1/(\Delta t_i)}{\sum_{i=0}^n 1/(\Delta t_i)}\%
\end{equation}
where $\Delta t_i = \Delta t_{i,obs}(1+z)^{-1}$
is the duration (corrected for redshift z) of a monitoring interval of a source of the selected class and $N_i$ is equal to 0 when the source is not variable at $\Delta t_i$, while $N_i$ is equal to 1 when the source is variable.
We adapted the calculation to our specific analysis and used the light curves  for the evaluation of the DC.
For each time bin in which the source is detected (i.e., TS $>$ 4 or N$\mathrm{_{pred}}>$ 3 or $\mathrm{\Delta}$Flux/Flux < 0.5), we compared the flux to the ``averaged sub-threshold flux'' (the flux averaged over 10 years using the evaluation of spectral index reported in 1FLT). If the flux in the monthly time bin is above the ``averaged sub-threshold flux'' we considered it in the calculation of the DC. 

In our case the DC can be evaluated as
DC\,\begin{math}=100\frac{\sum_{i=0}^n N_i }{120}
\end{math}
since $\Delta t_i$ is one month for each source. We can calculate the DC for each source class independently of its redshift.
We also calculated the duty cycle for each transient source class in the 4FGL using the same procedure adopted for the 1FLT. Since the light curves reported in the 4FGL have a time binning of 2 months, which is different from our sampling interval, we can only perform a qualitative comparison between the duty cycles of the 1FLT and 4FGL sources. In Figure \ref{figure:dc} we compare between the average value of the DC for each transient source class in the 1FLT and the 4FGL catalogs. The 1FLT sources are characterized by a very low duty cycle. 
Clearly there is a selection effect due to the way that the 1FLT catalog was built because a bias was introduced by the definition of a significant point in the LC extraction. We can, however, obtain some qualitative insights about the difference between the DC of the radio galaxies and the blazar-like objects. In fact the radio galaxy class has a greater DC than the FSRQ and BCU classes which are characterized by shorter flaring episodes. 

\begin{figure}
 \centering
 \includegraphics[width=0.7\textwidth]{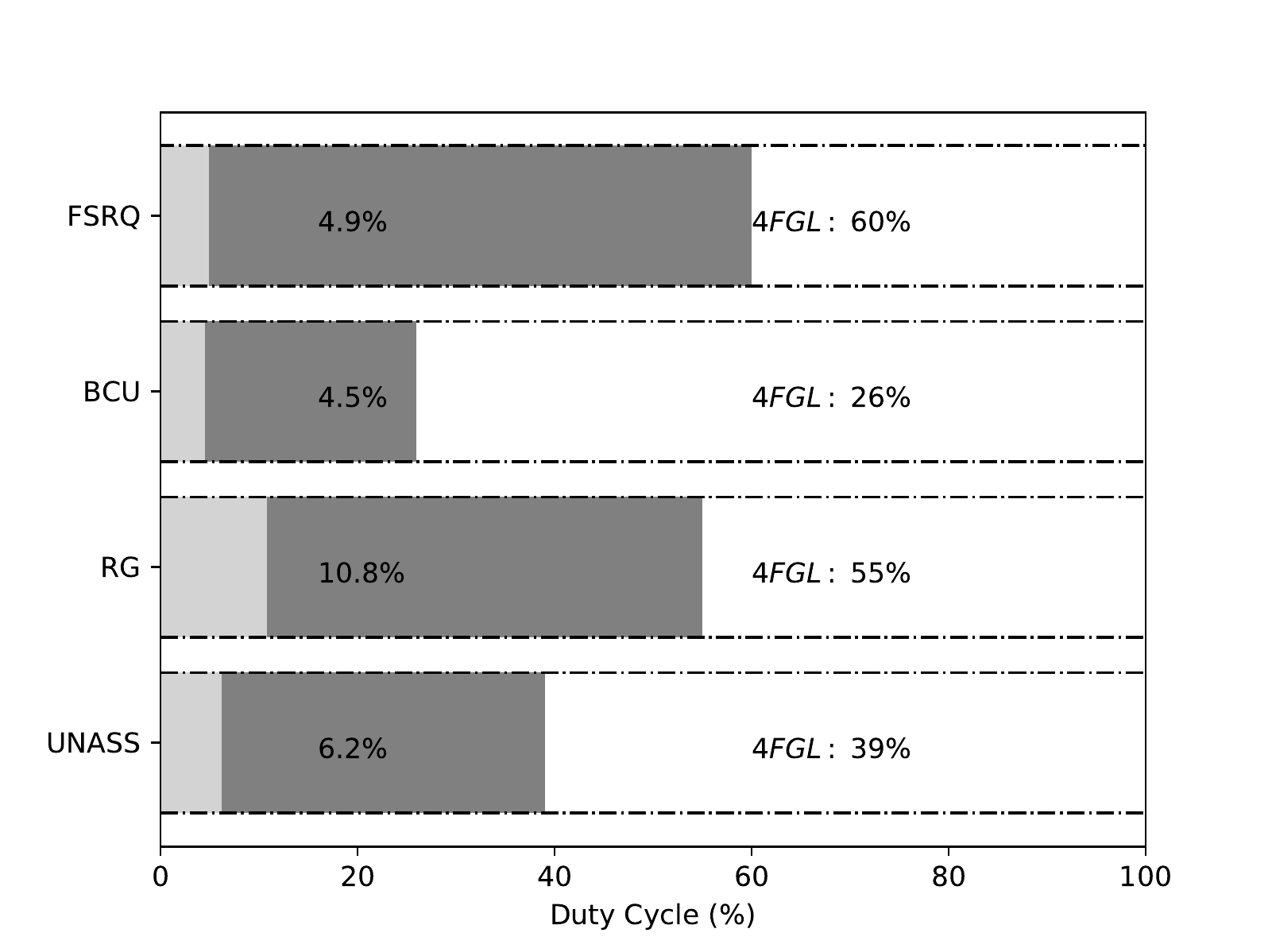}
 \caption{Average duty cycles for the different classes of objects from the 1FLT catalog compared to their values in the 4FGL catalog. The duty cycles of the 1FLT sources are shown in light grey with their numerical values in bold font. The duty cycles of the 4FGL sources are shown in dark grey with their numerical values labeled as such. The time binning of the light curves on which the duty-cycle calculation is based is not the same for 1FLT and 4FGL: the 1FLT light curves are binned in one-month intervals while the 4FGL light curves binning is two months.}
\end{figure}\label{figure:dc}

\subsection{Discussion of peculiar sources}
We discuss here some peculiar cases which deserve a detailed investigation and for which we built SEDs collecting all archival multi-wavelength data\footnote{\url{https://tools.ssdc.asi.it/SED/docs/SED_catalogs_reference.html}\label{fn:arc}} using the \textit{SSDC-SED Tool}\footnote{\url{https://tools.ssdc.asi.it/SED/}} and adding our data points for each of the TBINs.
%%------------%%%

\subsubsection{Soft $\gamma$-ray blazars detected with the {\it Swift} Burst Alert Telescope}
Two 1FLT sources were also detected with the {\it Swift} Burst Alert Telescope \citep{krimm2013swift}.
The source 1FLT J0010$+$1056 is associated with a Bayesian probability of 0.96 with Mrk 1501 which is reported as SWIFT J0010.5$+$1057 in the \textit{Swift-BAT 105-Month Hard X-ray Survey}\footnote{\url{https://swift.gsfc.nasa.gov/results/bs105mon/}}. Mrk 1501 is a FSRQ located in the local universe at redshift of $z\,=\,0.089338$ \citep{sargent1970isolated}. This source is the nearest FSRQ in the whole 1FLT sample. It was also reported by \citet{arsioli2018complete}. They searched for sources with short-lived $\gamma$-ray emission and found a TS$\sim26$ for this source in just one monthly bin, that from June 2010. We detected this source in the TBIN 21.5 (2010 May 20 to 2010 June 20) thus confirming this short $\gamma$-ray flare from Mrk 1501.

\begin{figure}[hbt!]
    \centering
    \includegraphics[width=0.7\textwidth]{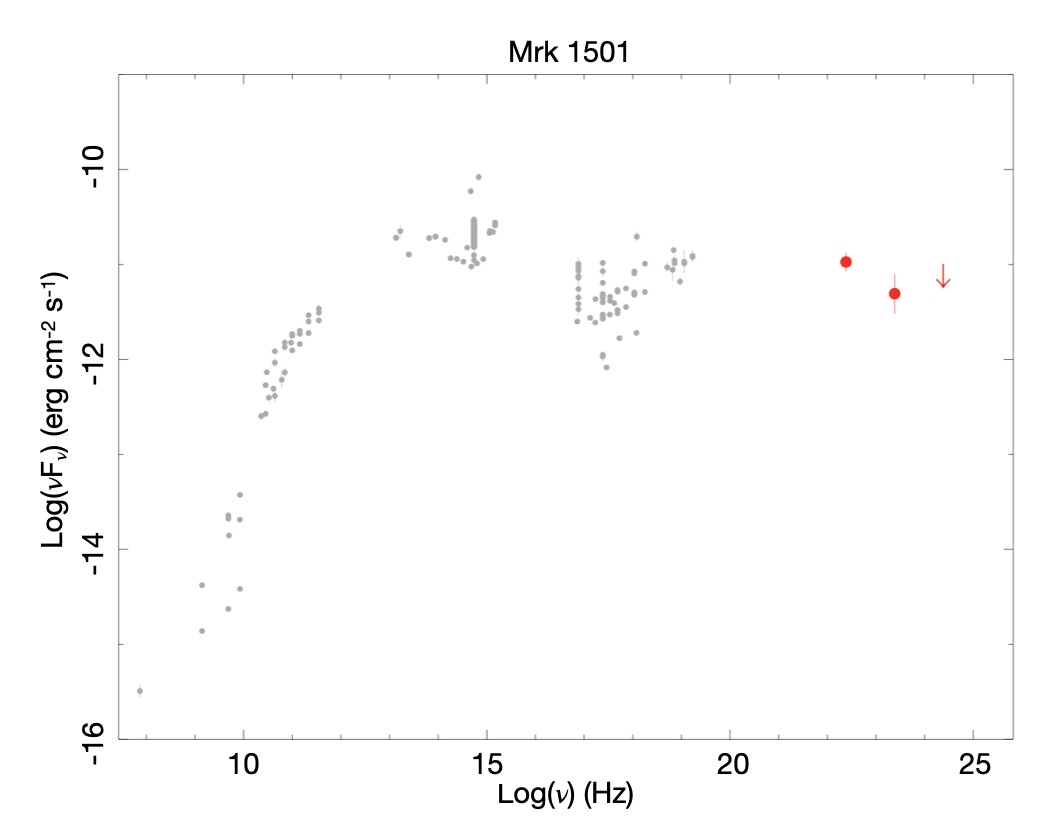}
    \caption{The SED of Mrk 1501 associated with 1FLT J0010+1056 detected in TBIN 21.5. The archival data points\footref{fn:arc} at other wavelengths are shown in gray. The $\gamma$-ray spectral data points for the corresponding 1FLT source are shown in red.}\label{figure:MRK1501}
\end{figure}

The SED of Mrk 1501 is shown in Figure \ref{figure:MRK1501}. The typical two-humped, blazar-like SED is clearly reproduced. Although the source was in a $\gamma$-ray flaring state, the Compton Dominance ($CD_{\gamma}$), defined as the ratio of the inverse Compton to the synchrotron peak luminosity) is, in logarithmic scale, less than 1 as is typical of high synchrotron-peaked BL Lacs (\citealt{abdo2010spectral}, see Figure 22 in  \citealt{giommi2012simultaneous}).

The source 1FLT J2010$-$2523 is associated with a Bayesian probability of 0.99 with PMN J2010$-$2524, a well-known FSRQ ($z= 0.825$, \citealt{massaro20155th}). It is also reported in the \textit{Swift-BAT 105-Month Hard X-ray Survey} as SWIFT J2010.6$-$2521 and as a $\gamma$-ray emitter by \citet{2019ApJ...881..154P}.
We detected the source in several temporal bins. The first detection was in TBIN 6 (2009 February  3 to 2009  March 5) followed by a detection in TBINs 68 and 68.5 (2014 April  5 to  2014  May 20) and then from TBIN 72 to TBIN 73.5 (2014 August 5 to 2014  October 20). The flaring activity of the source is quite evident in the light curve reported in Appendix \ref{sec:light}.

\begin{figure}[hbt!]
    \centering
    \includegraphics[width=0.7\textwidth]{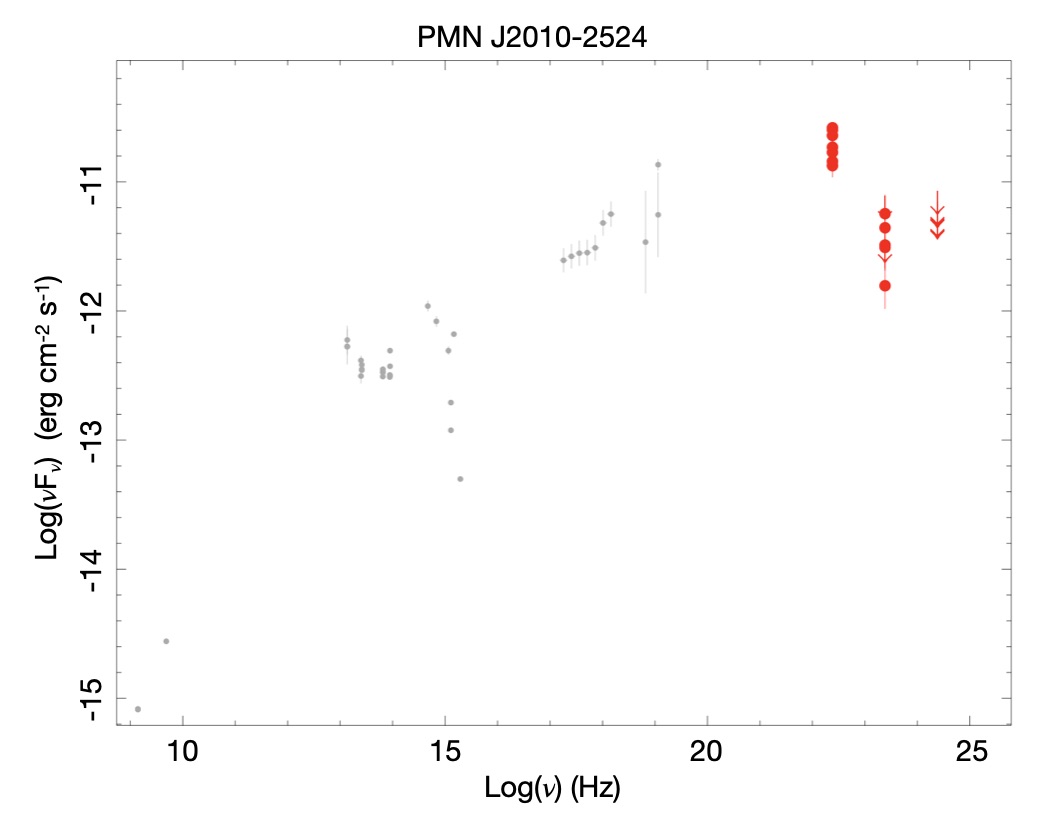}
    \caption{The SED of PMN J2010-2524 associated with 1FLT J2010-2523 detected in seven TBINs 6, 68, 68.5, 72, 72.5, 73 and 73.5. The archival data points\footref{fn:arc} at other wavelengths are shown in gray. The $\gamma$-ray spectral data points for the corresponding 1FLT (and for its different TBIN detections) source are shown in red.
    }\label{figure:PMN2010}
\end{figure}

In contrast to Mrk 1501, the SED of PMN J2010$-$2524 has a $CD_{\gamma}$ greater then 1, which is fairly common among FSRQs. The $CD_{\gamma}$ in Figure \ref{figure:PMN2010}.

\subsubsection{Compact Steep Spectrum source}
\begin{figure}[hbt!]
    \centering
    \includegraphics[width=0.7\textwidth]{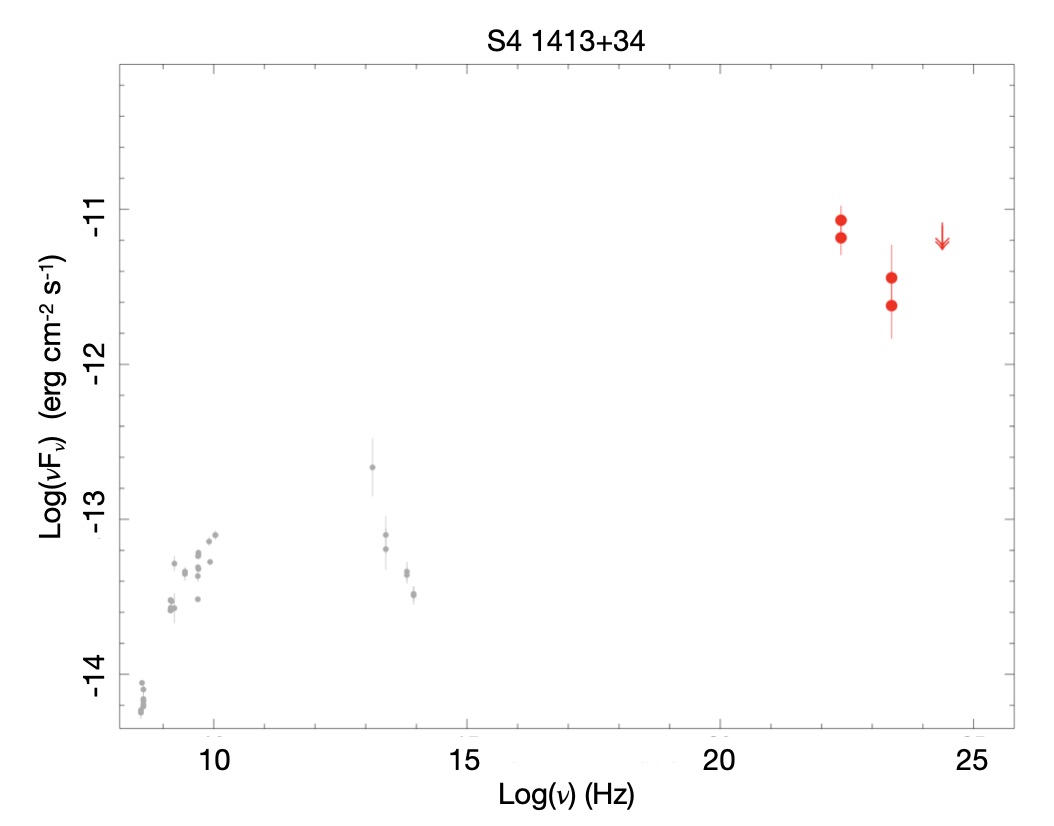}
    \caption{The SED of S4 1413+34 associated with 1FLT J1416+3447 detected in two different TBINs 111.5 and 112. The archival data points\footref{fn:arc} at other wavelengths are shown in gray. The $\gamma$-ray spectral data points for the corresponding 1FLT source (and for its different TBIN detections) are shown in red.}\label{figure:S41413}
\end{figure}

1FLT J1416$+$3447 was detected in two consecutive and overlapping TBINs, TBIN 111.5 and TBIN 112 (2017 November 19 to 2018 January 04), which strengthens the reliability of the detection. It is associated, through the Bayesian method, with a probability of 0.85, with a radio Compact Steep Spectrum (CSS) source, S4 1413+34. The SED of S4 1413+34 is shown in Figure \ref{figure:S41413}.\\
CSS sources are thought to represent the first stages of evolution of the radio sources that eventually physically expand as Fanaroff-Riley I (FRI) or Fanaroff-Riley II (FRII) radio galaxies. $\gamma$-ray emission in CSS sources was predicted by \citet{stawarz2008gamma} but these sources have proved to be elusive in the xFGL catalogs. 
Recently, \citet{shultz} reported on the results of a dedicated analysis of PKS 2004$-$447, which is associated with a $\gamma$-ray loud young radio source that has a radio spectrum resembling that of a CSS and an optical spectrum which is like that of a narrow-line Seyfert 1.
S4 1413+34, which does not have an optical identification,  shows a core-jet structure at 5~GHz. The  radio morphology (see Figure 9 of \citealt{Dallacasa2013}) shows a well-collimated jet, but the radio emission is dominated by the brightest compact component.\\
\citet{abdo2010magn} report on eleven misaligned AGN detected at $\gamma$-ray energies, which, with the exception of NGC 1275, show no evidence for variability. \citet{abdo2010magn} suggested that sources with a larger CD, which are observed with a line of sight closer to the jet  axis, are those preferentially  detected  with  the \fermilat (in contrast with what was found by \citealt{angioni}). We evaluated the CD of S4 1413+34 at 5~GHz, and obtained a value of CD = 0.28, which  is consistent with the values previously found for the other misaligned AGN. We conclude, therefore, that the observed $\gamma$-ray emission should be produced in the compact core region.

\subsubsection{NGC 4278}
\begin{figure}[hbt!]
    \centering
    \includegraphics[width=0.7\textwidth]{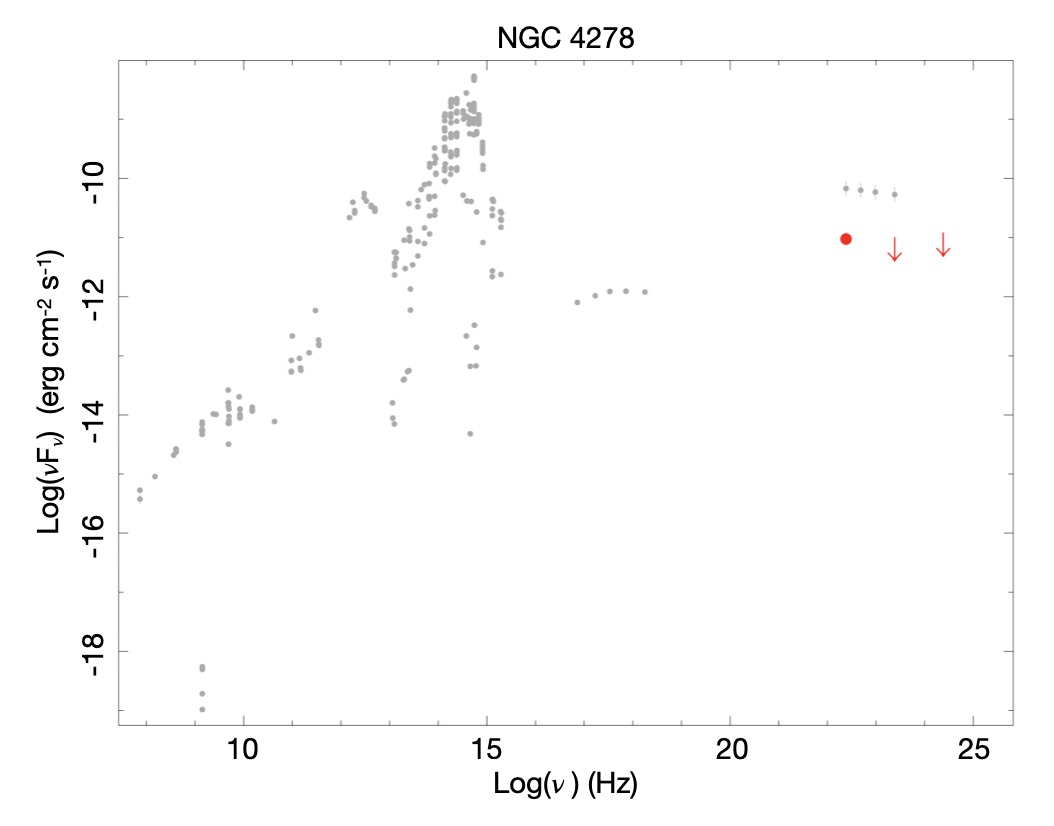}
    \caption{The SED of NGC 4278 associated with 1FLT J1219+2907 detected in TBIN 7. The archival data points\footref{fn:arc} at other wavelengths are shown in gray. The $\gamma$-ray spectral data points for the corresponding 1FLT source are shown in red. Gray data points at $\gamma$-ray frequency are AGILE data points.}\label{figure:NGC4278}
\end{figure}

1FLT J1219$+$2907 was detected in TBIN 7 (2009 March 5 to 2009 April 5) of our analysis and, using the positional association procedure, we found that NGC 4278 lies  within its error ellipse. The SED of NGC 4278 is shown in Figure \ref{figure:NGC4278}. NGC 4278 (z = 0.0021) is a nearby galaxy identified as a likely radio compact symmetric object (CSO) with a FRI  morphology. It shows a dominant, bright, flat-spectrum component identified as the core (see Figures 2M and 3M in \citealt{tremblay2016compact}). From the north and south ends of the core, diffuse jets extend out to the west and east respectively, creating an ‘S-shaped’ symmetry as observed in other CSO radio sources. We calculated a CD of 0.71 at 366 5~GHz using the data reported in \citet{schilizzi1983vlbi}. This is a greater than average value of CD for a FRI although it is reasonable for a $\gamma$-ray emitter.

\citet{giroletti2005low} reported that the source is oriented at a very small viewing angle ($2^{\circ}<\theta<4^{\circ}$) with respect to the line of sight, although they did not
find compelling evidence for strong beaming. They also proposed an alternative scenario in which the source could be oriented at a larger angle, with asymmetries related to the jet's interaction with the surrounding medium. NGC 4278 is optically classified as a low-luminosity and low-accreting AGN with a low-ionization nuclear emission-line region (LINER).
All of these characteristics make NGC 4278 a very intriguing new type of $\gamma$-ray emitter. Until now, the detection of two CSOs, PKS 1718$-$649 \citep{migliori2016first} and PMN J1603$-$4904 \citep{krauss}, had been reported at $\gamma$-ray energies. \citet{stawarz2008momentum} postulate, in their dynamical radiative model of young radio sources, that the $\gamma$-ray emission is produced via inverse-Comptonization of circum-nuclear (IR–to–UV) photon fields off relativistic  electrons in compact, expanding lobes.

\subsubsection{3C 226}
\begin{figure}[hbt!]
    \centering
   \includegraphics[width=0.7\textwidth]{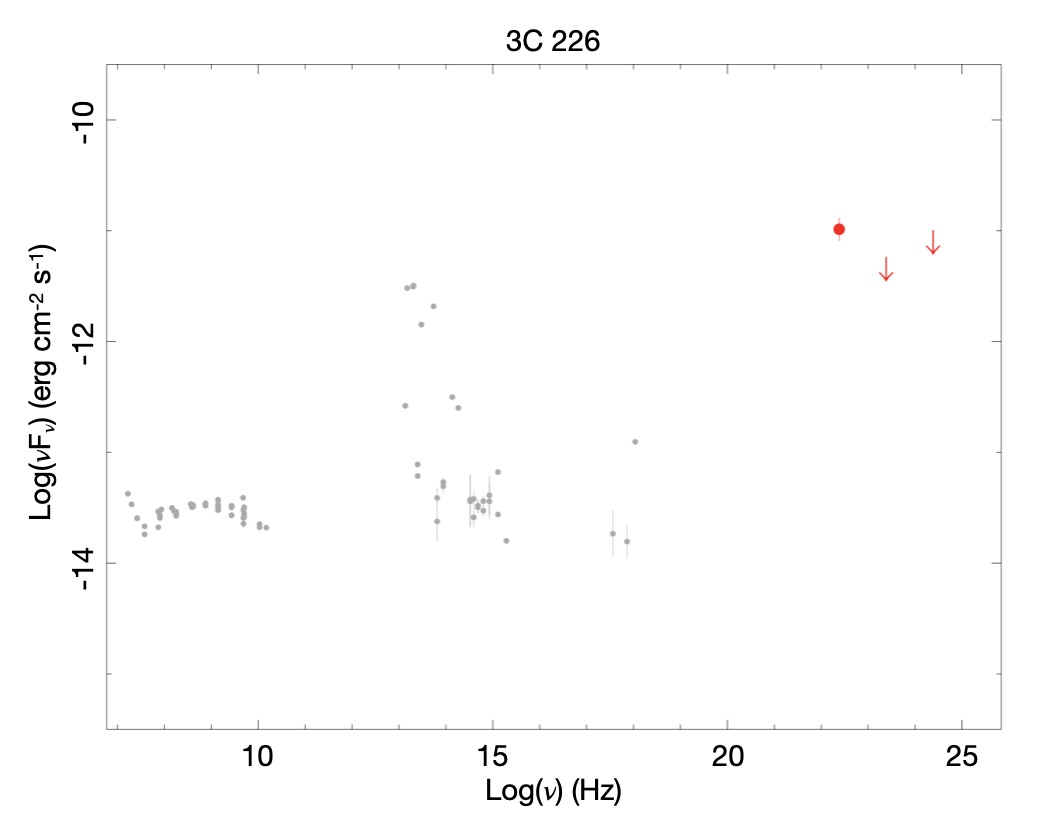}
\caption{The SED of 3C 226 associated with 1FLT J0943+0940 detected in TBIN 44. The archival data points\footref{fn:arc} at other wavelengths are shown in gray. The $\gamma$-ray spectral data points for the corresponding 1FLT source are shown in red.}
    \label{figure:3C226}
\end{figure}

1FLT J0943$+$0940 is positionally consistent with 3C 226 and was detected in the TBIN 44 (2012 April 05 to 2012  May 05) of our analysis. 3C 226, located at $z=0.818$ \citep[][]{hewitt1991optical}, is an FRII radio galaxy \citep{laing1983bright} with an optical spectrum that shows the presence of narrow emission lines (NLRG). The FRII radio morphology is clearly seen over a span of 260 kpc \citep{best1997hst}. The source has an asymmetric radio structure, having the southeastern lobe closer to the radio core than the northwestern lobe. However, when the radio and optical observations are combined, the source shows a large opening angle. It has a  low CD of -2 which makes 3C 266 a unfavored candidate for $\gamma$-ray emission. Furthermore, it can be seen from the SED, which is shown in Figure \ref{figure:3C226}, that there is a discrepancy between the archival data and the $\gamma$-ray data. FRII radio galaxies are, however, found to be variable from radio to $\gamma$-ray energies, e.g., 3C 111 in \citet{grandi2012gamma} whose variability has been postulated to be due to a new knot component ejected from the core. A similar scenario could be at play in 3C 226 which would strengthen the case for its association with 1FLT J0943+0940.

\subsubsection{The highest redshift FSRQ: PMN J2219$-$2719}
\begin{figure}[hbt!]
    \centering
    \includegraphics[width=0.7\textwidth]{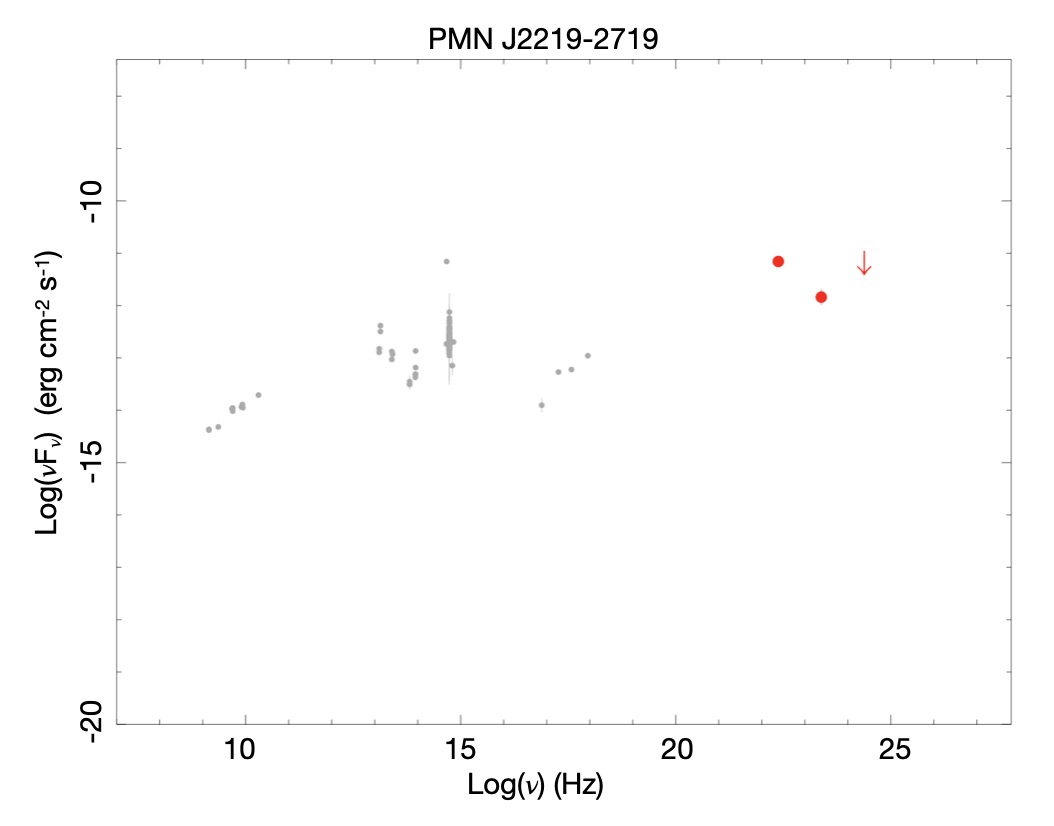}
    \caption{The SED of PMN J2219-2719 associated with 1FLT J2219-2732 detected in TBIN 96. The archival data points\footref{fn:arc} at other wavelengths are shown in gray. The $\gamma$-ray spectral data points for the corresponding 1FLT source are shown in red.}\label{figure:PMNJ2219-2719}
\end{figure}

%%%---------------
1FLT J2219$-$2732 is  positionally consistent with PMN J2219$-$2719 and was detected in TBIN 96 (2016 August 04 to 2016 September 04) of our analysis. PMN J2219$-$2719 is a well-known FSRQ belonging to the Fifth Roma-BZCAT catalog \citep{massaro2014optical} at a redshift of $z$ = 3.634 \citep{hook2002discovery}. This source has a very low duty cycle (see section \ref{DC}) of about 2.5\% and it lies at the highest redshift of the entire 1FLT catalog. Examination of the SED (see Figure  \ref{figure:PMNJ2219-2719}) reveals a  large $CD_{\gamma}$ of $\sim 1.8$ during a period of flaring activity. This behavior is typical of FSRQs and is usually interpreted as an external Compton process (see Figure 11 of \citealt{abdo2010pks}). PMN J2219$-$2719 is also identified as the most distant TS $>$ 25 blazar with monthly variability detected in \citet{kreter2020search}. The other source with high redshift reported in \citet{kreter2020search}, i.e., 5BZQ J0422$-$3844, is not detected with our method. The blind search undertaken with PGWave found several monthly seeds that were positionally compatible with this source but were discarded because they were located within 50 arcmin of a 4FGL source.

\section{Conclusions}\label{sec:conc}
%%%--------
In this work we have investigated the \fermilat sky on monthly timescales to search for sources which are not detected when integrating on timescales of years and/or that are not already hosted in the \fermilat xFGL general catalogs.
We found that:
\begin{itemize}
\item The 1FLT catalog is mainly populated by soft sources with an average spectral index of $\sim$ 2.7 compared to $\sim$ 2.3 for the 4FGL-DR2. This confirms that these soft sources are not distinguishable from the diffuse $\gamma$-ray background when considered over multi-year integration times. 
\item This study found that there are more FSRQs found when using a monthly binning compared to catalogs using data averaged over longer intervals. This is due both to BL Lacs that are in general less variable around 1 GeV and to FSRQs whose strong gamma-ray activity is mainly seen only during flaring events. If the flare intensity is fainter than the integrated background over years, we lose the capability to detect these FSRQs in long-time integrated catalogs.
\item In the 1FLT we detected sources only when they were in an active flaring phase. For this reason they show a high $\gamma$-ray luminosity with respect to the 4LAC sources whose fluxes were calculated over a longer integration time (see Figure \ref{figure:lumsp}).
\item The 1FLT sources show a fainter radio flux than the 4LAC FSRQs (see Figure \ref{figure:histr}), and they also have  a very low duty cycle  relative to 4FGL sources. This correlation between the radio-loudness and variability was already described in \citet{romero1999optical} in the optical wave band.
\end{itemize}

In conclusion, the 1FLT catalog comprises 142 $\gamma$-ray sources of which $\sim$ 70\% are associated with soft AGN-type counterparts, principally BCU and FSRQ. Approximately 28\% of 1FLT sources remain unassociated. This fraction is similar to the percentage of unassociated sources found in the \fermilat general catalogs. The 72 1FLT sources that have TS between 25 and 30 and are detected in only one of the monthly time bins should be regarded as low confidence, given their $\sim$ 34\% probability of being statistical fluctuations. 
Out of the 142 monthly detections we found only 4 correspondences with the 2FAV catalog whose analysis is on a one-week time interval. This shows that integrating over different time intervals does not reproduce the same $\gamma$-ray sky and that the discovery space of \fermilat remains large.
The 1FLT catalog will be available online for an easy and fast visualization at \url{www.ssdc.asi.it/fermi1flt}.

\acknowledgments

The \textit{Fermi} LAT Collaboration acknowledges generous ongoing support from a number of agencies and institutes that have supported both the development and the operation of the LAT as well as scientific data analysis. These include the National Aeronautics and Space Administration and the Department of Energy in the United States, the Commissariat \`a l'Energie Atomique and the Centre National de la Recherche Scientifique / Institut National de Physique Nucl\'eaire et de Physique des Particules in France, the Agenzia Spaziale Italiana and the Istituto Nazionale di Fisica Nucleare in Italy, the Ministry of Education, Culture, Sports, Science and Technology (MEXT), High Energy Accelerator Research Organization (KEK) and Japan Aerospace Exploration Agency (JAXA) in Japan, and the K.~A.~Wallenberg Foundation, the Swedish Research Council and the Swedish National Space Board in Sweden.
 
Additional support for science analysis during the operations phase is gratefully acknowledged from the Istituto Nazionale di Astrofisica in Italy and the Centre National d'\'Etudes Spatiales in France. This work performed in part under DOE Contract DE-AC02-76SF00515.
Sara Buson acknowledges financial support by the European Research Council for the ERC Starting grant MessMapp, under contract no. 949555.
G. Tosti acknowledges support from the Grant ASI/INAF n.2015-023-R.1. 
This work has been partially supported by the EOSC-hub EU project G.A 777536 and a special thank goes to Daniele Spiga (INFN Perugia) and Mirko Mariotti (University of Perugia) for the  help with the computing infrastructure. \\
Additional support for science analysis during the operations phase is gratefully acknowledged from Space Science Data Center Tools\footnote{\url{https://www.asdc.asi.it/}}, Open Universe portal  and VOU-Blazar Tool\footref{fn:open} (\citealt{giommi2020open}, \citealt{giommi2019open}).

\vspace{5mm}
\facilities{\fermilat}
\software{PGWave tool \citep{ciprini20071d}, Fermipy (v0.18.0; \citealt{wood2017fermipy}), Fermitools (v1.2.1; Fermi Science Support Development Team 2019)}

\clearpage
\newpage
\medskip

\bibliography{1FLT}

\clearpage
\newpage
\appendix
\setcounter{table}{0}

\section{Description of the FITS version of the 1FLT Catalog}\label{sec:catdesc}
\renewcommand{\thetable}{A\arabic{table}}

\begin{table}[hbt!]
\centering
\caption{Catalog Table \texttt{SOURCES} Description}
%%\begin{table}
%%\centering
\tablecaption{Catalog Description}
\begin{tabular}{ll} 
%%\toprule
\hline
\hline
  %%\multicolumn{2}{c}{Catalog Description} \\
%%\cmidrule(r){1-2}
%%\hline
  \multicolumn{1}{l}{Column}&
  \multicolumn{1}{l}{Description}\\
%%\midrule
\hline
  Source\_Name  & 1FLT JHHMM+DDMM, according to IAU Specifications for Nomenclature.\\
  SRCNUM & Source number. It is the same for multiple flares of the same source\\
  TBIN\_1m & Time Bin in which the source was detected with the greatest TS, where \\ & 0-119 stand for first 120 months starting August 04, 2018 UTC (MET 239557417), \\ & 0.5-119.5 stands for 120 15-day shifted months staring August 19, 2018 UTC (MET 240846000)\\
  Flares & Number of flares of this source\\
  RAJ2000  & Right Ascension, J2000, in degrees\\
  DEJ2000 & Declination, J2000, in degrees\\
  GLON & Galactic longitude, in degrees\\
  GLAT & Galactic latitude, in degrees\\
  Conf\_95\_SemiMajor & Semimajor axis of the error ellipse at 95\% confidence, in degrees\\
  Conf\_95\_SemiMinor & Semiminor axis of the error ellipse at 95\% confidence, in degrees\\
  Conf\_95\_PosAng & The position angle of the 95$\%$-confidence semi-major axis, from celestial North, positive toward\\ 
  & increasing R.A. (eastward), in degrees\\
  Test\_Statistic & Likelihood test statistic for 100 MeV\textendash 300 GeV analysis\\
  Npred & The number of predicted events in the model\\
  PL\_Index & The photon index for the PowerLaw fit\\
  Unc\_PL\_Index &  The 1$\sigma$ error on PL\_Index\\ 
  Flux & The integral photon flux for 0.1 to 300 GeV, in photons/cm$^2$/s\\
  Unc\_Flux &  The 1$\sigma$ error on Flux, in photons/cm$^2$/s\\
  Energy\_Flux & The energy flux in MeV/cm$^2$/s, in the 100 MeV to 300 GeV range obtained by spectral fitting\\ 
  & from 100 MeV to 300 GeV\\
  Unc\_Energy\_Flux & The 1$\sigma$ error on Energy\_Flux, in MeV/cm$^2$/s\\
  ASSOC\_FERMI & Correspondence to \fermilat xFGL and FL8Y catalog\\
  ASSOC\_GAM & Correspondence to $\gamma$-ray source catalog\\
  ASSOC\_FAVA & Correspondence to 2FAV source catalog\\ 
  LMC & True if the source is in the LMC region\\ 
  CenA & True if the source is near Cen A\\ 
  Class & Class of the most likely counterpart. See also sec. \ref{sec:class}\\
  Assoc$\_$Name & The designation of the most likely associated counterpart\\
  Assoc$\_$Prob$\_$Bay & The probability of association according to the Bayesian method. It is set to 0 for point sources with\\ 
  & only positional association (see Sect. \ref{sec:assoc})\\
  Assoc$\_$RA & The most likely counterpart R.A., J2000, in degrees\\
  Assoc$\_$Dec & The most likely counterpart Declination, J2000, in degrees\\
  Redshift & The most likely counterpart redshift\\
  Radio$\_$Flux & The 1.4 GHz most likely counterpart flux density; Units: mJy\\
  log10(Nu$\_$Peak) & Log10 Synchrotron peak frequency measured with a third degree polynomial fit function; Units: Hz\\
  & \citep{ackermann20153LAC}\\
  Low\_Confidence & True if a source has a TS < 30 and only one monthly detection\\
\hline
\end{tabular}
%%\end{table}

\end{table}\label{table:catalogdescription}

\begin{table}
\centering
\caption{Catalog Table \texttt{FLARES} Description}
\tablecaption{SED points table description}
\begin{tabular}{ll} 
%%\toprule
\hline
\hline
  %%\multicolumn{2}{c}{SED table Description} \\
%%\cmidrule(r){1-2}
%%\hline
  \multicolumn{1}{l}{Column}&
  \multicolumn{1}{l}{Description}\\
%%\midrule
\hline
  SRCNUM & Multiple monthly flares of the same 1FLT source have the same assigned number\\
  repROI & True if the source is detected in multiple overlapping ROI\\
  repTBIN &  True if the source is detected in more than one TBIN\\
  TBIN\_1m & See Table \ref{table:catalogdescription}\\
  TSTART & Beginning of observation time window in MET seconds\\
  TSTOP & End of observation time window in MET seconds\\
  RAJ2000  & See Table \ref{table:catalogdescription}\\
  DEJ2000 & See Table \ref{table:catalogdescription}\\
  GLON & See Table \ref{table:catalogdescription}\\
  GLAT & See Table \ref{table:catalogdescription}\\
  Conf\_95\_SemiMajor & See Table \ref{table:catalogdescription}\\
  Conf\_95\_SemiMinor & See Table \ref{table:catalogdescription}\\
  Conf\_95\_PosAng & See Table \ref{table:catalogdescription}\\ 
  Test\_Statistic & See Table \ref{table:catalogdescription}\\
  Npred & See Table \ref{table:catalogdescription}\\
  PL\_Index & See Table \ref{table:catalogdescription}\\
  Unc\_PL\_Index &  See Table \ref{table:catalogdescription}\\
  Flux &  See Table \ref{table:catalogdescription}\\
  Unc\_Flux &  See Table \ref{table:catalogdescription}\\
  Energy\_Flux & See Table \ref{table:catalogdescription}\\ 
  Unc\_Energy\_Flux &  See Table \ref{table:catalogdescription}\\
  nuFnu\_band &  SED data points in three energy bands, as reported in table \ref{table:EB}, in MeV/cm$^2$/s\\
  nuFnu\_ul\_band & SED data points upper limits at 2$\sigma$-confidence in three energy bands, as reported in table \ref{table:EB},\\
   &in MeV/cm$^2$/s\\
  Unc\_nuFnu\_band &  The 1$\sigma$ error on SED data points in three energy bands, as reported in table \ref{table:EB},\\
   & in MeV/cm$^2$/s\\
  Test\_Statistic\_band &  SED data points test statistic in three energy bands, as reported in table \ref{table:EB}\\
%%\bottomrule
\hline
\end{tabular}

\end{table}\label{table:SED}

\begin{table}
\centering
\caption{Catalog Table \texttt{EnergyBounds} Description}
\tablecaption{EnergyBounds table description}
\begin{tabular}{ll} 
%%\toprule
\hline
\hline
  %%\multicolumn{2}{c}{SED table Description} \\
%%\cmidrule(r){1-2}
%%\hline
  \multicolumn{1}{l}{Column}&
  \multicolumn{1}{l}{Description}\\
%%\midrule
\hline
  LowerEnergy &  Value of the minimum energy, in MeV\\
  UpperEnergy &  Value of the maximum energy, in MeV\\
  ENumBins & Bins per decade in energy\\
  EvType &  LAT event class\\
  ZenithCut &  Zenith angle cut, in deg\\
  PixelSize &  Binning in space, in deg\\
\hline
\end{tabular}
\end{table}\label{table:EB}

The FITS format version of the 1FLT catalog has three binary table extensions. The extension \texttt{SOURCES} has all of the information about the sources. Its format is described in Table \ref{table:catalogdescription}.

The extension \texttt{FLARES} contains the information about all the monthly detections, including multiple monthly flares of the same 1FLT source reported with the same assigned number. It includes also SED data points. Its format is described in Table \ref{table:SED}.

The extension \texttt{EnergyBounds} contains the definitions of the bands in which the SED data points were computed, and the settings
of the analysis. Its format is described in Table \ref{table:EB}.

\section{Source Lists}\label{sec:srclist}
\setcounter{table}{0}
\renewcommand{\thetable}{B\arabic{table}}
Table \ref{table:extract} contains the complete 1FLT source list but only a subset of the columns reported in the extension \texttt{SOURCES} of the FITS format version of the catalog. Tables \ref{table:sun} and \ref{table:grb} contain respectively the Sun and the GRB detection descriptions.

\begin{longrotatetable}
\begin{deluxetable*}{|l|c|c|c|c|c|c|c|c|c|c|c|l|l|c|}
\tablecaption{Catalog table: see the description of the columns in Table \ref{table:catalogdescription}.\label{table:extract}}
\tablewidth{700pt}
\tabletypesize{\scriptsize}
\tablehead{
  \multicolumn{1}{|c|}{1FLT} &
  \multicolumn{1}{c|}{SRC} &
  \multicolumn{1}{c|}{TBIN} &
  \multicolumn{1}{c|}{Flares} &
  \multicolumn{1}{c|}{RAJ2000} &
  \multicolumn{1}{c|}{DEJ2000} &
  \multicolumn{1}{c|}{Conf 95} &
  \multicolumn{1}{c|}{Conf 95} &
  \multicolumn{1}{c|}{Conf 95} &
  \multicolumn{1}{c|}{TS} &
  \multicolumn{1}{c|}{PL} &
  \multicolumn{1}{c|}{Flux} &
  \multicolumn{1}{c|}{Class} &
  \multicolumn{1}{c|}{Assoc Name}&
  \multicolumn{1}{c|}{Low}\\
  \multicolumn{1}{|c|}{Source Name} &
  \multicolumn{1}{c|}{NUM} &
  \multicolumn{1}{c|}{1m} &
  \multicolumn{1}{c|}{} &
  \multicolumn{1}{c|}{} &
  \multicolumn{1}{c|}{} &
  \multicolumn{1}{c|}{SemiMaj} &
  \multicolumn{1}{c|}{SemiMin} &
  \multicolumn{1}{c|}{PosAng} &
  \multicolumn{1}{c|}{} &
  \multicolumn{1}{c|}{index} &
  \multicolumn{1}{c|}{} &
  \multicolumn{1}{c|}{} &
  \multicolumn{1}{c|}{} &
  \multicolumn{1}{c|}{Conf}\\
  \multicolumn{1}{|c|}{} &
  \multicolumn{1}{c|}{} &
  \multicolumn{1}{c|}{} &
  \multicolumn{1}{c|}{} &
  \multicolumn{1}{c|}{(degree)} &
  \multicolumn{1}{c|}{(degree)} &
  \multicolumn{1}{c|}{(degree)} &
  \multicolumn{1}{c|}{(degree)} &
  \multicolumn{1}{c|}{(degree)} &
  \multicolumn{1}{c|}{} &
  \multicolumn{1}{c|}{} &
  \multicolumn{1}{c|}{${\rm phcm^{-2}s^{-1}}$} &
  \multicolumn{1}{c|}{} &
  \multicolumn{1}{c|}{} &
  \multicolumn{1}{c|}{}
}
\startdata
\hline
 J0002$-$2148 & 1 & 118.5 & 1 & 0.5 & -21.81 & 0.18 & 0.14 & 115 & 43 & 1.8$\pm$0.2 & 1.5E-8$\pm$6.1E-9 & bcu & PKS 2359$-$221 & false\\
 J0010+1056 & 2 & 21.5 & 1 & 2.53 & 10.94 & 0.31 & 0.24 & 140 & 38 & 2.5$\pm$0.2 & 8.3E-8$\pm$2.3E-8 & fsrq & Mrk 1501 & false\\
 J0010+3905 & 3 & 16 & 1 & 2.7 & 39.09 & 0.57 & 0.46 & 91 & 33 & 3.4$\pm$0.3 & 1.1E-7$\pm$2.4E-8 & bcu & GB6 J0008+3856 & false\\
 J0112+0711 & 4 & 34 & 1 & 18.21 & 7.2 & 0.46 & 0.29 & 2 & 25 & 2.3$\pm$0.3 & 4.1E-8$\pm$1.7E-8 & unass &  & true\\
 J0115+5230 & 5 & 59 & 1 & 18.86 & 52.51 & 0.23 & 0.19 & 138 & 26 & 2.1$\pm$0.2 & 3.7E-8$\pm$1.8E-8 & unass &  & true\\
 J0115$-$2237 & 6 & 19.5 & 1 & 18.96 & -22.63 & 0.3 & 0.25 & 4 & 30 & 2.1$\pm$0.2 & 3.7E-8$\pm$1.8E-8 & unass &  & false\\
 J0121+2602 & 7 & 73.5 & 2 & 20.4 & 26.04 & 0.56 & 0.45 & 105 & 33 & 3$\pm$0.3 & 1.3E-7$\pm$3.4E-8 & fsrq & TXS 0120+259 & false\\
 J0125+2212 & 8 & 55.5 & 1 & 21.46 & 22.22 & 0.62 & 0.49 & 129 & 26 & 3.1$\pm$0.3 & 1E-7$\pm$2.9E-8 & bcu & TXS 0122+220 & true\\
 J0136+1505 & 9 & 56.5 & 1 & 24.16 & 15.09 & 0.65 & 0.41 & 139 & 27 & 2.9$\pm$0.3 & 7.7E-8$\pm$2.1E-8 & bcu & TXS 0133+146 & true\\
 J0139+1744 & 10 & 10.5 & 1 & 24.83 & 17.74 & 0.25 & 0.19 & 138 & 28 & 2.1$\pm$0.2 & 2.6E-8$\pm$1.2E-8 & fsrq & PKS 0136+176 & true\\
 J0141$-$1330 & 11 & 106.5 & 2 & 25.42 & -13.51 & 0.2 & 0.18 & 132 & 54 & 2.1$\pm$0.2 & 3.7E-8$\pm$1.1E-8 & bcu & TXS 0139$-$138 & false\\
 J0149+2602 & 12 & 51 & 1 & 27.34 & 26.04 & 0.83 & 0.58 & 67 & 31 & 3.1$\pm$0.3 & 1E-7$\pm$2.6E-8 & fsrq & TXS 0145+256 & false\\
 J0202+3910 & 13 & 33.5 & 1 & 30.6 & 39.17 & 0.58 & 0.52 & 134 & 26 & 3$\pm$0.3 & 7.6E-8$\pm$2.1E-8 & unass &  & true\\
 J0209+7459 & 14 & 4.5 & 1 & 32.38 & 74.99 & 0.85 & 0.68 & 93 & 27 & 3.2$\pm$0.2 & 1.4E-7$\pm$3.7E-8 & bcu & GB6 J0214+7437 & true\\
 J0230+1423 & 15 & 85.5 & 1 & 37.64 & 14.39 & 0.67 & 0.64 & 79 & 28 & 3.4$\pm$0.4 & 1.4E-7$\pm$3.4E-8 & unass &  & true\\
 J0240$-$4657 & 16 & 21.5 & 2 & 40.16 & -46.95 & 0.62 & 0.42 & 73 & 35 & 2.8$\pm$0.3 & 9.9E-8$\pm$2.6E-8 & unass &  & false\\
 J0240$-$4739 & 17 & 101 & 1 & 40.22 & -47.66 & 1.55 & 0.64 & 44 & 29 & 3.6$\pm$0.4 & 7.5E-8$\pm$1.7E-8 & bcu & PMN J0240$-$4822 & true\\
 J0253+1801 & 18 & 9 & 1 & 43.38 & 18.02 & 0.7 & 0.58 & 82 & 27 & 2.9$\pm$0.2 & 1.1E-7$\pm$2.7E-8 & bcu & PKS 0250+178 & true\\
 J0259-5406 & 19 & 64.5 & 1 & 44.84 & -54.1 & 0.45 & 0.44 & 86 & 28 & 2.9$\pm$0.3 & 5.5E-8$\pm$1.6E-8 & unass &  & true\\
 J0321$-$1136 & 20 & 85.5 & 1 & 50.29 & -11.61 & 0.49 & 0.32 & 61 & 30 & 2.6$\pm$0.2 & 6.7E-8$\pm$2E-8 & bcu & TXS 0318$-$115 & false\\
 J0332$-$1733 & 21 & 111 & 1 & 53.03 & -17.56 & 0.27 & 0.24 & 107 & 27 & 2.5$\pm$0.3 & 5.1E-8$\pm$1.9E-8 & bcu & PMN J0332$-$1734 & true\\
 J0342$-$1807 & 22 & 45 & 1 & 55.53 & -18.12 & 0.58 & 0.84 & 83 & 28 & 3.1$\pm$0.3 & 7.6E-8$\pm$1.9E-8 & bcu & TXS 0340$-$182 & true\\
 J0343+1200 & 23 & 116 & 1 & 55.76 & 12.0 & 0.46 & 0.34 & 29 & 29 & 2.5$\pm$0.2 & 1.9E-7$\pm$6.3E-8 & unass &  & true\\
 J0350+0512 & 24 & 90 & 1 & 57.68 & 5.2 & 0.34 & 0.22 & 56 & 31 & 2.3$\pm$0.2 & 7.1E-8$\pm$2.4E-8 & bcu & TXS 0348+049 & false\\
 J0357$-$1509 & 25 & 50 & 1 & 59.4 & -15.15 & 0.92 & 0.67 & 148 & 28 & 3.8$\pm$0.6 & 9E-8$\pm$2.2E-8 & unass &  & true\\
 J0402$-$1227 & 26 & 72 & 1 & 60.51 & -12.46 & 0.55 & 0.47 & 4 & 26 & 3$\pm$0.3 & 9E-8$\pm$2.8E-8 & unass &  & true\\
 J0402+1431 & 27 & 100.5 & 1 & 60.66 & 14.52 & 0.77 & 0.66 & 44 & 34 & 4.1$\pm$0.6 & 1.6E-7$\pm$3.5E-8 & unass &  & false\\
 J0408+6839 & 28 & 73 & 1 & 62.15 & 68.66 & 0.34 & 0.28 & 49 & 27 & 2.4$\pm$0.2 & 7.3E-8$\pm$2.2E-8 & bcu & TXS 0402+682 & true\\
 J0409$-$5530 & 29 & 83 & 1 & 62.29 & -55.5 & 0.93 & 0.73 & 145 & 30 & 3.5$\pm$0.4 & 8.1E-8$\pm$1.9E-8 & unass &  & false\\
 J0448$-$3424 & 30 & 100.5 & 1 & 72.15 & -34.41 & 0.17 & 0.15 & 16 & 49 & 2.1$\pm$0.2 & 3.7E-8$\pm$1.3E-8 & bcu & TXS 0447$-$345 & false\\
 J0457$-$6830 & 31 & 63 & 1 & 74.3 & -68.51 & 0.36 & 0.29 & 136 & 29 & 2.8$\pm$0.3 & 8.1E-8$\pm$2.2E-8 & unass &  & true\\
 J0459$-$7909 & 32 & 80 & 2 & 75.0 & -79.16 & 0.84 & 0.8 & 30 & 31 & 4.1$\pm$0.6 & 1E-7$\pm$2.3E-8 & bcu & PKS 0509$-$792 & false\\
 J0512$-$7007 & 33 & 30 & 1 & 78.05 & -70.13 & 0.52 & 0.43 & 117 & 32 & 2.8$\pm$0.3 & 9.6E-8$\pm$2.7E-8 & unass &  & false\\
 J0519$-$3709 & 34 & 116 & 1 & 79.94 & -37.16 & 0.81 & 0.36 & 72 & 60 & 3$\pm$0.2 & 1E-7$\pm$1.8E-8 & bcu & SUMSS J051941$-$371449 & false\\
 J0520$-$6726 & 35 & 35.5 & 1 & 80.04 & -67.44 & 0.66 & 0.43 & 42 & 41 & 2.7$\pm$0.2 & 1.1E-7$\pm$2.6E-8 & unass &  & false\\
 J0527$-$3747 & 36 & 37.5 & 1 & 81.8 & -37.79 & 1.16 & 0.89 & 105 & 33 & 4.1$\pm$0.5 & 1.2E-7$\pm$2.6E-8 & unass &  & false\\
 J0546+8247 & 37 & 41 & 2 & 86.54 & 82.79 & 0.42 & 0.28 & 40 & 31 & 2.2$\pm$0.2 & 3.8E-8$\pm$1.2E-8 & bcu & S5 0532+82 & false\\
 J0620$-$1226 & 38 & 97.5 & 1 & 95.14 & -12.44 & 0.63 & 0.44 & 84 & 26 & 2.7$\pm$0.2 & 8.1E-8$\pm$2.6E-8 & agn & TXS 0617$-$123 & true\\
 J0642$-$3237 & 39 & 48 & 1 & 100.52 & -32.63 & 0.87 & 0.51 & 93 & 31 & 3.4$\pm$0.3 & 1.1E-7$\pm$2.5E-8 & bcu & PKS 0641$-$331 & false\\
 J0645+7222 & 40 & 6.5 & 1 & 101.27 & 72.38 & 1.0 & 0.83 & 15 & 36 & 4$\pm$0.6 & 1.2E-7$\pm$2.4E-8 & bcu & GB6 J0642+7215 & false\\
 J0653$-$3649 & 41 & 100.5 & 1 & 103.47 & -36.83 & 0.26 & 0.21 & 139 & 25 & 2.3$\pm$0.3 & 4.6E-8$\pm$2.1E-8 & bcu & NVSS J065356$-$365202 & true\\
 J0657+4619 & 42 & 112.5 & 1 & 104.39 & 46.33 & 0.76 & 0.56 & 63 & 28 & 3.4$\pm$0.4 & 9.8E-8$\pm$2.4E-8 & rg & B3 0653+464 & true\\
 J0657$-$5348 & 43 & 104 & 2 & 104.44 & -53.8 & 0.19 & 0.17 & 151 & 60 & 2.3$\pm$0.2 & 6.8E-8$\pm$1.7E-8 & bcu & PMN J0657$-$5406 & false\\
 J0724+7027 & 44 & 100.5 & 1 & 111.01 & 70.45 & 0.34 & 0.25 & 96 & 47 & 2.5$\pm$0.2 & 6.3E-8$\pm$1.6E-8 & bcu & 87GB 071829.7+701845 & false\\
 J0734$-$6137 & 45 & 56 & 2 & 113.66 & -61.62 & 1.02 & 0.66 & 117 & 27 & 3.5$\pm$0.4 & 1.1E-7$\pm$2.7E-8 & bcu & PMN J0734$-$6122 & false\\
 J0734+1721 & 46 & 109 & 1 & 113.7 & 17.36 & 0.5 & 0.5 & 0 & 29 & 2.5$\pm$0.2 & 7.4E-8$\pm$2.2E-8 & unass &  & true\\
 J0804+8311 & 47 & 66 & 1 & 121.25 & 83.2 & 0.96 & 0.72 & 124 & 34 & 3.5$\pm$0.3 & 8.4E-8$\pm$1.9E-8 & fsrq & S5 0740+82 & false\\
 J0837+2500 & 48 & 112.5 & 1 & 129.43 & 25.01 & 0.21 & 0.17 & 0 & 41 & 1.8$\pm$0.2 & 2E-8$\pm$8.9E-9 & fsrq & B2 0834+25 & false\\
 J0845+5040 & 49 & 35.5 & 1 & 131.33 & 50.67 & 0.5 & 0.5 & 0 & 25 & 3.4$\pm$0.4 & 7.9E-8$\pm$2E-8 & bcu & TXS 0843+510 & true\\
 J0855$-$1233 & 50 & 111 & 1 & 133.75 & -12.56 & 0.94 & 0.6 & 178 & 25 & 3.3$\pm$0.3 & 8.3E-8$\pm$2.1E-8 & unass &  & true\\
 J0902+3323 & 51 & 62.5 & 1 & 135.66 & 33.39 & 0.6 & 0.3 & 130 & 25 & 2.6$\pm$0.3 & 4.9E-8$\pm$1.9E-8 & bcu & VLSS J0900.5+3330* & true\\
 J0943+0940 & 52 & 44 & 1 & 145.81 & 9.68 & 0.92 & 1.2 & 81 & 31 & 3.4$\pm$0.4 & 1.2E-7$\pm$2.8E-8 & rg & 3C 226 & false\\
 J0953$-$3006 & 53 & 116 & 5 & 148.37 & -30.11 & 0.24 & 0.18 & 138 & 39 & 2.3$\pm$0.2 & 4.1E-8$\pm$1.2E-8 & bcu & PMN J0952$-$3006 & false\\
 J1008+1319 & 54 & 34 & 1 & 152.1 & 13.33 & 0.5 & 0.34 & 5 & 29 & 2.3$\pm$0.2 & 5.7E-8$\pm$2.1E-8 & agn & GB6 J1009+1322 & true\\
 J1008+0528 & 55 & 66.5 & 1 & 152.14 & 5.47 & 1.17 & 0.5 & 84 & 27 & 2.8$\pm$0.3 & 1E-7$\pm$3.1E-8 & fsrq & TXS 1006+056 & true\\
 J1016+0959 & 56 & 87.5 & 1 & 154.04 & 9.99 & 0.83 & 0.7 & 61 & 25 & 3.9$\pm$0.5 & 9.1E-8$\pm$2.3E-8 & bcu & 4C +09.36 & true\\
 J1019$-$3216 & 57 & 48.5 & 1 & 154.95 & -32.28 & 0.55 & 0.43 & 143 & 30 & 2.8$\pm$0.2 & 8.8E-8$\pm$2.4E-8 & ssrq & TXS 1018$-$319 & true\\
 J1022$-$3140 & 58 & 74.5 & 1 & 155.74 & -31.67 & 0.32 & 0.26 & 176 & 25 & 2.1$\pm$0.3 & 3.1E-8$\pm$1.5E-8 & unass &  & true\\
 J1053+1409 & 59 & 89 & 1 & 163.35 & 14.16 & 0.7 & 0.63 & 144 & 33 & 3.1$\pm$0.3 & 1.1E-7$\pm$2.5E-8 & fsrq & TXS 1051+147 & false\\
 J1053+1529 & 60 & 19.5 & 1 & 163.46 & 15.5 & 0.66 & 0.6 & 84 & 25 & 3.2$\pm$0.4 & 8.4E-8$\pm$2.4E-8 & fsrq & GB6 J1054+1507 & true\\
 J1055+6509 & 61 & 80.5 & 2 & 163.97 & 65.16 & 0.18 & 0.17 & 18 & 57 & 2.4$\pm$0.2 & 5.3E-8$\pm$1.4E-8 & bcu & GB6 J1055+6509 & false\\
 J1112$-$0934 & 62 & 108 & 1 & 168.08 & -9.57 & 0.26 & 0.24 & 100 & 29 & 2.3$\pm$0.3 & 5.3E-8$\pm$2.4E-8 & bcu & TXS 1109$-$093 & true\\
 J1113$-$0537 & 63 & 83 & 1 & 168.36 & -5.63 & 0.49 & 0.43 & 114 & 25 & 3$\pm$0.3 & 8.5E-8$\pm$2.4E-8 & unass &  & true\\
 J1117+8436 & 64 & 89.5 & 1 & 169.28 & 84.61 & 0.52 & 0.47 & 61 & 25 & 2.9$\pm$0.3 & 6.7E-8$\pm$1.9E-8 & unass &  & true\\
 J1117$-$4839 & 65 & 14 & 2 & 169.39 & -48.66 & 0.19 & 0.18 & 111 & 62 & 2.4$\pm$0.2 & 1.1E-7$\pm$2.8E-8 & bcu & CRATES J1117$-$4838 & false\\
 J1117$-$1604 & 66 & 74 & 1 & 169.45 & -16.08 & 0.52 & 0.31 & 26 & 27 & 2.4$\pm$0.2 & 6E-8$\pm$2E-8 & bcu & PMN J1117$-$1609 & true\\
 J1133$-$1052 & 67 & 7.5 & 1 & 173.43 & -10.87 & 0.2 & 0.17 & 128 & 36 & 2$\pm$0.2 & 2.7E-8$\pm$1.1E-8 & bcu & NVSS J113347$-$105640 & false\\
 J1137+1010 & 68 & 54 & 1 & 174.35 & 10.17 & 0.27 & 0.25 & 147 & 26 & 2.3$\pm$0.3 & 4.1E-8$\pm$1.8E-8 & bcu & TXS 1136+106 & true\\
 J1141+0023 & 69 & 51.5 & 1 & 175.41 & 0.39 & 0.23 & 0.19 & 125 & 36 & 2.1$\pm$0.2 & 3.2E-8$\pm$1.2E-8 & fsrq & PMN J1141+0022 & false\\
 J1145$-$4044 & 70 & 67.5 & 1 & 176.27 & -40.75 & 0.62 & 0.48 & 90 & 26 & 2.8$\pm$0.3 & 9.5E-8$\pm$2.8E-8 & bcu & SUMSS J114604$-$405932 & true\\
 J1145$-$1531 & 71 & 41 & 2 & 176.39 & -15.53 & 0.28 & 0.22 & 140 & 39 & 2.4$\pm$0.2 & 6.3E-8$\pm$1.8E-8 & bcu & TXS 1143$-$152 & false\\
 J1146$-$0926 & 72 & 102 & 3 & 176.6 & -9.45 & 0.18 & 0.16 & 96 & 110 & 2.5$\pm$0.1 & 1.4E-7$\pm$2.4E-8 & bcu & PMN J1146$-$0932 & false\\
 J1151+3938 & 73 & 98.5 & 1 & 177.91 & 39.64 & 0.46 & 0.35 & 29 & 25 & 2.1$\pm$0.2 & 3E-8$\pm$1.4E-8 & fsrq & GB6 J1151+4008 & true\\
 J1153$-$3658 & 74 & 74.5 & 1 & 178.42 & -36.97 & 0.26 & 0.25 & 109 & 26 & 2.3$\pm$0.2 & 5.5E-8$\pm$2.2E-8 & bcu & PKS 1150$-$369 & true\\
 J1201$-$2658 & 75 & 104 & 1 & 180.43 & -26.98 & 0.64 & 0.58 & 133 & 30 & 3.1$\pm$0.3 & 7.6E-8$\pm$1.7E-8 & bcu & TXS 1200$-$261 & true\\
 J1202$-$2144 & 76 & 46 & 1 & 180.73 & -21.74 & 0.54 & 0.81 & 69 & 25 & 2.8$\pm$0.3 & 7.4E-8$\pm$2.1E-8 & unass &  & true\\
 J1210+3708 & 77 & 13 & 1 & 182.58 & 37.14 & 0.38 & 0.38 & 37 & 29 & 2.7$\pm$0.3 & 6.4E-8$\pm$2.3E-8 & fsrq & B3 1206+374 & true\\
 J1219+2907 & 78 & 7 & 1 & 184.9 & 29.12 & 0.5 & 0.5 & 0 & 25 & 3.3$\pm$0.4 & 9.8E-8$\pm$3.1E-8 & rg & NGC 4278 & true\\
 J1223$-$0847 & 79 & 26 & 1 & 185.93 & -8.78 & 0.37 & 0.3 & 147 & 28 & 2.5$\pm$0.2 & 8.7E-8$\pm$3.2E-8 & bcu & PMN J1223$-$0907 & true\\
 J1224$-$0536 & 80 & 8 & 2 & 186.16 & -5.6 & 0.44 & 0.23 & 48 & 42 & 2.2$\pm$0.2 & 4.7E-8$\pm$1.4E-8 & bcu & NVSS J122430$-$053030 & false\\
 J1226+0232 & 81 & 47 & 1 & 186.51 & 2.54 & 0.37 & 0.26 & 50 & 50 & 2.9$\pm$0.2 & 1.6E-7$\pm$3.9E-8 & bcu & PMN J1225+0235 & false\\
 J1237$-$4430 & 82 & 104 & 1 & 189.48 & -44.51 & 0.48 & 0.34 & 57 & 29 & 2.8$\pm$0.2 & 9E-8$\pm$2.6E-8 & unass &  & true\\
 J1259+5444 & 83 & 58.5 & 1 & 194.96 & 54.74 & 0.5 & 0.5 & 0 & 27 & 3$\pm$0.3 & 6.9E-8$\pm$1.7E-8 & fsrq & VLSS J1259.3+5432 & true\\
 J1322$-$4521 & 84 & 6 & 1 & 200.7 & -45.36 & 1.22 & 1.07 & 131 & 26 & 3.1$\pm$0.3 & 1.3E-7$\pm$3.3E-8 & unass &  & true\\
 J1323$-$4439 & 85 & 86.5 & 2 & 200.9 & -44.67 & 0.31 & 0.23 & 146 & 26 & 2.4$\pm$0.2 & 6.5E-8$\pm$2.4E-8 & fsrq & PKS 1320$-$446 & false\\
 J1331+1346 & 86 & 103.5 & 1 & 202.84 & 13.77 & 0.26 & 0.2 & 108 & 38 & 2$\pm$0.2 & 2.9E-8$\pm$1.2E-8 & bcu & NVSS J133134+135227 & false\\
 J1356$-$1545 & 87 & 21.5 & 2 & 209.1 & -15.77 & 0.36 & 0.31 & 139 & 29 & 2.6$\pm$0.2 & 6.8E-8$\pm$1.9E-8 & fsrq & PMN J1357$-$1527 & false\\
 J1402$-$1547 & 88 & 96 & 1 & 210.72 & -15.79 & 0.72 & 0.59 & 125 & 29 & 3.1$\pm$0.2 & 1.3E-7$\pm$3.6E-8 & bcu & PMN J1404$-$1538 & true\\
 J1412+7641 & 89 & 27.5 & 1 & 213.2 & 76.69 & 0.45 & 0.31 & 0 & 33 & 2.6$\pm$0.2 & 5.1E-8$\pm$1.3E-8 & bcu & NVSS J141134+763515 & false\\
 J1416+3447 & 90 & 111.5 & 2 & 214.07 & 34.8 & 0.24 & 0.21 & 159 & 47 & 2.5$\pm$0.2 & 6.4E-8$\pm$1.9E-8 & css & S4 1413+34 & false\\
 J1419+3708 & 91 & 59.5 & 2 & 214.76 & 37.14 & 0.71 & 0.62 & 65 & 36 & 3.2$\pm$0.4 & 7.7E-8$\pm$1.7E-8 & fsrq & B3 1417+375 & false\\
 J1421$-$2222 & 92 & 37.5 & 1 & 215.41 & -22.38 & 0.23 & 0.2 & 64 & 27 & 2.3$\pm$0.2 & 4.4E-8$\pm$1.8E-8 & bcu & PMN J1421$-$2221 & true\\
 J1422$-$3254 & 93 & 28 & 1 & 215.67 & -32.9 & 1.0 & 0.91 & 27 & 27 & 2.6$\pm$0.2 & 7.9E-8$\pm$2.2E-8 & bcu & PKS 1421$-$322 & true\\
 J1426+3139 & 94 & 113 & 1 & 216.75 & 31.66 & 0.7 & 0.53 & 118 & 26 & 2.3$\pm$0.3 & 5.8E-8$\pm$2.1E-8 & bcu & TXS 1423+323 & true\\
 J1502$-$2420 & 95 & 107.5 & 1 & 225.73 & -24.34 & 0.31 & 0.23 & 140 & 33 & 2.5$\pm$0.2 & 8.6E-8$\pm$2.4E-8 & unass &  & false\\
 J1513$-$2830 & 96 & 70 & 2 & 228.4 & -28.5 & 0.17 & 0.21 & 161 & 66 & 2.4$\pm$0.2 & 8.6E-8$\pm$2E-8 & bcu & PMN J1512$-$2828 & false\\
 J1522+1535 & 97 & 27 & 1 & 230.6 & 15.6 & 0.51 & 0.31 & 23 & 26 & 2.5$\pm$0.2 & 7E-8$\pm$2.5E-8 & fsrq & SDSS J152247.54+153520.8 & true\\
 J1528$-$1348 & 98 & 91.5 & 1 & 232.22 & -13.8 & 0.41 & 0.27 & 38 & 42 & 2.8$\pm$0.2 & 1.2E-7$\pm$2.8E-8 & bcu & TXS 1527$-$135 & false\\
 J1533$-$2130 & 99 & 98 & 3 & 233.45 & -21.5 & 0.18 & 0.16 & 118 & 37 & 2.1$\pm$0.2 & 5.2E-8$\pm$1.9E-8 & unass &  & false\\
 J1544+2705 & 100 & 77.5 & 1 & 236.04 & 27.1 & 1.35 & 0.78 & 137 & 30 & 2.4$\pm$0.2 & 5E-8$\pm$1.8E-8 & unass &  & false\\
 J1554$-$0252 & 101 & 109 & 2 & 238.56 & -2.87 & 0.23 & 0.18 & 77 & 41 & 2.3$\pm$0.2 & 8.5E-8$\pm$2.4E-8 & bcu & PMN J1554$-$0239 & false\\
 J1605+7725 & 102 & 91.5 & 2 & 241.37 & 77.42 & 0.38 & 0.26 & 121 & 40 & 2.5$\pm$0.2 & 6E-8$\pm$1.5E-8 & bcu & VLSS J1606.2+7658 & false\\
 J1618$-$1016 & 103 & 101.5 & 1 & 244.57 & -10.28 & 0.3 & 0.22 & 61 & 27 & 2.2$\pm$0.2 & 5.2E-8$\pm$1.9E-8 & bcu & TXS 1615$-$102 & true\\
 J1622+0044 & 104 & 30 & 1 & 245.71 & 0.74 & 0.43 & 0.25 & 54 & 27 & 2.3$\pm$0.3 & 5.6E-8$\pm$2.4E-8 & unass &  & true\\
 J1626+5436 & 105 & 86.5 & 2 & 246.51 & 54.61 & 0.24 & 0.2 & 13 & 60 & 2.7$\pm$0.2 & 8.9E-8$\pm$2E-8 & fsrq & CRATES J1626+5442 & false\\
 J1628$-$1449 & 106 & 110.5 & 1 & 247.05 & -14.82 & 0.26 & 0.22 & 171 & 25 & 2.2$\pm$0.2 & 5.4E-8$\pm$2.5E-8 & bcu & PKS 1625$-$149 & true\\
 J1649+2725 & 107 & 4.5 & 1 & 252.49 & 27.42 & 0.94 & 0.85 & 169 & 28 & 4.2$\pm$0.7 & 9.3E-8$\pm$2.1E-8 & bll & B2 1645+27 & true\\
 J1701+2801 & 108 & 0.5 & 1 & 255.42 & 28.02 & 0.31 & 0.25 & 56 & 27 & 2.6$\pm$0.3 & 5.4E-8$\pm$1.8E-8 & unass &  & true\\
 J1708$-$1246 & 109 & 72 & 2 & 257.14 & -12.78 & 0.19 & 0.15 & 162 & 39 & 2.1$\pm$0.1 & 4.3E-8$\pm$1.3E-8 & bcu & NVSS J170822$-$124426 & false\\
 J1732+1510 & 110 & 71.5 & 2 & 263.2 & 15.17 & 0.31 & 0.26 & 152 & 46 & 2.9$\pm$0.2 & 8E-8$\pm$1.7E-8 & bcu & TXS 1731+152A & false\\
 J1734+5158 & 111 & 84.5 & 1 & 263.61 & 51.97 & 0.63 & 0.43 & 53 & 45 & 3.3$\pm$0.3 & 1.5E-7$\pm$3.1E-8 & bcu & TXS 1734+516 & false\\
 J1737+1854 & 112 & 42 & 1 & 264.27 & 18.9 & 0.29 & 0.23 & 100 & 28 & 2.4$\pm$0.3 & 5.4E-8$\pm$2.3E-8 & unass &  & true\\
 J1752+4355 & 113 & 119.5 & 1 & 268.2 & 43.92 & 0.28 & 0.27 & 94 & 26 & 2.5$\pm$0.3 & 4.1E-8$\pm$1.6E-8 & fsrq & S4 1751+44 & true\\
 J1801$-$7816 & 114 & 32.5 & 2 & 270.31 & -78.27 & 0.28 & 0.23 & 129 & 55 & 2.6$\pm$0.2 & 1.3E-7$\pm$2.8E-8 & bcu & PKS 1754$-$782 & false\\
 J1803+2523 & 115 & 13.5 & 1 & 270.84 & 25.4 & 0.35 & 0.28 & 32 & 32 & 2.6$\pm$0.2 & 8.9E-8$\pm$2.5E-8 & bcu & TXS 1801+253 & false\\
 J1815+2017 & 116 & 83.5 & 2 & 273.95 & 20.29 & 0.42 & 0.3 & 120 & 29 & 2.9$\pm$0.3 & 1.1E-7$\pm$3.2E-8 & unass &  & false\\
 J1823+4759 & 117 & 71.5 & 1 & 275.82 & 48.0 & 0.5 & 0.5 & 0 & 25 & 3$\pm$0.3 & 9.4E-8$\pm$2.4E-8 & bcu & TXS 1821+483 & true\\
 J1851$-$6844 & 118 & 5 & 1 & 282.83 & -68.74 & 0.25 & 0.19 & 156 & 47 & 2.4$\pm$0.2 & 8.7E-8$\pm$2.6E-8 & bcu & PKS 1847$-$688 & false\\
 J1918$-$2816 & 119 & 27.5 & 1 & 289.63 & -28.27 & 0.7 & 0.67 & 142 & 30 & 3.2$\pm$0.3 & 1.6E-7$\pm$4E-8 & bcu & PMN J1919$-$2823 & true\\
 J1919$-$4543 & 120 & 21 & 2 & 289.84 & -45.73 & 0.26 & 0.22 & 34 & 41 & 2.6$\pm$0.2 & 8.2E-8$\pm$2.1E-8 & fsrq & PKS 1915$-$458 & false\\
 J1922$-$6615 & 121 & 90.5 & 1 & 290.61 & -66.25 & 0.34 & 0.25 & 97 & 27 & 2.5$\pm$0.2 & 5.8E-8$\pm$2E-8 & unass &  & true\\
 J1929$-$7909 & 122 & 34.5 & 2 & 292.36 & -79.15 & 0.35 & 0.27 & 138 & 36 & 2.7$\pm$0.2 & 1E-7$\pm$2.8E-8 & bcu & PMN J1931$-$7930 & false\\
 J1933$-$5420 & 123 & 101 & 1 & 293.42 & -54.34 & 0.5 & 0.5 & 0 & 27 & 2.8$\pm$0.2 & 8.5E-8$\pm$2.2E-8 & unass &  & true\\
 J1936+5341 & 124 & 20.5 & 1 & 294.21 & 53.69 & 0.16 & 0.14 & 70 & 41 & 2$\pm$0.2 & 3E-8$\pm$1.1E-8 & bcu & TXS 1935+536 & false\\
 J1937$-$5509 & 125 & 99.5 & 2 & 294.28 & -55.15 & 0.18 & 0.15 & 81 & 80 & 2.2$\pm$0.1 & 8.1E-8$\pm$1.7E-8 & bcu & PMN J1936$-$5512 & false\\
 J1943$-$3034 & 126 & 40.5 & 1 & 295.91 & -30.57 & 0.8 & 0.47 & 148 & 28 & 3$\pm$0.2 & 1E-7$\pm$2.7E-8 & unass &  & true\\
 J1944$-$5135 & 127 & 99 & 1 & 296.12 & -51.6 & 0.34 & 0.28 & 106 & 26 & 2.6$\pm$0.2 & 6E-8$\pm$2.1E-8 & unass &  & true\\
 J1952+4942 & 128 & 110 & 2 & 298.01 & 49.7 & 0.43 & 0.33 & 42 & 30 & 2.6$\pm$0.2 & 8E-8$\pm$2.5E-8 & fsrq & TXS 1951+498 & false\\
 J2010$-$2523 & 129 & 72.5 & 7 & 302.56 & -25.4 & 0.17 & 0.15 & 109 & 177 & 2.9$\pm$0.1 & 2.3E-7$\pm$2.6E-8 & fsrq & PMN J2010$-$2524 & false\\
 J2013$-$1732 & 130 & 55.5 & 1 & 303.5 & -17.55 & 0.5 & 0.5 & 0 & 27 & 3.2$\pm$0.3 & 1.3E-7$\pm$3.2E-8 & unass &  & true\\
 J2035$-$2708 & 131 & 55 & 1 & 308.76 & -27.14 & 0.62 & 0.5 & 82 & 26 & 3$\pm$0.2 & 1.1E-7$\pm$3.4E-8 & bcu & CRATES J2035$-$2632 & true\\
 J2035+0955 & 132 & 73.5 & 1 & 308.9 & 9.92 & 0.5 & 0.5 & 0 & 30 & 2.8$\pm$0.2 & 1.2E-7$\pm$3.2E-8 & unass &  & false\\
 J2112$-$3207 & 133 & 0 & 1 & 318.1 & -32.12 & 0.74 & 0.43 & 94 & 43 & 2.9$\pm$0.2 & 1.2E-7$\pm$2.6E-8 & bcu & PKS 2108$-$326 & false\\
 J2113$-$2616 & 134 & 71.5 & 1 & 318.37 & -26.27 & 0.5 & 0.5 & 0 & 31 & 4.1$\pm$0.6 & 1.5E-7$\pm$3.4E-8 & unass &  & false\\
 J2114+1113 & 135 & 60.5 & 2 & 318.51 & 11.22 & 0.19 & 0.16 & 39 & 43 & 2.2$\pm$0.2 & 4.3E-8$\pm$1.3E-8 & bcu & TXS 2112+108 & false\\
 J2141+1615 & 136 & 45.5 & 1 & 325.44 & 16.26 & 0.7 & 0.43 & 32 & 28 & 2.9$\pm$0.2 & 9.8E-8$\pm$2.6E-8 & bcu & 87GB 213814.6+161434 & true\\
 J2142$-$2303 & 137 & 64.5 & 2 & 325.7 & -23.06 & 0.34 & 0.25 & 172 & 50 & 2.4$\pm$0.2 & 1.2E-7$\pm$2.9E-8 & bcu & PMN J2142$-$2303 & false\\
 J2144$-$3215 & 138 & 20.5 & 1 & 326.08 & -32.26 & 1.09 & 1.07 & 84 & 26 & 3.4$\pm$0.3 & 9.2E-8$\pm$2.5E-8 & fsrq & PKS 2142$-$319 & true\\
 J2219$-$2732 & 139 & 96 & 1 & 334.99 & -27.53 & 0.41 & 0.32 & 113 & 30 & 2.6$\pm$0.2 & 5.1E-8$\pm$1.5E-8 & fsrq & PMN J2219$-$2719 & true\\
 J2241$-$3550 & 140 & 86 & 2 & 340.42 & -35.84 & 0.72 & 0.39 & 79 & 31 & 2.9$\pm$0.3 & 7.6E-8$\pm$1.9E-8 & bcu & PMN J2241$-$3559 & false\\
 J2302+0457 & 141 & 47.5 & 1 & 345.53 & 4.97 & 0.55 & 0.46 & 90 & 28 & 2.8$\pm$0.3 & 8.3E-8$\pm$2.3E-8 & unass &  & true\\
 J2341$-$4205 & 142 & 86.5 & 1 & 355.38 & -42.1 & 0.29 & 0.23 & 113 & 25 & 2.2$\pm$0.3 & 3.2E-8$\pm$1.5E-8 & unass &  & true\\
\hline
\enddata
\end{deluxetable*}%\label{table:extract}
\end{longrotatetable}

\begin{longrotatetable}
\begin{deluxetable*}{|c|c|c|c|c|c|c|c|}
\tablecaption{Sun detection summary table: the columns show the detection time bin, the detection localization in Equatorial coordinates (J2000), a flag indicating whether the source is detected in overlapping ROIs, the test statistic, the flux with its 1$\sigma$ uncertainty, the power law index with its 1$\sigma$ uncertainty and the name of the associated solar flare as reported in The First Fermi-LAT Solar Flare Catalog \cite{2021ApJS..252...13A}. \label{table:sun}}
\tablewidth{700pt}
\tabletypesize{\scriptsize}
\tablehead{
  \multicolumn{1}{|c|}{TBIN 1m} &
  \multicolumn{1}{c|}{RAJ2000} &
  \multicolumn{1}{c|}{DEJ2000} &
  \multicolumn{1}{c|}{repROI} &
  \multicolumn{1}{c|}{Test Statistic} &
  \multicolumn{1}{c|}{Flux} &
  \multicolumn{1}{c|}{Power Law Index}&
  \multicolumn{1}{c|}{FLSF Name}\\
  \multicolumn{1}{|c|}{} &
  \multicolumn{1}{c|}{(degree)} &
  \multicolumn{1}{c|}{(degree)} &
  \multicolumn{1}{c|}{} &
  \multicolumn{1}{c|}{} &
  \multicolumn{1}{c|}{${\rm (phcm^{-2}s^{-1})}$} &
  \multicolumn{1}{c|}{} &
  \multicolumn{1}{c|}{} 
  }
\startdata
\hline
0 & 161.66 & 7.76 & false & 30 & 4.2E-8$\pm$1.6E-8 & 2.2 $\pm$ 0.2 & - \\
0 & 148.65 & 12.81 & true & 28 & 6.2E-8$\pm$2.7E-8 & 2.5 $\pm$ 0.3 & - \\
0 & 135.82 & 16.92 & true & 41 & 6.3E-8$\pm$2.1E-8 & 2.3 $\pm$ 0.2 & - \\
0.5 & 165.39 & 6.31 & false & 34 & 5.4E-8$\pm$1.7E-8 & 2.3 $\pm$ 0.2 & - \\
1 & 165.46 & 6.38 & false & 40 & 5.9E-8$\pm$1.7E-8 & 2.3 $\pm$ 0.2 & - \\
5.5 & 317.67 & -16.2 & false & 33 & 3E-8$\pm$1.2E-8 & 1.9 $\pm$ 0.2 & - \\
6 & 343.37 & -7.18 & true & 28 & 7.3E-8$\pm$2.3E-8 & 2.6 $\pm$ 0.2 & - \\
6 & 317.88 & -16.2 & true & 31 & 3.4E-8$\pm$1.5E-8 & 2.0 $\pm$ 0.2 & - \\
7 & 356.23 & -2.04 & false & 30 & 1.1E-7$\pm$2.7E-8 & 3.3 $\pm$ 0.4 & - \\
11 & 128.32 & 18.8 & false & 28 & 5.8E-8$\pm$2E-8 & 2.4 $\pm$ 0.2 & - \\
11.5 & 143.8 & 14.48 & false & 34 & 1.0E-7$\pm$2.6E-8 & 3.1 $\pm$ 0.3 & - \\
11.5 & 139.56 & 15.43 & false & 30 & 5.3E-8$\pm$1.7E-8 & 2.4 $\pm$ 0.2 & - \\
12 & 144.14 & 14.18 & false & 33 & 8.5E-8$\pm$2.5E-8 & 2.7 $\pm$ 0.3 & - \\
12 & 139.6 & 15.46 & false & 31 & 5.6E-8$\pm$1.7E-8 & 2.3 $\pm$ 0.2 & - \\
13 & 166.21 & 5.83 & true & 31 & 8.5E-8$\pm$2.7E-8 & 2.5 $\pm$ 0.2 & - \\
17 & 310.77 & -18.46 & false & 33 & 7E-8$\pm$2.3E-8 & 2.3 $\pm$ 0.2 & - \\
18.5 & 355.84 & -1.84 & false & 29 & 1.1E-7$\pm$3.2E-8 & 2.9 $\pm$ 0.3 & - \\
22.5 & 110.21 & 22.15 & false & 28 & 5.6E-8$\pm$2.4E-8 & 2.3 $\pm$ 0.3 & - \\
23 & 130.92 & 18.05 & false & 25 & 3.2E-8$\pm$2E-8 & 2.0 $\pm$ 0.3 & - \\
23.5 & 143.92 & 14.38 & false & 31 & 8.4E-7$\pm$2.5E-8 & 2.5 $\pm$ 0.2 & - \\
23.5 & 127.69 & 19.05 & false & 35 & 6.6E-8$\pm$2.3E-8 & 2.3 $\pm$ 0.2 & - \\
29 & 60.69 & 20.32 & false & 30 & 1E-7$\pm$2.7E-8 & 2.8 $\pm$ 0.2 & - \\
30.5 & 347.8 & -5.1 & false & 540 & 8E-7$\pm$5.4E-8 & 3.1 $\pm$ 0.1 &  2011$-$03$-$07    \\
31 & 347.77 & -5.08 & false & 533 & 5.9E-7$\pm$3.9E-8 & 3.1 $\pm$ 0.1 &  2011$-$03$-$07    \\
31.5 & 18.16 & 8.61 & false & 29 & 1E-7$\pm$2.9E-8 & 2.8 $\pm$ 0.2 & - \\
33.5 & 75.27 & 22.49 & false & 37 & 2E-7$\pm$4.2E-8 & 3.4 $\pm$ 0.3 &  2011$-$06$-$07 \\
35 & 134.3 & 16.51 & true & 30 & 1E-7$\pm$2.5E-8 & 3.1 $\pm$ 0.3 &  2011$-$08$-$04 \\
35 & 133.49 & 17.77 & true & 32 & 1E-7$\pm$2.6E-8 & 3.1 $\pm$ 0.3 &  2011$-$08$-$04 \\
36.5 & 165.03 & 6.51 & false & 1649 & 1.1E-6$\pm$4.7E-8 & 2.9 $\pm$ 0.1 &  2011$-$09$-$06    \\
37 & 167.86 & 5.43 & true & 39 & 7.5E-8$\pm$2.4E-8 & 2.3 $\pm$ 0.3 & - \\
37 & 165.04 & 6.49 & false & 1685 & 1.1E-6$\pm$5.5E-8 & 2.9 $\pm$ 0.1 &  2011$-$09$-$06    \\
41 & 304.7 & -19.37 & true & 263 & 5.7E-7$\pm$4.8E-8 & 3.5 $\pm$ 0.1 &  2012$-$01$-$23    \\
41.5 & 304.25 & -19.52 & false & 261 & 5.7E-7$\pm$4.8E-8 & 3.5 $\pm$ 0.1 &  2012$-$01$-$23    \\
42 & 347.92 & -5.06 & false & 161920 & 2.2E-5$\pm$1.7E-7 & 2.8 $\pm$ 0.1 &  2012$-$03$-$05    \\
42.5 & 347.91 & -5.0 & false & 74014 & 2.2E-5$\pm$1.8E-7 & 2.8 $\pm$ 0.1 &  2012$-$03$-$07    \\
43 & 347.91 & -5.0 & false & 168860 & 2.2E-5$\pm$1.7E-7 & 2.8 $\pm$ 0.1 &  2012$-$03$-$07    \\
46.5 & 106.04 & 22.28 & true & 178 & 3E-7$\pm$3.5E-8 & 3.0 $\pm$ 0.1 &  2012$-$07$-$06    \\
54 & 331.17 & -12.05 & true & 29 & 8.2E-8$\pm$2.3E-8 & 2.6 $\pm$ 0.2 & - \\
56 & 19.8 & 8.11 & false & 160 & 3.4E-7$\pm$3.8E-8 & 3.2 $\pm$ 0.1 &  2013$-$04$-$11    \\
56.5 & 51.1 & 18.81 & false & 949 & 8.7E-7$\pm$4.5E-8 & 3.1 $\pm$ 0.1 &  2013$-$05$-$13a$-$b \\
57 & 51.11 & 18.83 & false & 1000 & 9.8E-7$\pm$5E-8 & 3.2 $\pm$ 0.1 &  2013$-$05$-$13a$-$b \\
61.5 & 196.81 & -7.11 & false & 858 & 9.7E-7$\pm$5.2E-8 & 3.0 $\pm$ 0.1 &  2013$-$10$-$11    \\
62 & 196.84 & -7.12 & false & 812 & 7E-7$\pm$3.8E-8 & 2.9 $\pm$ 0.1 &  2013$-$10$-$11    \\
62 & 212.8 & -13.31 & false & 26 & 7.1E-8$\pm$2.1E-8 & 2.6 $\pm$ 0.2 &  2013$-$10$-$28c \\
66 & 338.18 & -9.22 & false & 8764 & 6.8E-6$\pm$1.2E-7 & 2.8 $\pm$ 0.1 &  2014$-$02$-$25    \\
66.5 & 338.18 & -9.2 & false & 8560 & 6.2E-6$\pm$1.1E-7 & 2.8 $\pm$ 0.1 &  2014$-$02$-$25    \\
72 & 155.99 & 10.14 & true & 31 & 9.9E-8$\pm$3.5E-8 & 2.4 $\pm$ 0.2 & - \\
72 & 160.59 & 8.42 & true & 2428 & 3.1E-6$\pm$9.5E-8 & 2.8 $\pm$ 0.1 &  2014$-$09$-$01    \\
72.5 & 160.61 & 8.37 & true & 2542 & 2.6E-6$\pm$7.9E-8 & 2.8 $\pm$ 0.1 &  2014$-$09$-$01    \\
90 & 347.9 & -5.41 & false & 26 & 5.1E-8$\pm$2E-8 & 2.3 $\pm$ 0.3 & - \\
96 & 161.07 & 8.21 & true & 45 & 4.9E-8$\pm$1.7E-8 & 2.1 $\pm$ 0.2 & - \\
96.5 & 161.13 & 8.06 & true & 38 & 4.8E-8$\pm$1.7E-8 & 2.1 $\pm$ 0.2 & - \\
98 & 218.0 & -14.7 & true & 28 & 5.8E-8$\pm$2E-8 & 2.2 $\pm$ 0.2 & - \\
98.5 & 218.01 & -14.7 & false & 30 & 5.2E-8$\pm$1.7E-8 & 2.2 $\pm$ 0.2 & - \\
101 & 315.58 & -16.09 & true & 29 & 1E-7$\pm$2.7E-8 & 3.0 $\pm$ 0.3 & - \\
103.5 & 0.84 & 0.33 & true & 28 & 3.2E-8$\pm$1.6E-8 & 2.0 $\pm$ 0.2 & - \\
104 & 19.17 & 8.0 & false & 37 & 8.9E-8$\pm$2.5E-8 & 2.6 $\pm$ 0.2 & - \\
108.5 & 165.38 & 6.33 & false & 114 & 4E-7$\pm$5E-8 & 2.9 $\pm$0.1 &  2017$-$09$-$06a$-$b \\
108.5 & 168.63 & 4.81 & true & 19735 & 7.8E-6$\pm$1.1E-7 & 2.6 $\pm$ 0.1 &  2017$-$09$-$10   \\
109 & 168.61 & 4.83 & true & 13639 & 7.4E-6$\pm$1.1E-7 & 2.6 $\pm$ 0.1 &  2017$-$09$-$10  \\
\hline
\enddata
\end{deluxetable*}%\label{table:sun}

%\label{table:sun}
\end{longrotatetable}

\begin{longrotatetable}
\begin{deluxetable*}{|c|c|c|c|c|c|c|c|c|c|}
\tablecaption{GRB detection summary table: the columns show the detection time bin, the detection localization in Equatorial coordinates (J2000), a flag indicating whether the source is detected in overlapping ROIs, the test statistic, the energy flux with its 1$\sigma$ uncertainty, the flux with its 1$\sigma$ uncertainty, the power-law index with its 1$\sigma$ uncertainty, the GRB trigger number and the GRB name as reported in \cite{2019fermi10GRB}.\label{table:grb}}
\tablewidth{700pt}
\tabletypesize{\scriptsize}
\tablehead{
  \multicolumn{1}{|c|}{TBIN 1m} &
  \multicolumn{1}{c|}{RAJ2000} &
  \multicolumn{1}{c|}{DEJ2000} &
  \multicolumn{1}{c|}{repROI} &
  \multicolumn{1}{c|}{Test Statistic} &
  \multicolumn{1}{c|}{Energy Flux} &
  \multicolumn{1}{c|}{Flux} &
  \multicolumn{1}{c|}{Power Law Index} &
  \multicolumn{1}{c|}{Trigger} &
  \multicolumn{1}{c|}{GRB name} \\
  \multicolumn{1}{|c|}{} &
  \multicolumn{1}{c|}{(degree)} &
  \multicolumn{1}{c|}{(degree)} &
  \multicolumn{1}{c|}{} &
  \multicolumn{1}{c|}{} &
  \multicolumn{1}{c|}{$\mathrm{(MeVcm^{-2}s^{-1})}$} &
  \multicolumn{1}{c|}{${\rm (phcm^{-2}s^{-1})}$} &
  \multicolumn{1}{c|}{} &
  \multicolumn{1}{c|}{} & 
  \multicolumn{1}{c|}{}
}
\startdata
\hline
7 & 90.82 & -41.96 & false & 65 & 4.4E-5 $\pm$ 1.8E-5 & 4.3E-8 $\pm$ 1.5E-8 & 1.9 $\pm$ 0.2 & 90328401 & 090328A\\
7.5 & 90.69 & -41.96 & true & 50 & 2.8E-5 $\pm$ 1.0E-5 & 3.9E-8 $\pm$ 1.4E-8 & 2.0 $\pm$ 0.2 & 90328401 & 090328A\\
8.5 & 333.55 & -26.58 & false & 223 & 6.4E-5 $\pm$ 1.2E-5 & 1.0E-7 $\pm$ 1.6E-8 & 2.1 $\pm$ 0.1 & 90510016 & 090510A\\
9 & 333.52 & -26.6 & true & 168 & 5.5E-5 $\pm$ 1.2E-5 & 8.6E-8 $\pm$ 1.7E-8 & 2.1 $\pm$ 0.1 & 90510016 & 090510A\\
12 & 264.92 & 27.35 & true & 471 & 1.3E-4 $\pm$ 1.9E-5 & 1.9E-7 $\pm$ 2.1E-8 & 2.1 $\pm$ 0.1 & 90902462 & 090902B\\
12.5 & 264.93 & 27.33 & true & 473 & 1.2E-4 $\pm$ 1.7E-5 & 1.9E-7 $\pm$ 2.2E-8 & 2.1 $\pm$ 0.1 & 90902462 & 090902B\\
13 & 353.52 & -66.34 & true & 251 & 6.9E-5 $\pm$ 1.1E-5 & 1.5E-7 $\pm$ 2.1E-8 & 2.2 $\pm$ 0.1 & 90926181 & 090926A\\
13.5 & 353.56 & -66.35 & true & 271 & 7.1E-5 $\pm$ 1.0E-5 & 1.7E-7 $\pm$ 2.2E-8 & 2.3 $\pm$ 0.1 & 90926181 & 090926A\\
19.5 & 192.15 & 8.74 & false & 30 & 2.8E-5 $\pm$ 1.6E-5 & 1.9E-8 $\pm$ 1.2E-8 & 1.8 $\pm$ 0.3 & 100414097 & 100414A\\
20 & 192.17 & 8.72 & true & 31 & 2.8E-5 $\pm$ 1.5E-5 & 2.0E-8 $\pm$ 1.0E-8 & 1.8 $\pm$ 0.2 & 100414097 & 100414A\\
46 & 170.71 & 8.91 & true & 61 & 3.2E-5 $\pm$ 5.5E-6 & 1.3E-7 $\pm$ 2.7E-8 & 2.7 $\pm$ 0.2 & 120624933 & 120624B\\
46.5 & 170.83 & 8.94 & true & 60 & 3.3E-5 $\pm$ 5.7E-6 & 1.4E-7 $\pm$ 2.8E-8 & 2.7 $\pm$ 0.2 & 120624933 & 120624B\\
56 & 173.13 & 27.71 & true & 895 & 2.4E-4 $\pm$ 3.6E-5 & 2.0E-7 $\pm$ 1.9E-8 & 1.9 $\pm$ 0.1 & 130427324 & 130427A\\
56.5 & 173.13 & 27.72 & true & 861 & 2.2E-4 $\pm$ 3.0E-5 & 2.3E-7 $\pm$ 2.2E-8 & 1.9 $\pm$ 0.1 & 130427324 & 130427A\\
61 & 211.99 & 1.25 & true & 26 & 2.1E-5 $\pm$ 5.0E-6 & 1.1E-7 $\pm$ 3.1E-8 & 3.0 $\pm$ 0.3 & 140102887 & 140102A\\
62.5 & 156.67 & 9.67 & false & 31 & 1.8E-5 $\pm$ 4.2E-6 & 6.7E-8 $\pm$ 1.9E-8 & 2.6 $\pm$ 0.2 & 131108862 & 131108A\\
63 & 156.67 & 9.68 & false & 47 & 2.2E-5 $\pm$ 4.7E-6 & 7.6E-8 $\pm$ 1.9E-8 & 2.5 $\pm$ 0.2 & 131108862 & 131108A\\
64 & 10.58 & -1.74 & false & 87 & 1.1E-4 $\pm$ 5.6E-5 & 4.9E-8 $\pm$ 1.8E-8 & 1.7 $\pm$ 0.2 & 131231198 & 131231A\\
64.5 & 10.57 & -1.72 & false & 71 & 7.4E-5 $\pm$ 4.1E-5 & 2.4E-8 $\pm$ 1.1E-8 & 1.6 $\pm$ 0.2 & 131231198 & 131231A\\
65.5 & 315.25 & -8.73 & false & 74 & 4.0E-5 $\pm$ 8.3E-6 & 9.5E-8 $\pm$ 2.6E-8 & 2.3 $\pm$ 0.2 & 140206275 & 140206B\\
66 & 315.26 & -8.73 & false & 71 & 3.3E-5 $\pm$ 6.1E-6 & 9.1E-8 $\pm$ 2.4E-8 & 2.4 $\pm$ 0.2 & 140206275 & 140206B\\
94 & 308.5 & 6.92 & false & 103 & 4.3E-5 $\pm$ 7.9E-6 & 1.0E-7 $\pm$ 2.4E-8 & 2.3 $\pm$ 0.1 & 160625945 & 160625B\\
94.5 & 308.52 & 6.93 & false & 102 & 4.5E-5 $\pm$ 8.4E-6 & 1.0E-7 $\pm$ 2.3E-8 & 2.2 $\pm$ 0.1 & 160625945 & 160625B\\
101.5 & 256.27 & -1.86 & true & 73 & 5.0E-5 $\pm$ 8.0E-6 & 1.7E-7 $\pm$ 3.6E-8 & 2.5 $\pm$ 0.2 & 170214649 & 170214A\\
102 & 256.3 & -1.78 & true & 59 & 4.8E-5 $\pm$ 8.0E-6 & 1.8E-7 $\pm$ 3.9E-8 & 2.6 $\pm$ 0.2 & 170214649 & 170214A\\
118.5 & 0.62 & -2.95 & true & 48 & 2.2E-5 $\pm$ 5.9E-6 & 4.7E-8 $\pm$ 1.4E-8 & 2.2 $\pm$ 0.2 & 180720598 & 180720B\\
119 & 0.58 & -2.92 & true & 59 & 3.4E-5 $\pm$ 9.3E-6 & 6.8E-8 $\pm$ 1.7E-8 & 2.2 $\pm$ 0.2 & 180720598 & 180720B\\
\hline
\enddata
\end{deluxetable*}
%\label{table:grb}
\end{longrotatetable}

\clearpage
\newpage
\section{Light curves}\label{sec:light}
The light curves were produced as described in Sect. \ref{sec:SEDLC}. Figure \ref{fig:lightccurve} contains the computed light curves for each 1FLT source.
\begin{figure}[!b]
  \centering
\setlength\tabcolsep{0.0pt}
\begin{tabular}{ccc}
  \includegraphics[width=0.35\textwidth]{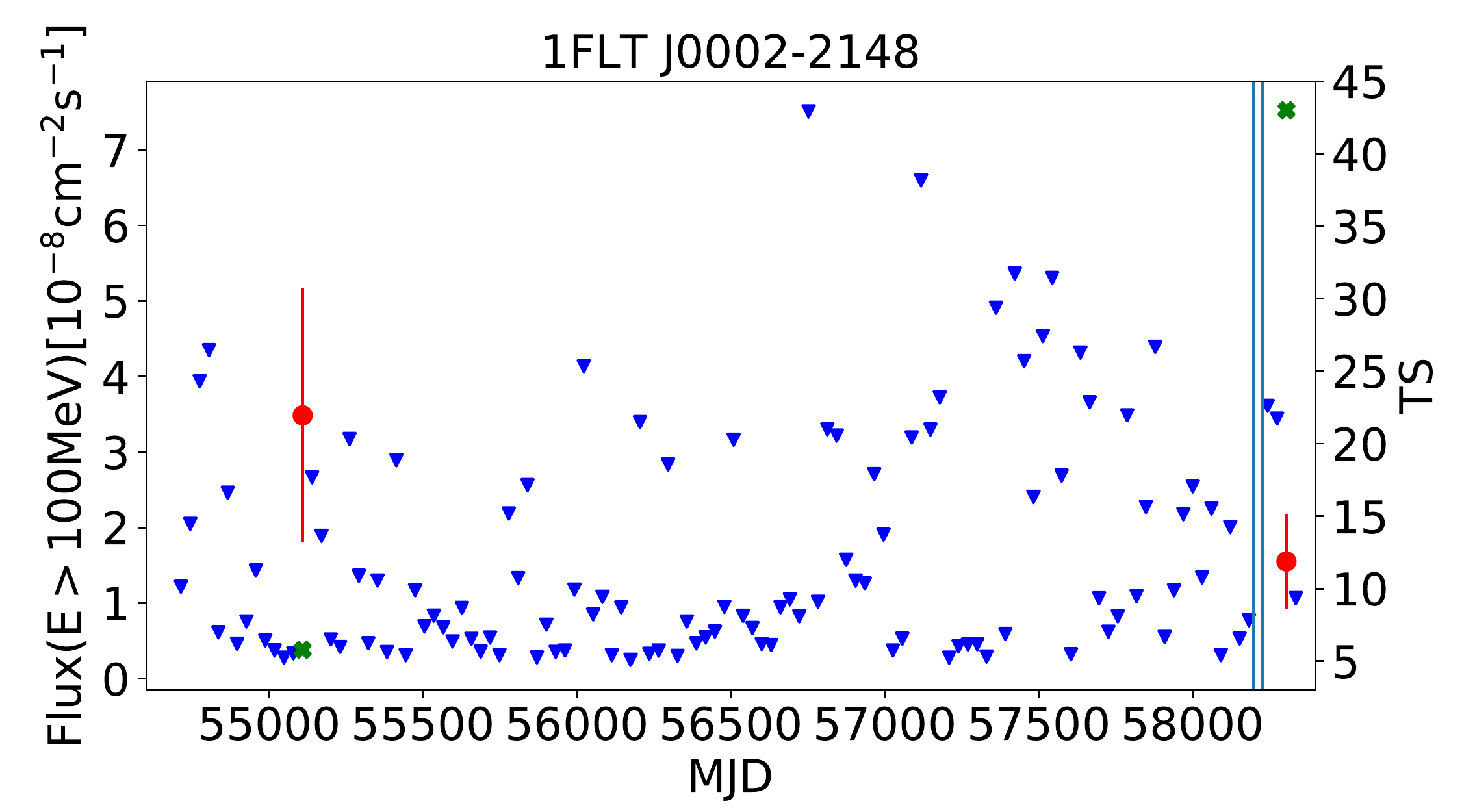}\label{fig:1FLTJ0002-2148}&
  \includegraphics[width=0.35\textwidth]{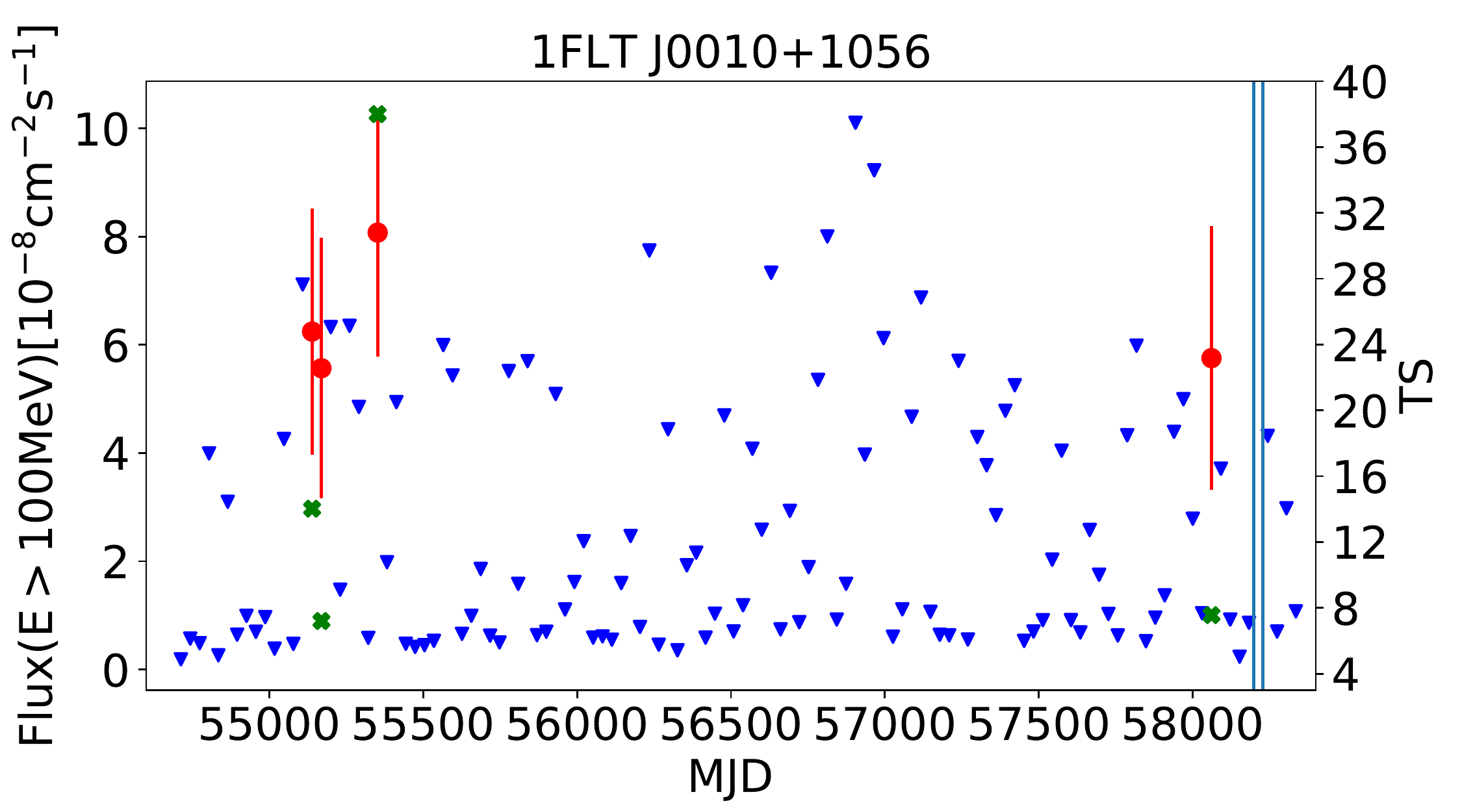}\label{fig:1FLTJ0010+1056}&
  \includegraphics[width=0.35\textwidth]{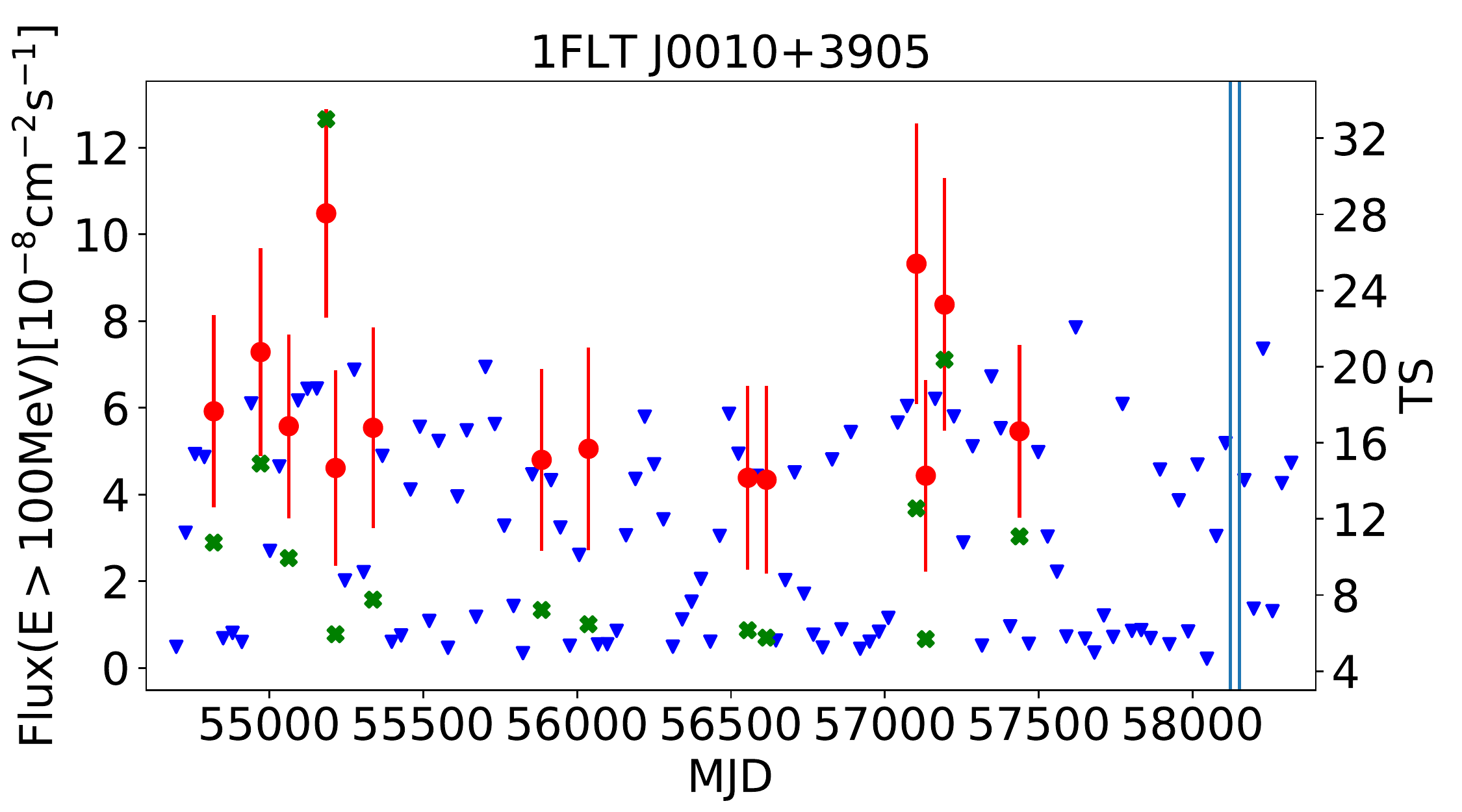}\label{fig:1FLTJ0010+3905}\\
  %[1FLTJ0002-2148]&%[1FLTJ0010+1056]&%[1FLTJ0010+3905]\\
  \includegraphics[width=0.35\textwidth]{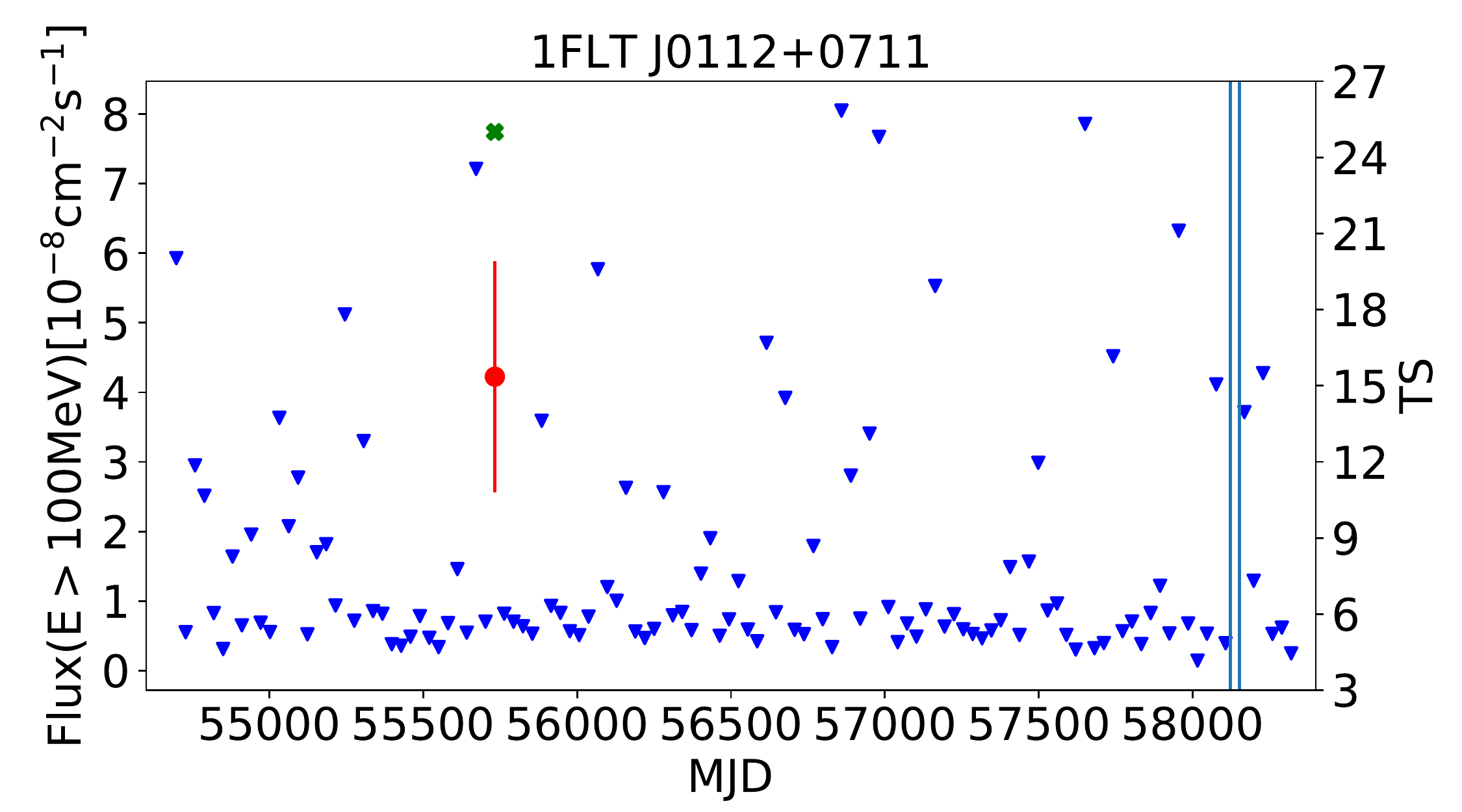}\label{fig:1FLTJ0112+0711}&
  \includegraphics[width=0.35\textwidth]{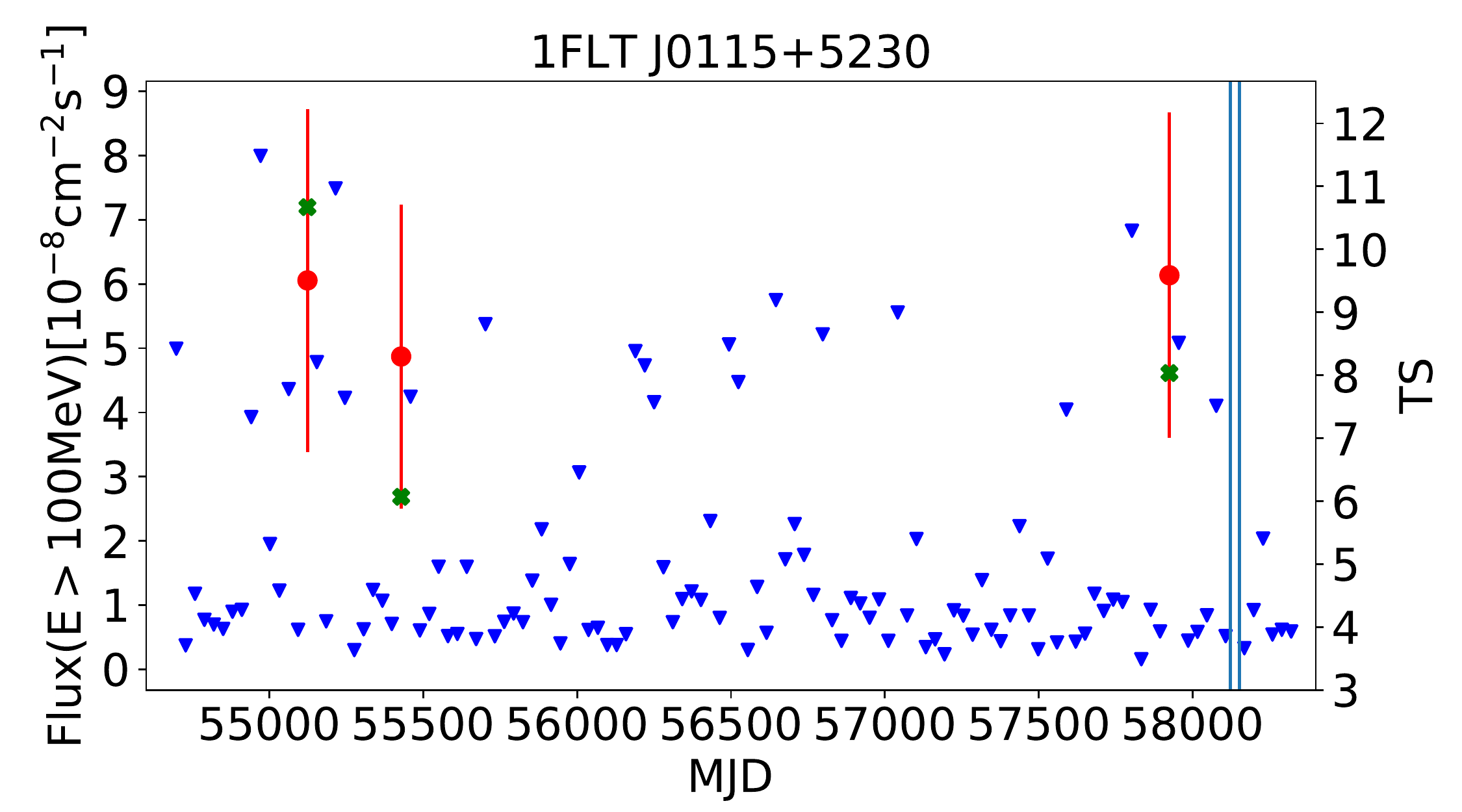}\label{fig:1FLTJ0115+5230}&
  \includegraphics[width=0.35\textwidth]{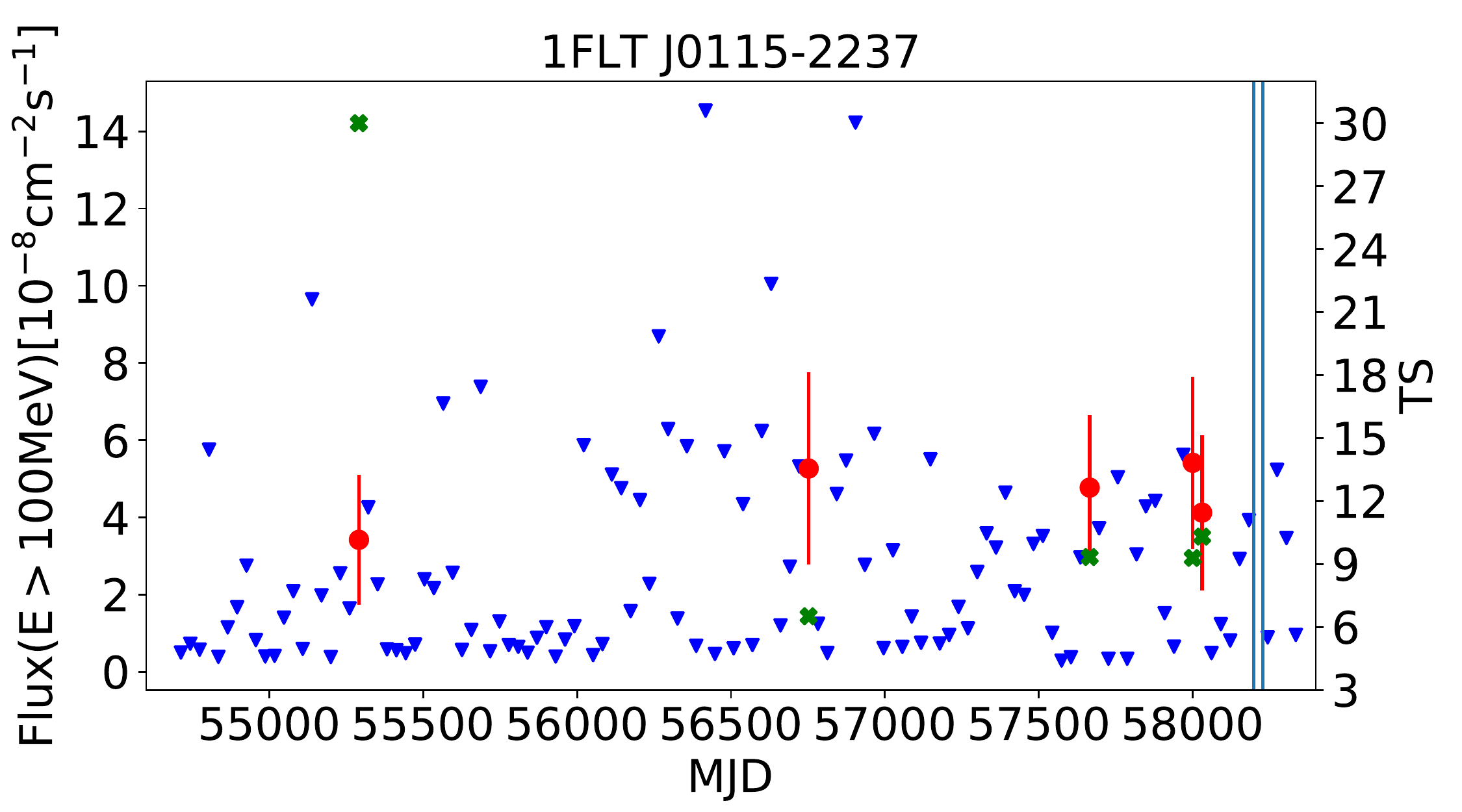}\label{fig:1FLTJ0115-2237}\\
  %[1FLTJ0112+0711]&%[1FLTJ0115+5230]&%[1FLTJ0115-2237]\\
  \includegraphics[width=0.35\textwidth]{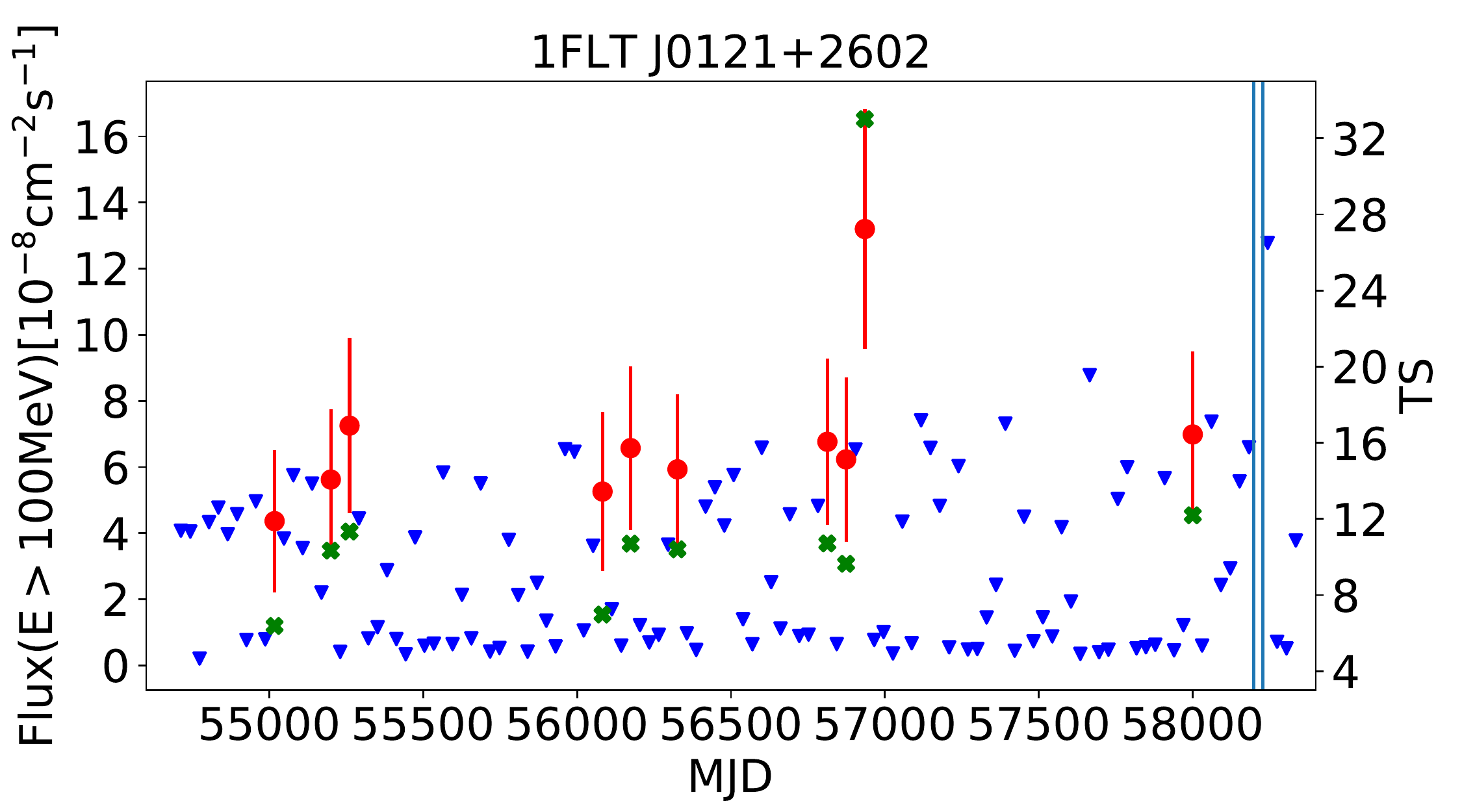}\label{fig:1FLTJ0121+2602}&
  \includegraphics[width=0.35\textwidth]{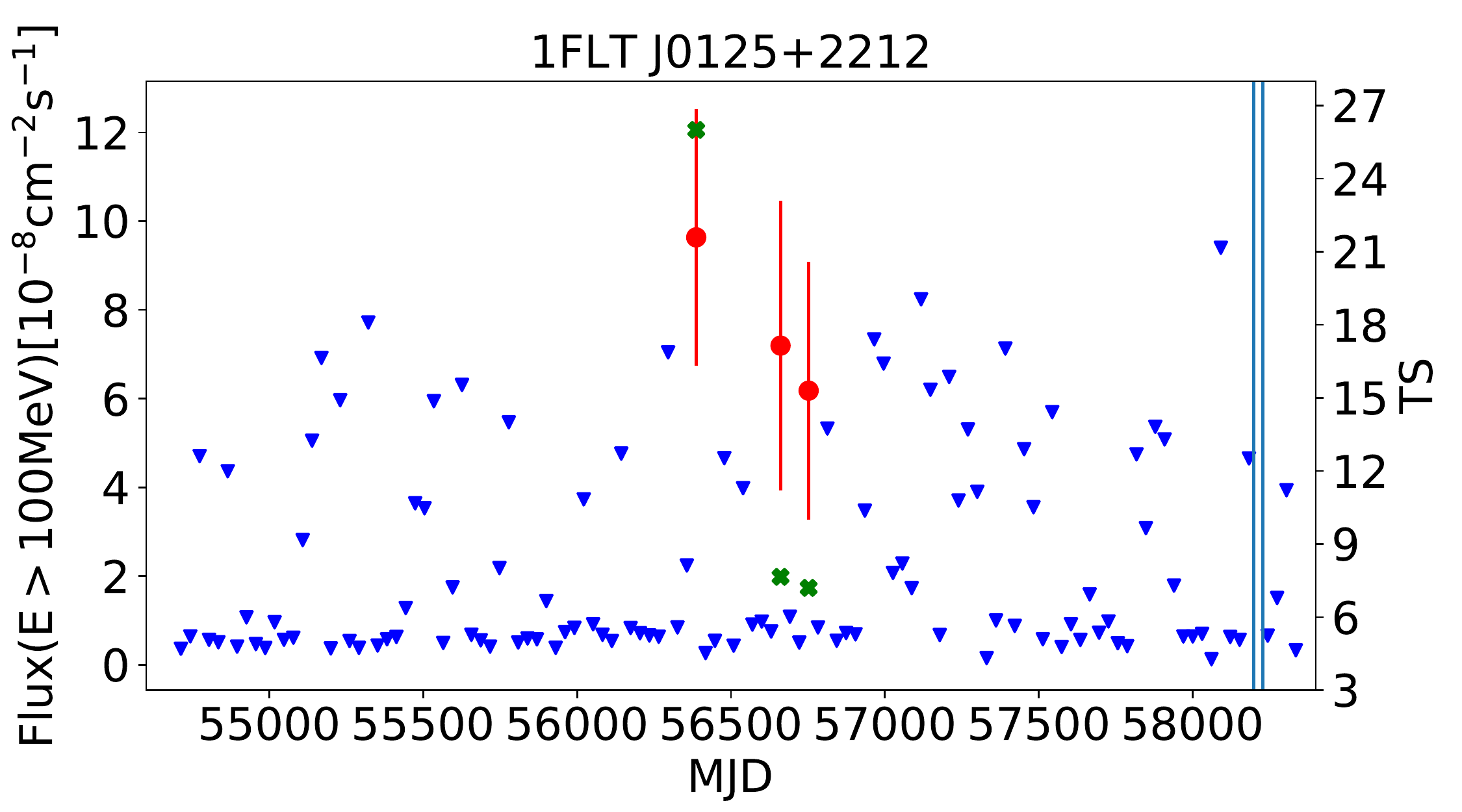}\label{fig:1FLTJ0125+2212}&
  \includegraphics[width=0.35\textwidth]{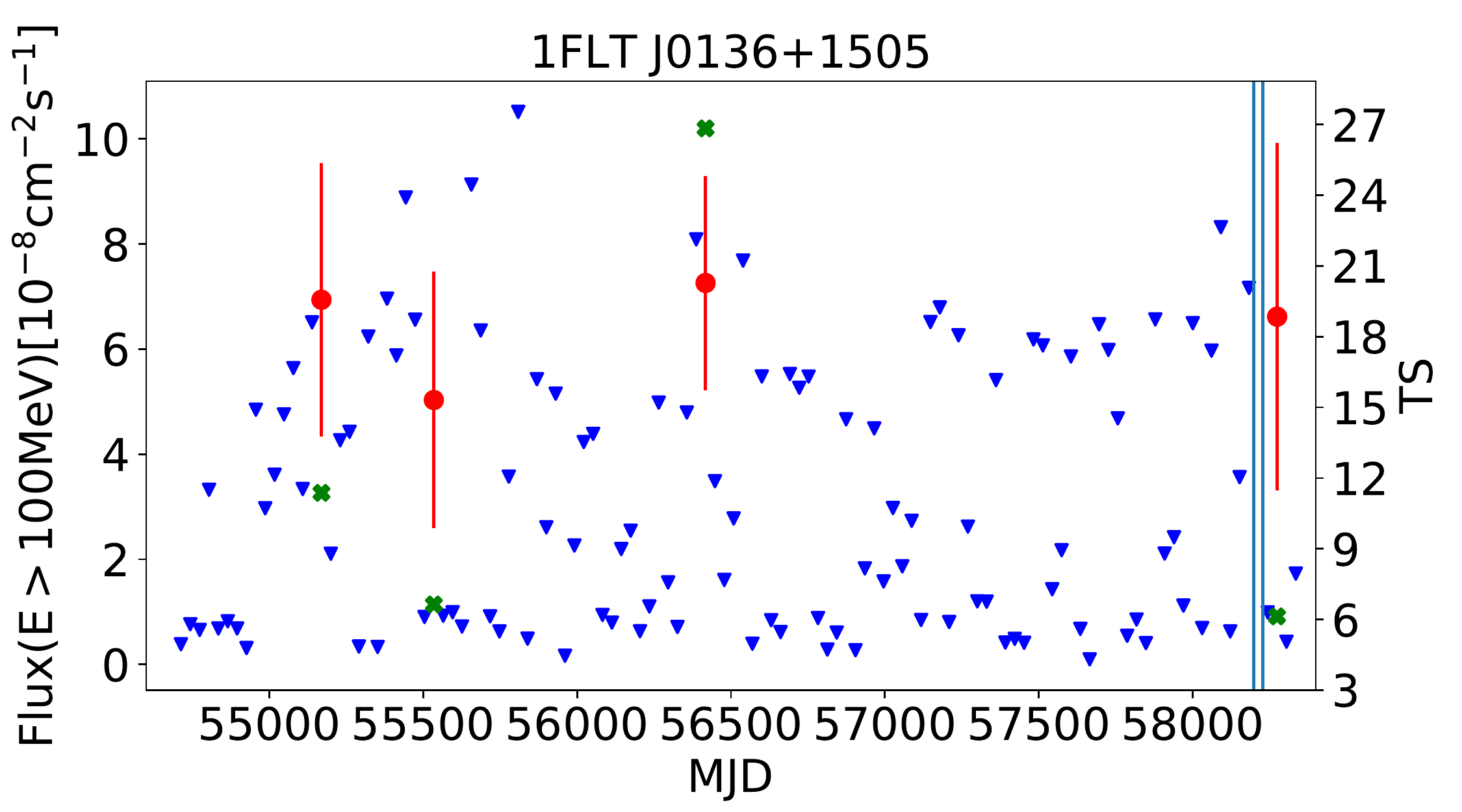}\label{fig:1FLTJ0136+1505}\\
  %[1FLTJ0121+2602]&%[1FLTJ0125+2212]&%[1FLTJ0136+1505]\\
  \includegraphics[width=0.35\textwidth]{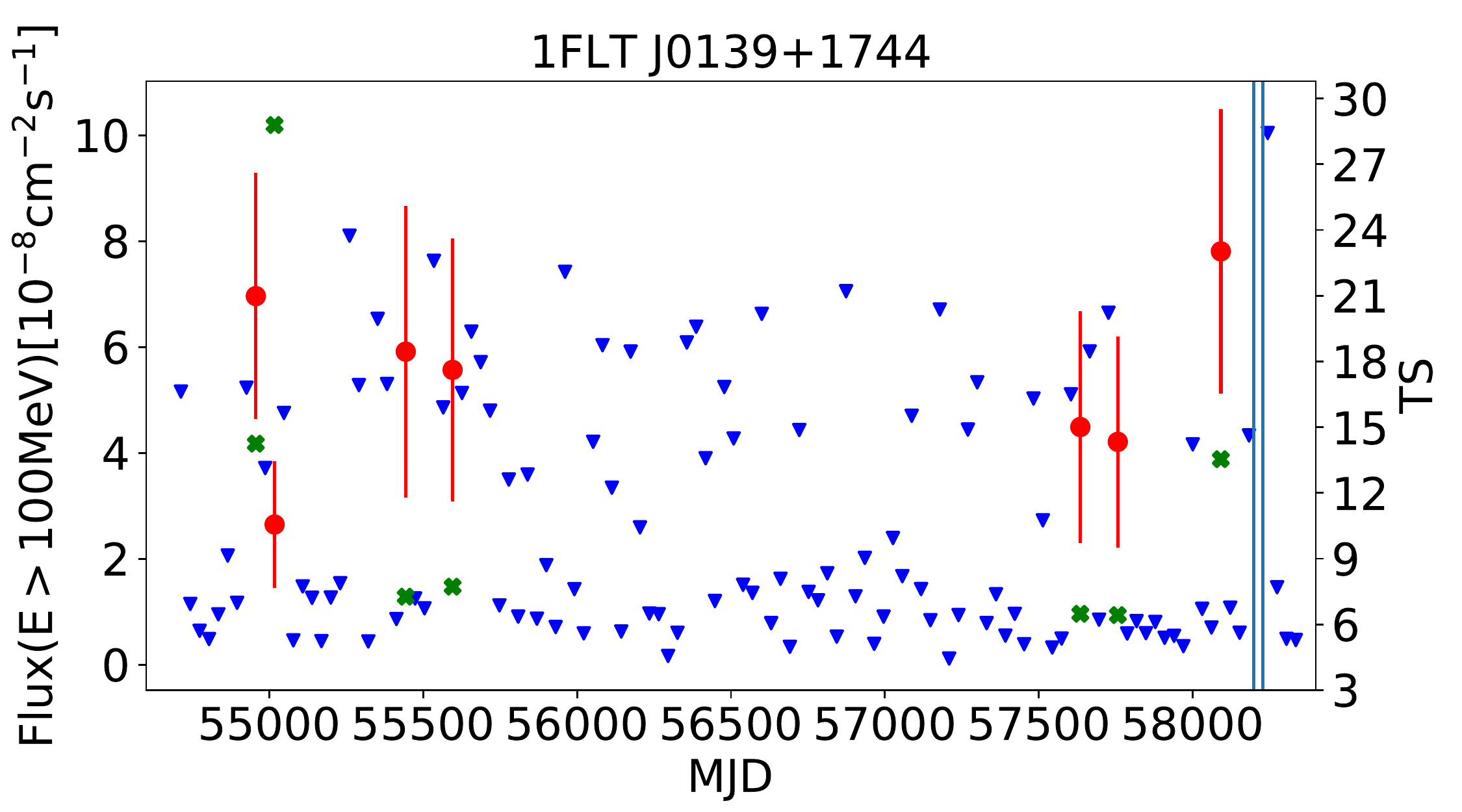}\label{fig:1FLTJ0139+1744}&
  \includegraphics[width=0.35\textwidth]{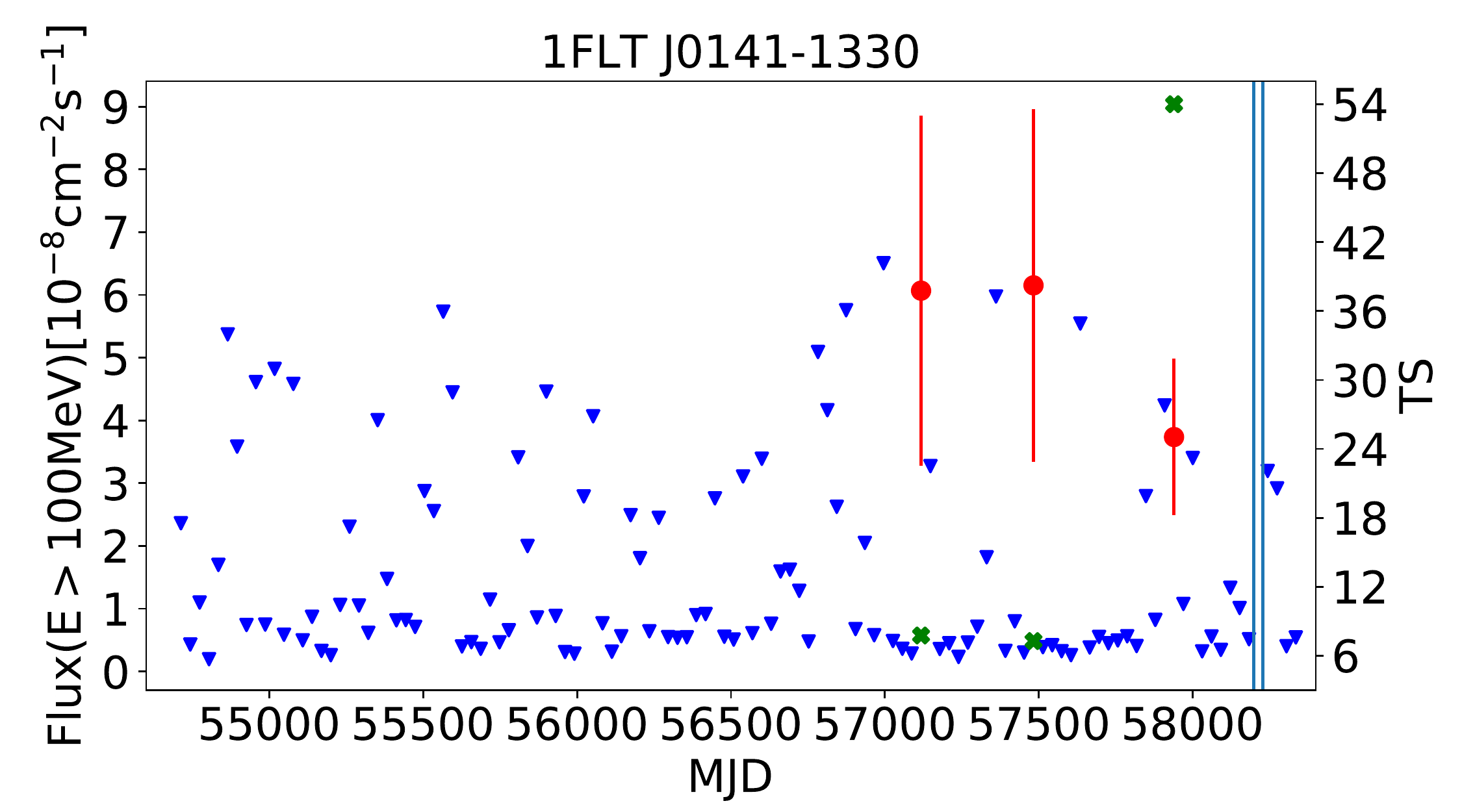}\label{fig:1FLTJ0141-1330}&
  \includegraphics[width=0.35\textwidth]{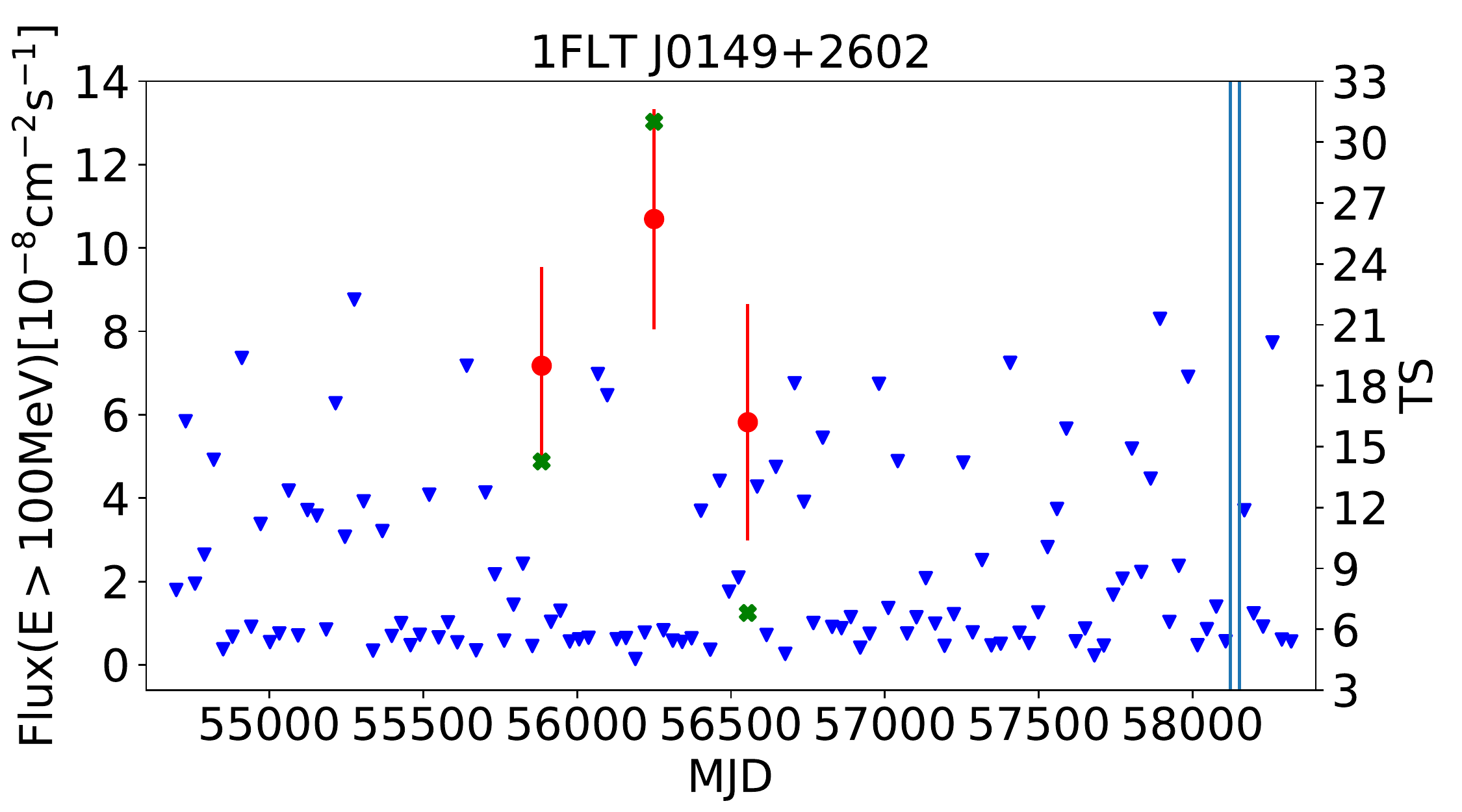}\label{fig:1FLTJ0149+2602}\\
  %[1FLTJ0139+1744]&%[1FLTJ0141-1330]&%[1FLTJ0149+2602]\\
  \includegraphics[width=0.35\textwidth]{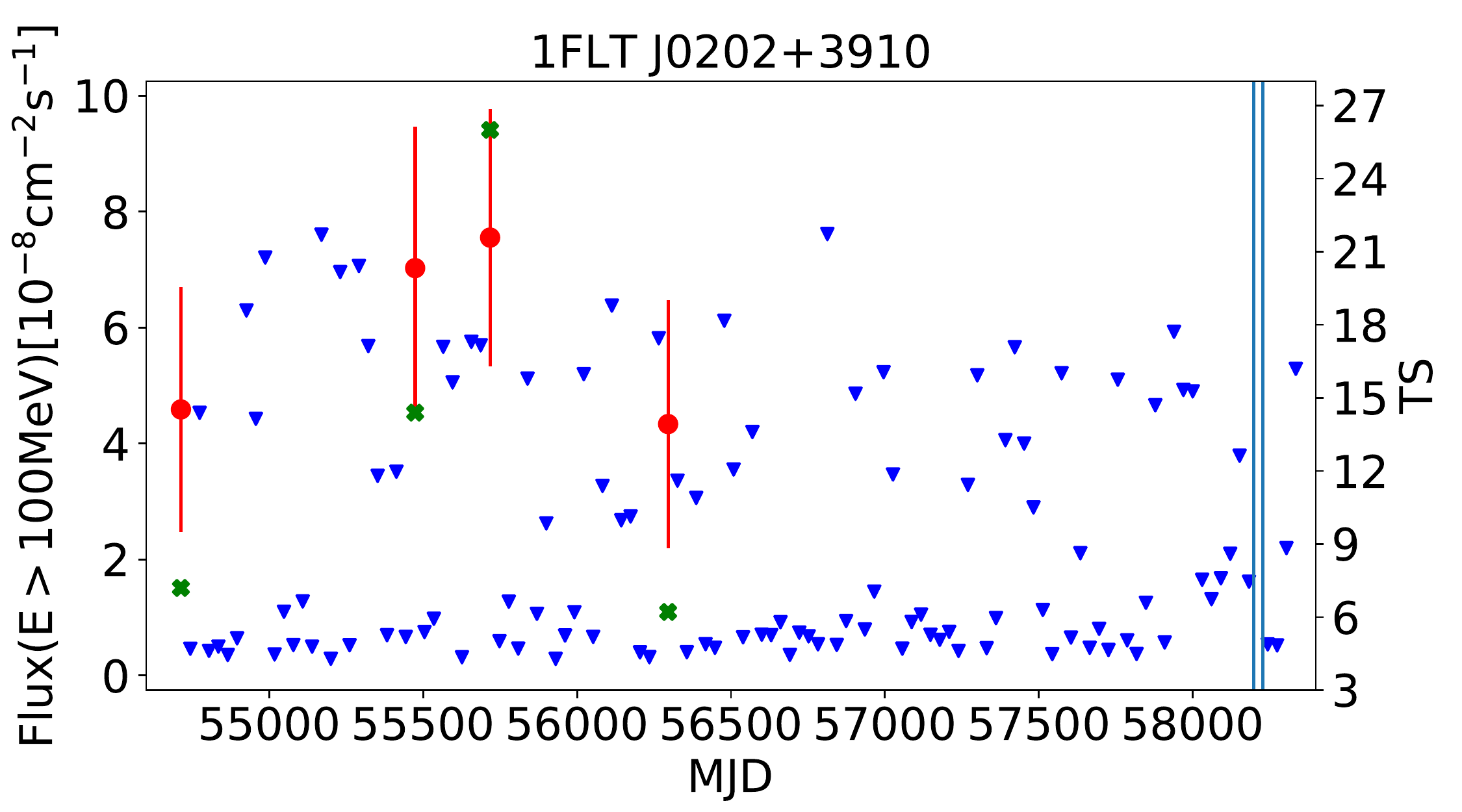}\label{fig:1FLTJ0202+3910}&
  \includegraphics[width=0.35\textwidth]{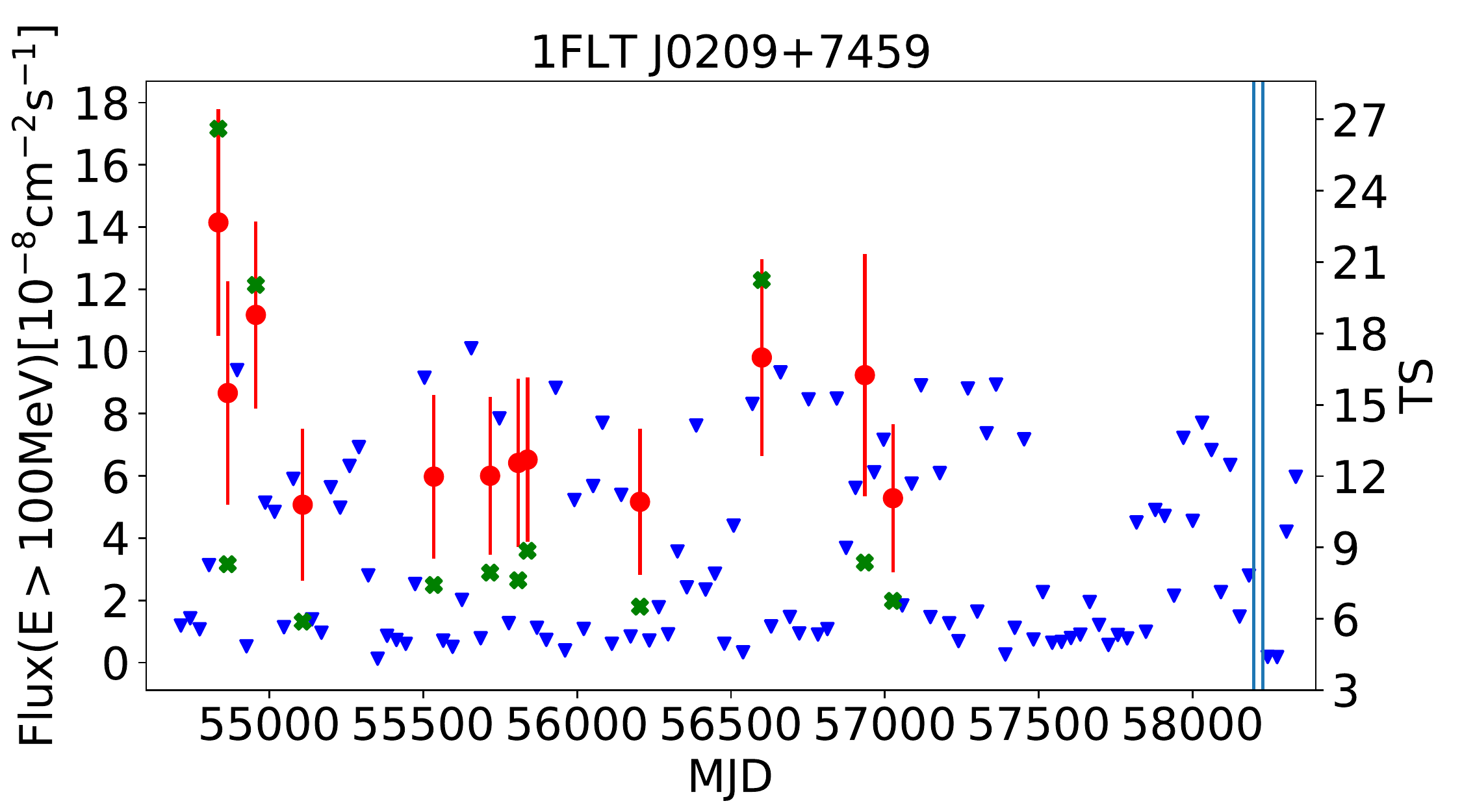}\label{fig:1FLTJ0209+7459}&
  \includegraphics[width=0.35\textwidth]{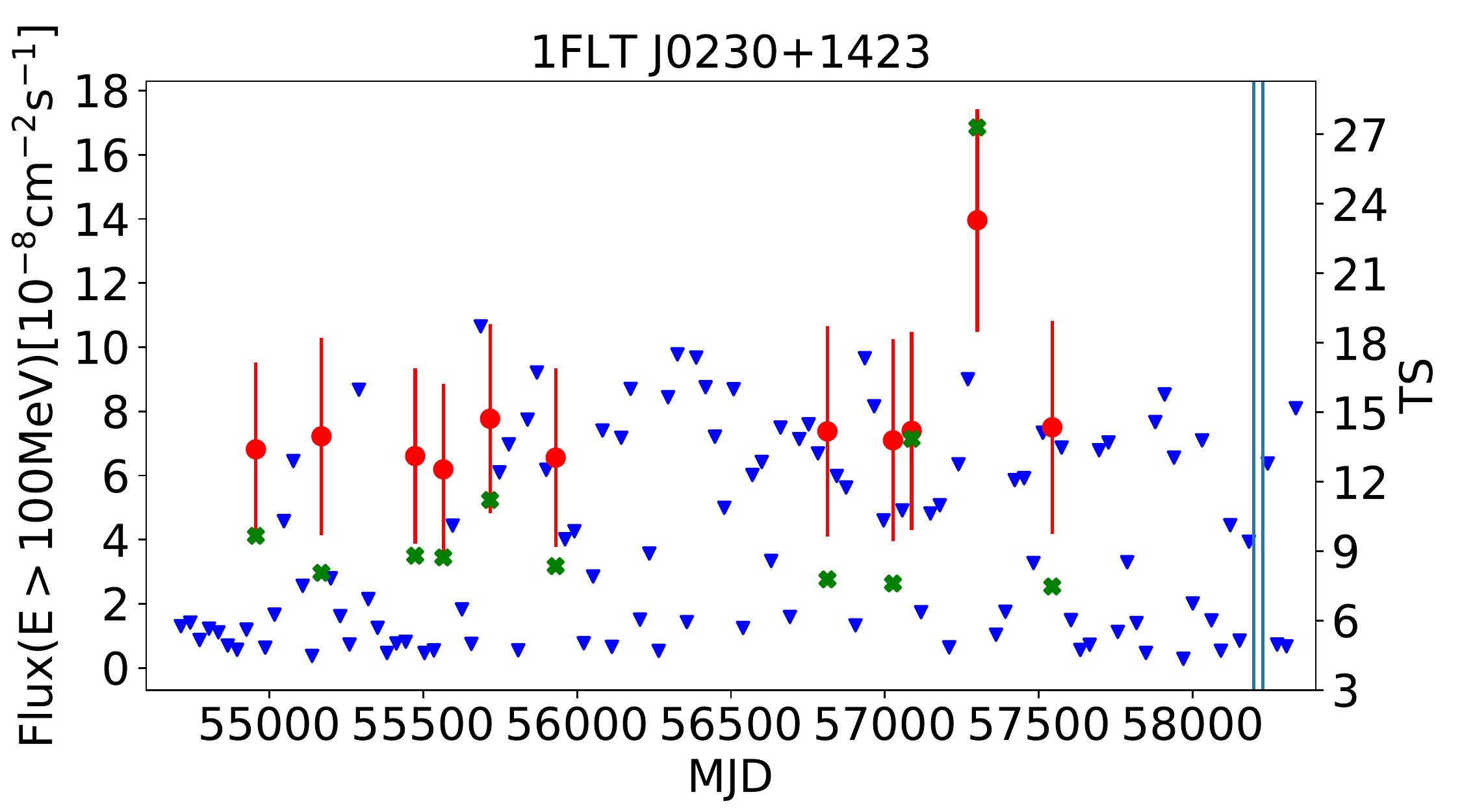}\label{fig:1FLTJ0230+1423}\\
  %[1FLTJ0202+3910]&%[1FLTJ0209+7459]&%[1FLTJ0230+1423]\\
\end{tabular}  
      \caption{%
   	 Light curves: red points with error bars represent monthly fluxes with the associated error, blue inverted triangles represent $2\sigma$ upper limits (when TS $<$ 4 or $\mathrm{\Delta}$Flux/Flux > 0.5 or N$\mathrm{_{pred}<}$ 3). The green crosses are the value of TS in correspondence of the flux point. The two light blue lines highlight TBIN 115 (nominal and shifted) which corresponds to a gap in the \fermilat data during the “safe hold” mode in March 2018 when the instrument was powered off. Light curves of sources in the LMC and Cen~A regions are not reported, except for 1FLT J0512$-$7007.}
\end{figure}\label{fig:lightccurve}
\begin{figure}[!t]%
      \centering      
      \ContinuedFloat% 
\setlength\tabcolsep{0.0pt}
\begin{tabular}{ccc}  
  \includegraphics[width=0.35\textwidth]{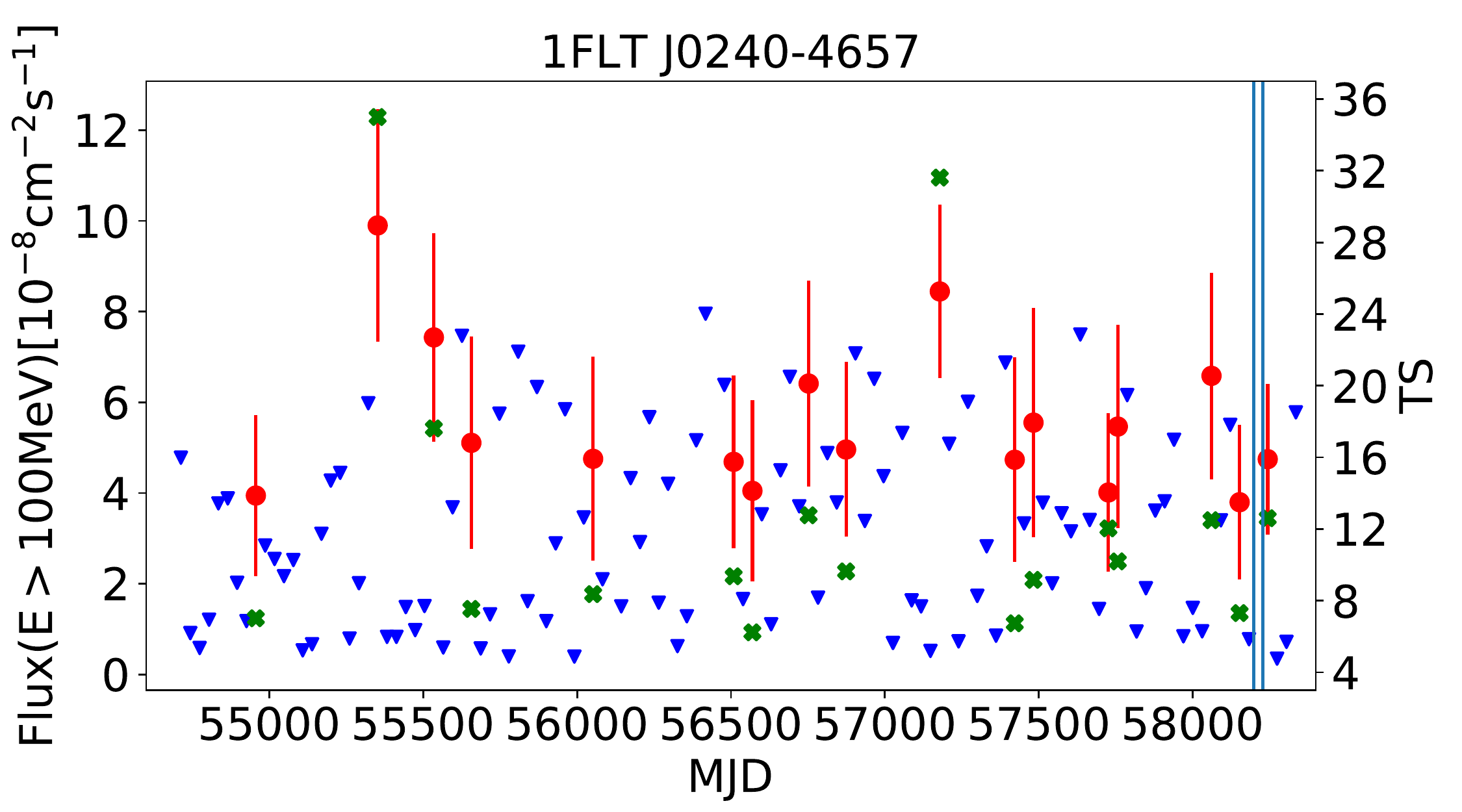}\label{fig:1FLTJ0240-4657}&
  \includegraphics[width=0.35\textwidth]{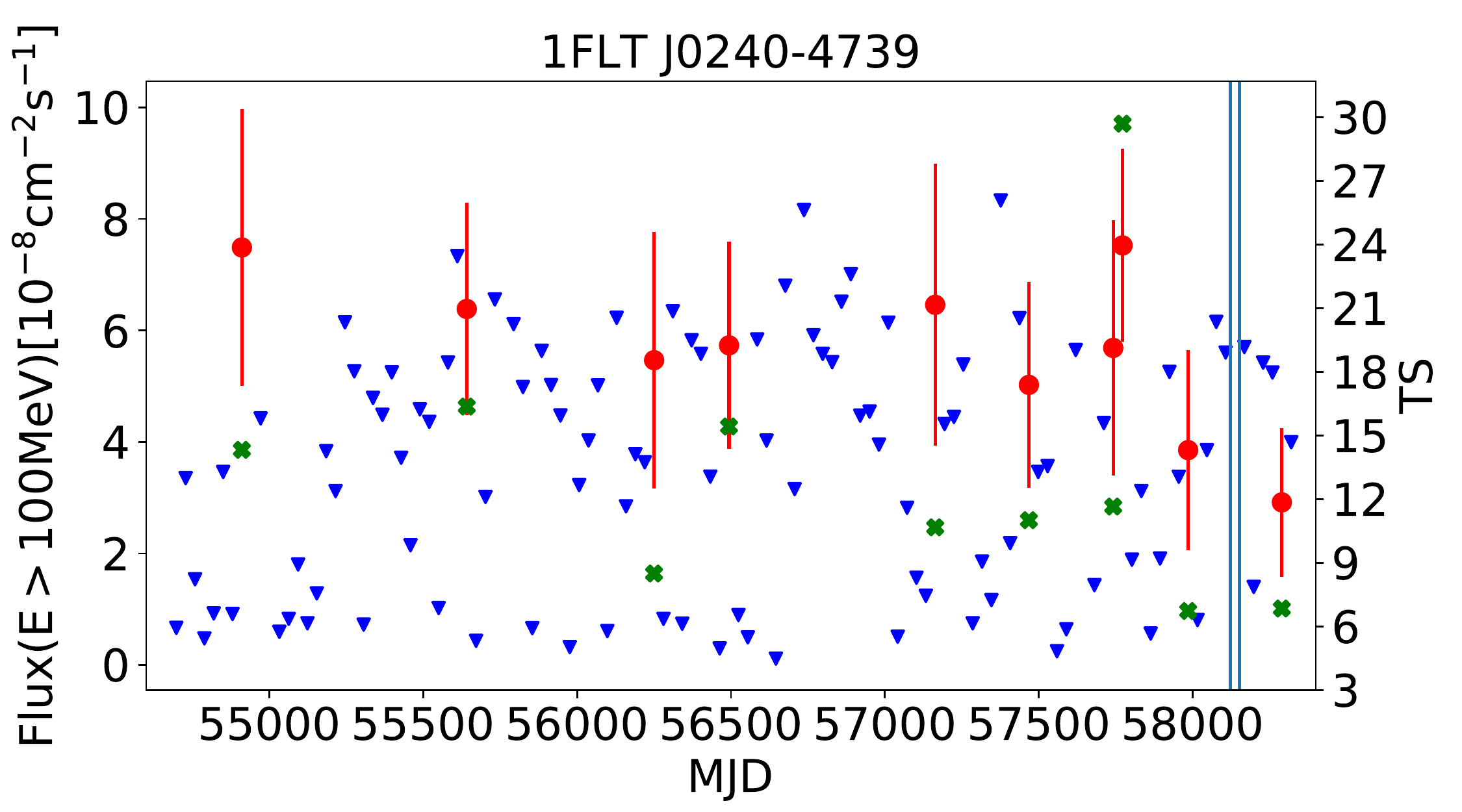}\label{fig:1FLTJ0240-4739}&
  \includegraphics[width=0.35\textwidth]{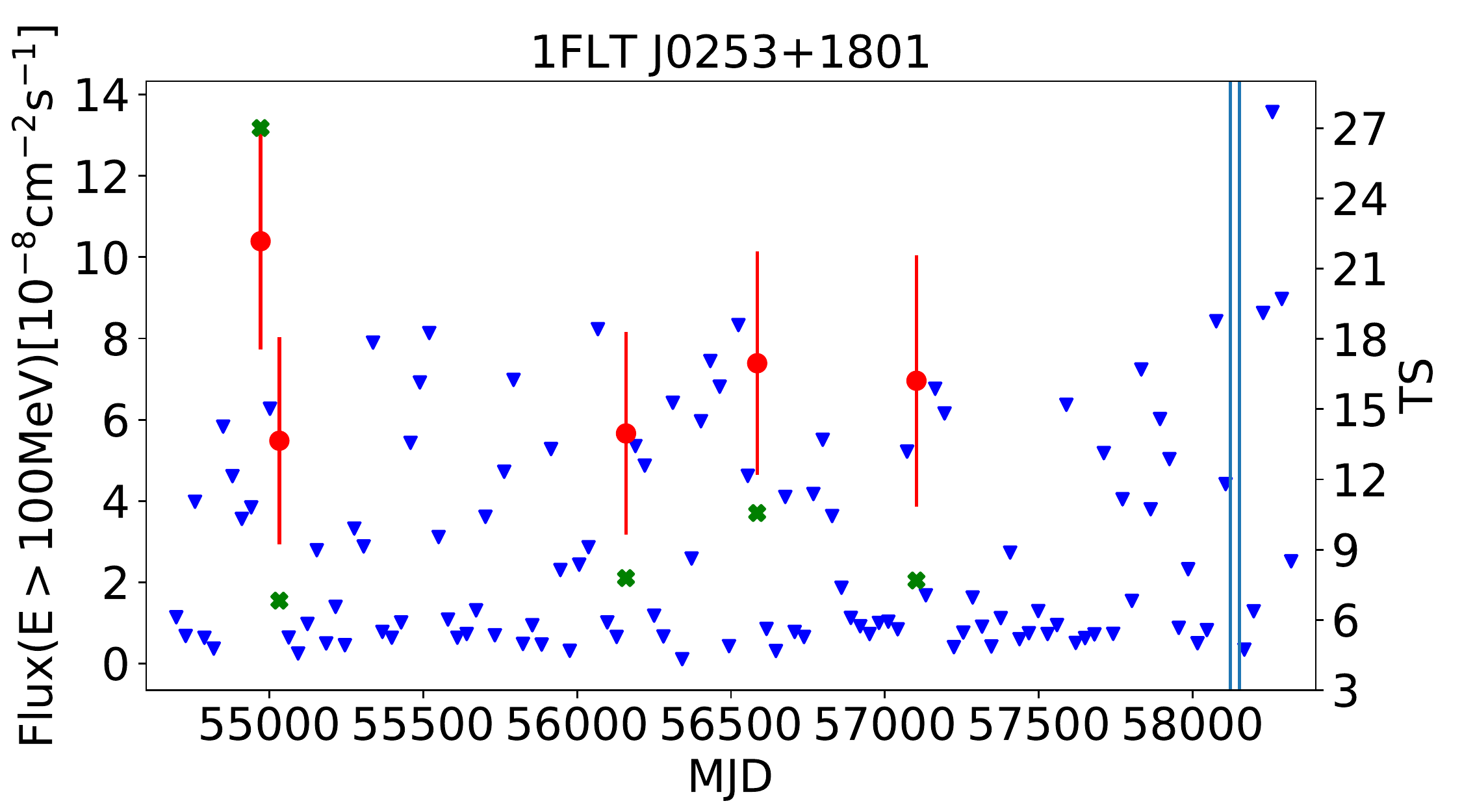}\label{fig:1FLTJ0253+1801}\\
  %[1FLTJ0240-4657]&%[1FLTJ0240-4739]&%[1FLTJ0253+1801]
  \includegraphics[width=0.35\textwidth]{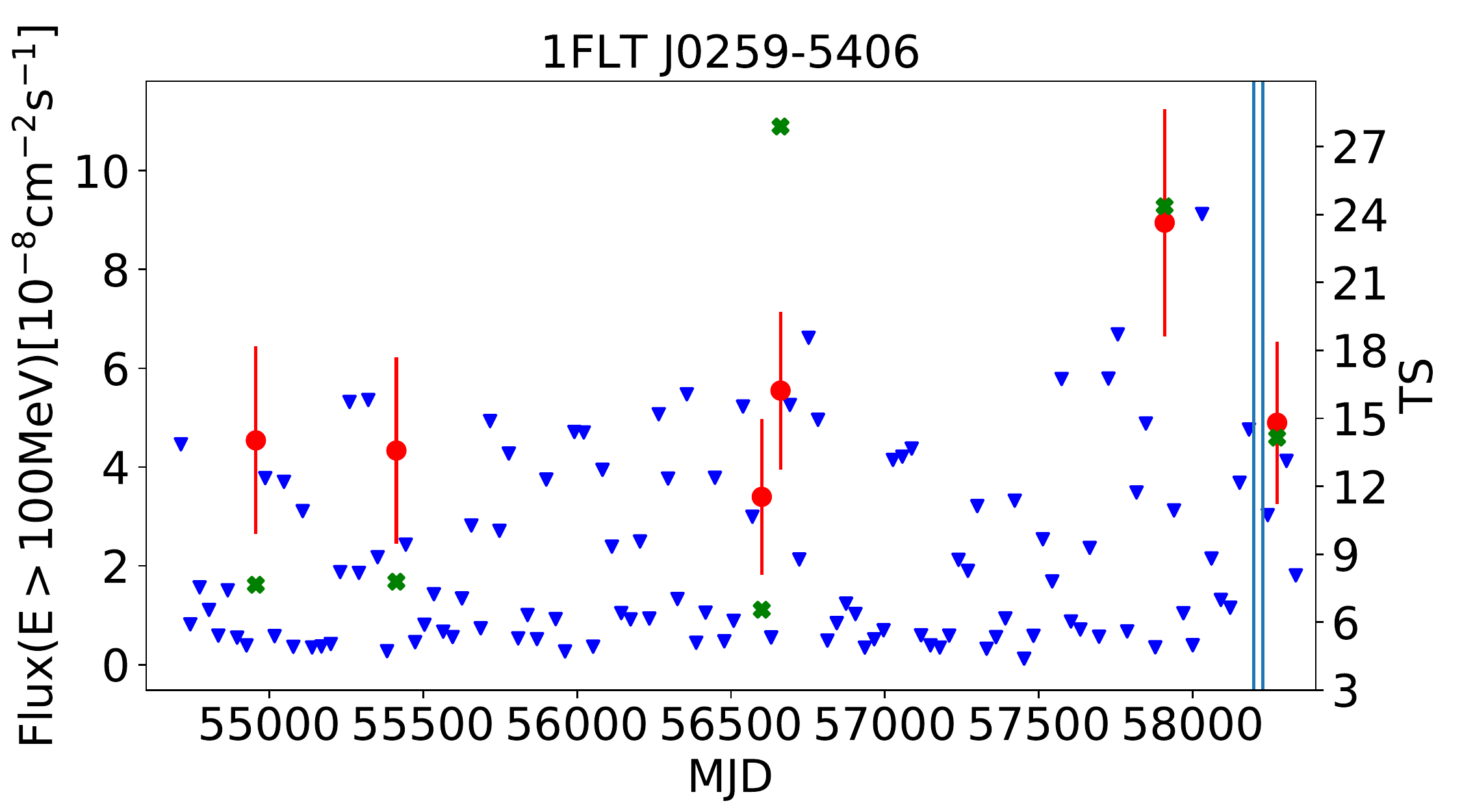}\label{fig:1FLTJ0259-5406}&
  \includegraphics[width=0.35\textwidth]{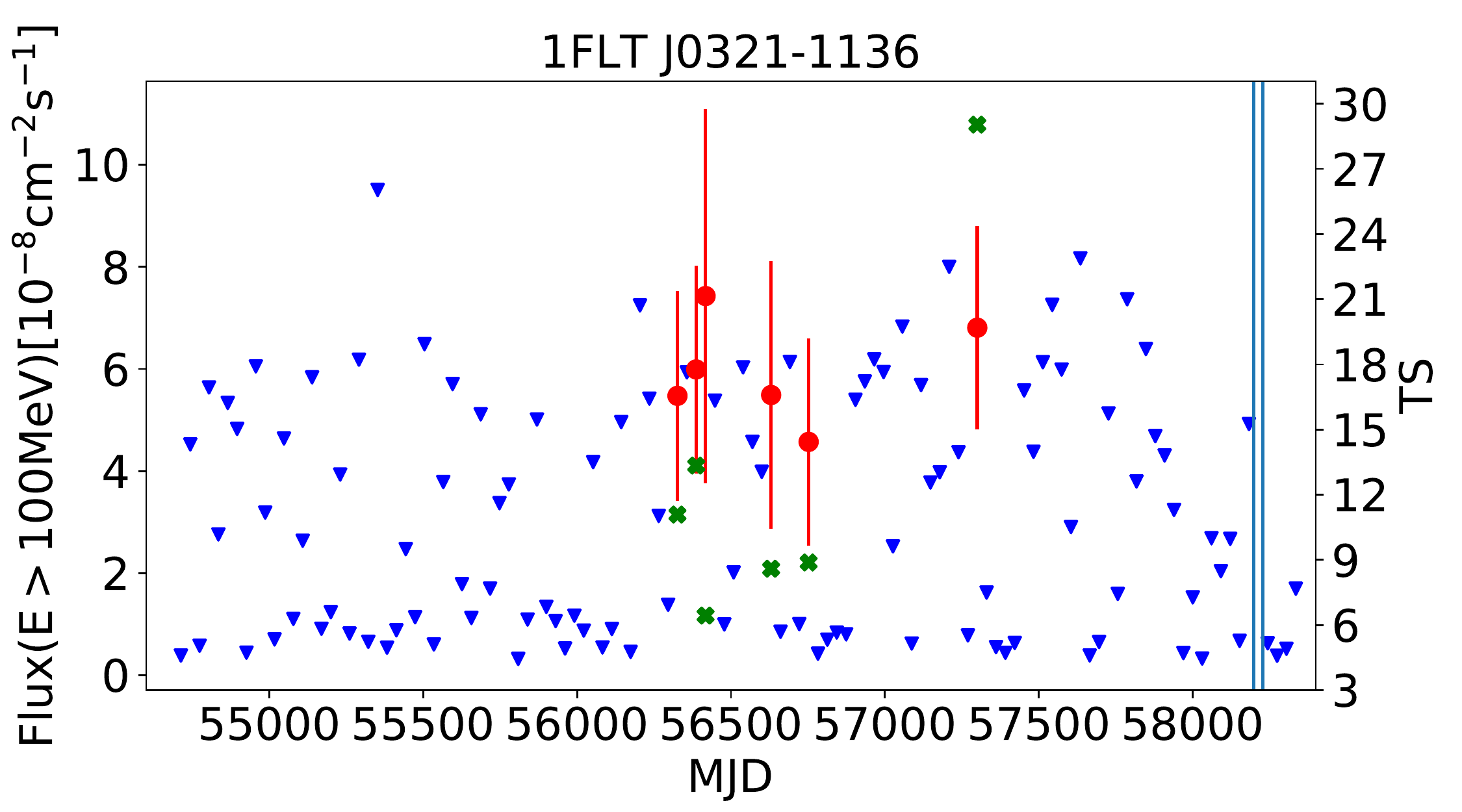}\label{fig:1FLTJ0321-1136}&
  \includegraphics[width=0.35\textwidth]{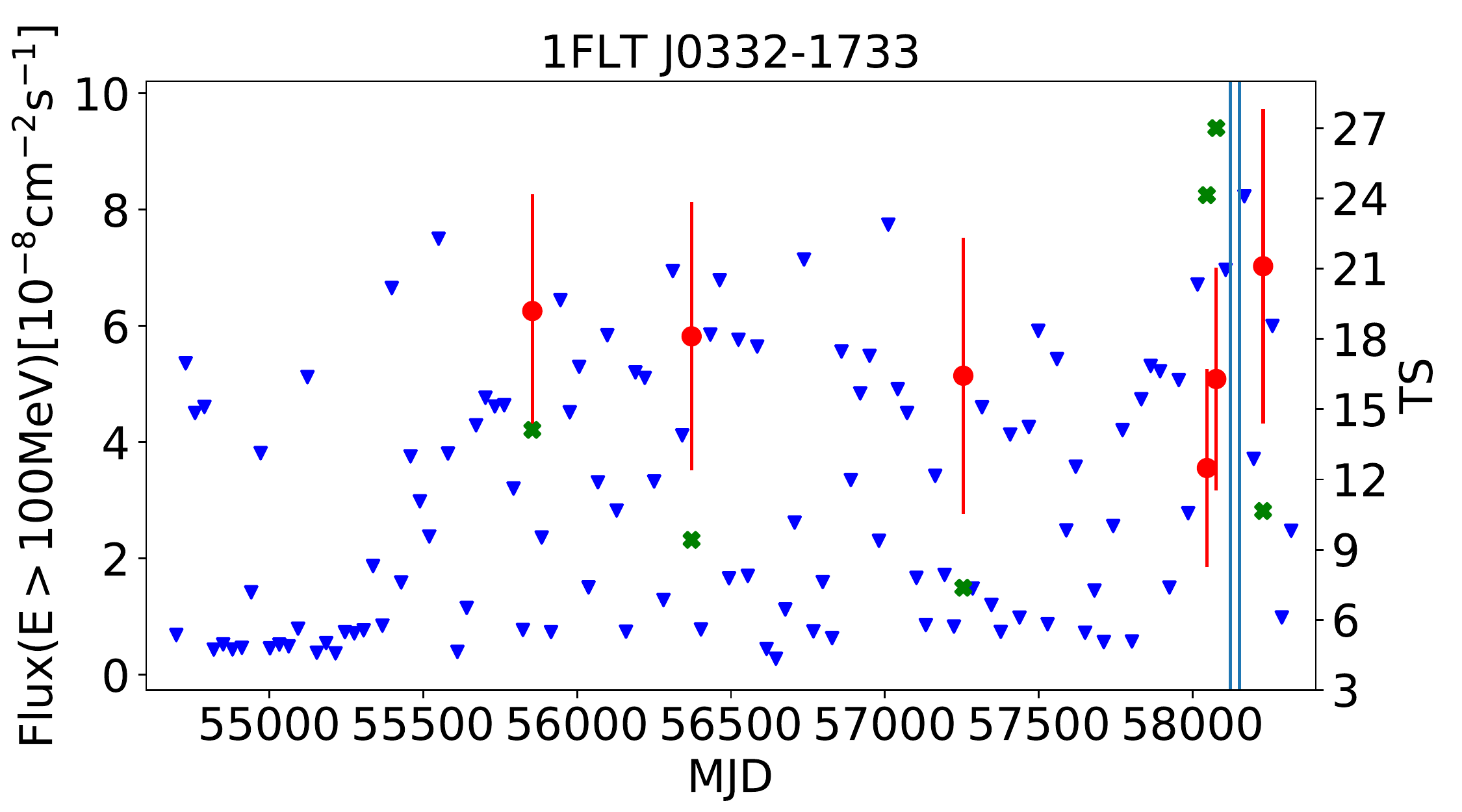}\label{fig:1FLTJ0332-1733}\\
  %[1FLTJ0259-5406]&%[1FLTJ0321-1136]&%[1FLTJ0332-1733]\\
  \includegraphics[width=0.35\textwidth]{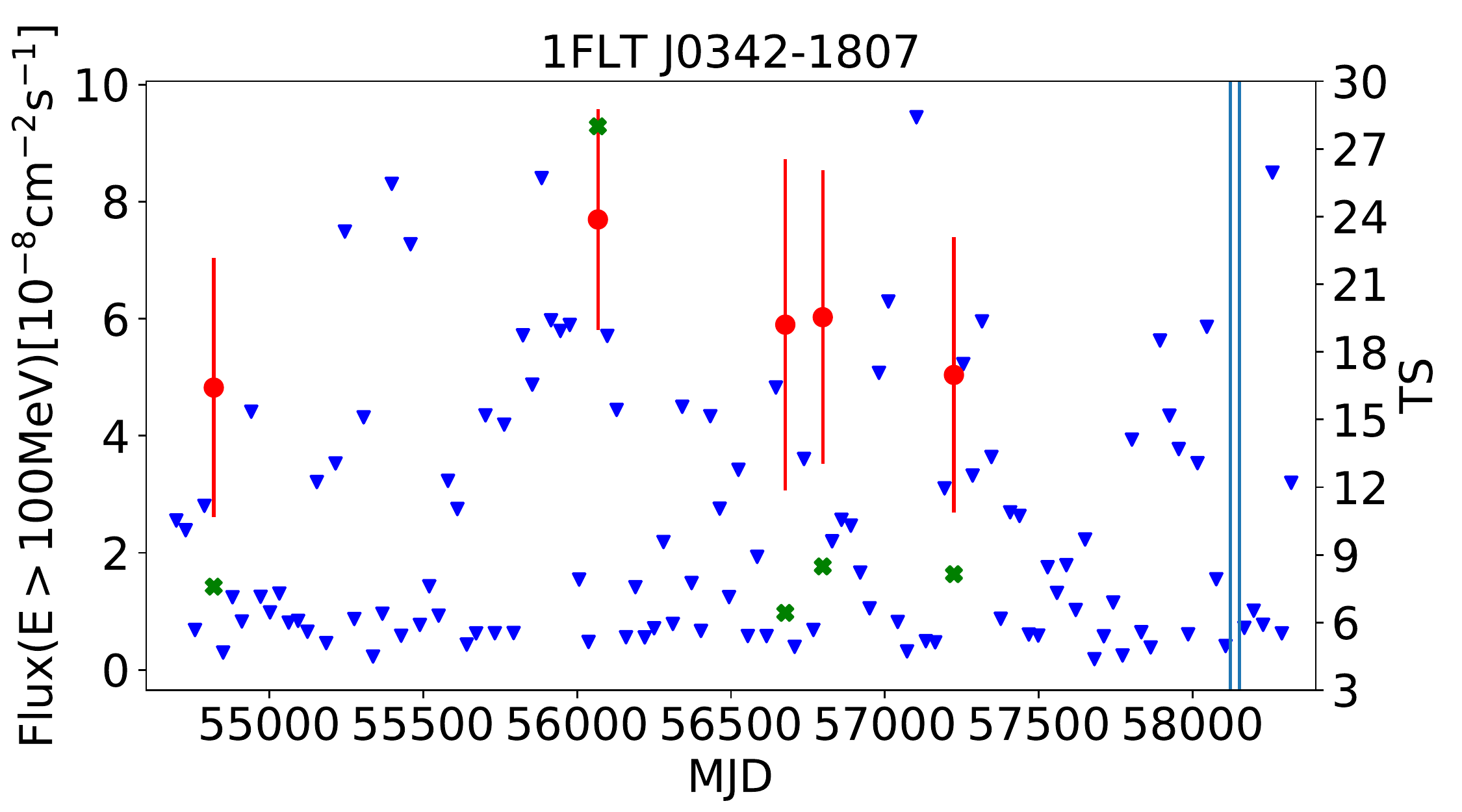}\label{fig:1FLTJ0342-1807}&
  \includegraphics[width=0.35\textwidth]{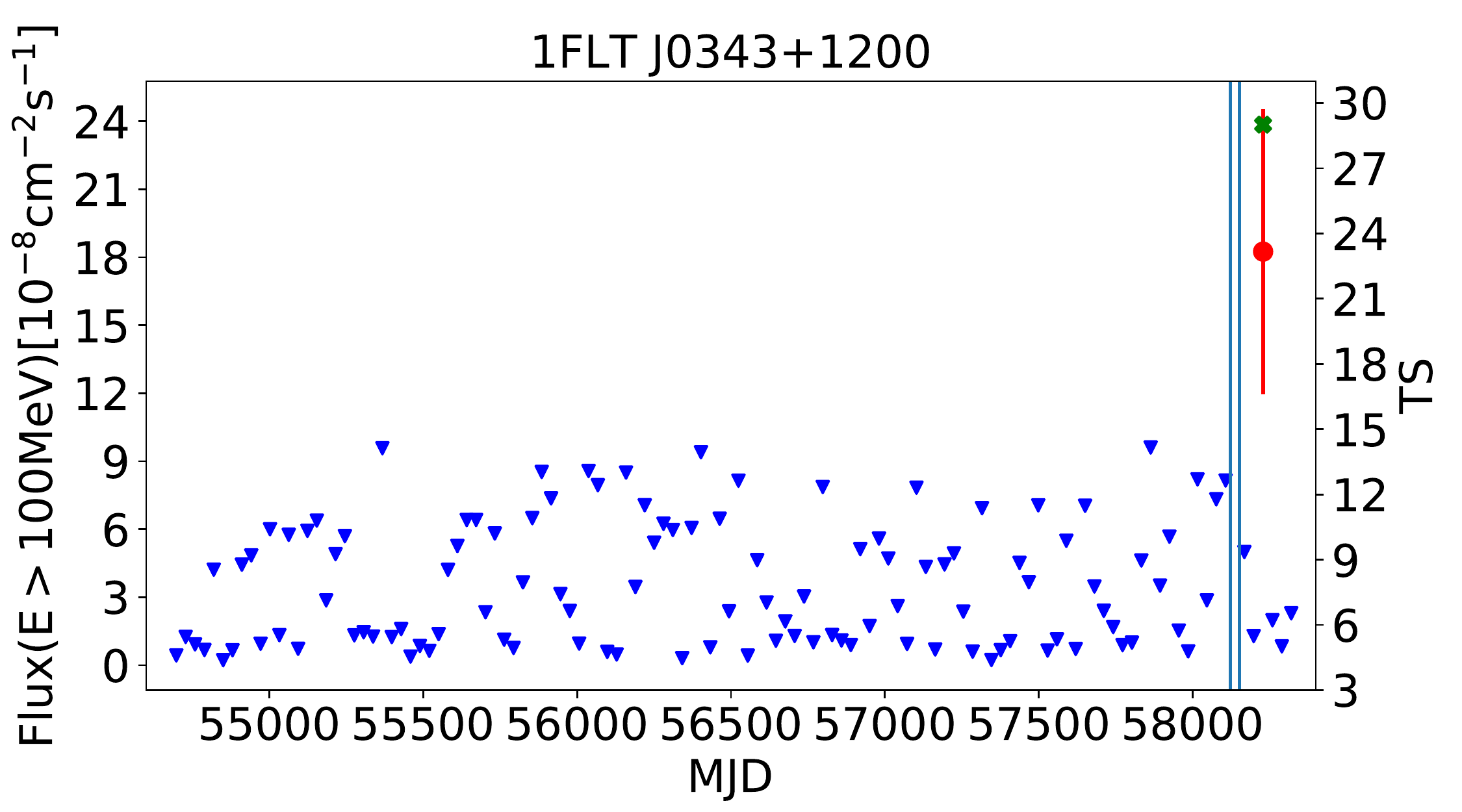}\label{fig:1FLTJ0343+1200}&
  \includegraphics[width=0.35\textwidth]{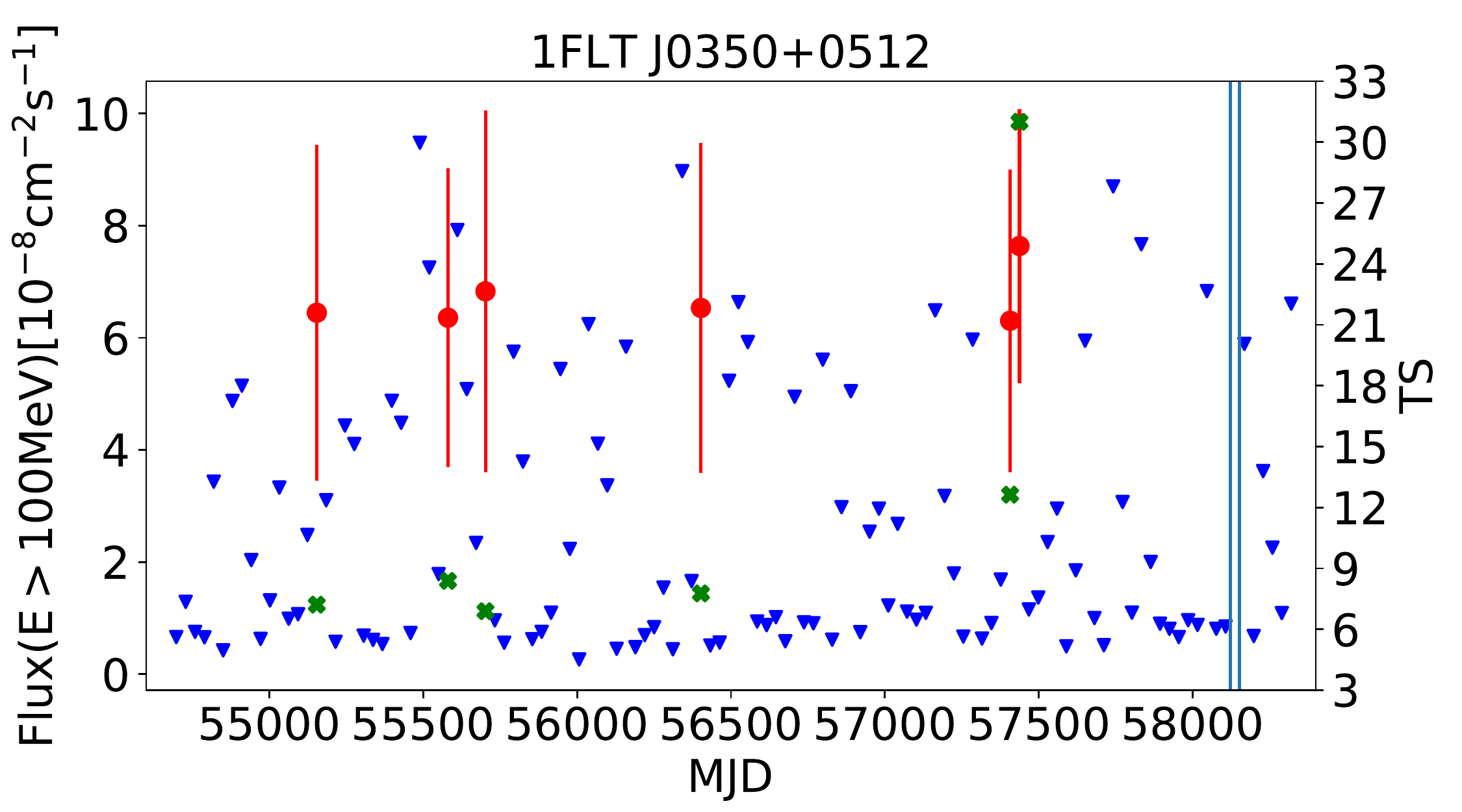}\label{fig:1FLTJ0350+0512}\\
  %[1FLTJ0342-1807]&%[1FLTJ0343+1200]&%[1FLTJ0350+0512]\\
  \includegraphics[width=0.35\textwidth]{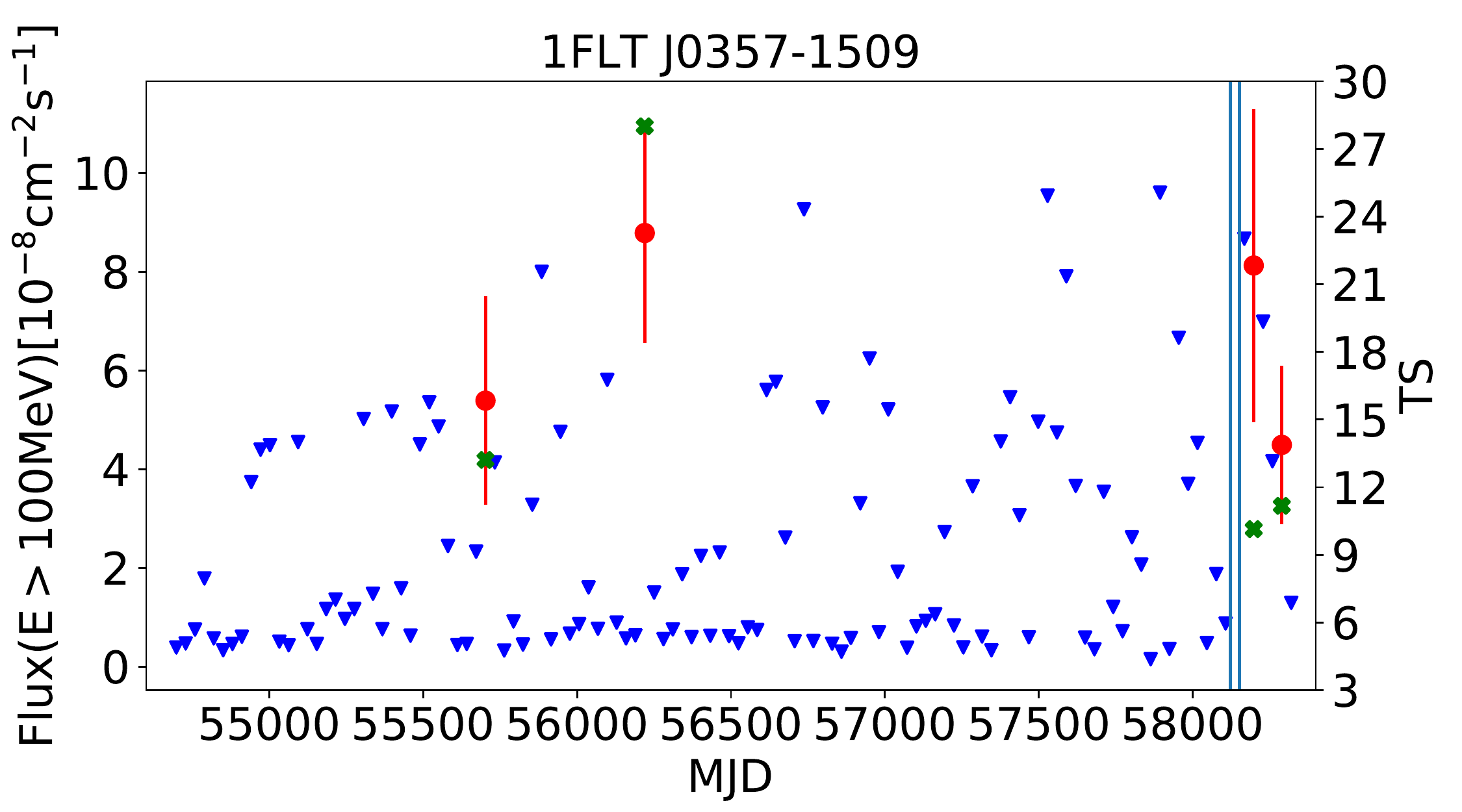}\label{fig:1FLTJ0357-1509}&
  \includegraphics[width=0.35\textwidth]{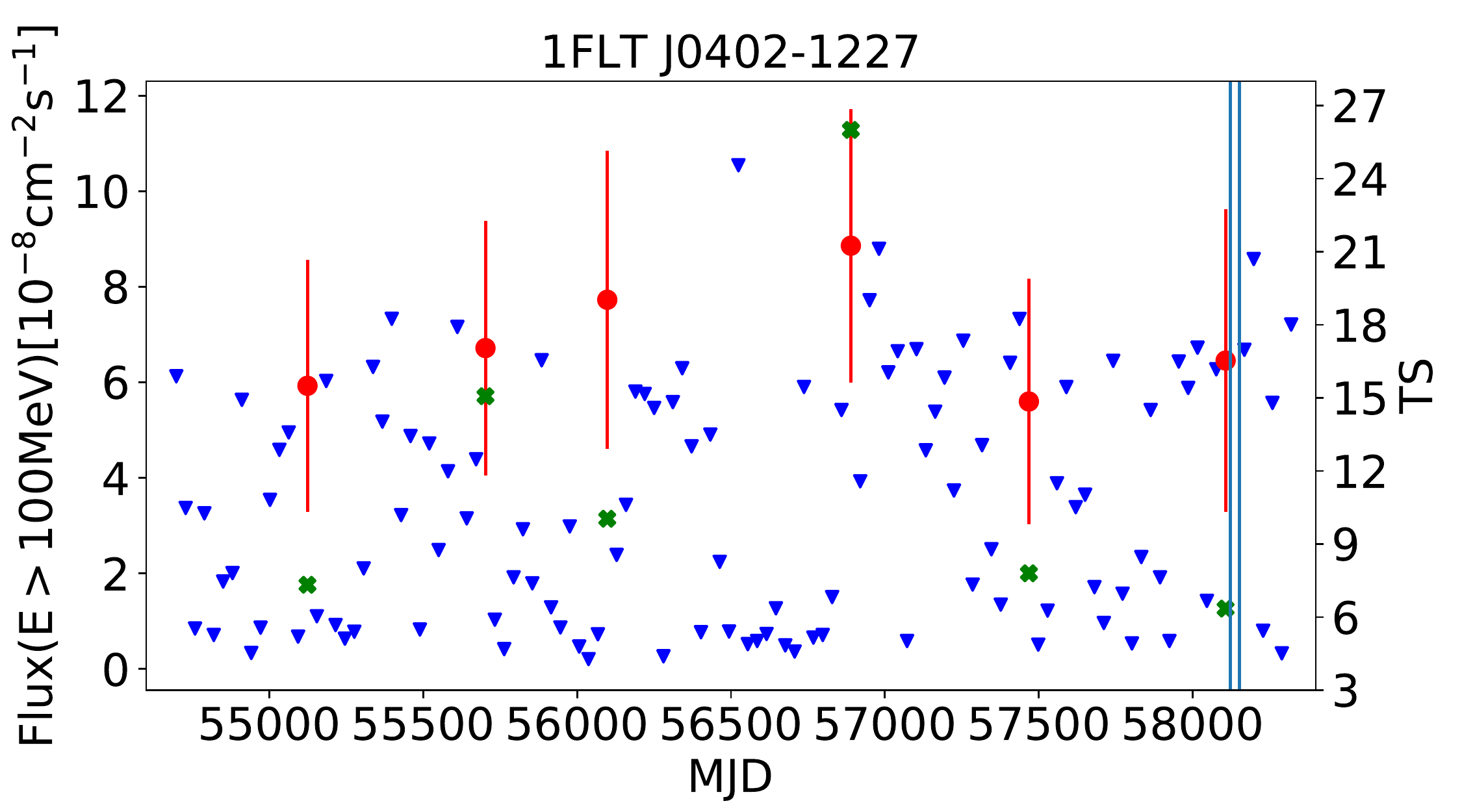}\label{fig:1FLTJ0402-1227}&
  \includegraphics[width=0.35\textwidth]{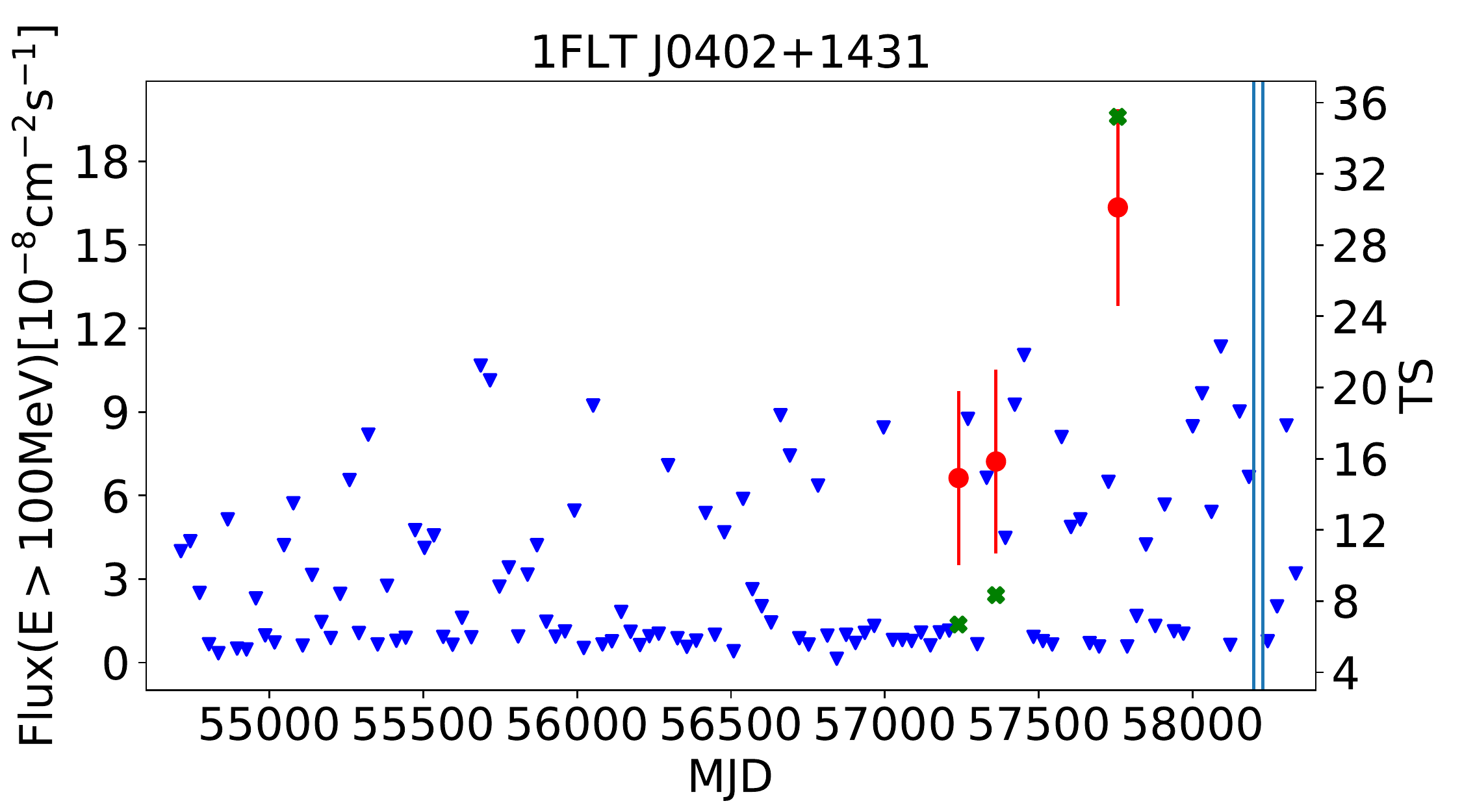}\label{fig:1FLTJ0402+1431}\\
  %[1FLTJ0357-1509]&%[1FLTJ0402-1227]&%[1FLTJ0402+1431]\\
  \includegraphics[width=0.35\textwidth]{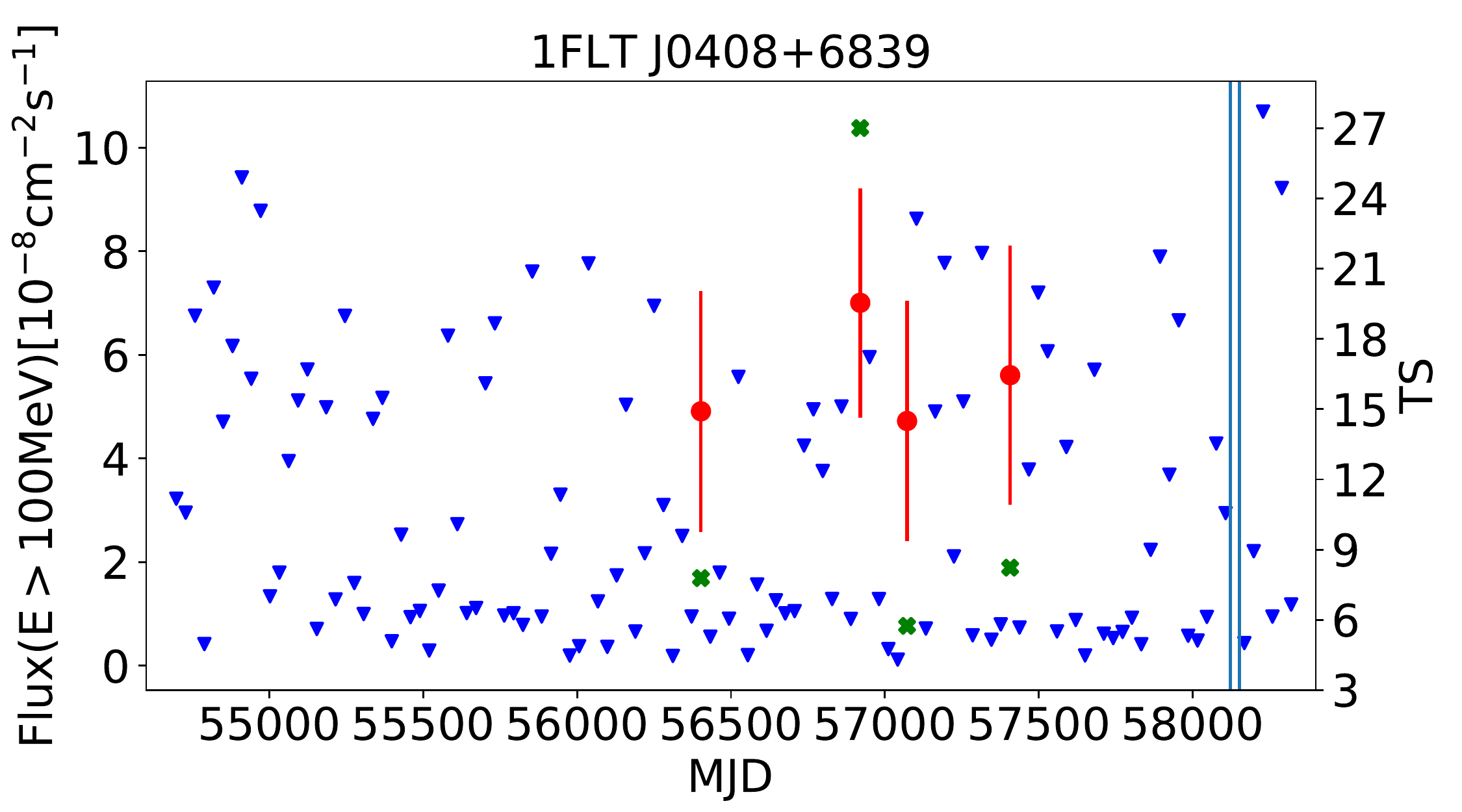}\label{fig:1FLTJ0408+6839}&
  \includegraphics[width=0.35\textwidth]{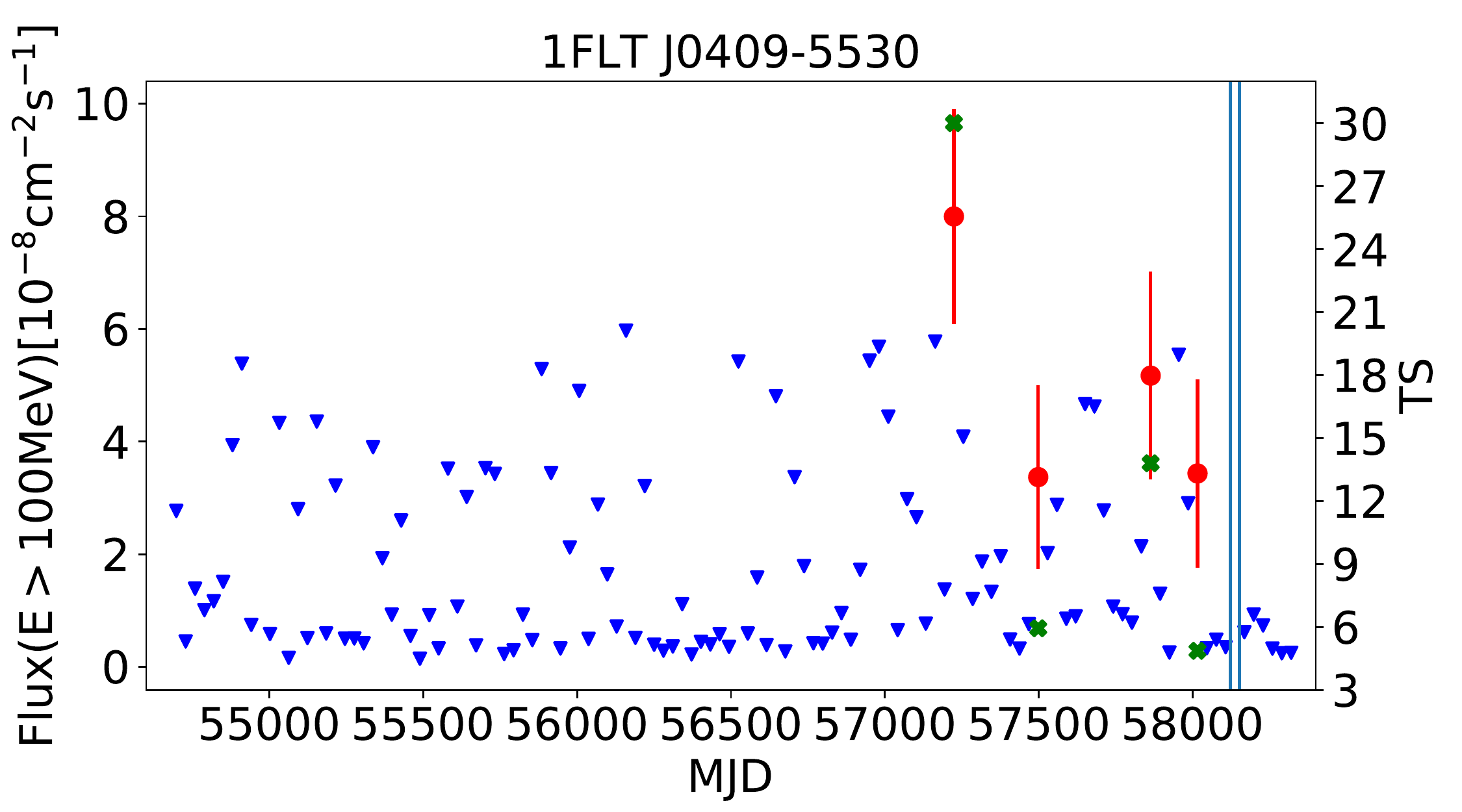}\label{fig:1FLTJ0409-5530}&
  \includegraphics[width=0.35\textwidth]{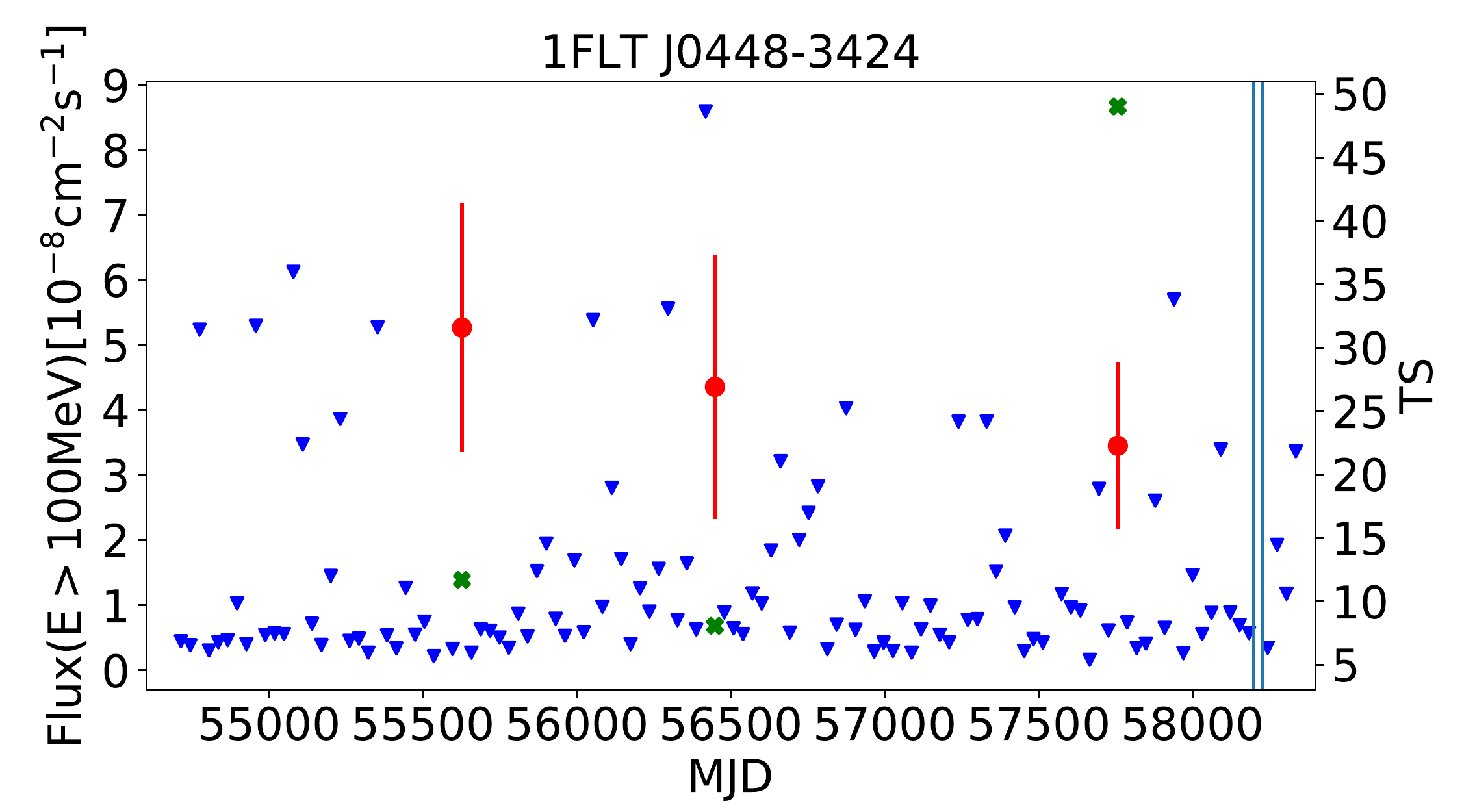}\label{fig:1FLTJ0448-3424}\\
  %[1FLTJ0408+6839]&%[1FLTJ0409-5530]&%[1FLTJ0448-3424]\\
  \includegraphics[width=0.35\textwidth]{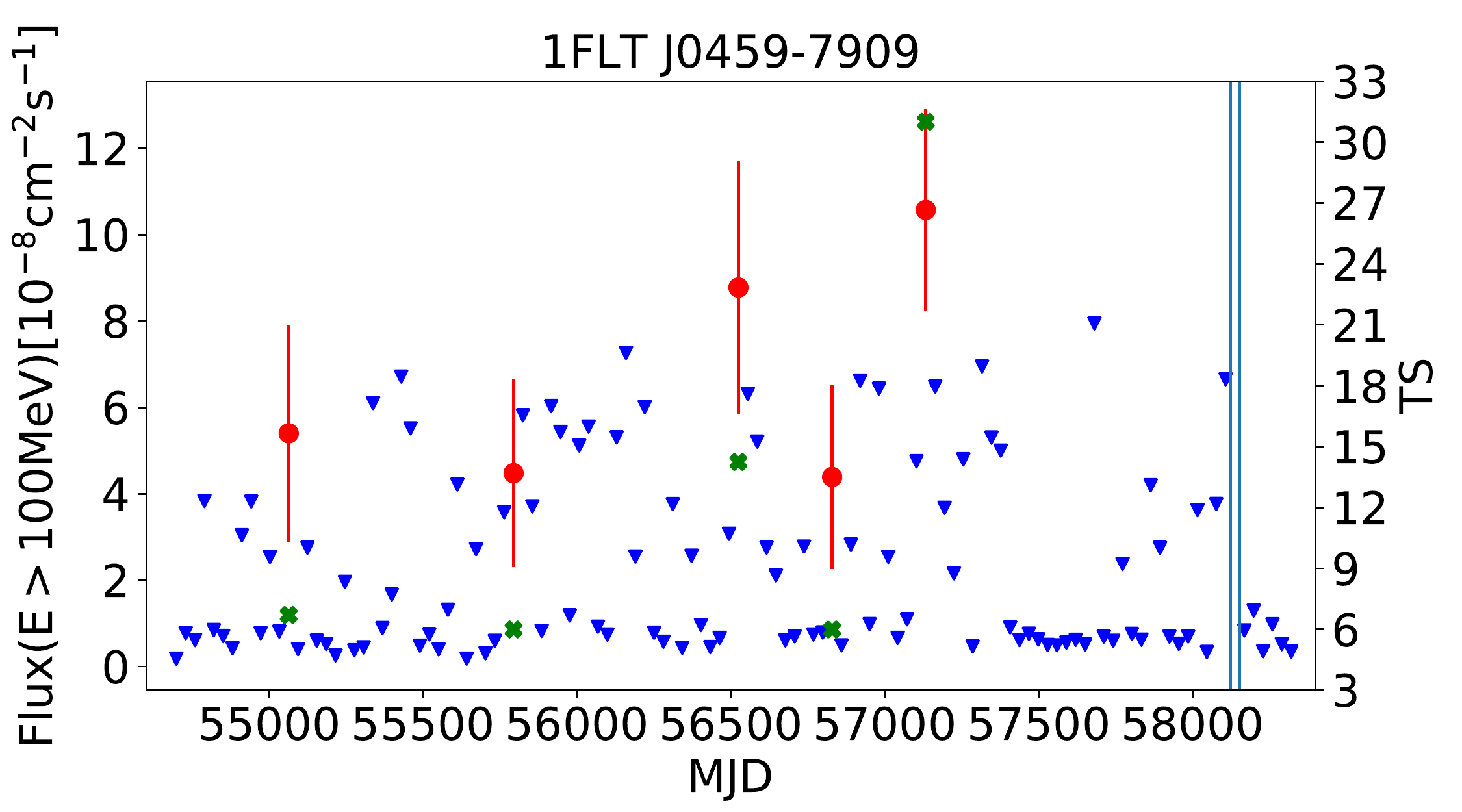}\label{fig:1FLTJ0459-7909}&
  \includegraphics[width=0.35\textwidth]{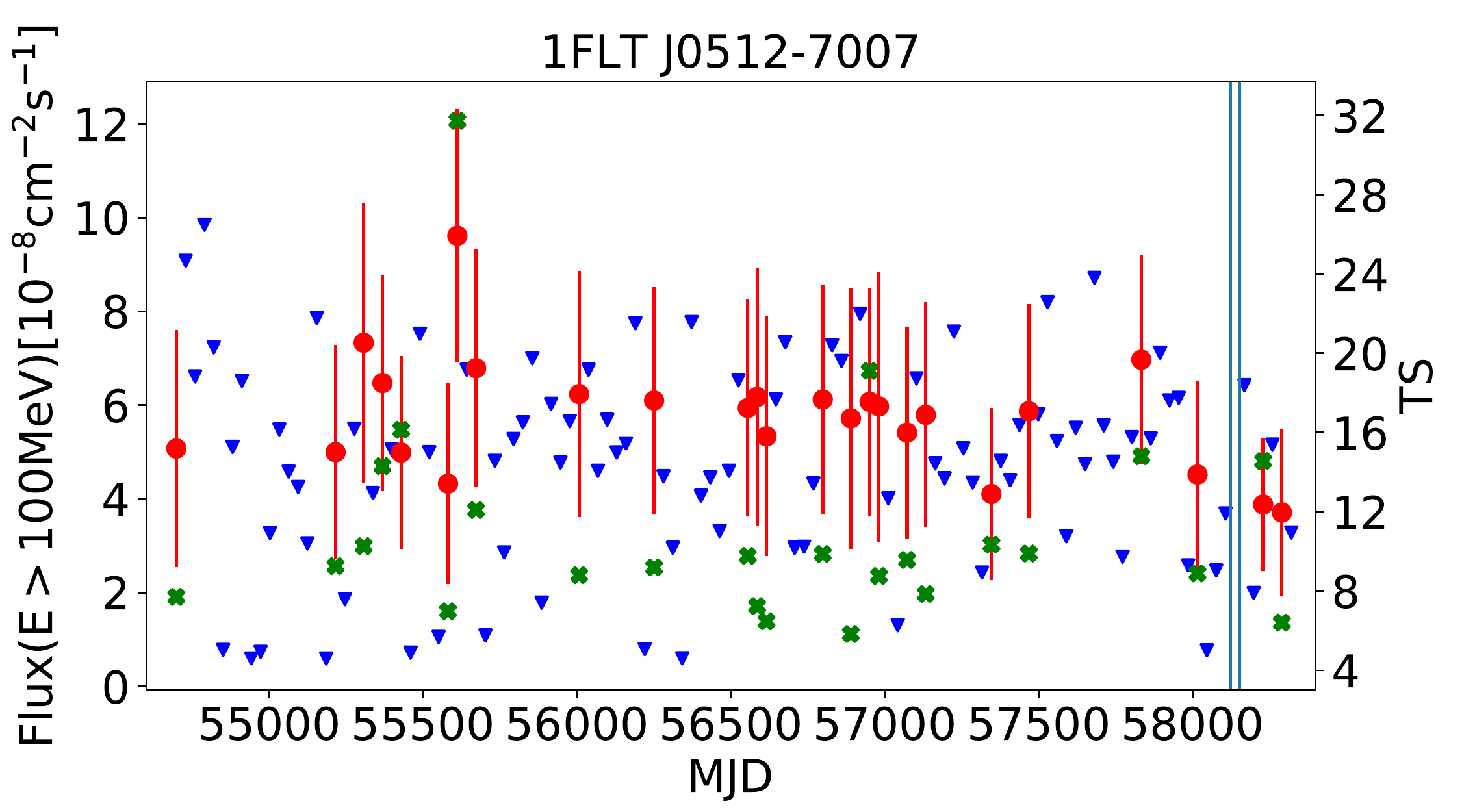}\label{fig:1FLTJ0512-7007}&
  \includegraphics[width=0.35\textwidth]{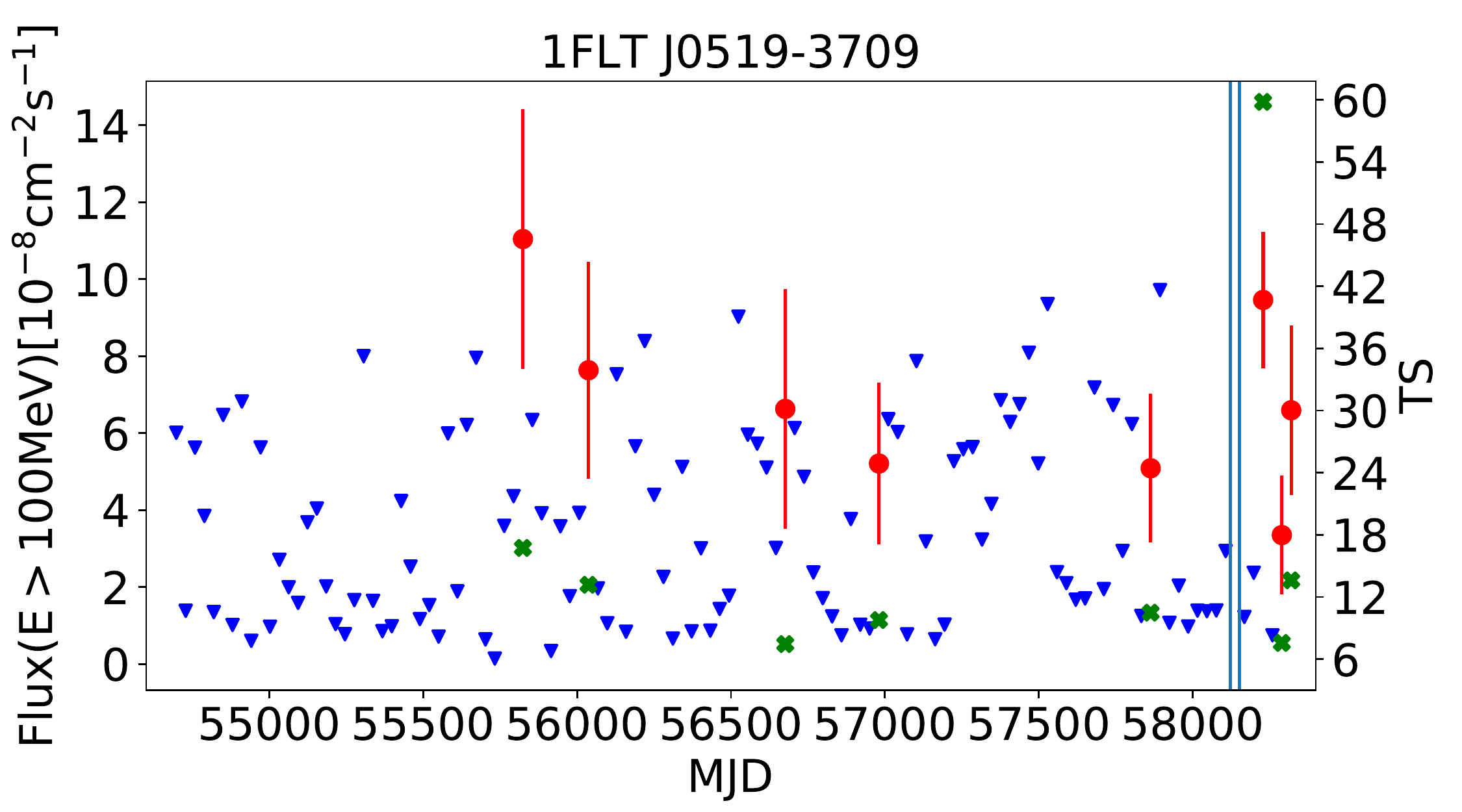}\label{fig:1FLTJ0519-3709}\\
  %[1FLTJ0459-7909]&%[1FLTJ0512-7007]&%[1FLTJ0519-3709]\\
\end{tabular}
\end{figure}
\begin{figure}[!t]
	\centering            
	\ContinuedFloat 
\setlength\tabcolsep{0.0pt}
\begin{tabular}{ccc} 
  \includegraphics[width=0.35\textwidth]{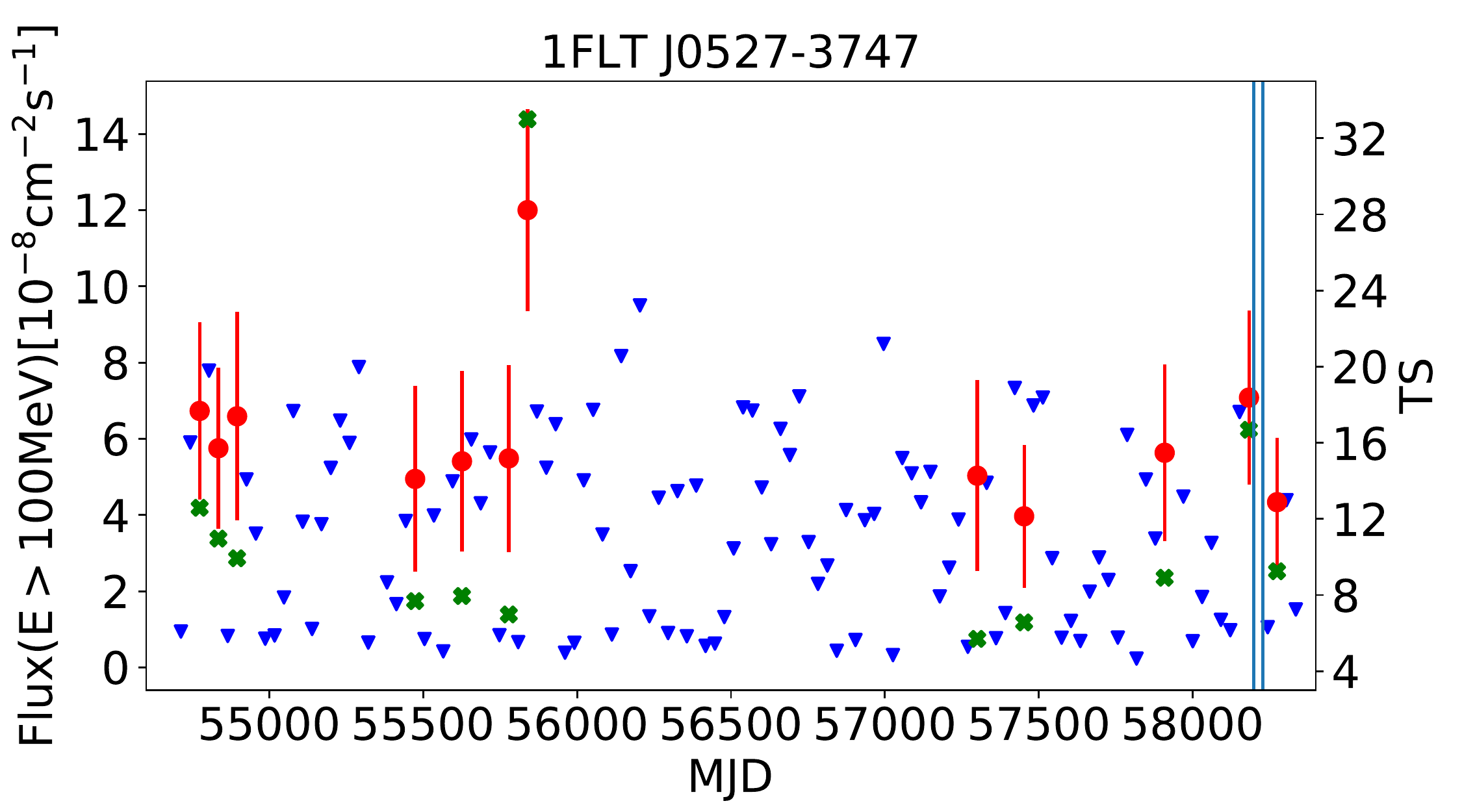}\label{fig:1FLTJ0527-3747}&
  \includegraphics[width=0.35\textwidth]{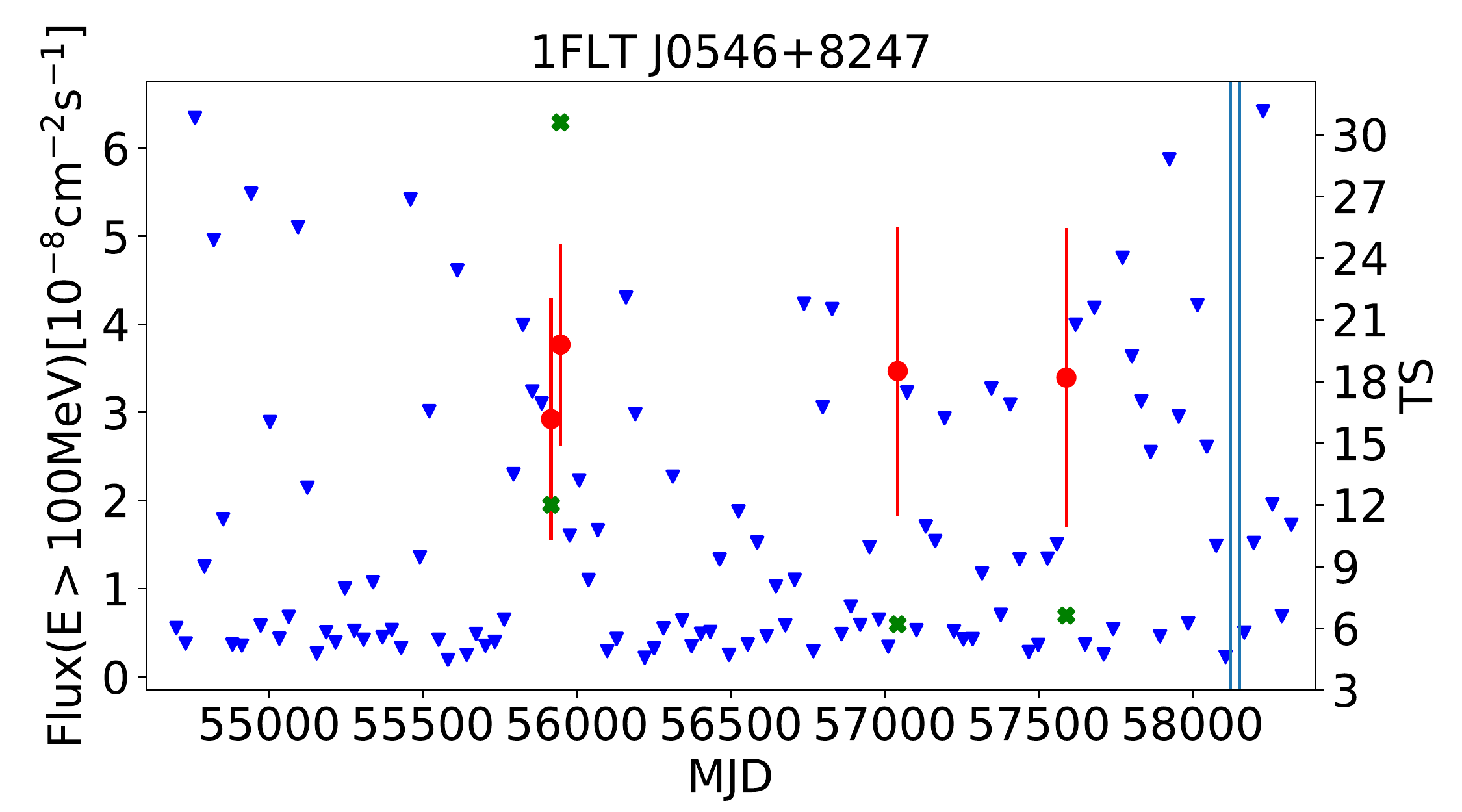}\label{fig:1FLTJ0546+8247}&
  \includegraphics[width=0.35\textwidth]{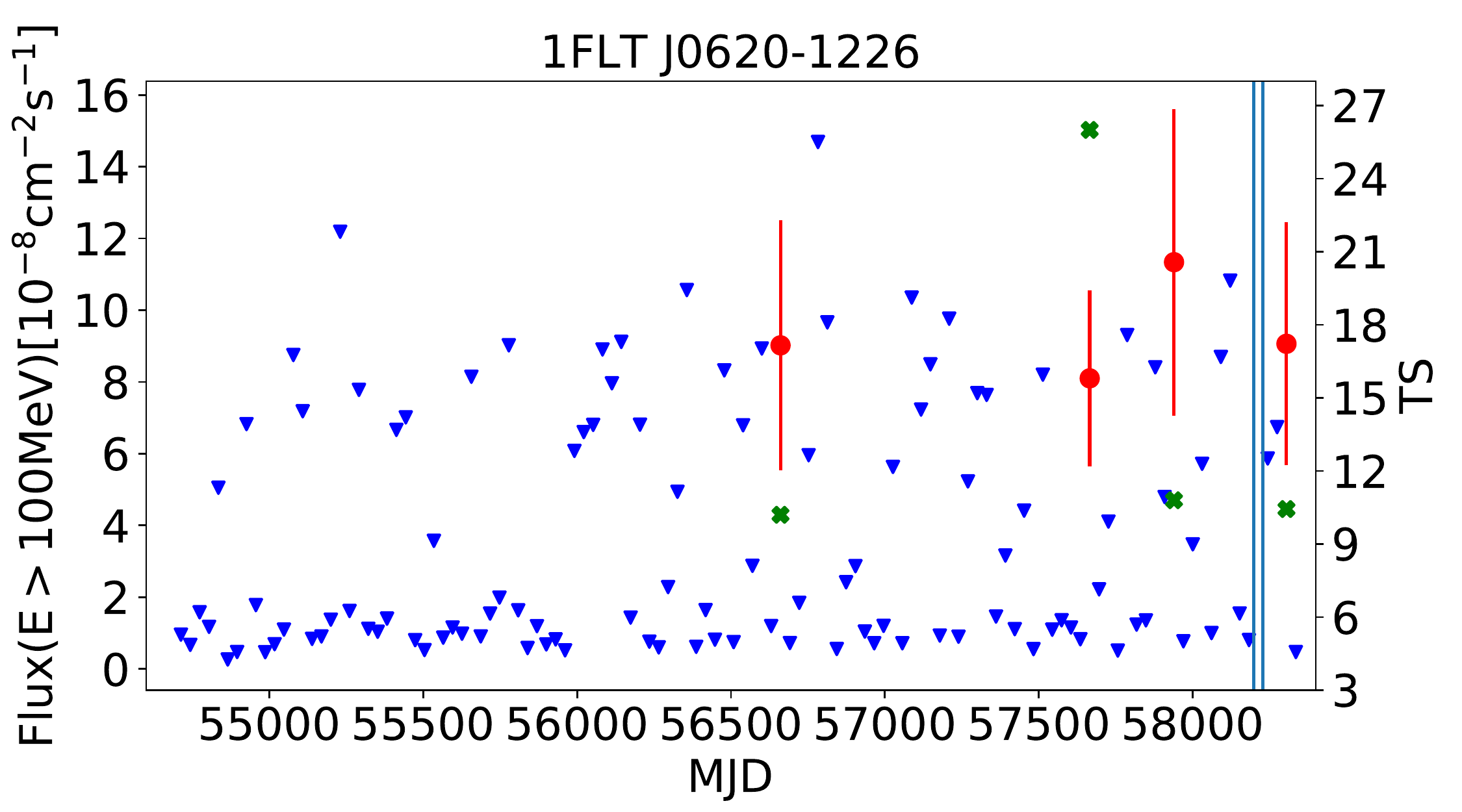}\label{fig:1FLTJ0620-1226}\\
  %[1FLTJ0527-3747]&%[1FLTJ0546+8247]&%[1FLTJ0620-1226]
  \includegraphics[width=0.35\textwidth]{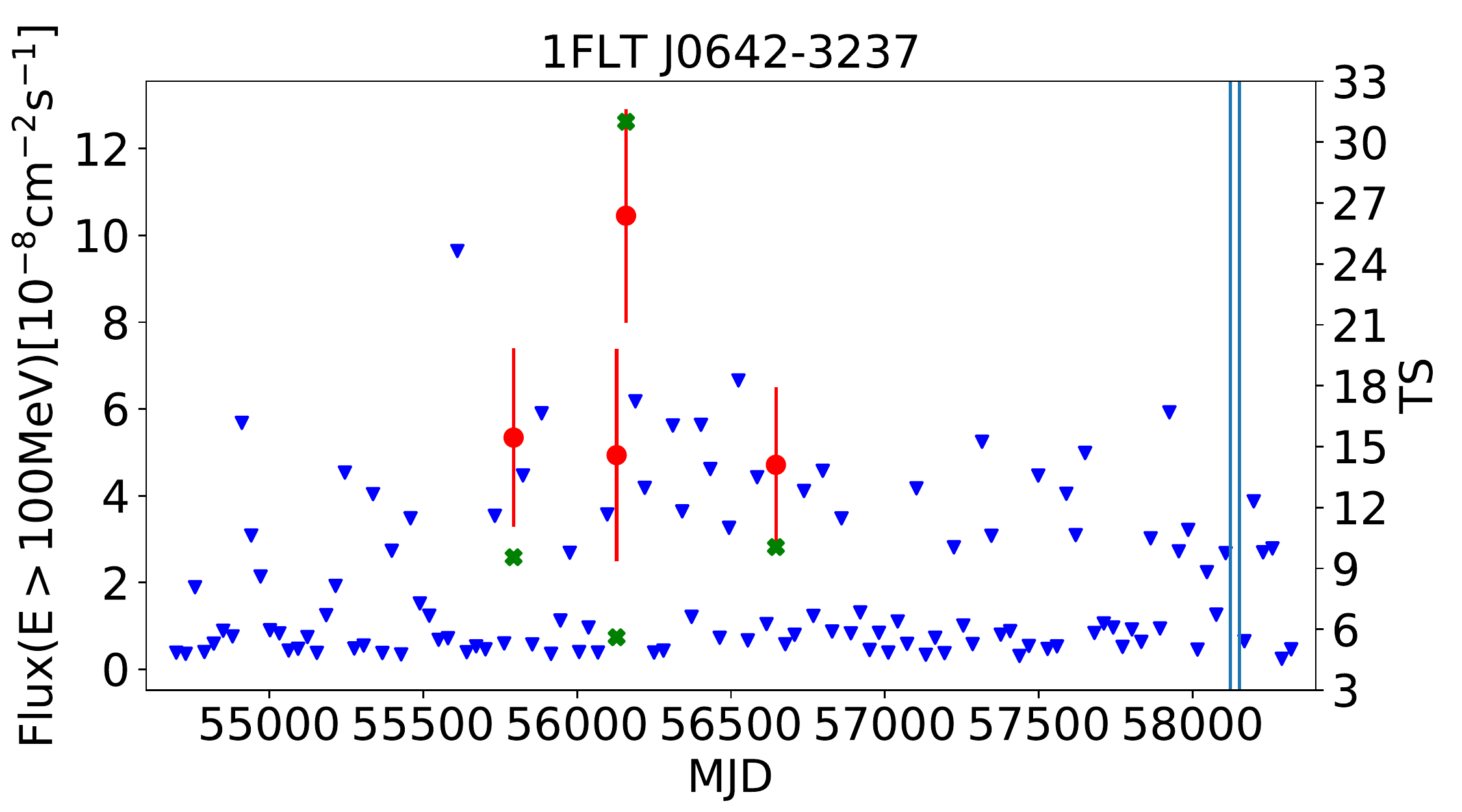}\label{fig:1FLTJ0642-3237}&
  \includegraphics[width=0.35\textwidth]{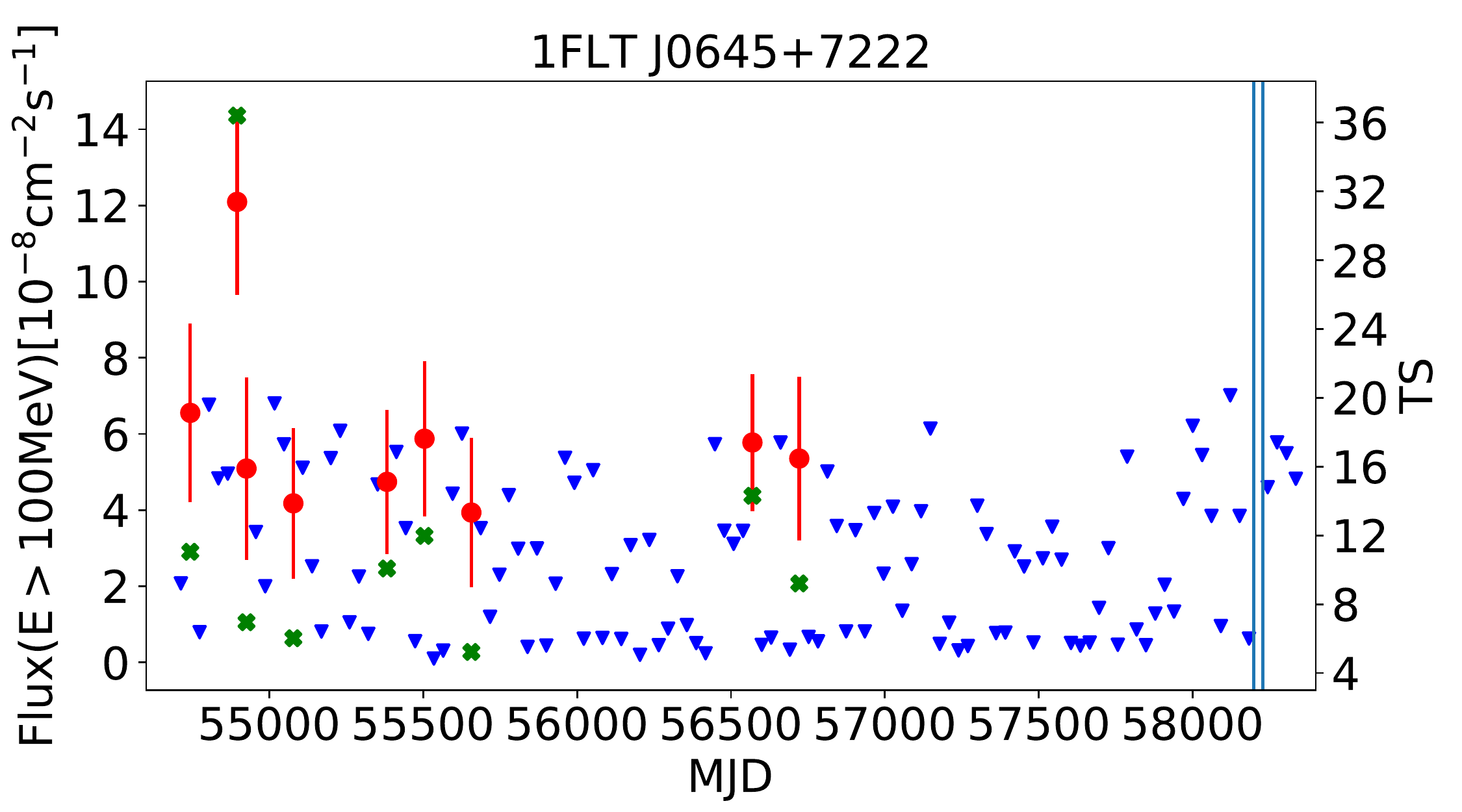}\label{fig:1FLTJ0645+7222}&
  \includegraphics[width=0.35\textwidth]{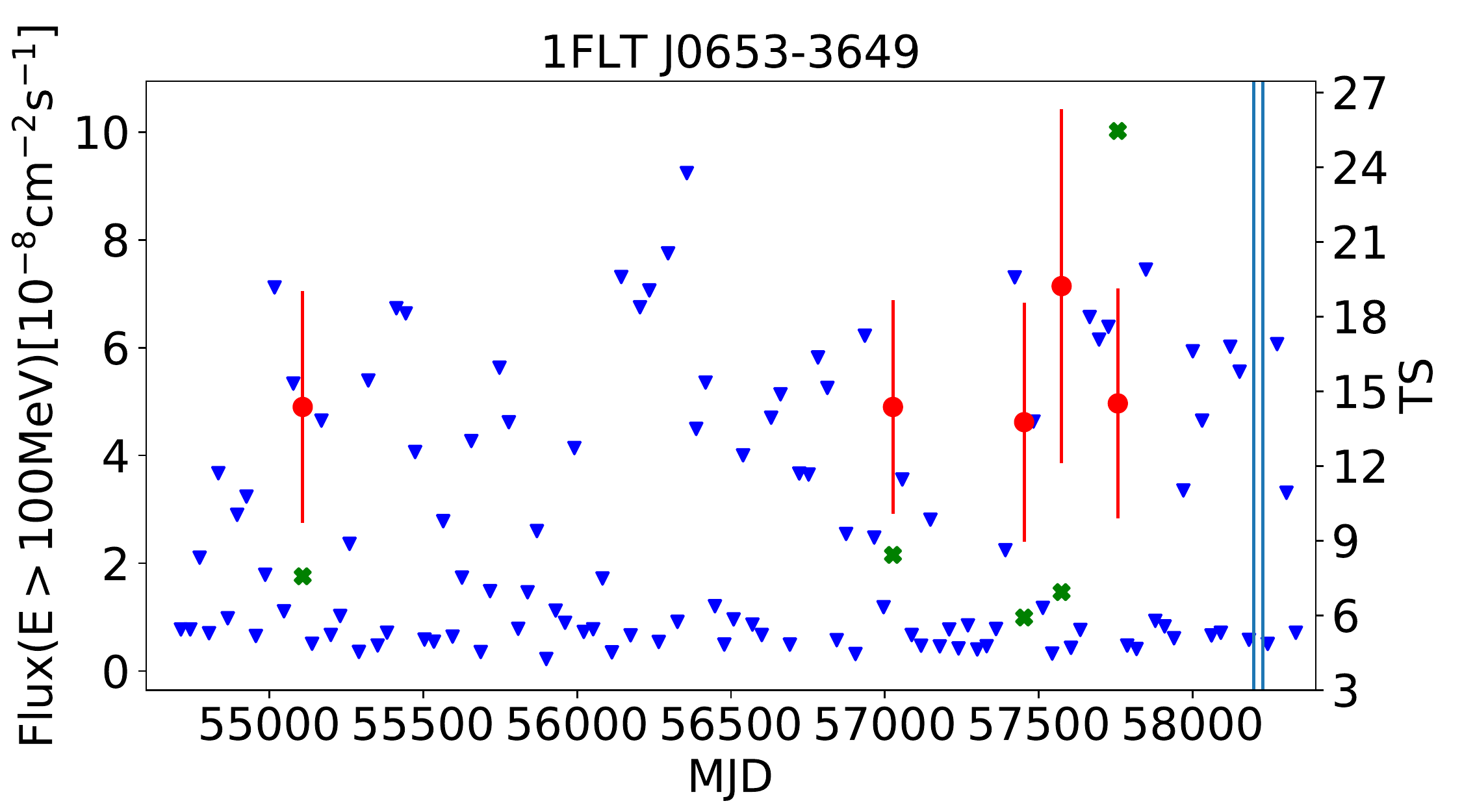}\label{fig:1FLTJ0653-3649}\\
  %[1FLTJ0642-3237]&%[1FLTJ0645+7222]&%[1FLTJ0653-3649]\\
  \includegraphics[width=0.35\textwidth]{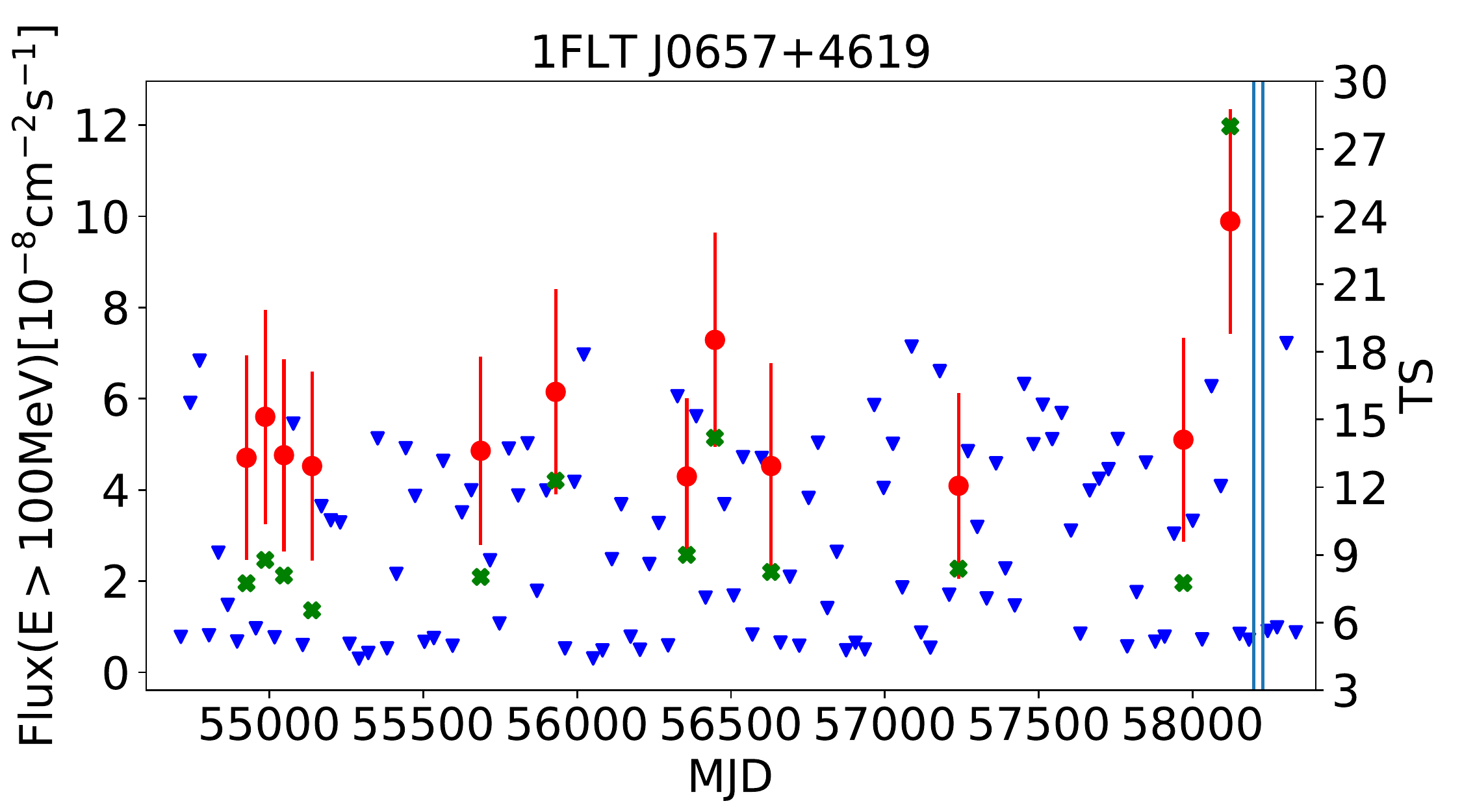}\label{fig:1FLTJ0657+4619}&
  \includegraphics[width=0.35\textwidth]{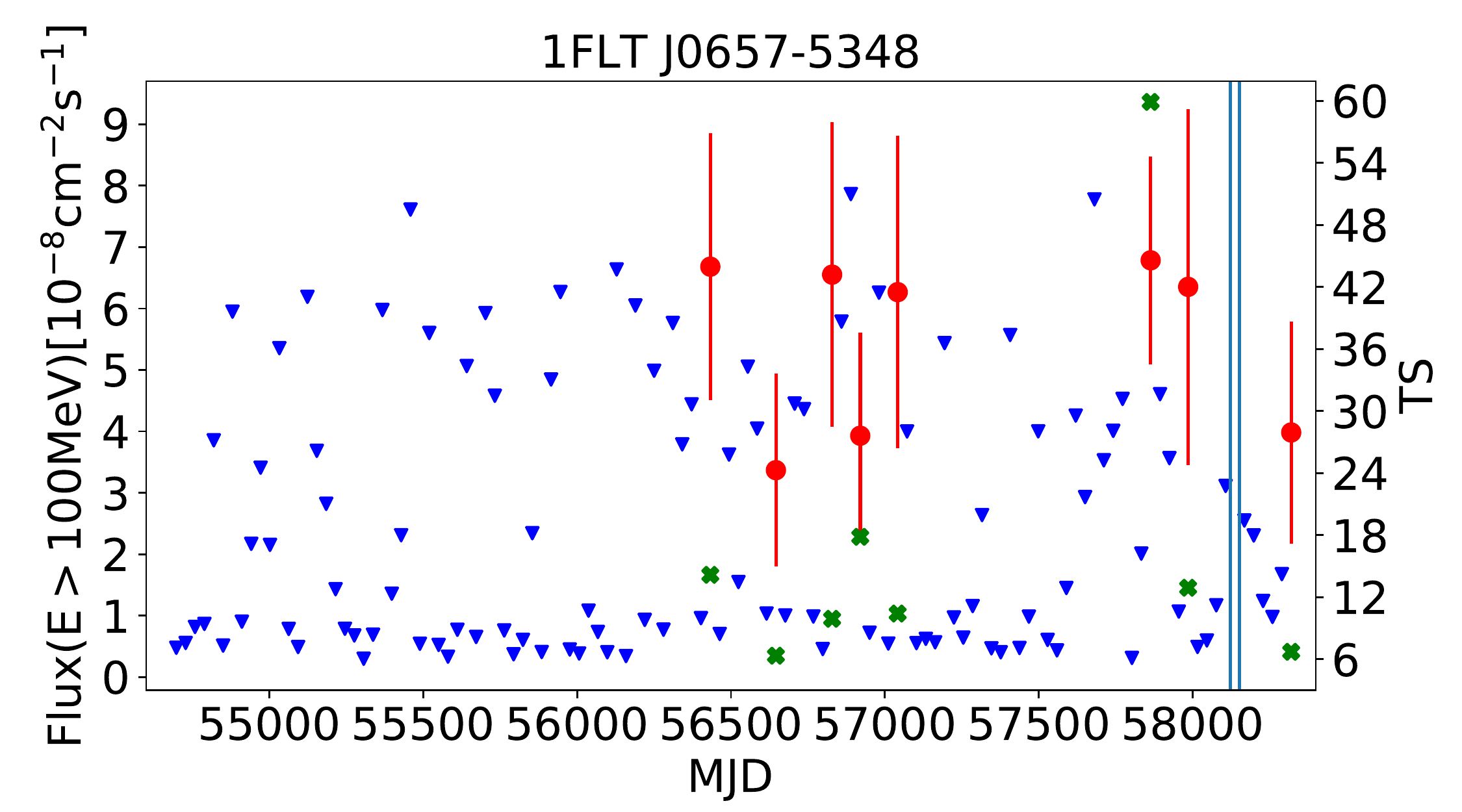}\label{fig:1FLTJ0657-5348}&
  \includegraphics[width=0.35\textwidth]{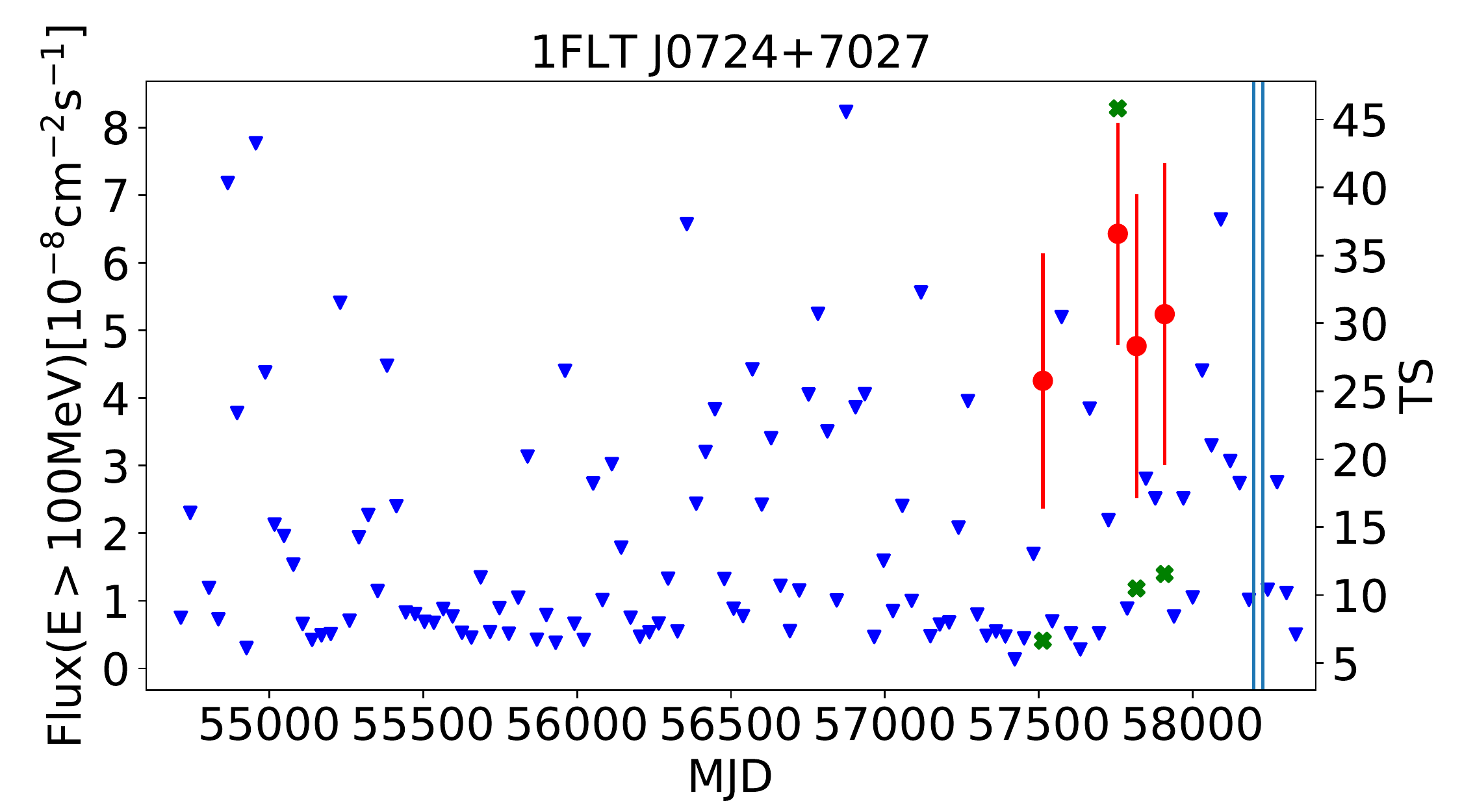}\label{fig:1FLTJ0724+7027}\\
  %[1FLTJ0657+4619]&%[1FLTJ0657-5348]&%[1FLTJ0724+7027]\\
  \includegraphics[width=0.35\textwidth]{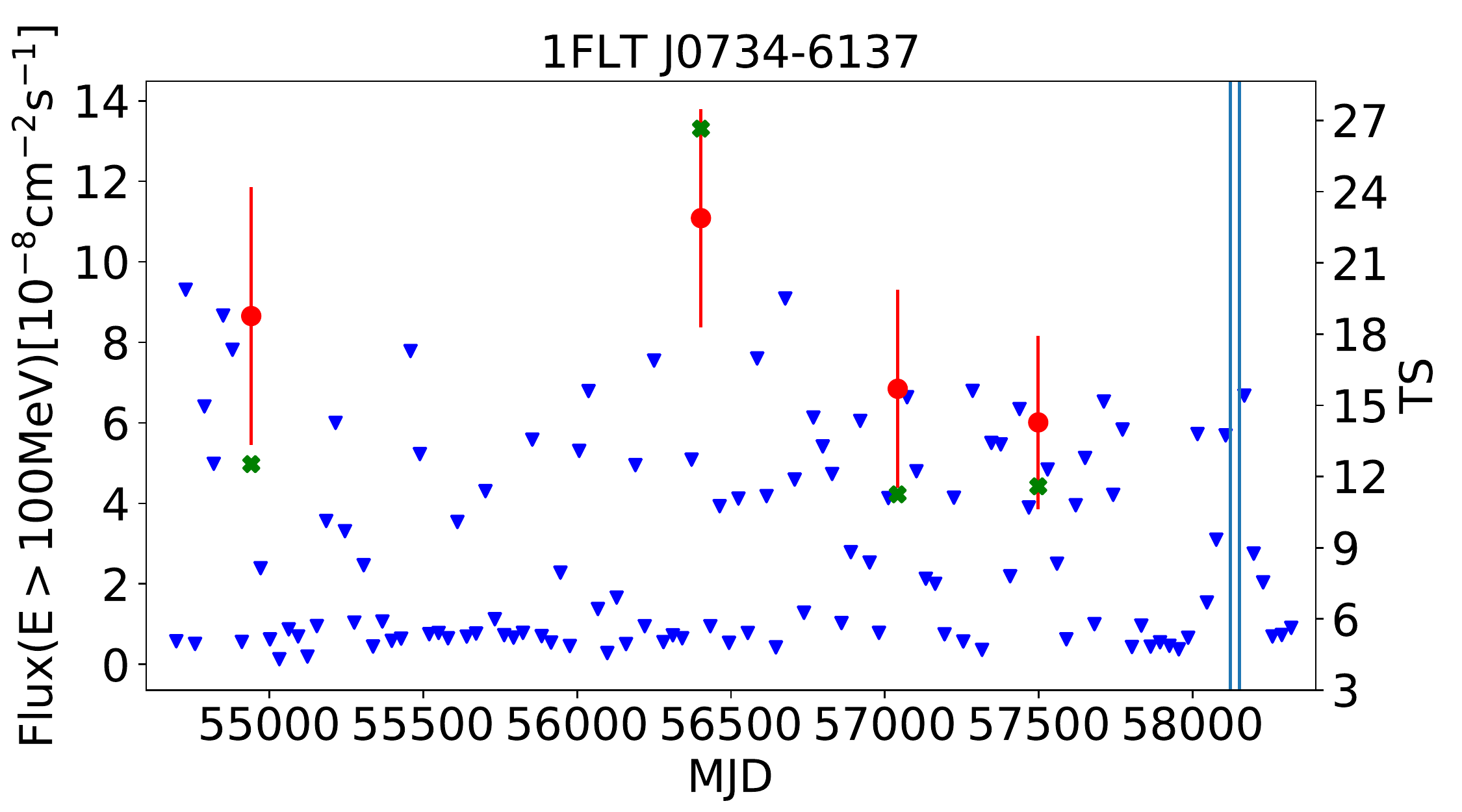}\label{fig:1FLTJ0734-6137}&
  \includegraphics[width=0.35\textwidth]{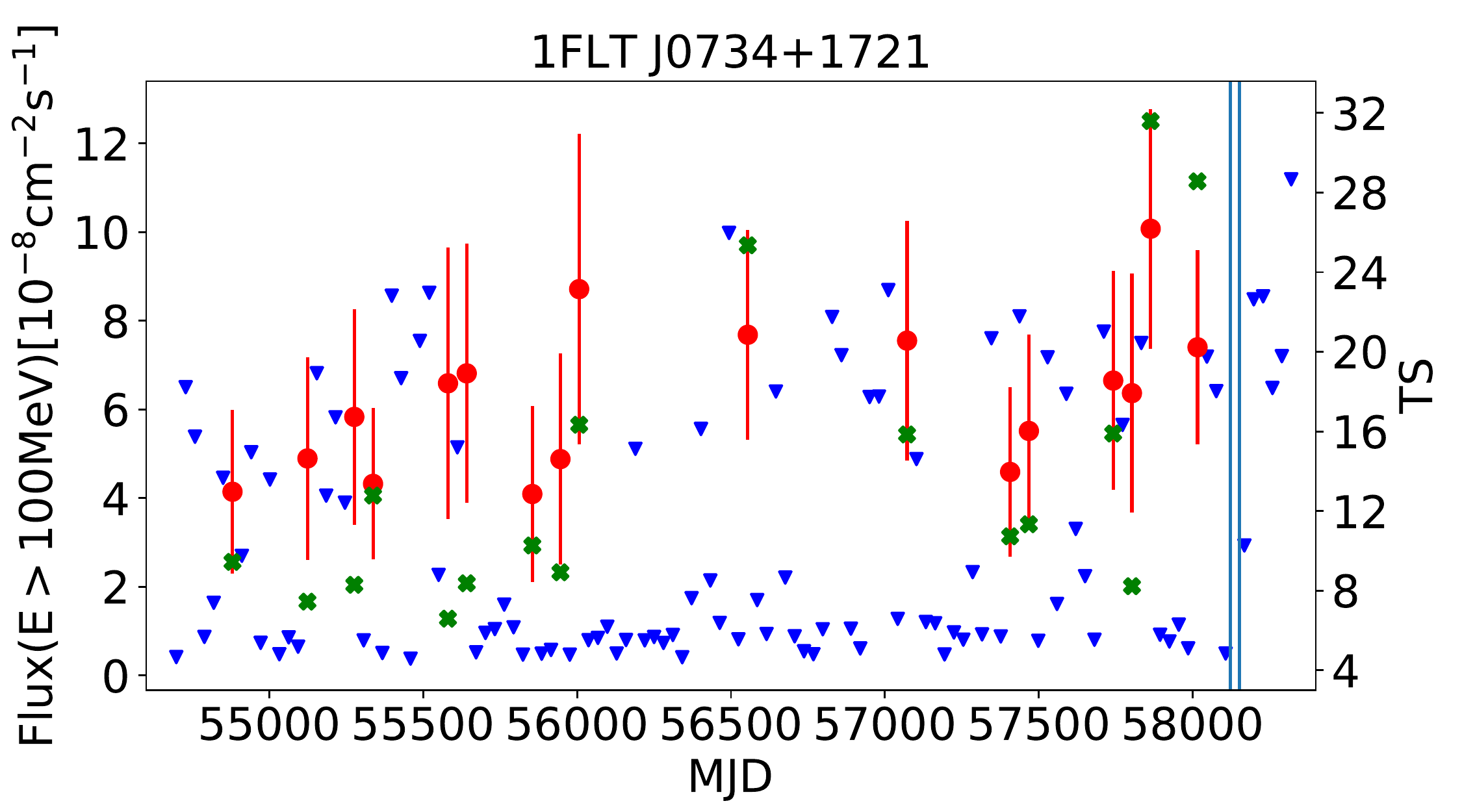}\label{fig:1FLTJ0734+1721}&
  \includegraphics[width=0.35\textwidth]{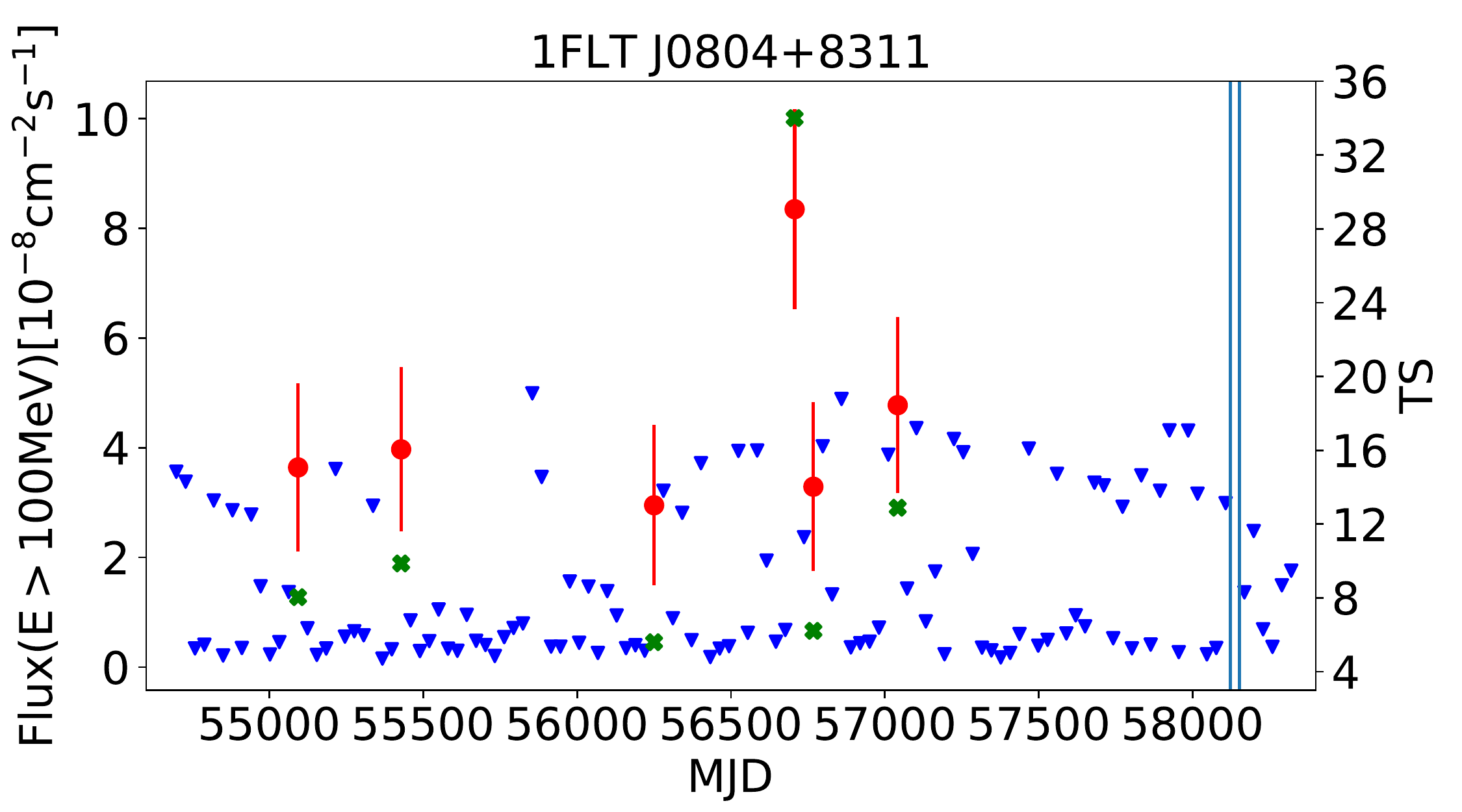}\label{fig:1FLTJ0804+8311}\\
  %[1FLTJ0734-6137]&%[1FLTJ0734+1721]&%[1FLTJ0804+8311]\\
  \includegraphics[width=0.35\textwidth]{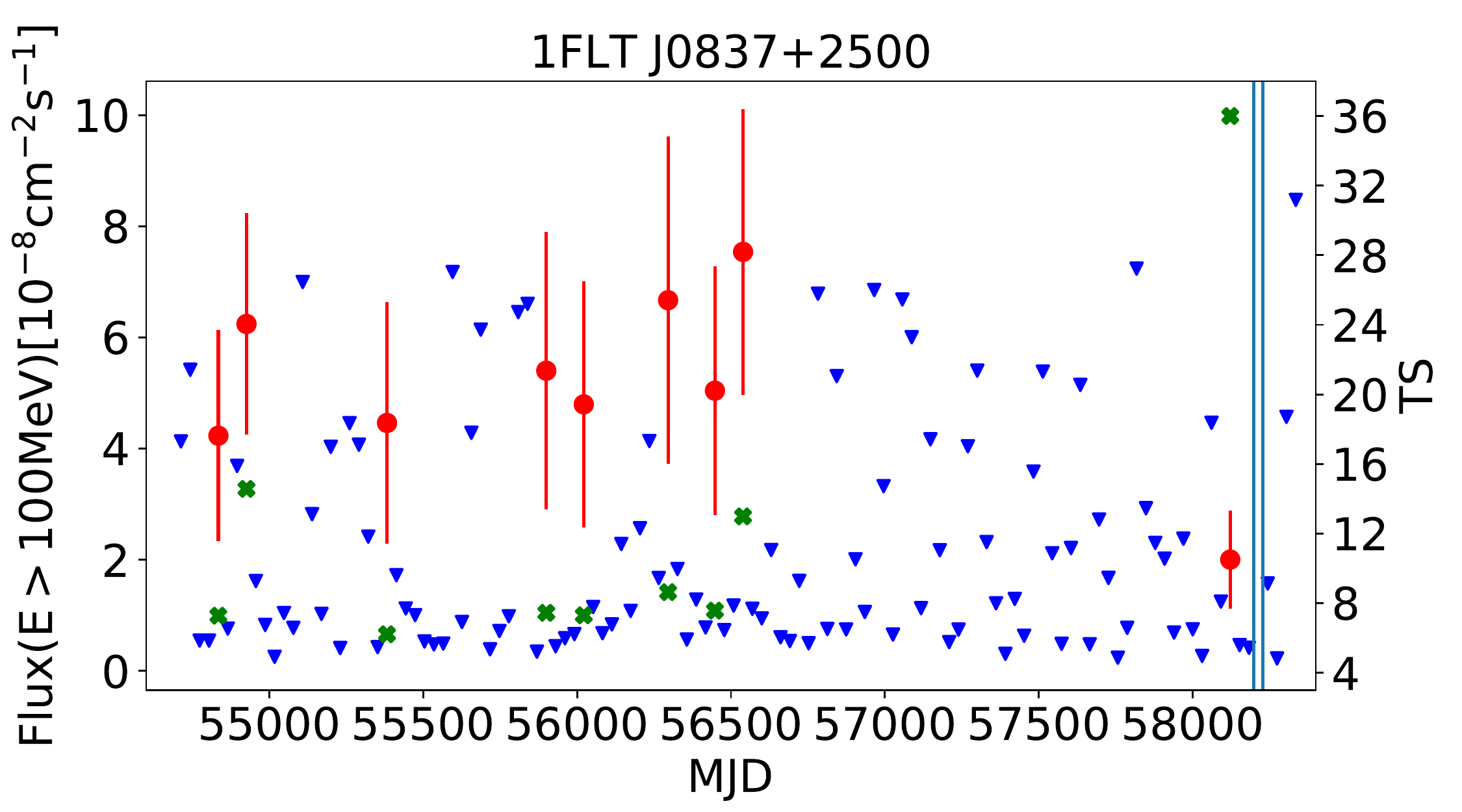}\label{fig:1FLTJ0837+2500}&
  \includegraphics[width=0.35\textwidth]{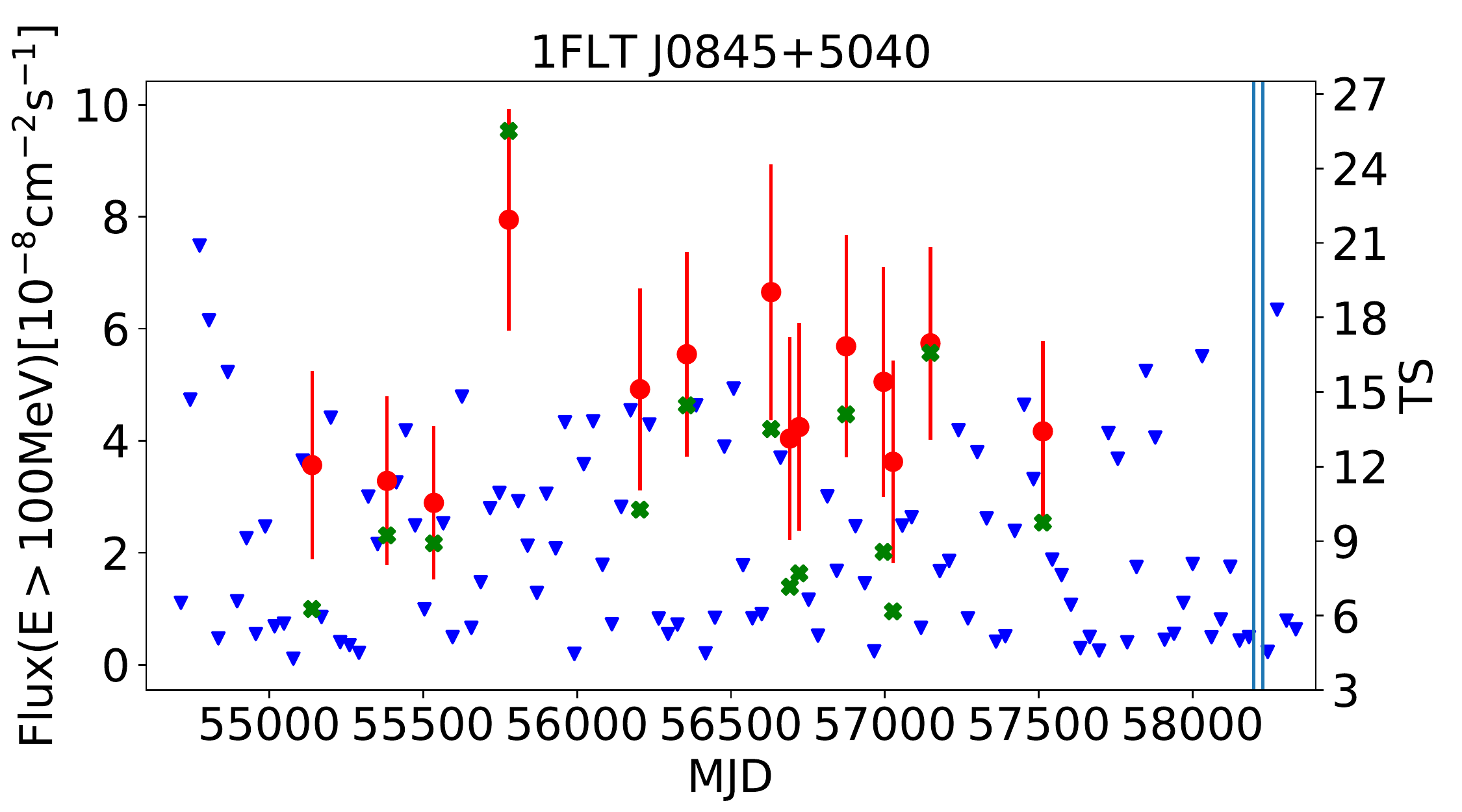}\label{fig:1FLTJ0845+5040}&
  \includegraphics[width=0.35\textwidth]{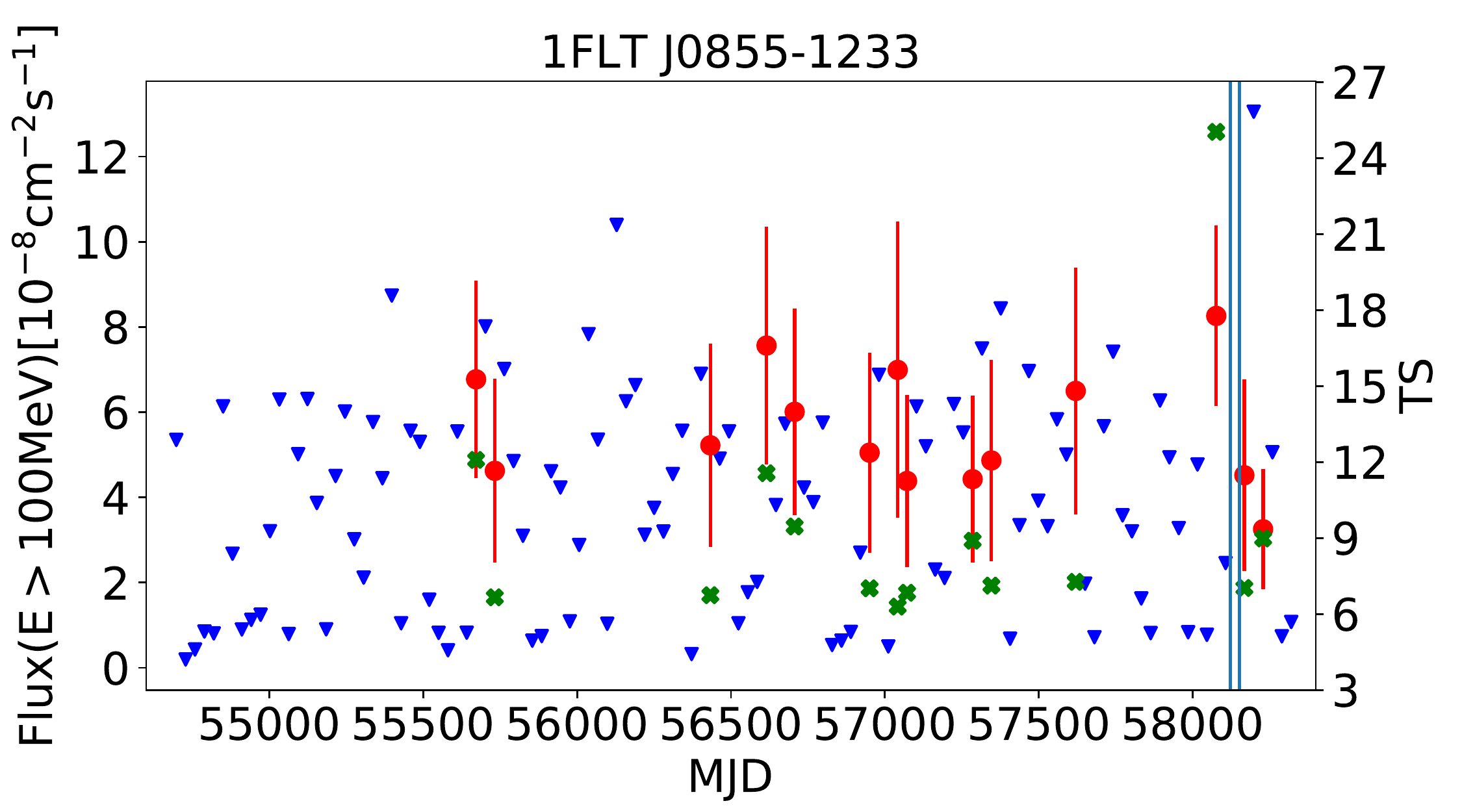}\label{fig:1FLTJ0855-1233}\\
  %[1FLTJ0837+2500]&%[1FLTJ0845+5040]&%[1FLTJ0855-1233]\\
  \includegraphics[width=0.35\textwidth]{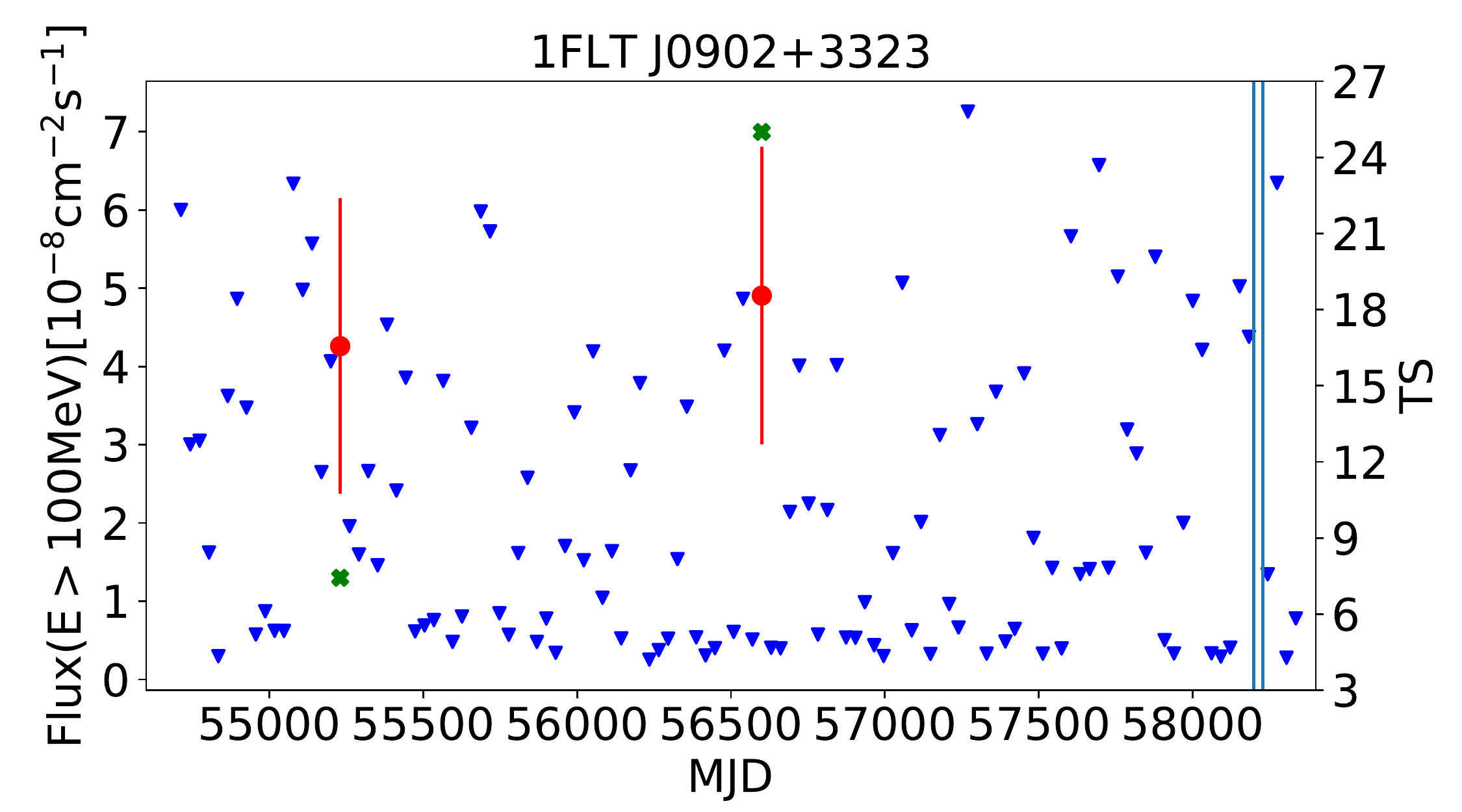}\label{fig:1FLTJ0902+3323}&
  \includegraphics[width=0.35\textwidth]{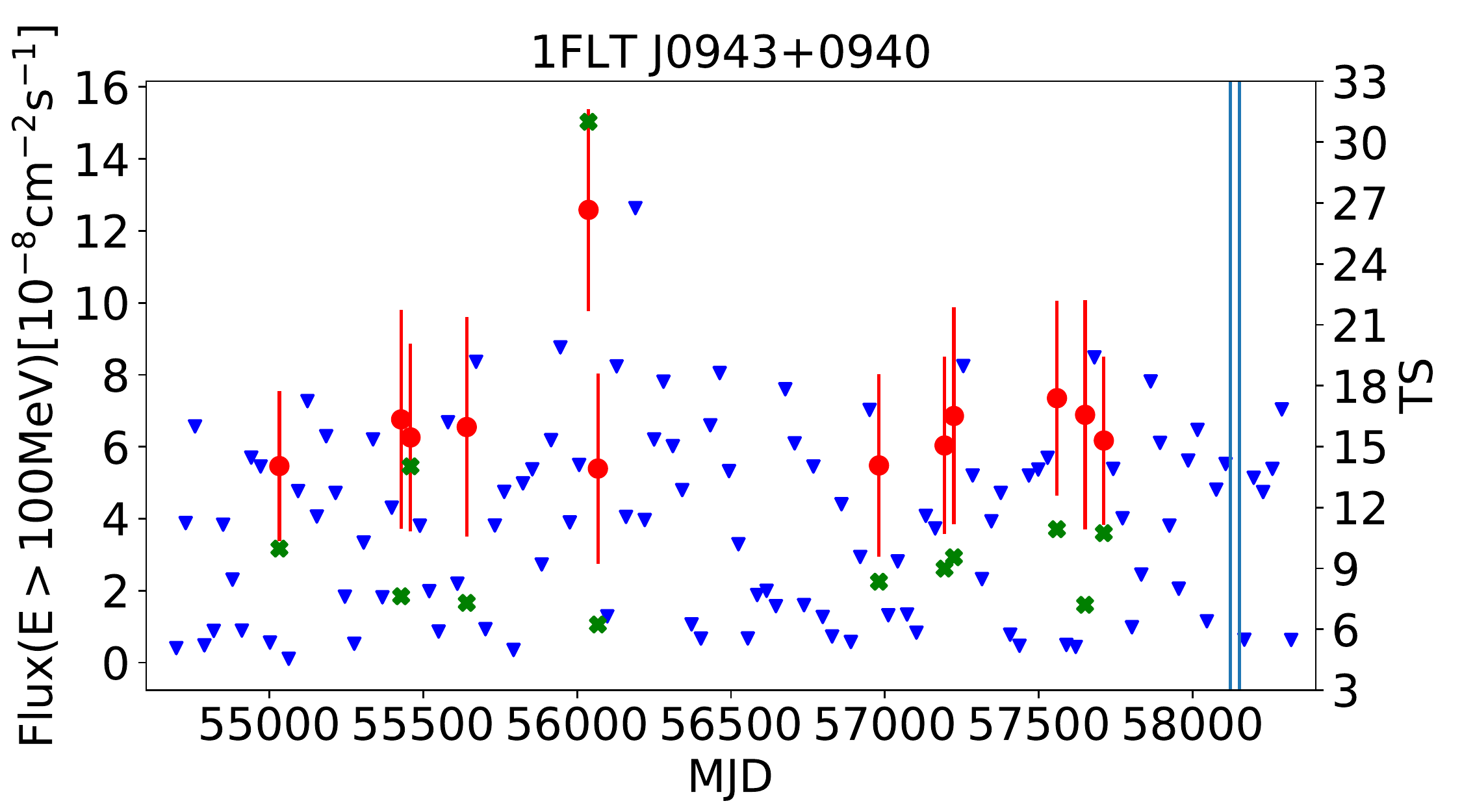}\label{fig:1FLTJ0943+0940}&
  \includegraphics[width=0.35\textwidth]{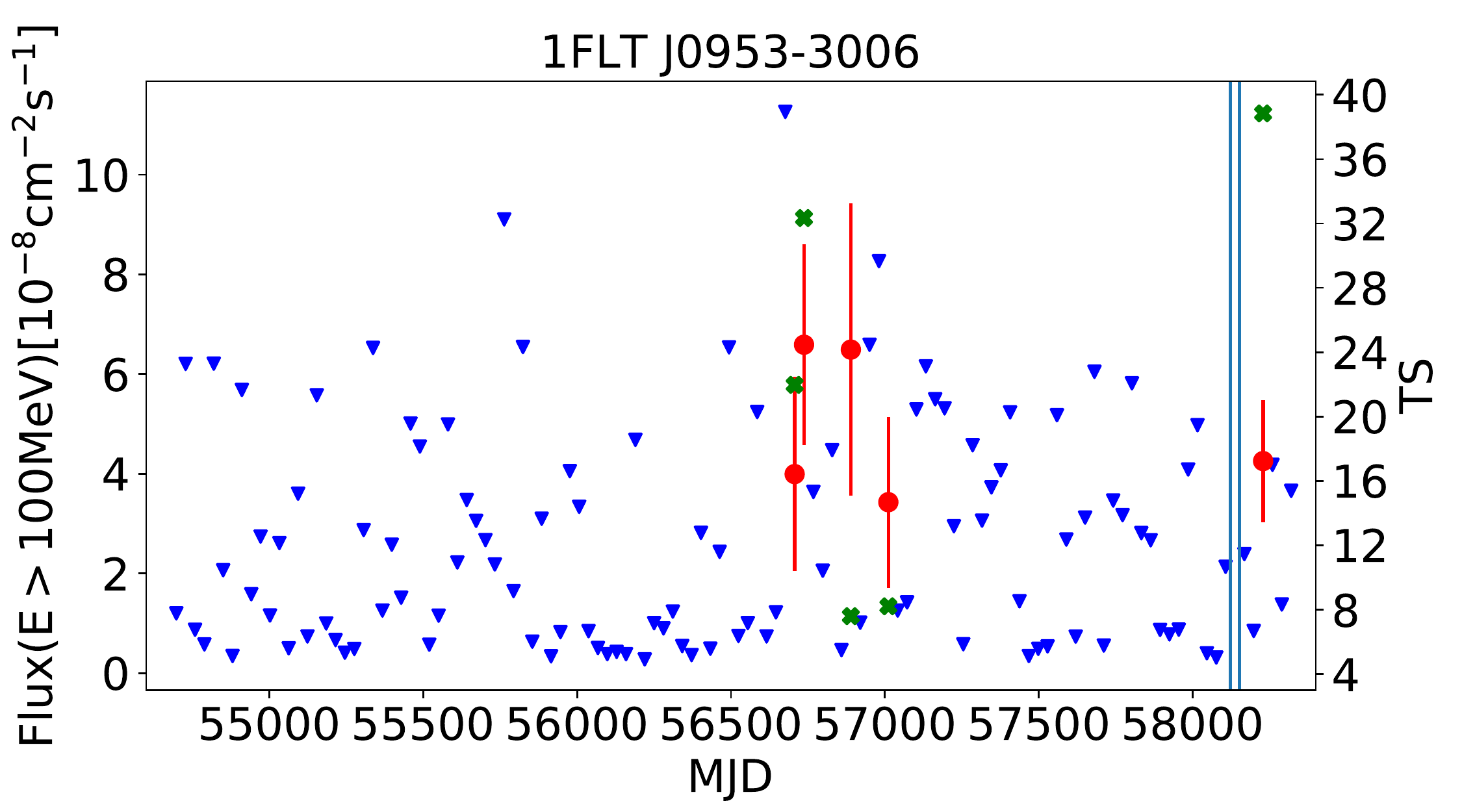}\label{fig:1FLTJ0953-3006}\\
  %[1FLTJ0902+3323]&%[1FLTJ0943+0940]&%[1FLTJ0953-3006]\\
\end{tabular}
\end{figure}
\begin{figure}[!t]
	\centering            
	\ContinuedFloat
\setlength\tabcolsep{0.0pt}
\begin{tabular}{ccc} 
  \includegraphics[width=0.35\textwidth]{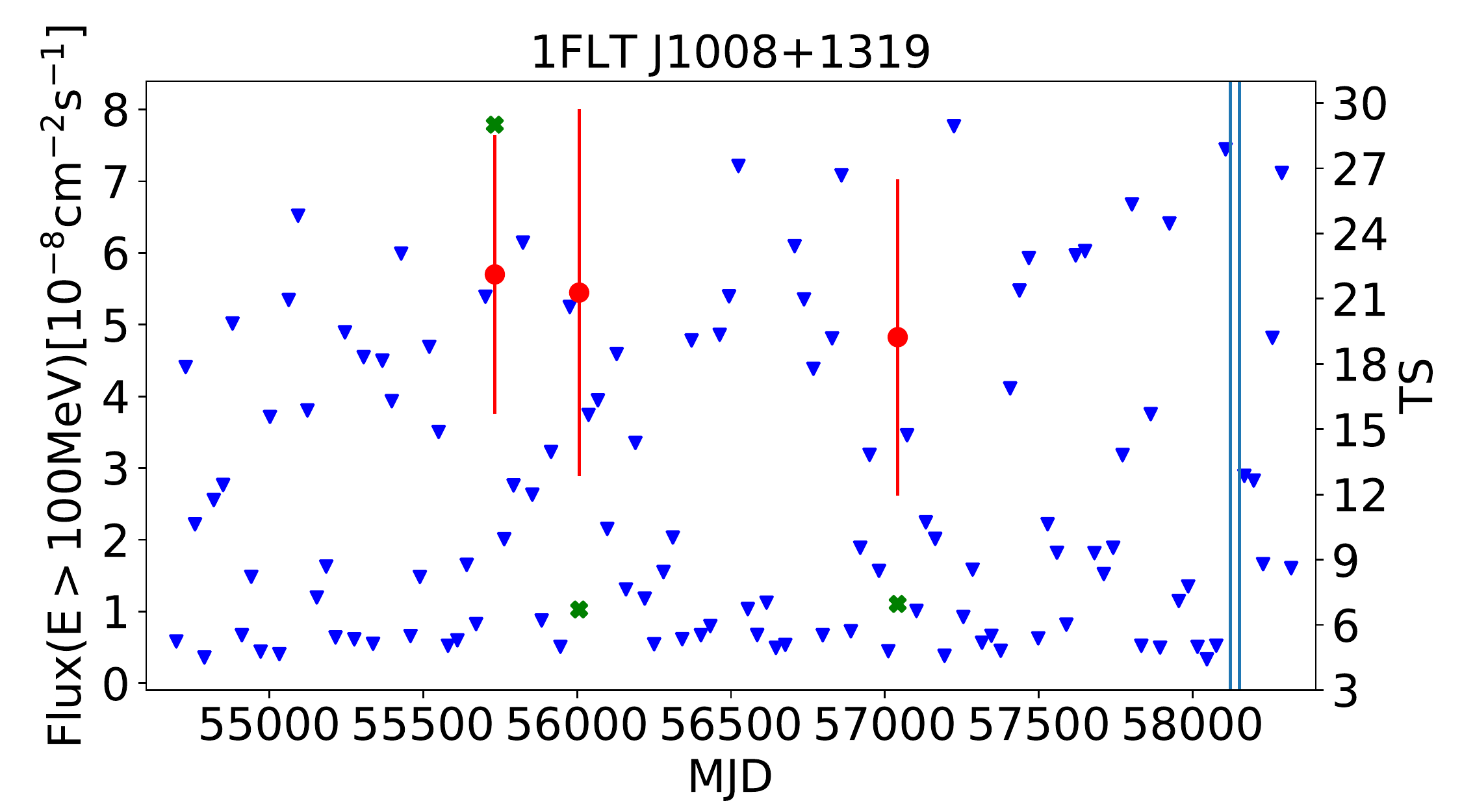}\label{fig:1FLTJ1008+1319}&
  \includegraphics[width=0.35\textwidth]{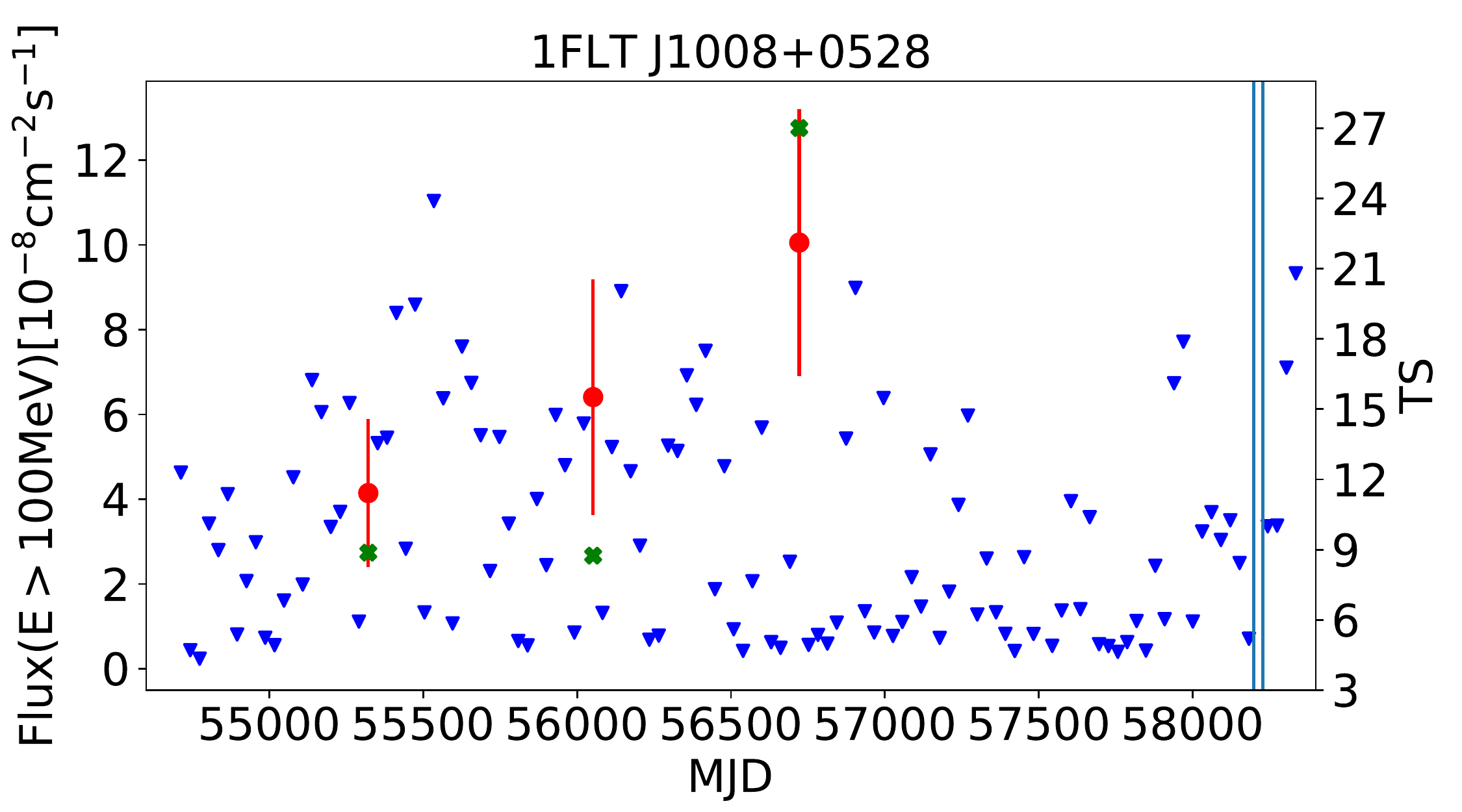}\label{fig:1FLTJ1008+0528}&
  \includegraphics[width=0.35\textwidth]{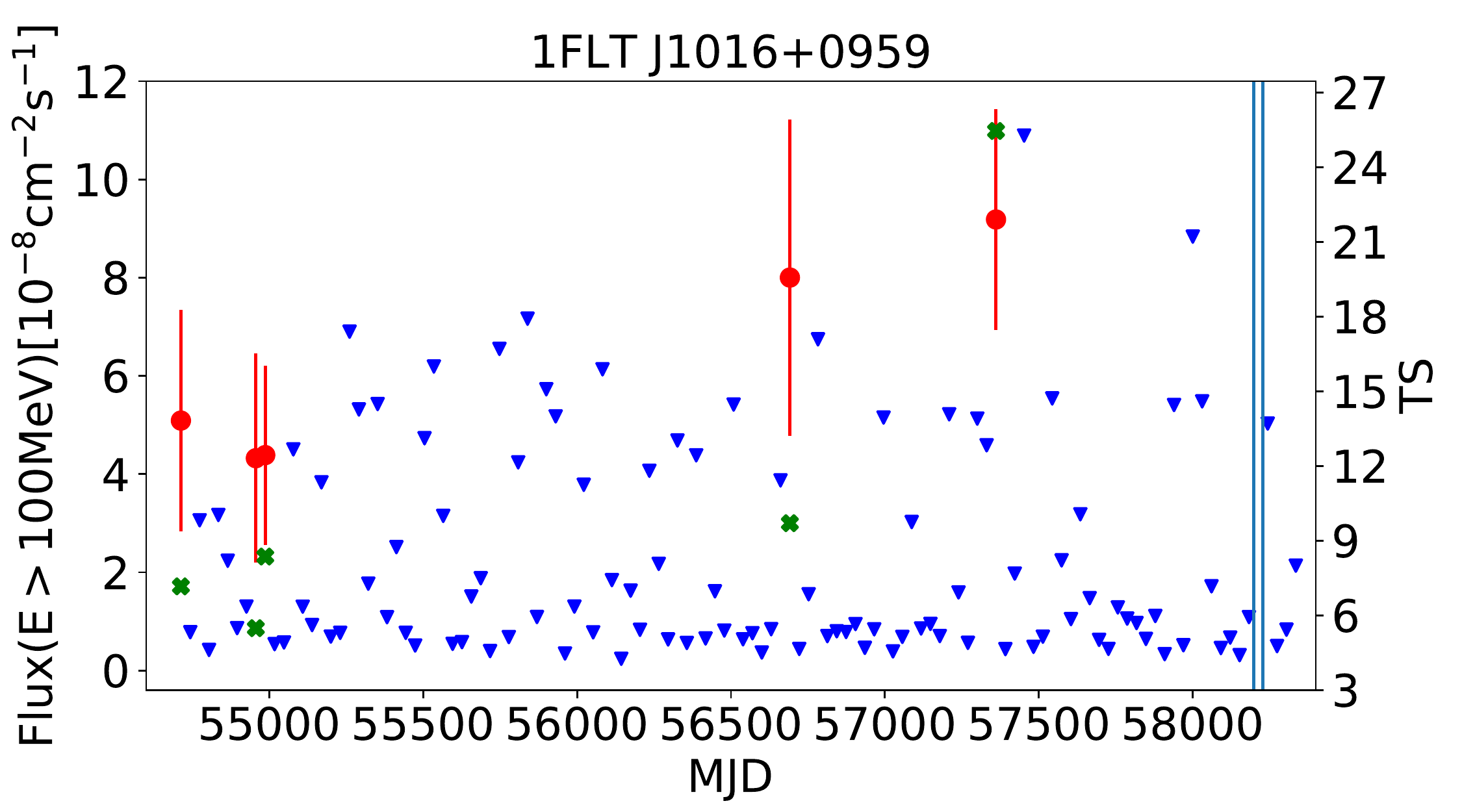}\label{fig:1FLTJ1016+0959}\\
  %[1FLTJ1008+1319]&%[1FLTJ1008+0528]&%[1FLTJ1016+0959]
  \includegraphics[width=0.35\textwidth]{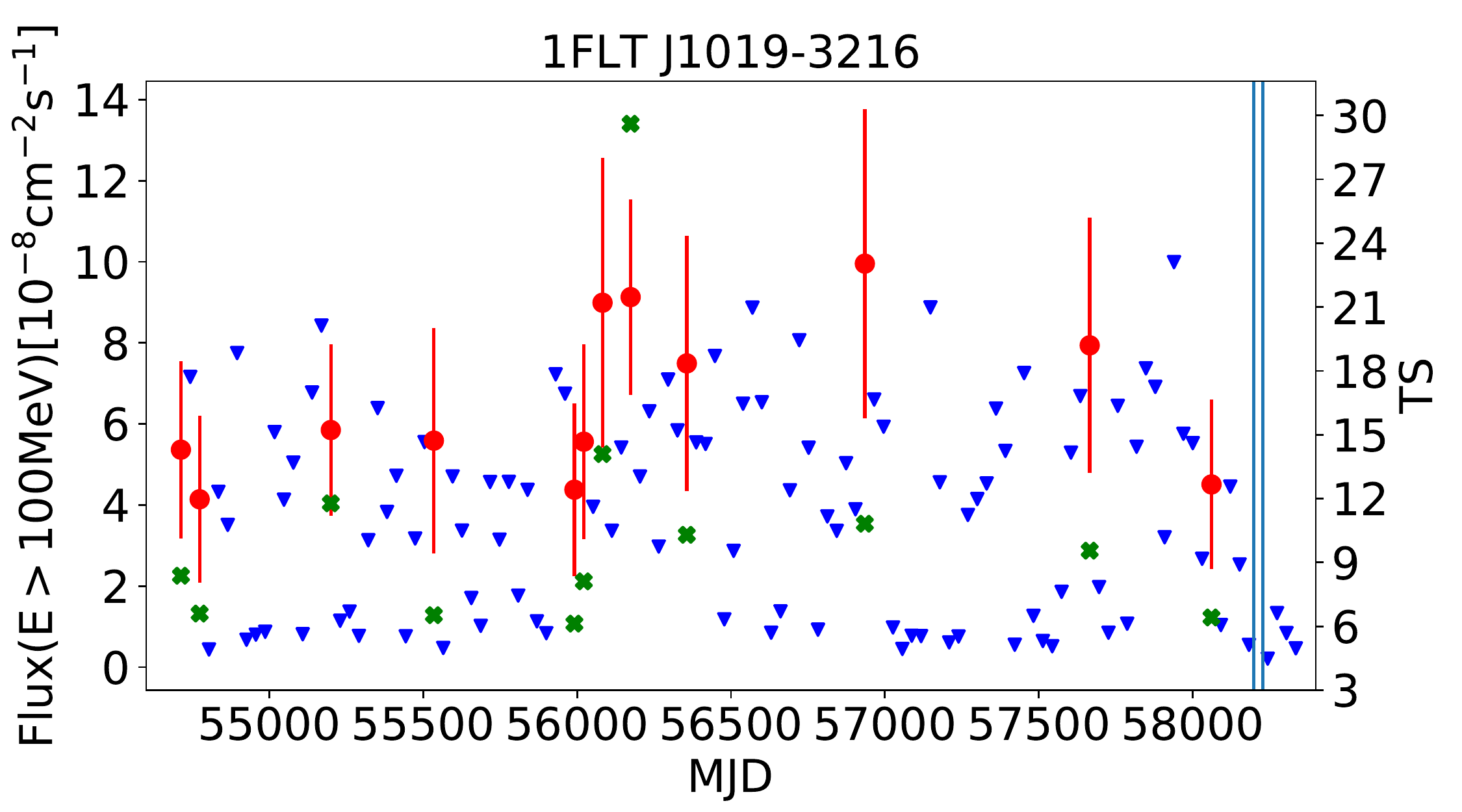}\label{fig:1FLTJ1019-3216}&
  \includegraphics[width=0.35\textwidth]{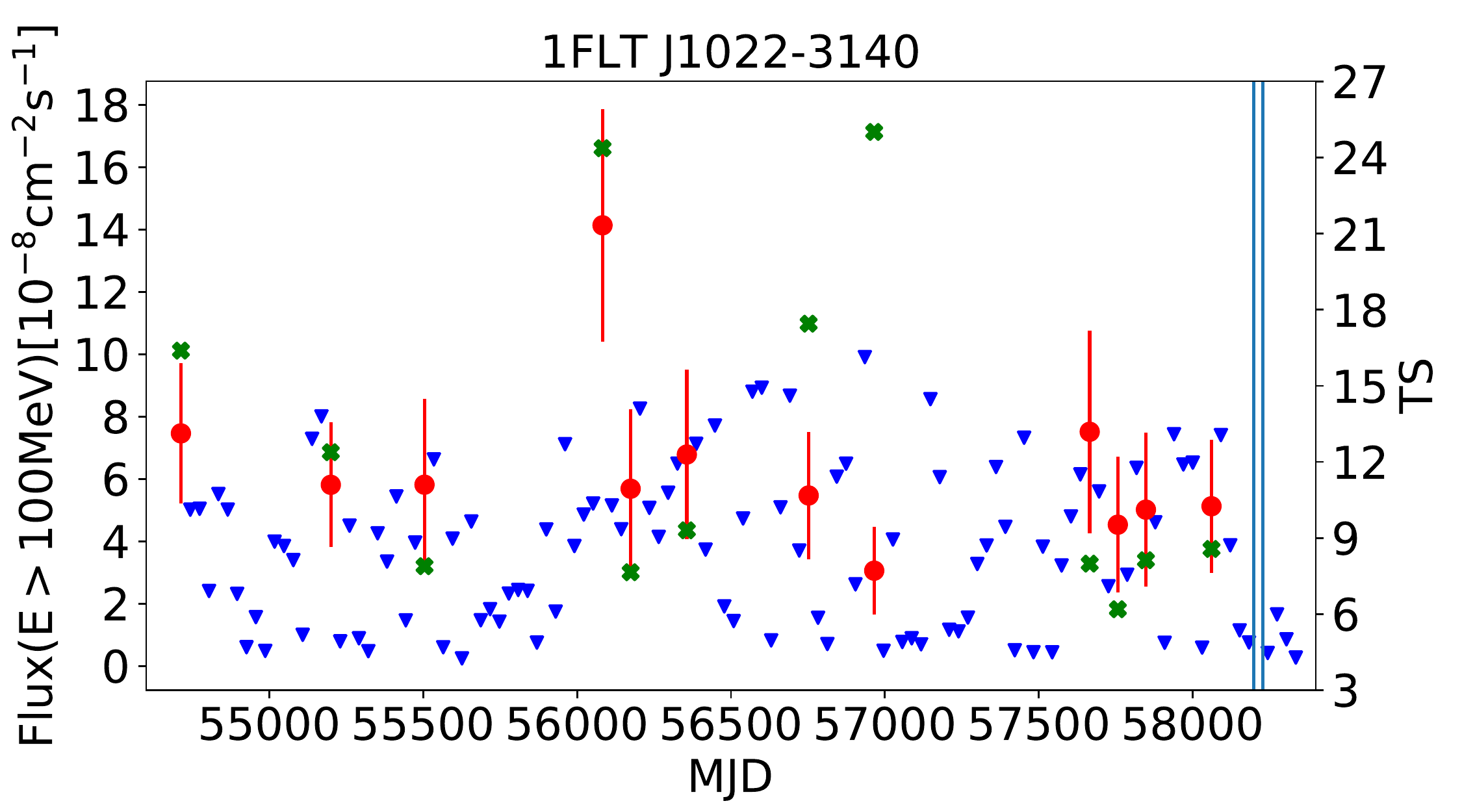}\label{fig:1FLTJ1022-3140}&
  \includegraphics[width=0.35\textwidth]{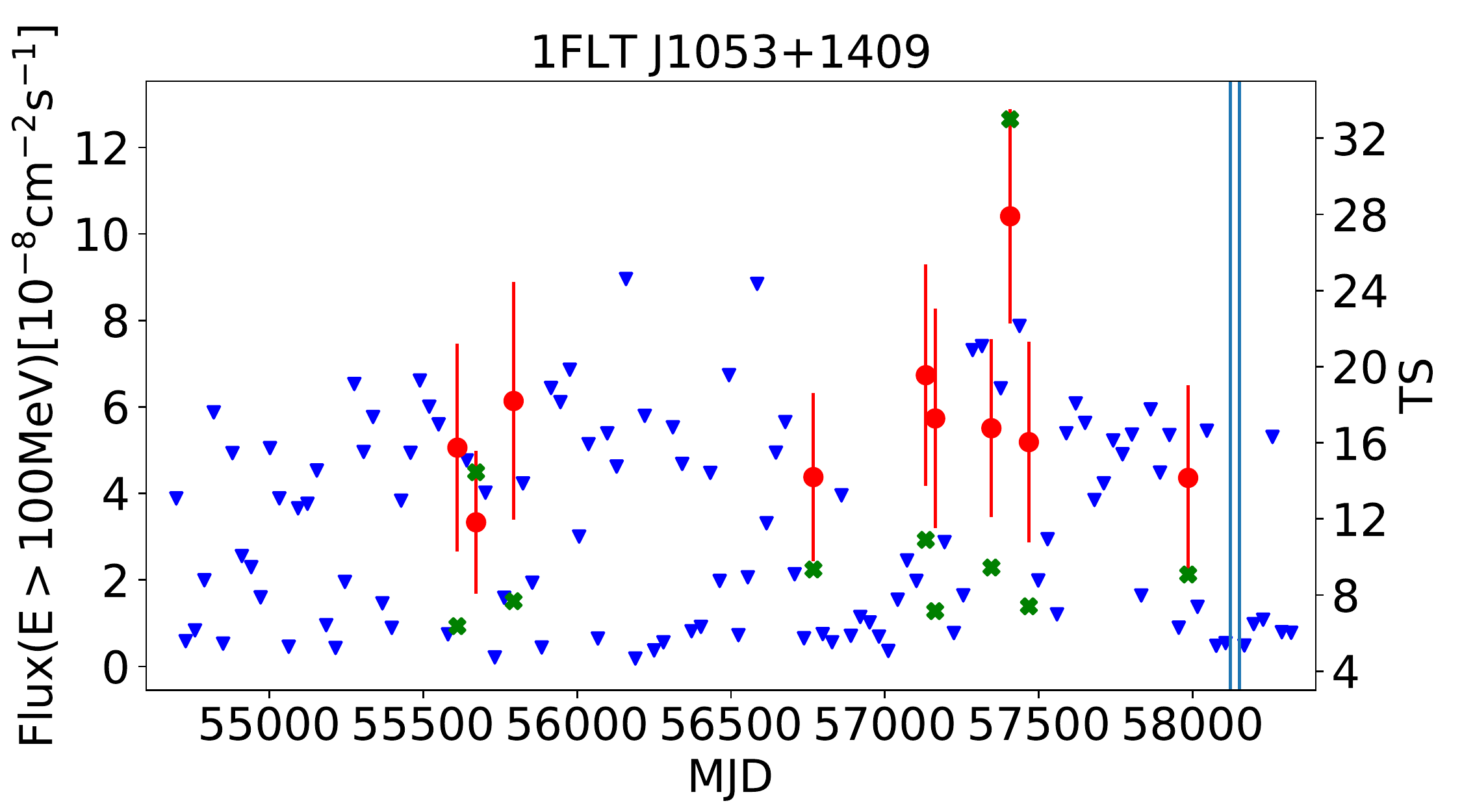}\label{fig:1FLTJ1053+1409}\\
  %[1FLTJ1019-3216]&%[1FLTJ1022-3140]&%[1FLTJ1053+1409]\\
  \includegraphics[width=0.35\textwidth]{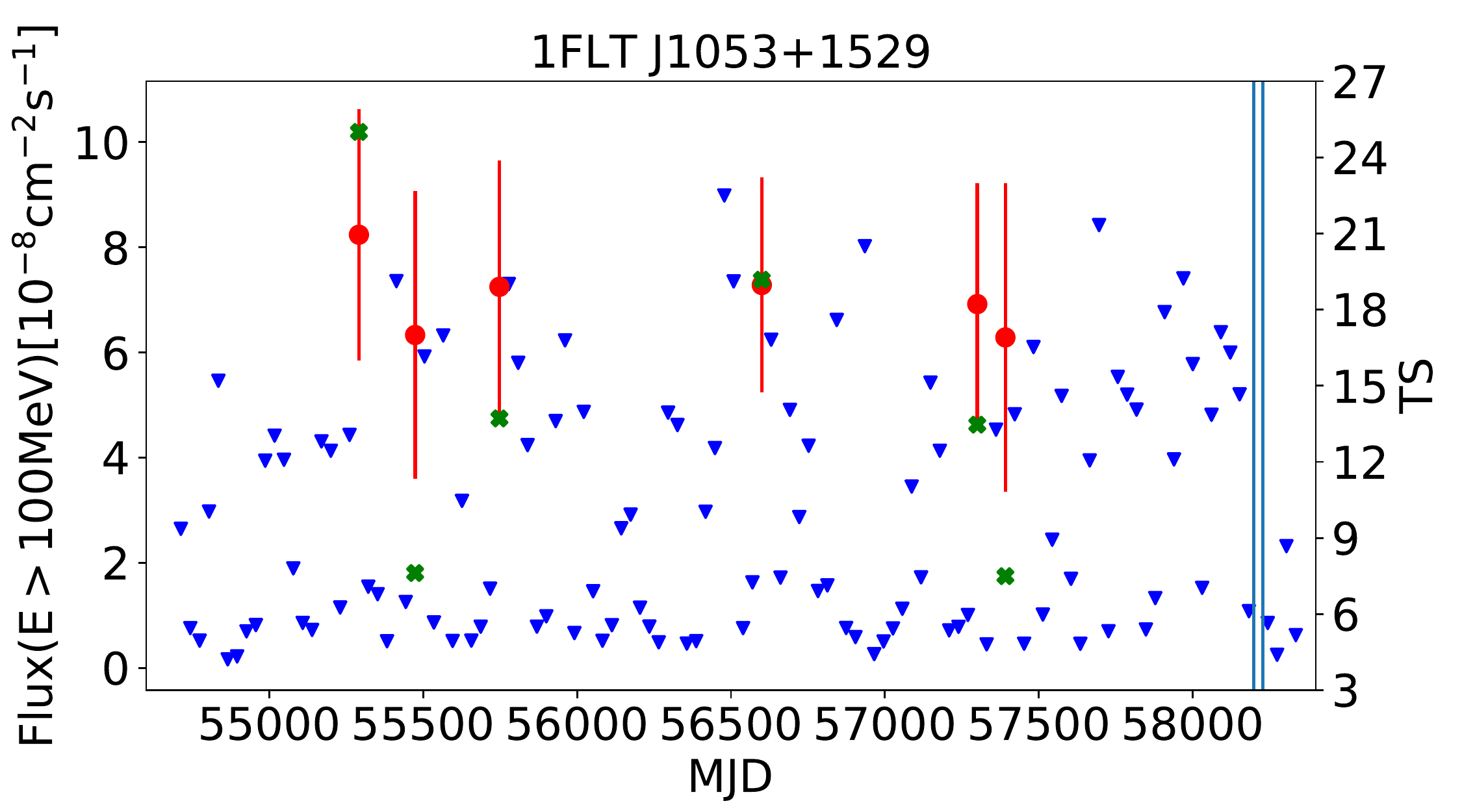}\label{fig:1FLTJ1053+1529}&
  \includegraphics[width=0.35\textwidth]{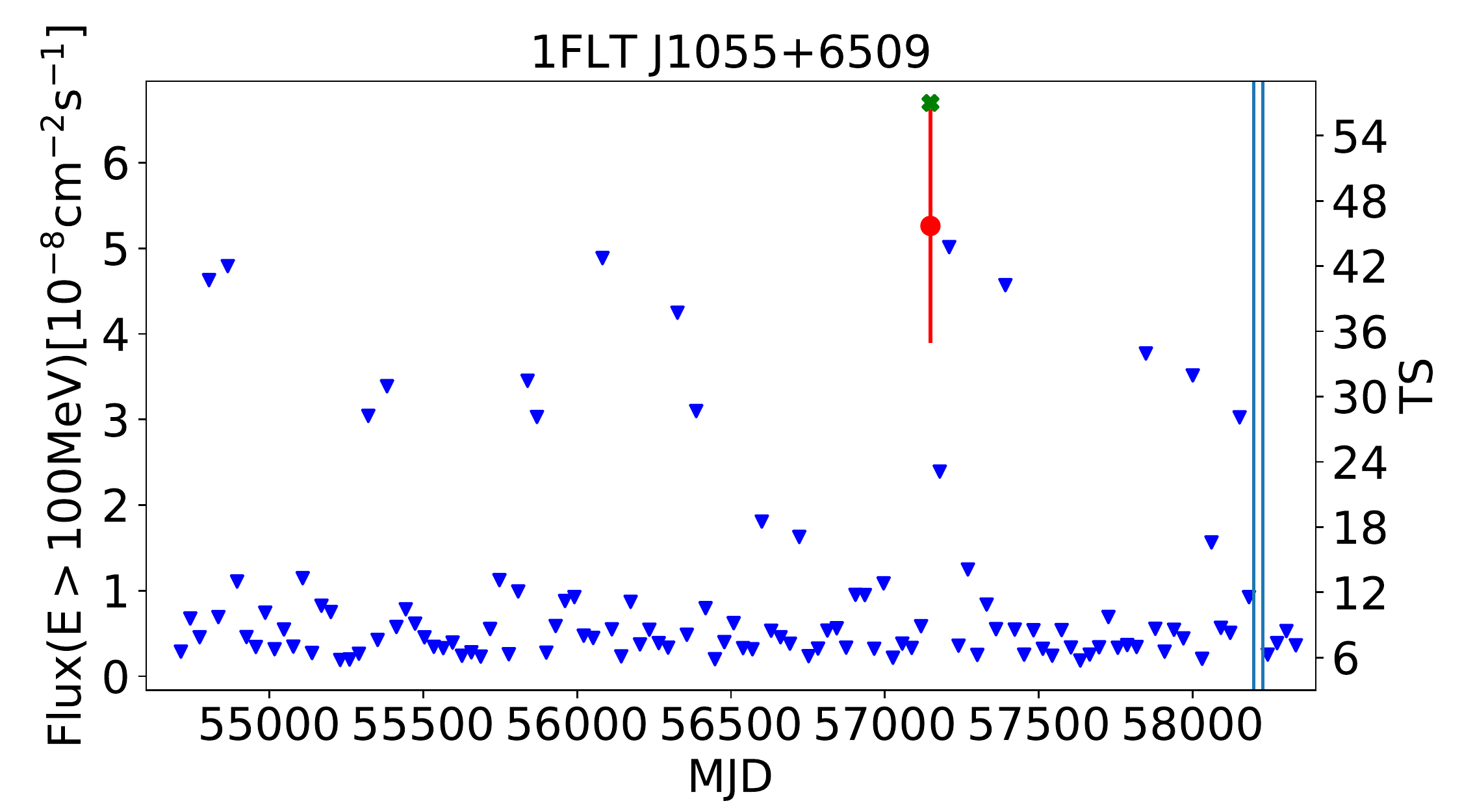}\label{fig:1FLTJ1055+6509}&
  \includegraphics[width=0.35\textwidth]{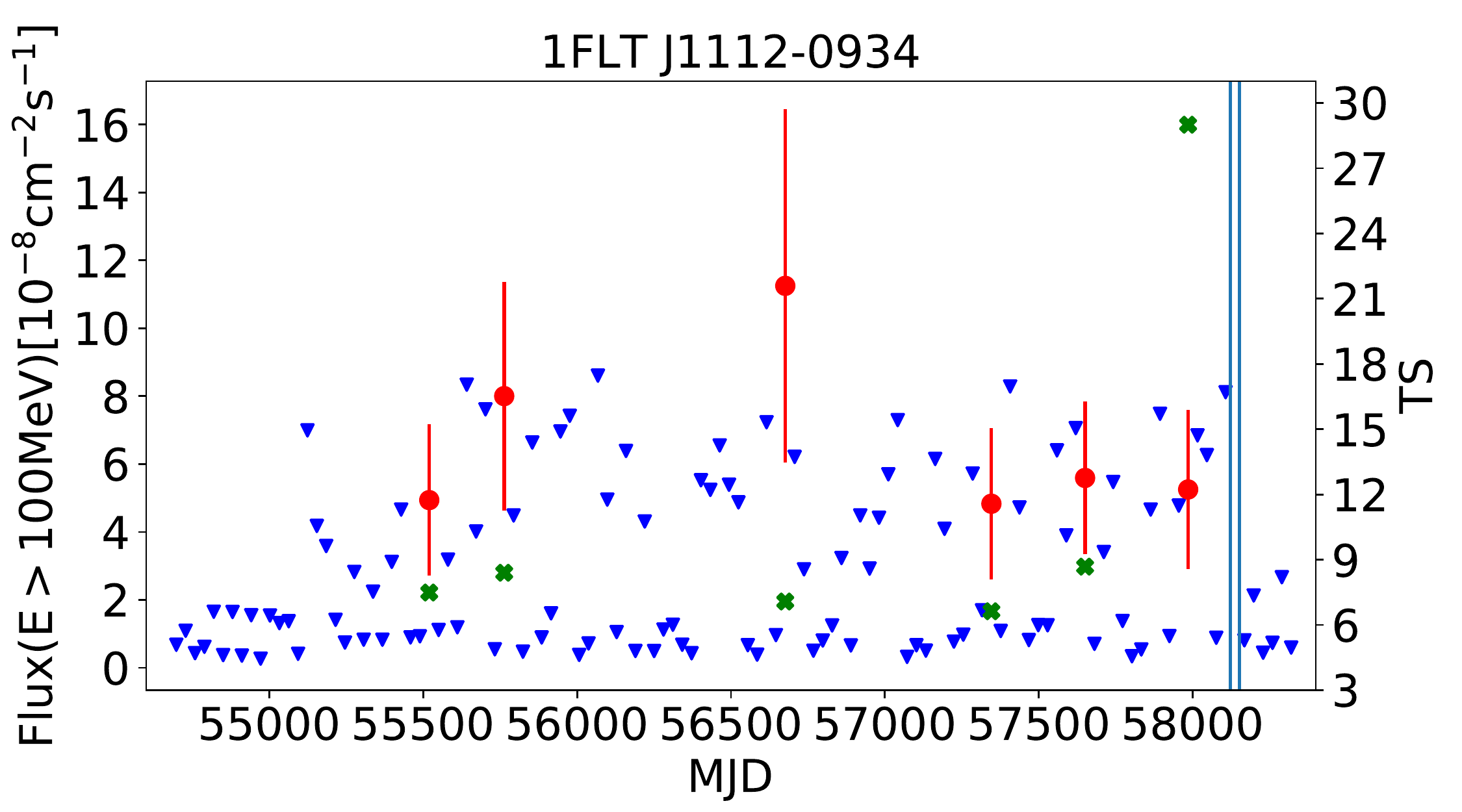}\label{fig:1FLTJ1112-0934}\\
  %[1FLTJ1053+1529]&%[1FLTJ1055+6509]&%[1FLTJ1112-0934]\\
  \includegraphics[width=0.35\textwidth]{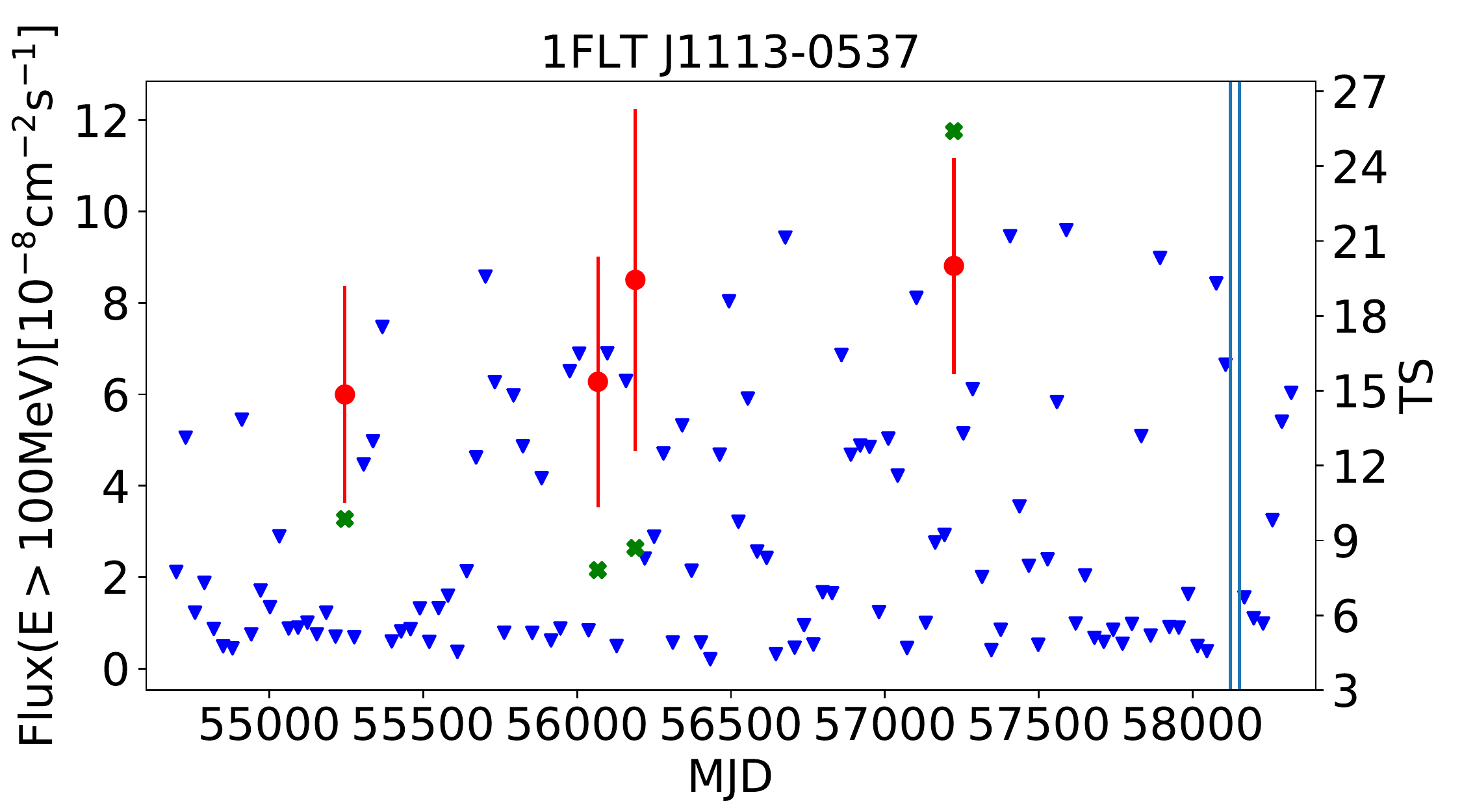}\label{fig:1FLTJ1113-0537}&
  \includegraphics[width=0.35\textwidth]{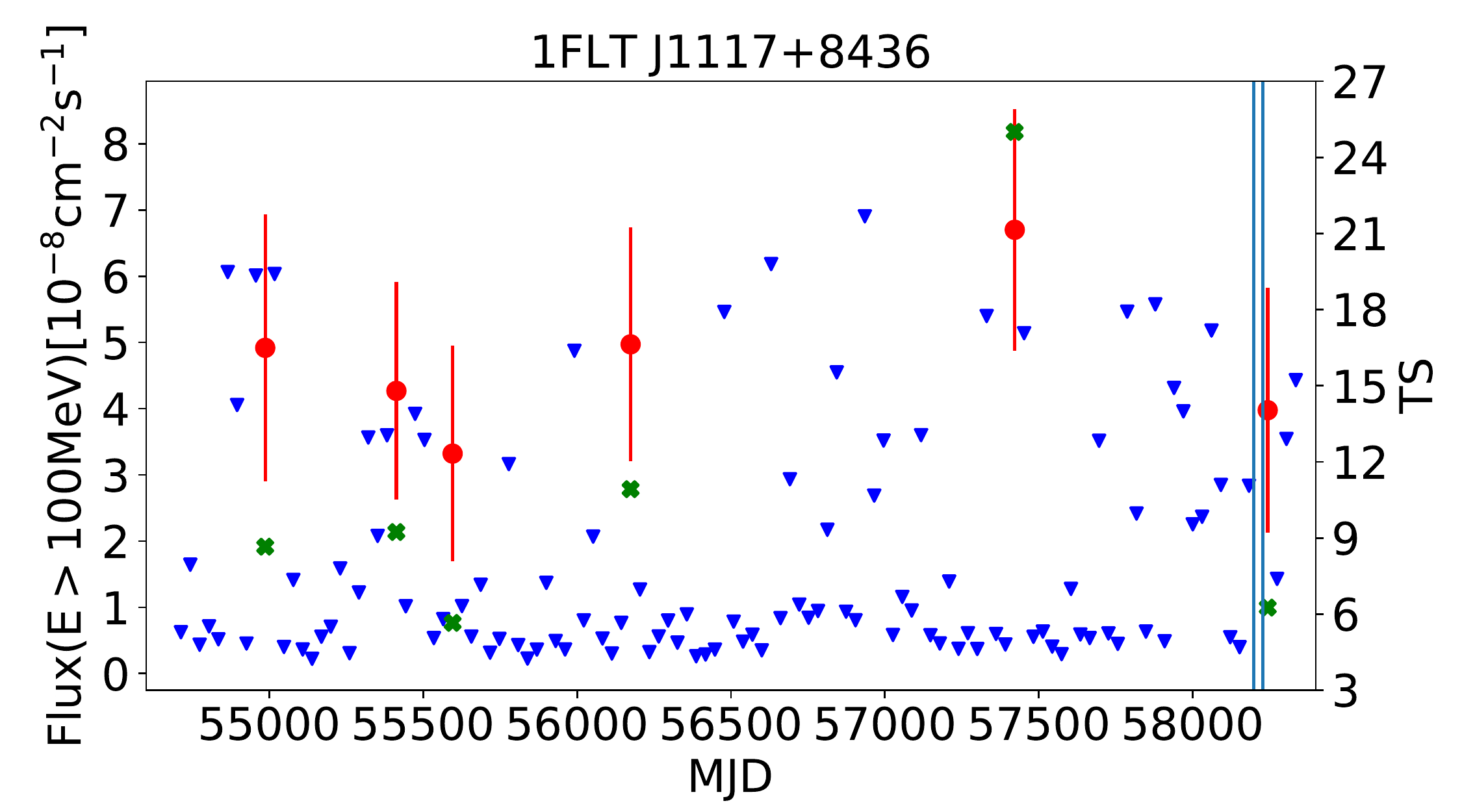}\label{fig:1FLTJ1117+8436}&
  \includegraphics[width=0.35\textwidth]{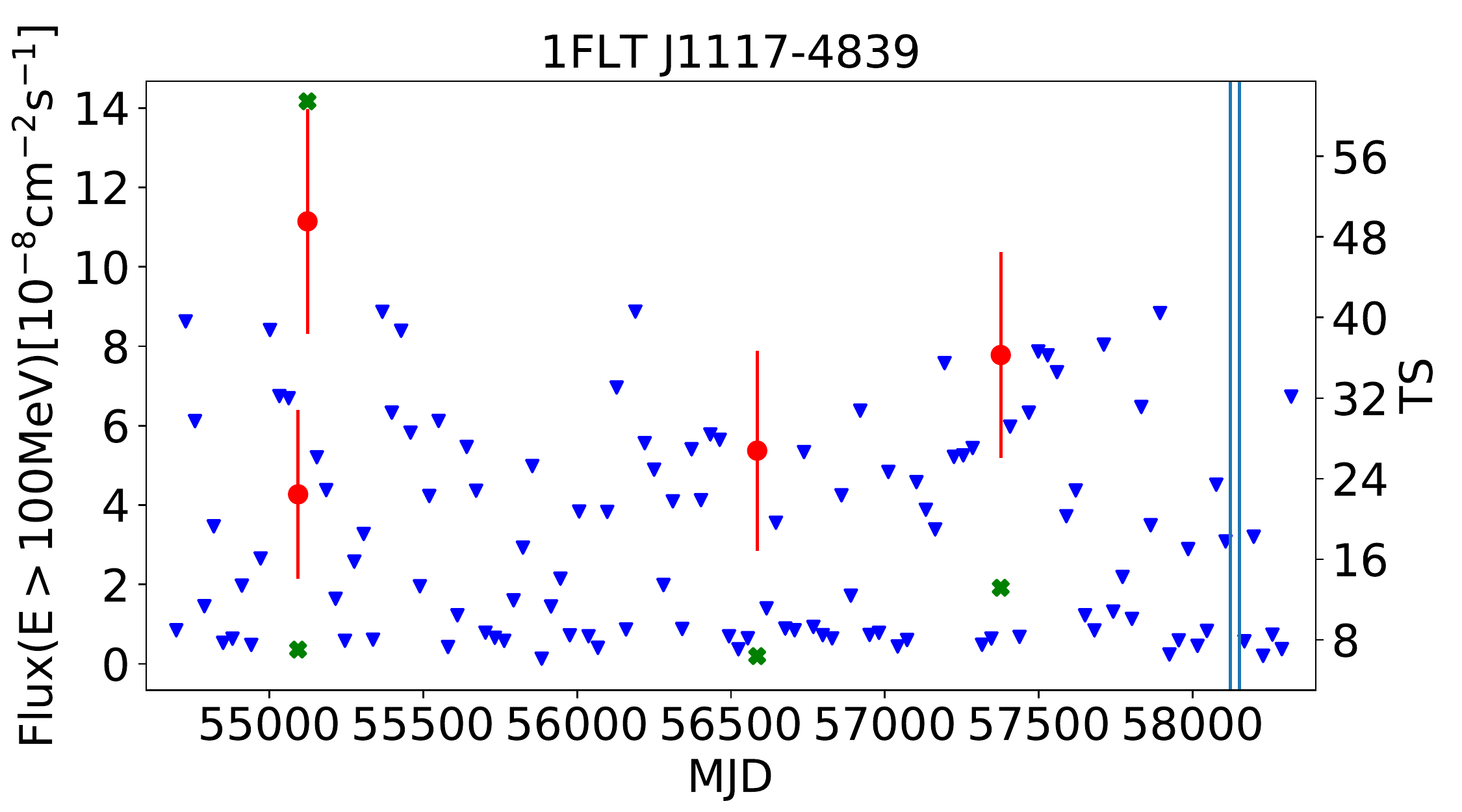}\label{fig:1FLTJ1117-4839}\\
  %[1FLTJ1113-0537]&%[1FLTJ1117+8436]&%[1FLTJ1117-4839]\\
  \includegraphics[width=0.35\textwidth]{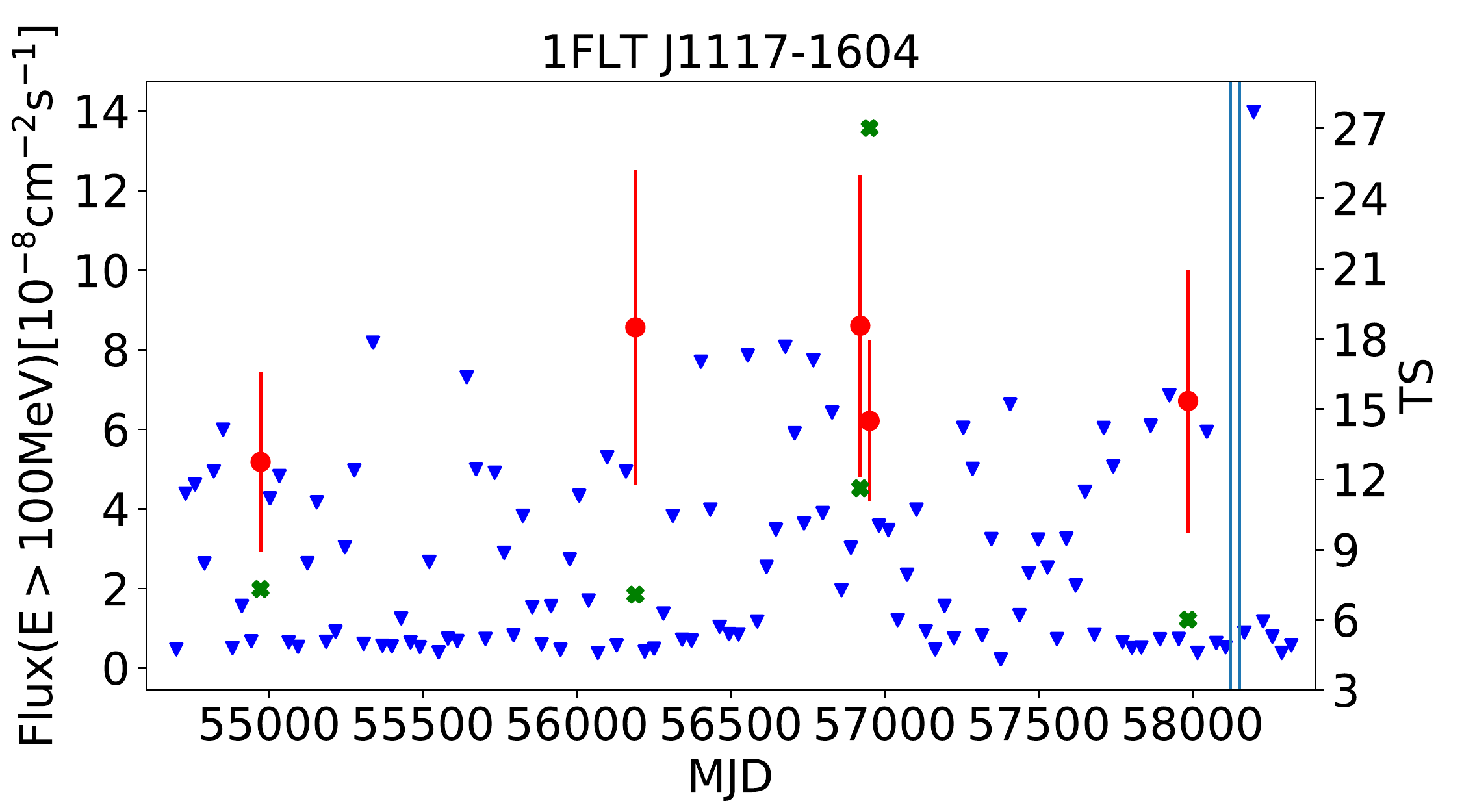}\label{fig:1FLTJ1117-1604}&
  \includegraphics[width=0.35\textwidth]{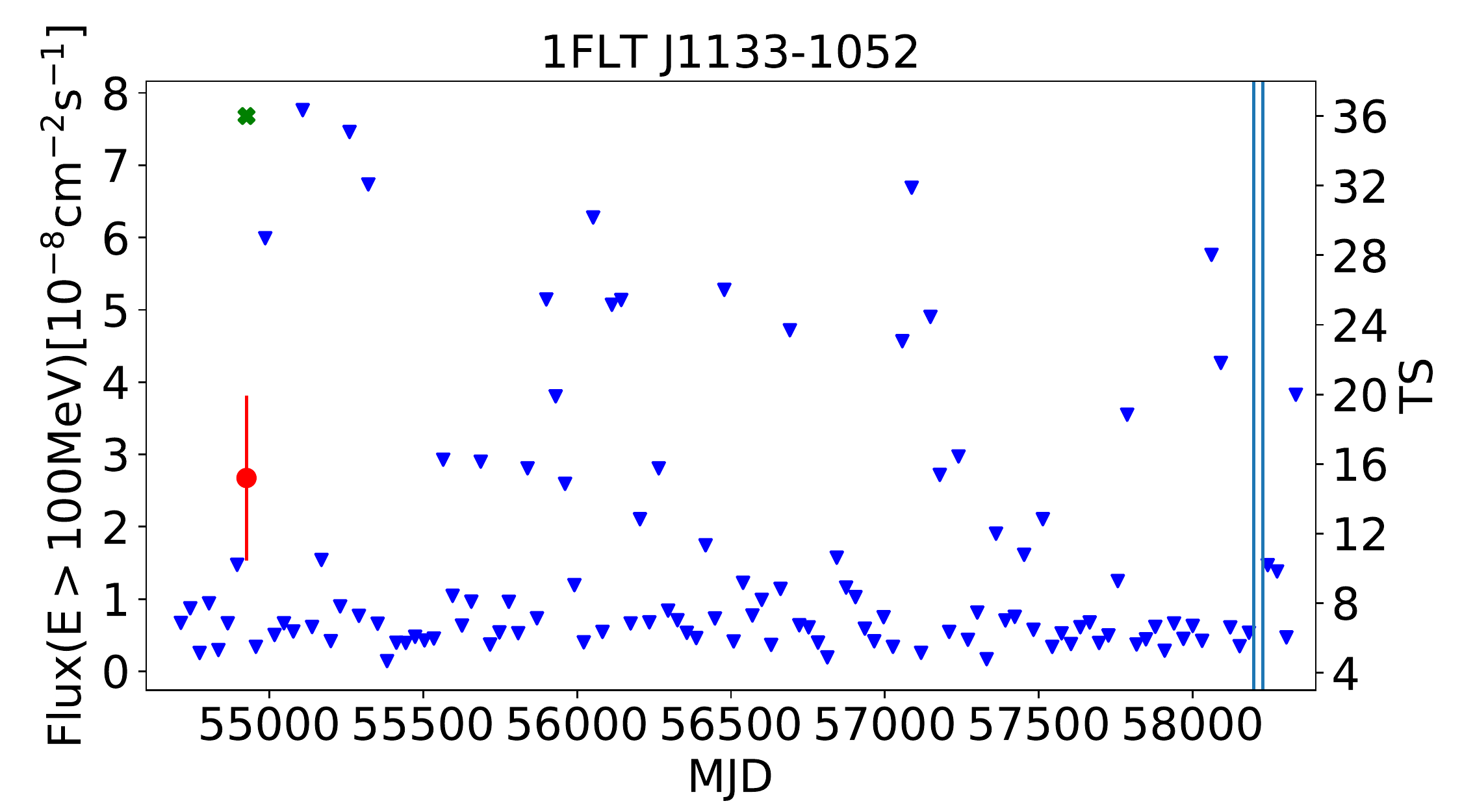}\label{fig:1FLTJ1133-1052}&
  \includegraphics[width=0.35\textwidth]{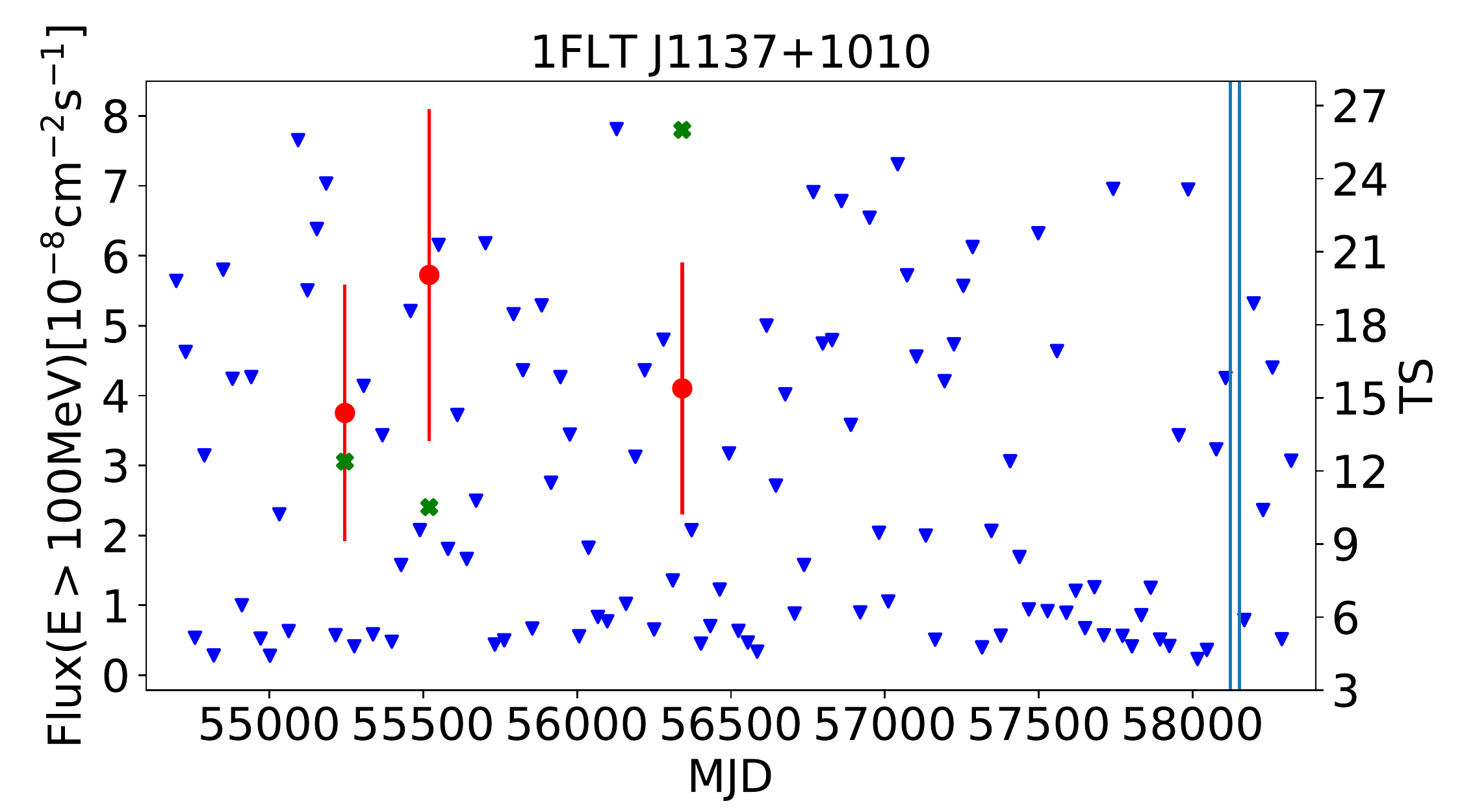}\label{fig:1FLTJ1137+1010}\\
  %[1FLTJ1117-1604]&%[1FLTJ1133-1052]&%[1FLTJ1137+1010]\\
  \includegraphics[width=0.35\textwidth]{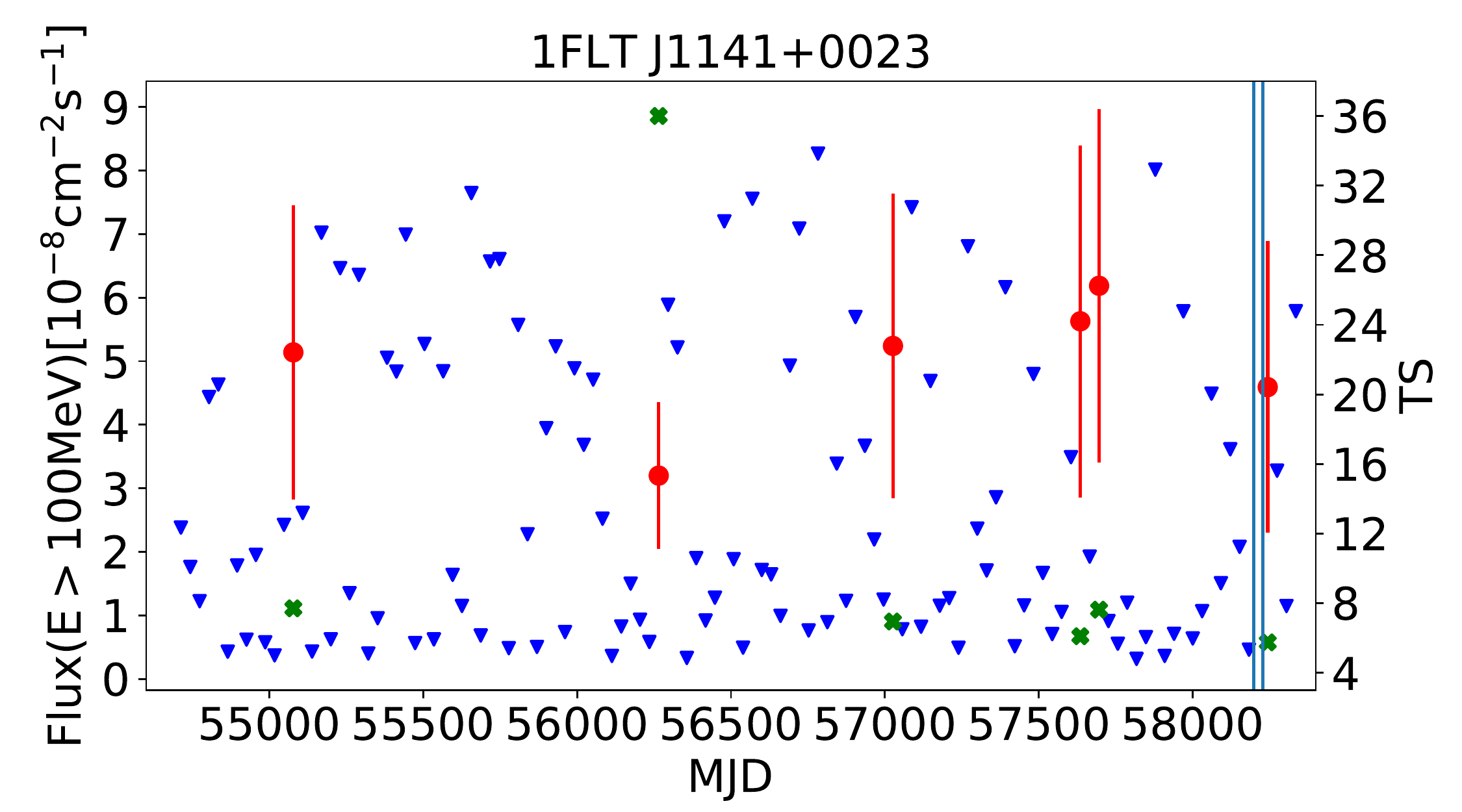}\label{fig:1FLTJ1141+0023}&
  \includegraphics[width=0.35\textwidth]{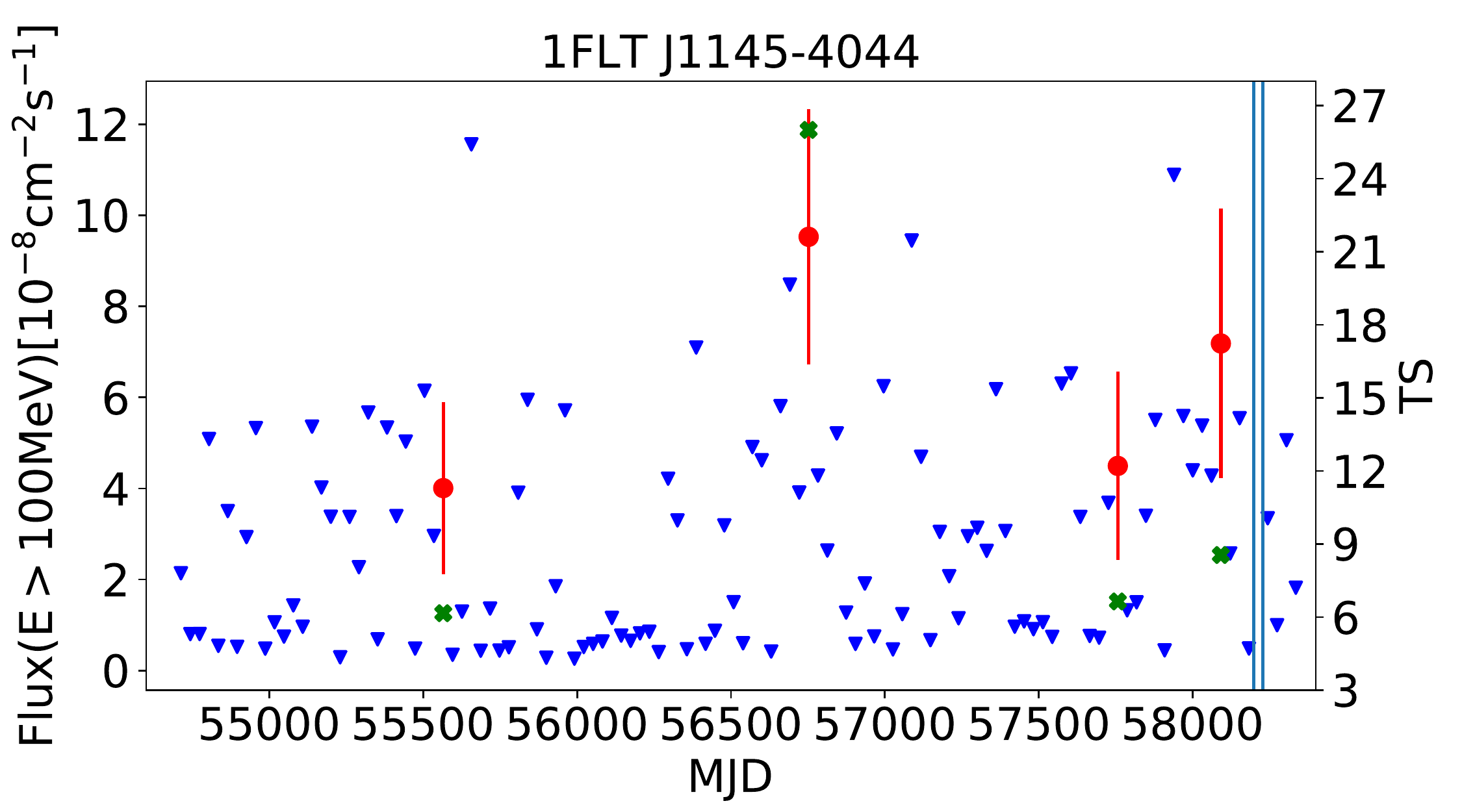}\label{fig:1FLTJ1145-4044}&
  \includegraphics[width=0.35\textwidth]{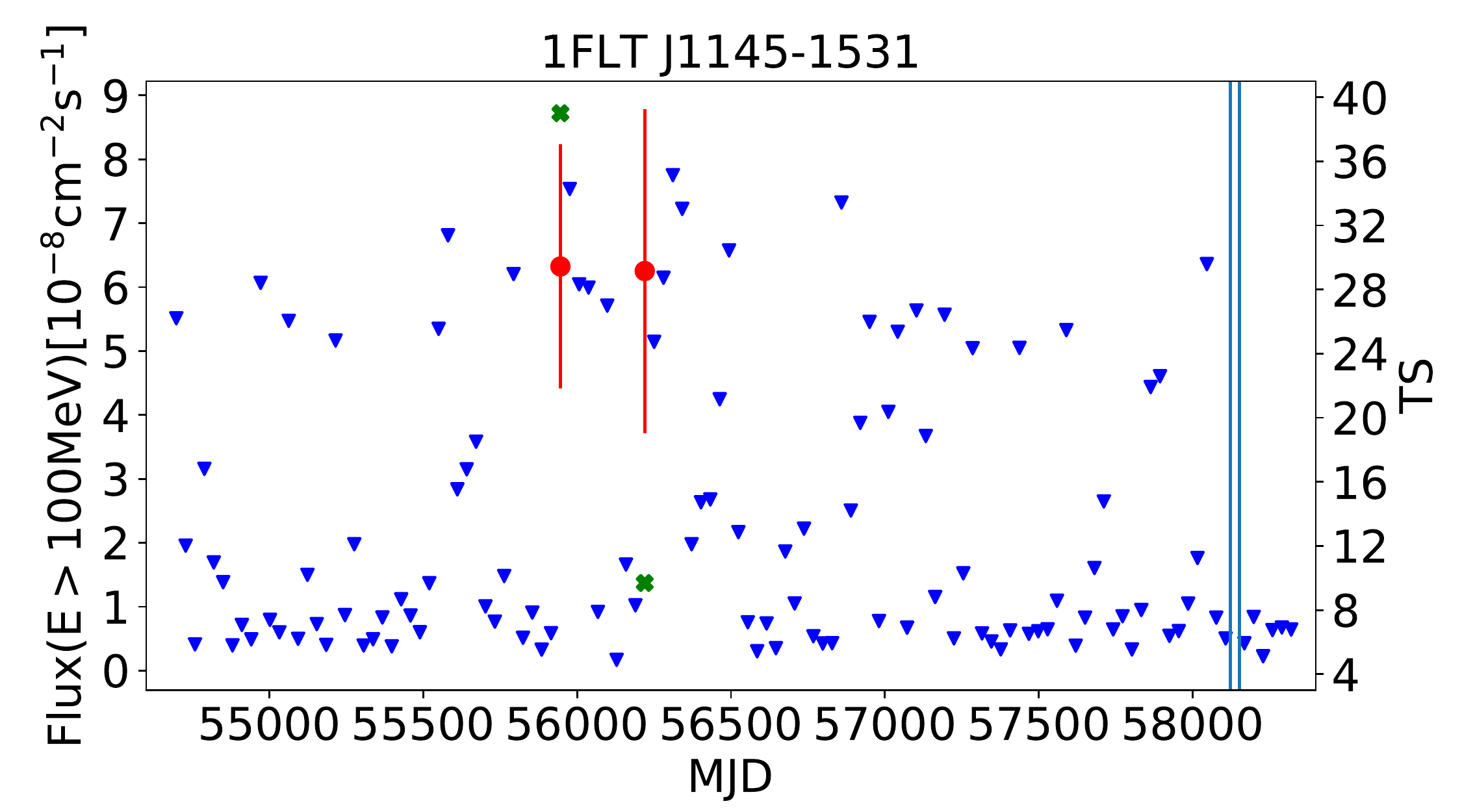}\label{fig:1FLTJ1145-1531}\\
  %[1FLTJ1141+0023]&%[1FLTJ1145-4044]&%[1FLTJ1145-1531]\\
\end{tabular}
\end{figure}
\begin{figure}[!t]
	\centering            
	\ContinuedFloat
\setlength\tabcolsep{0.0pt}
\begin{tabular}{ccc}
  \includegraphics[width=0.35\textwidth]{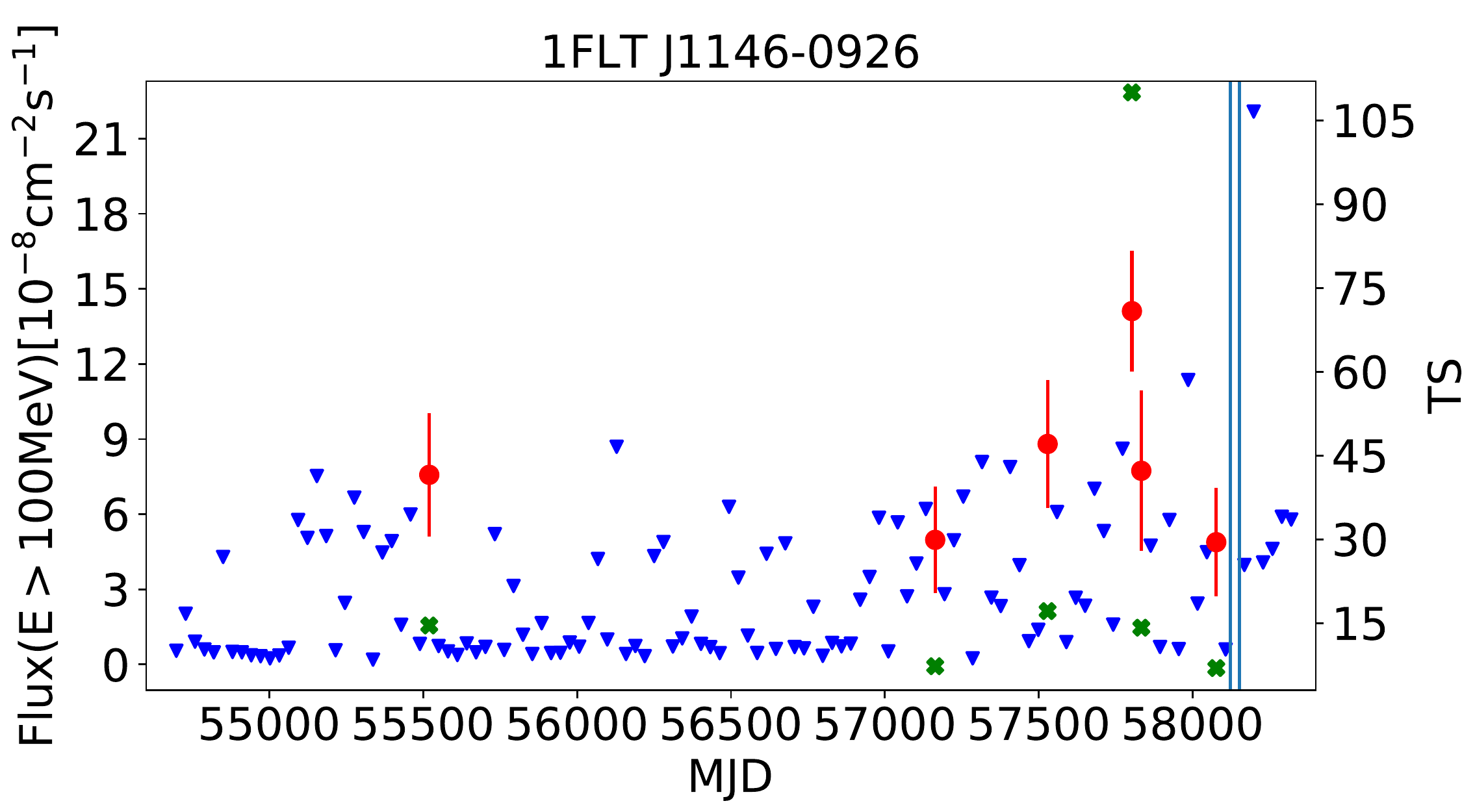}\label{fig:1FLTJ1146-0926}&
  \includegraphics[width=0.35\textwidth]{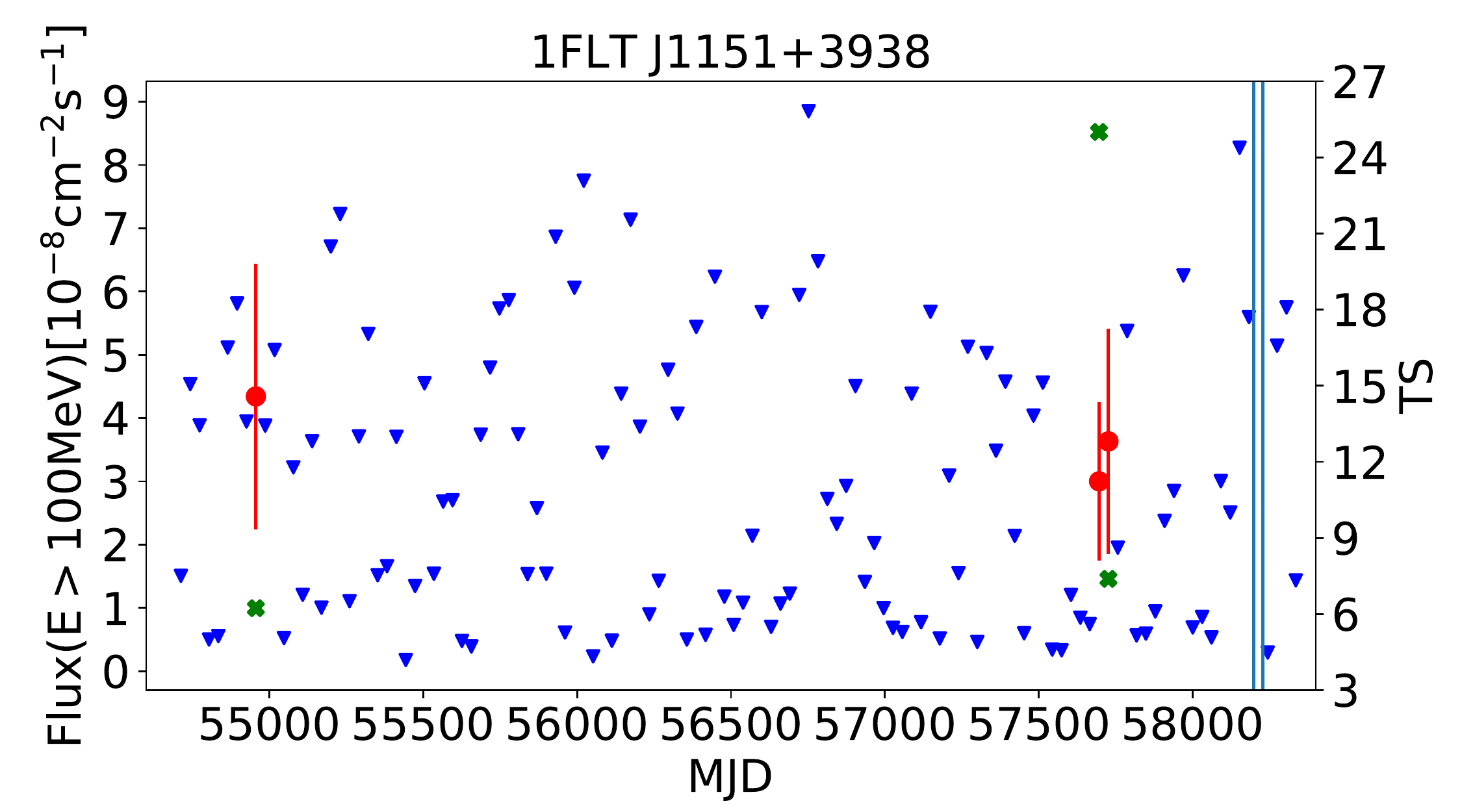}\label{fig:1FLTJ1151+3938}&
  \includegraphics[width=0.35\textwidth]{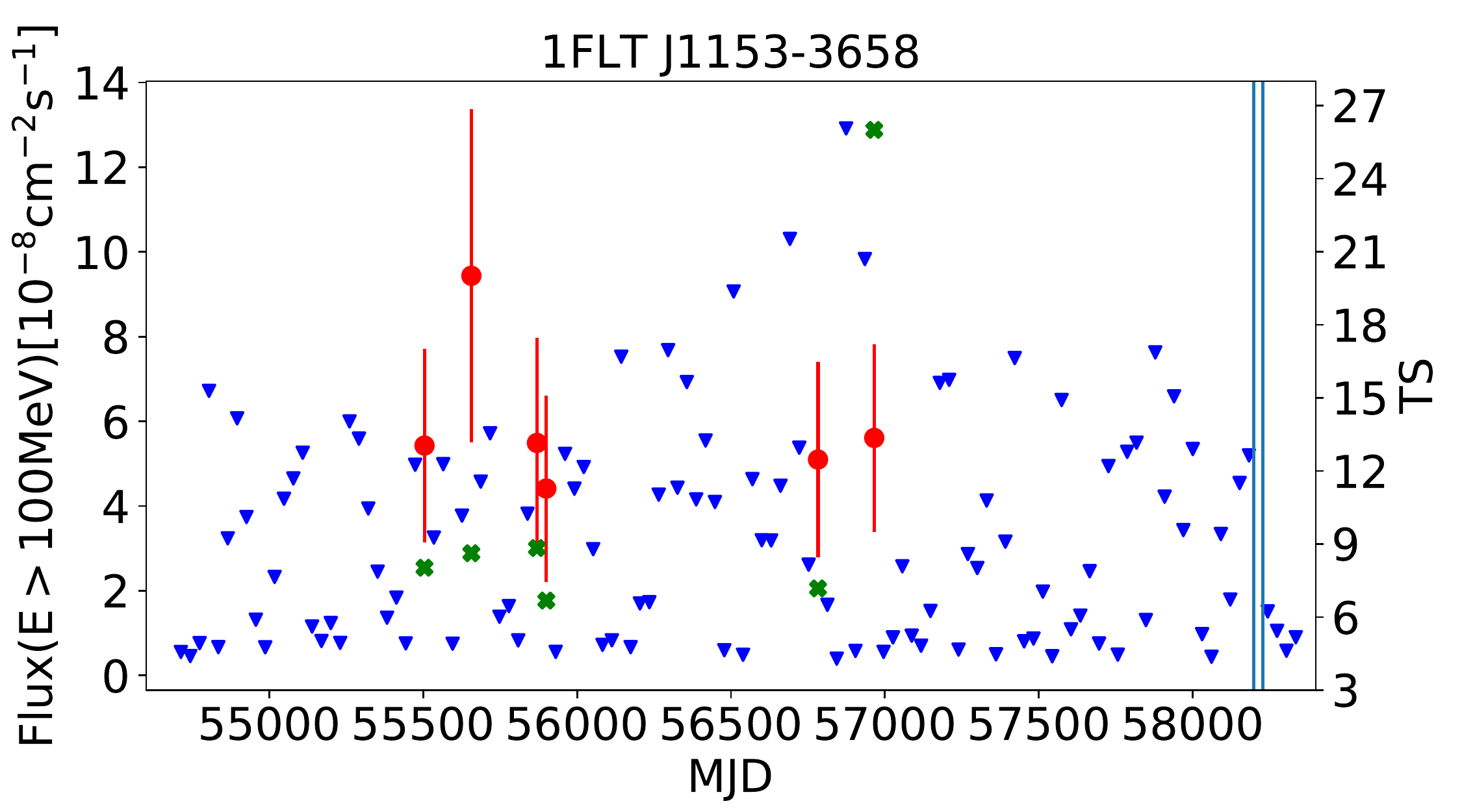}\label{fig:1FLTJ1153-3658}\\
  %[1FLTJ1146-0926]&%[1FLTJ1151+3938]&%[1FLTJ1153-3658]
  \includegraphics[width=0.35\textwidth]{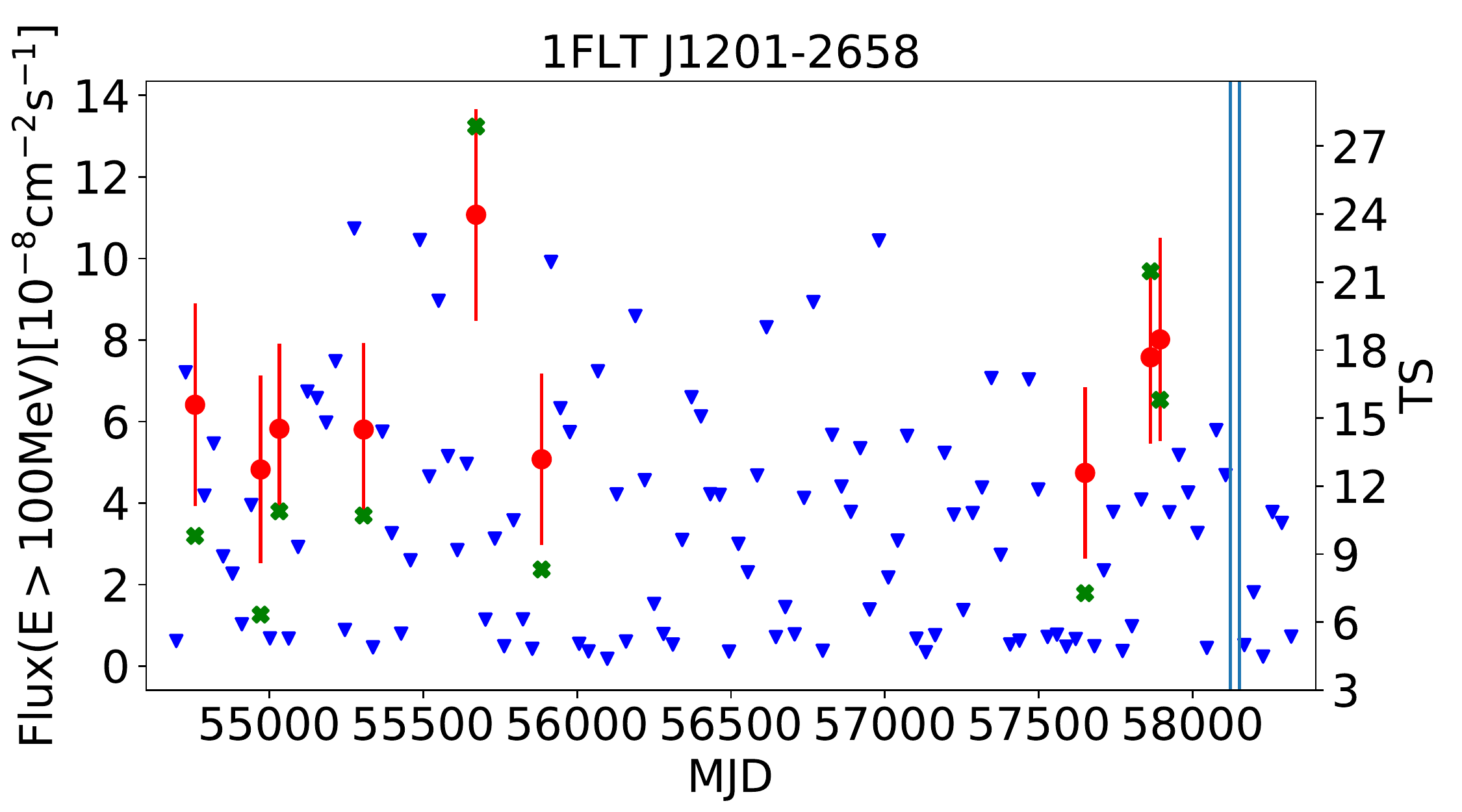}\label{fig:1FLTJ1201-2658}&
  \includegraphics[width=0.35\textwidth]{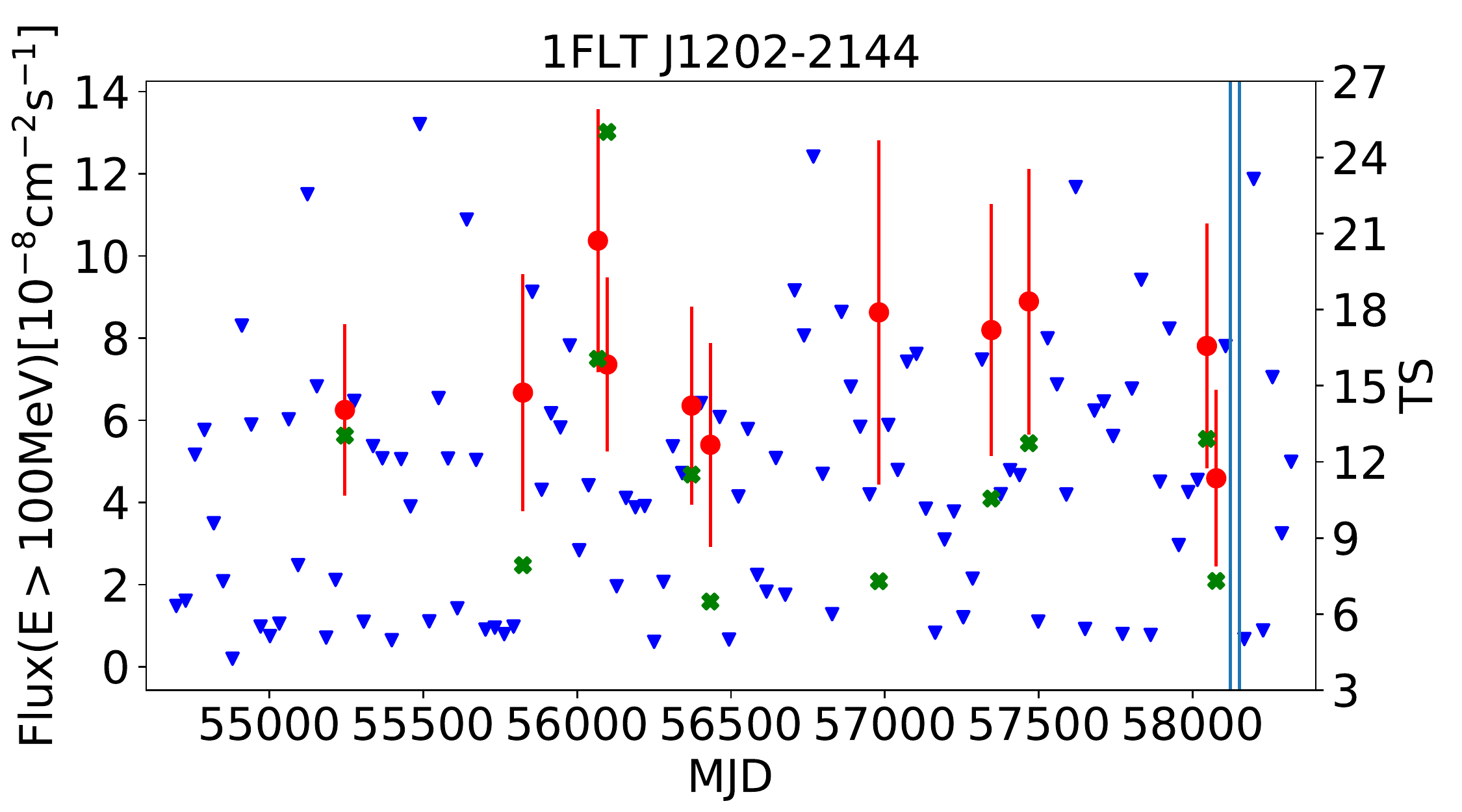}\label{fig:1FLTJ1202-2144}&
  \includegraphics[width=0.35\textwidth]{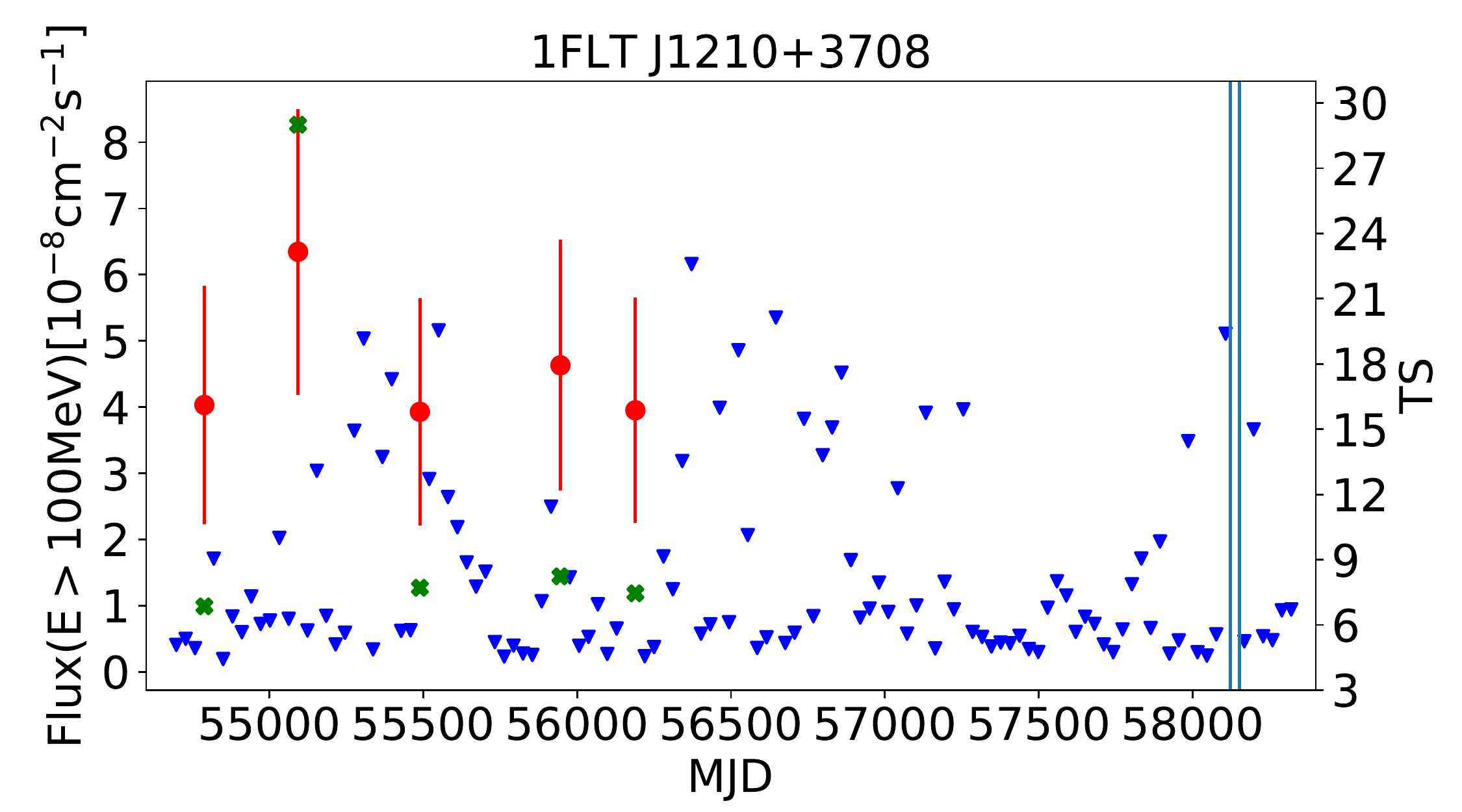}\label{fig:1FLTJ1210+3708}\\
  %[1FLTJ1201-2658]&%[1FLTJ1202-2144]&%[1FLTJ1210+3708]\\
  \includegraphics[width=0.35\textwidth]{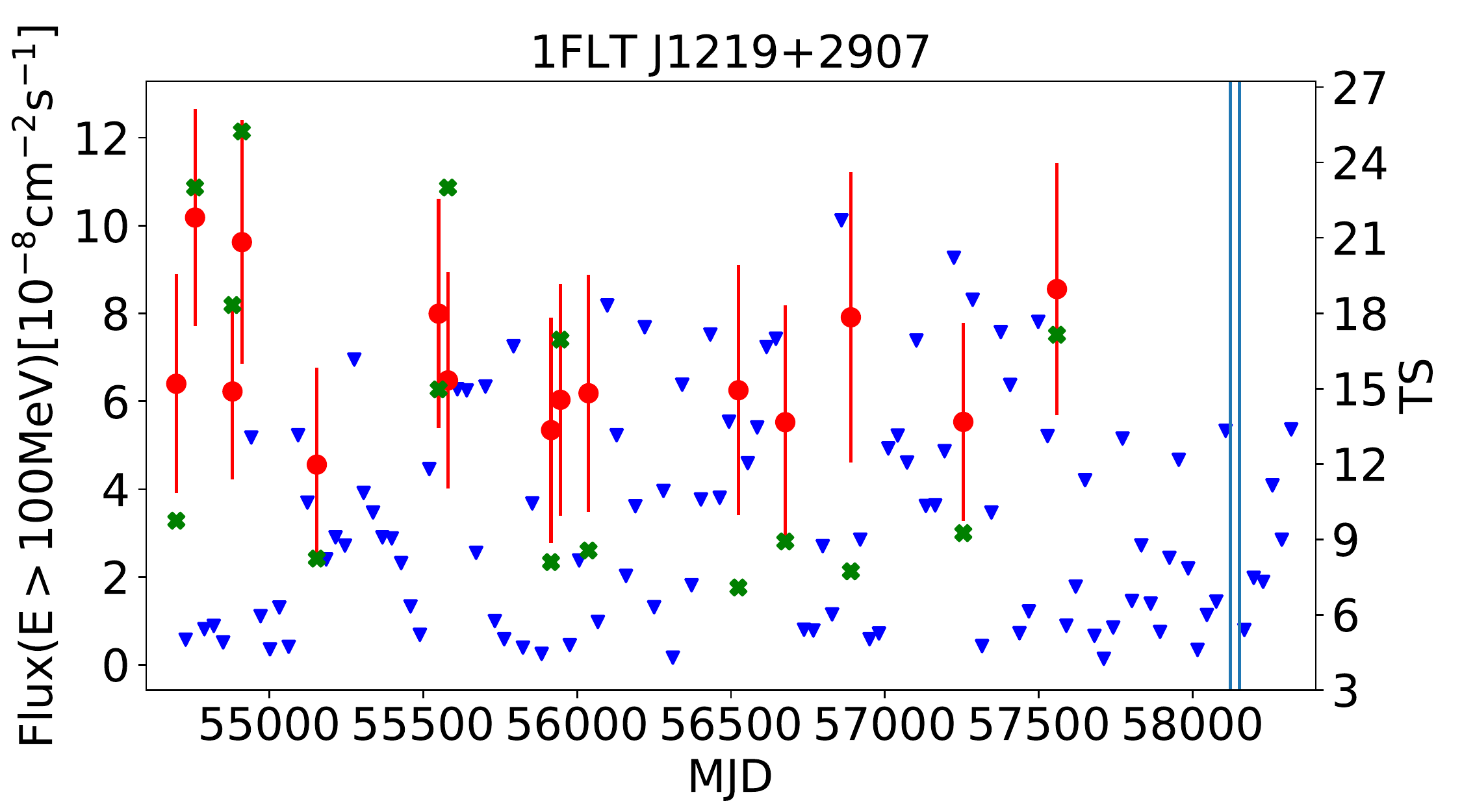}\label{fig:1FLTJ1219+2907}&
  \includegraphics[width=0.35\textwidth]{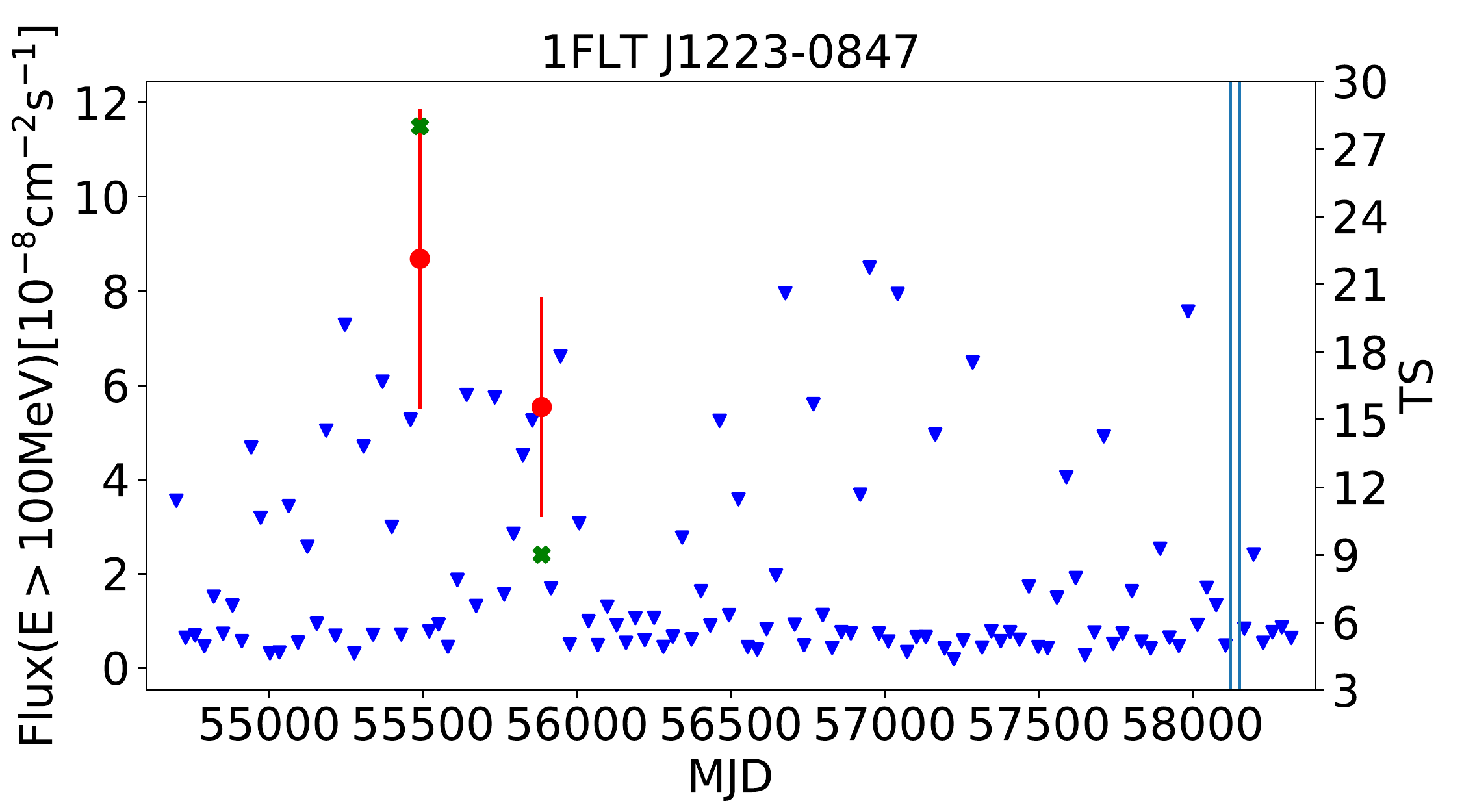}\label{fig:1FLTJ1223-0847}&
  \includegraphics[width=0.35\textwidth]{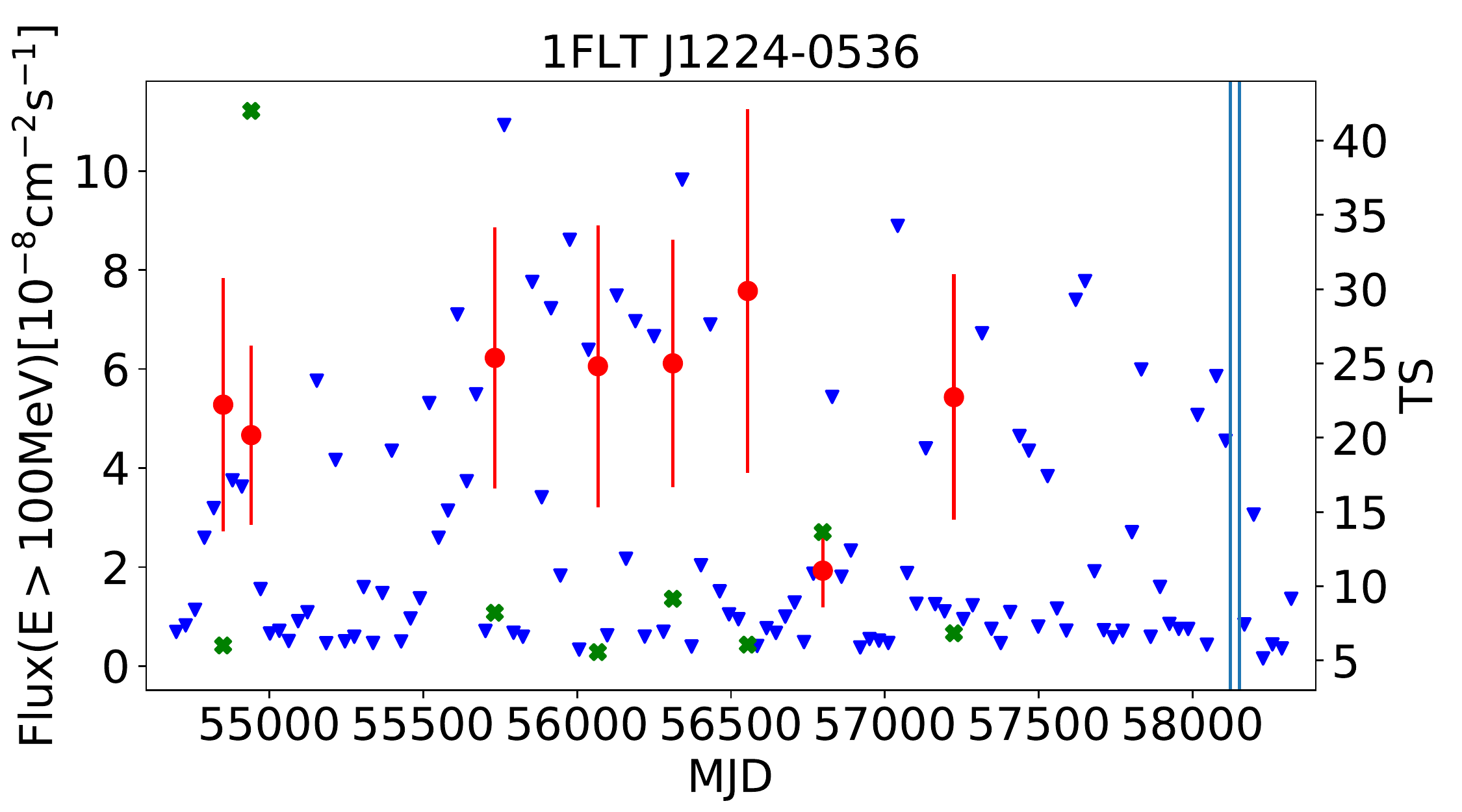}\label{fig:1FLTJ1224-0536}\\
  %[1FLTJ1219+2907]&%[1FLTJ1223-0847]&%[1FLTJ1224-0536]\\
  \includegraphics[width=0.35\textwidth]{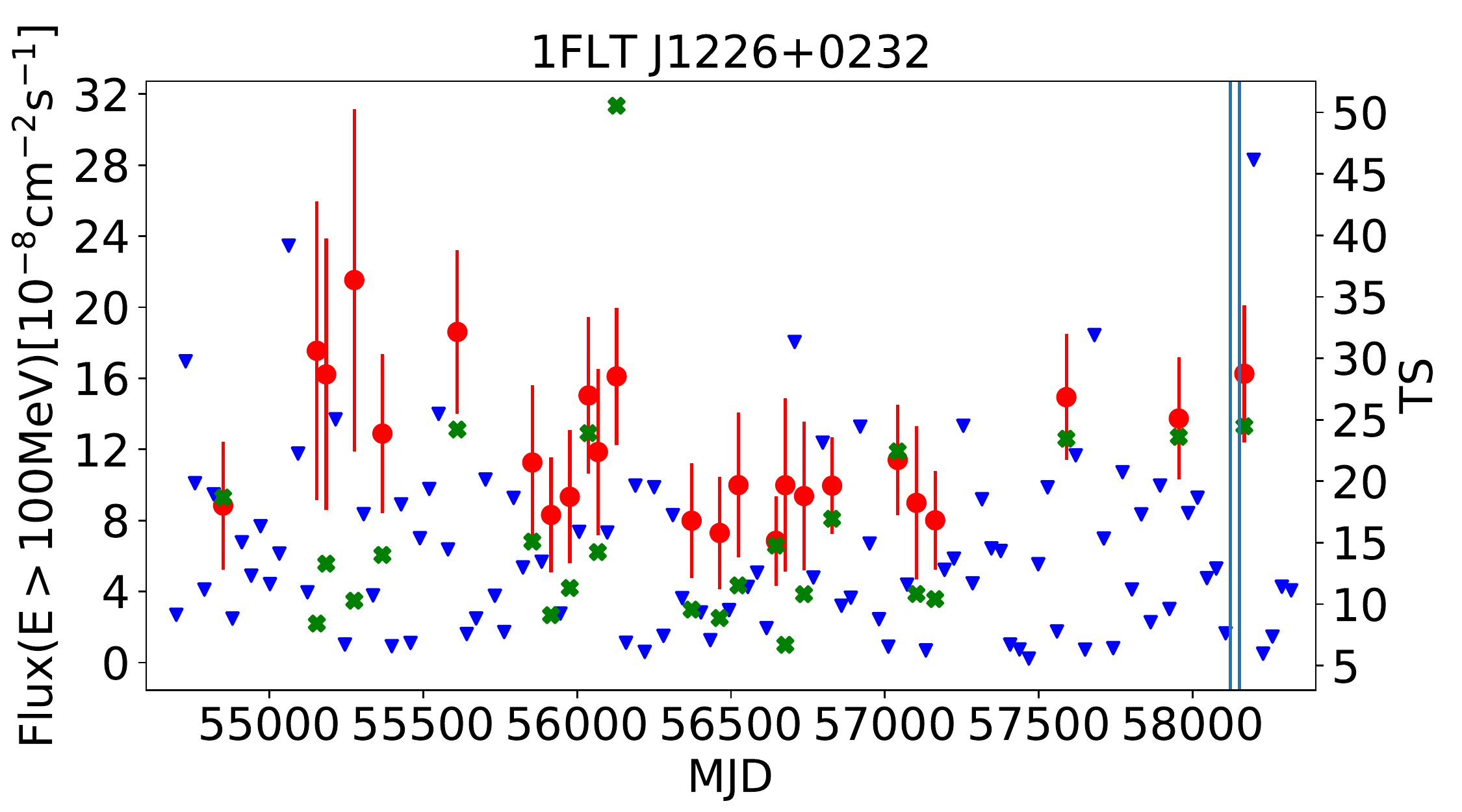}\label{fig:1FLTJ1226+0232}&
  \includegraphics[width=0.35\textwidth]{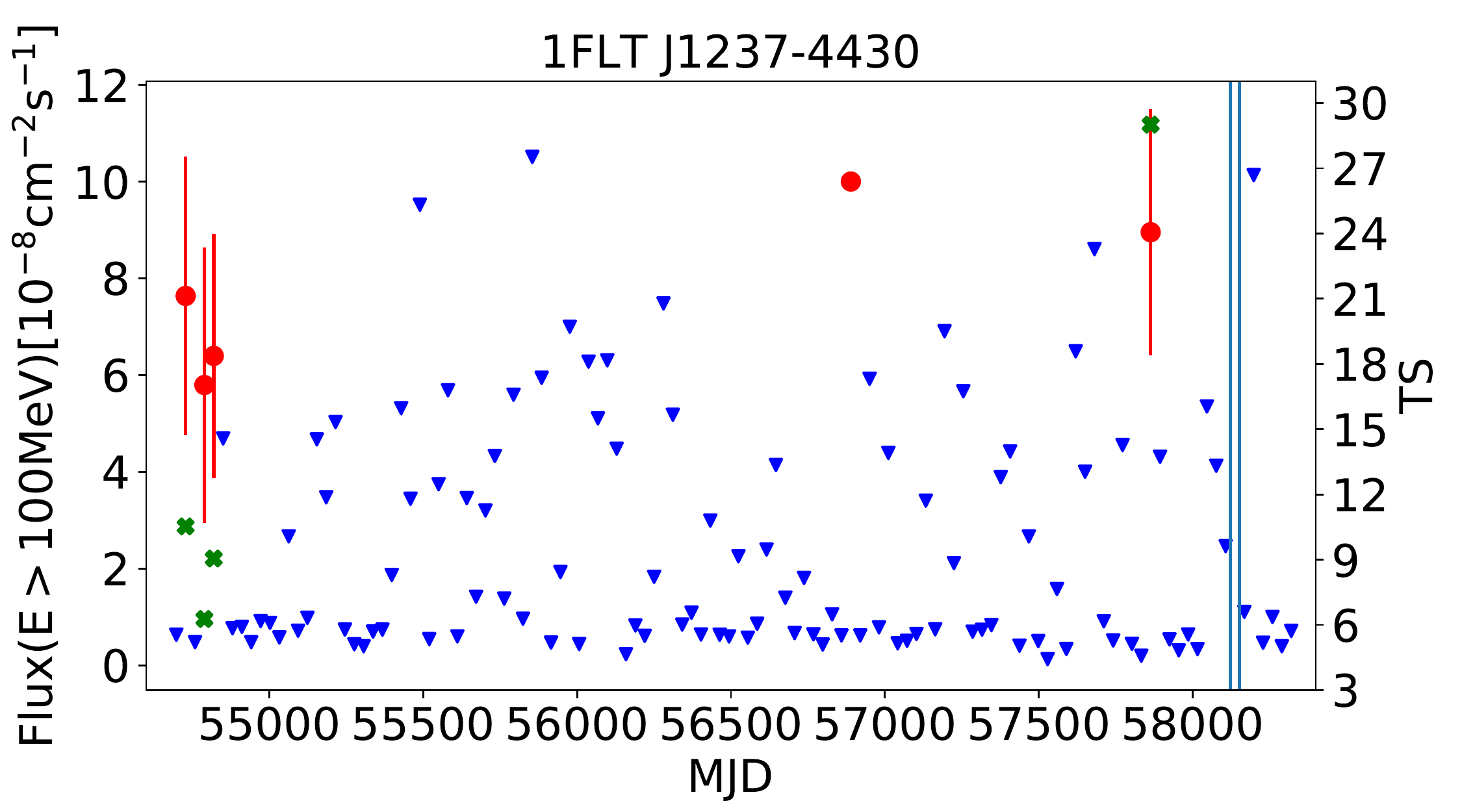}\label{fig:1FLTJ1237-4430}&
  \includegraphics[width=0.35\textwidth]{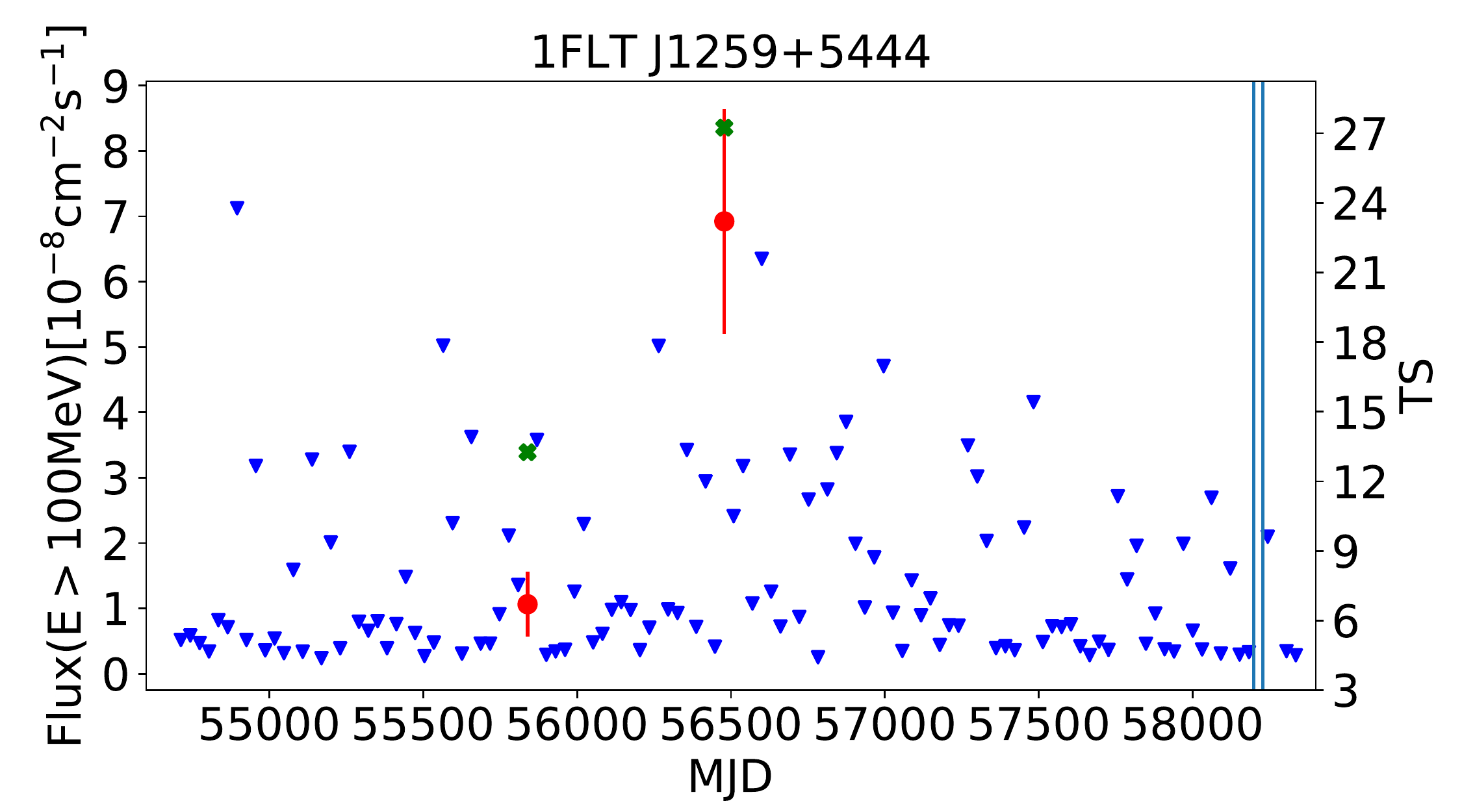}\label{fig:1FLTJ1259+5444}\\
  %[1FLTJ1226+0232]&%[1FLTJ1237-4430]&%[1FLTJ1259+5444]\\
  \includegraphics[width=0.35\textwidth]{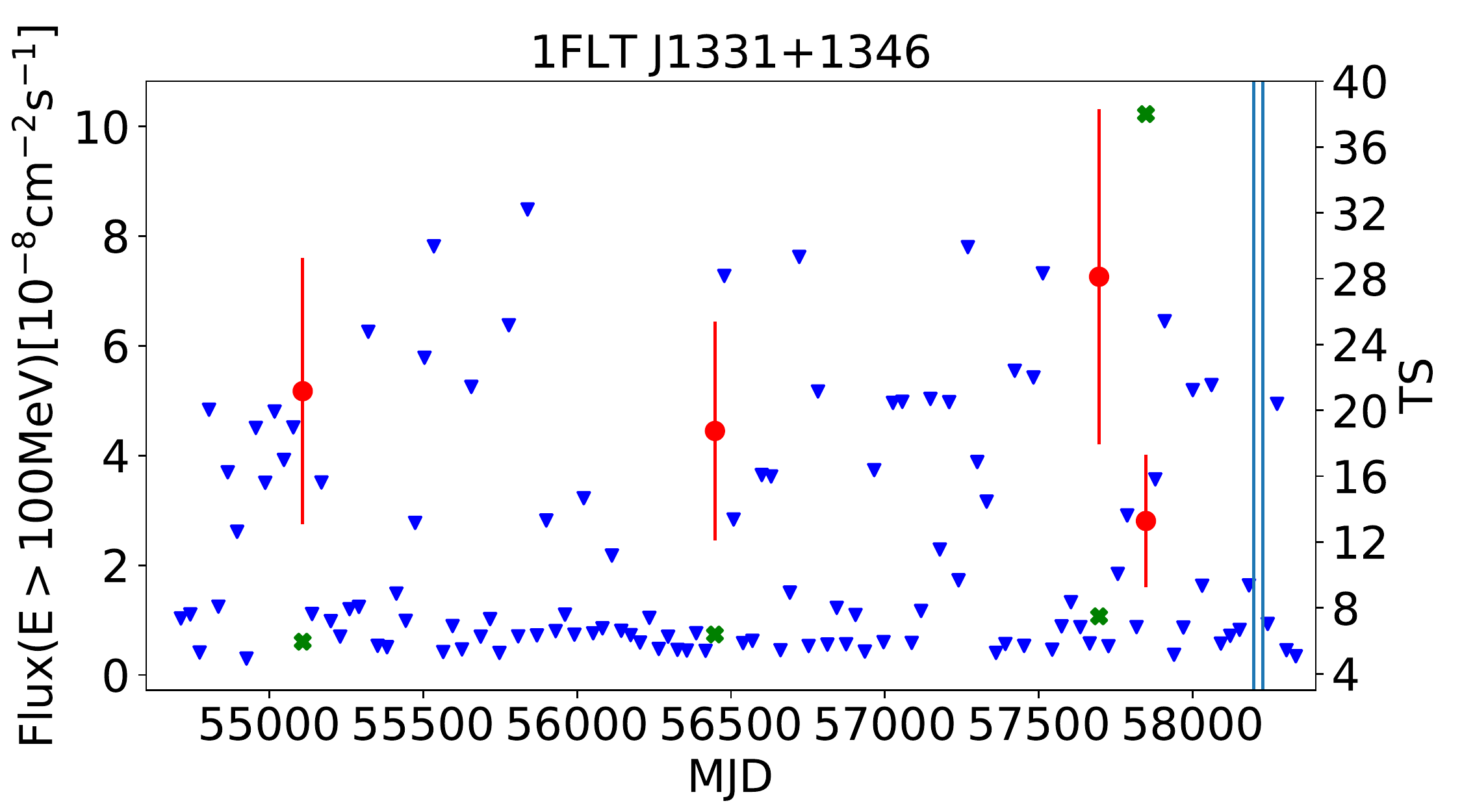}\label{fig:1FLTJ1331+1346}&
  \includegraphics[width=0.35\textwidth]{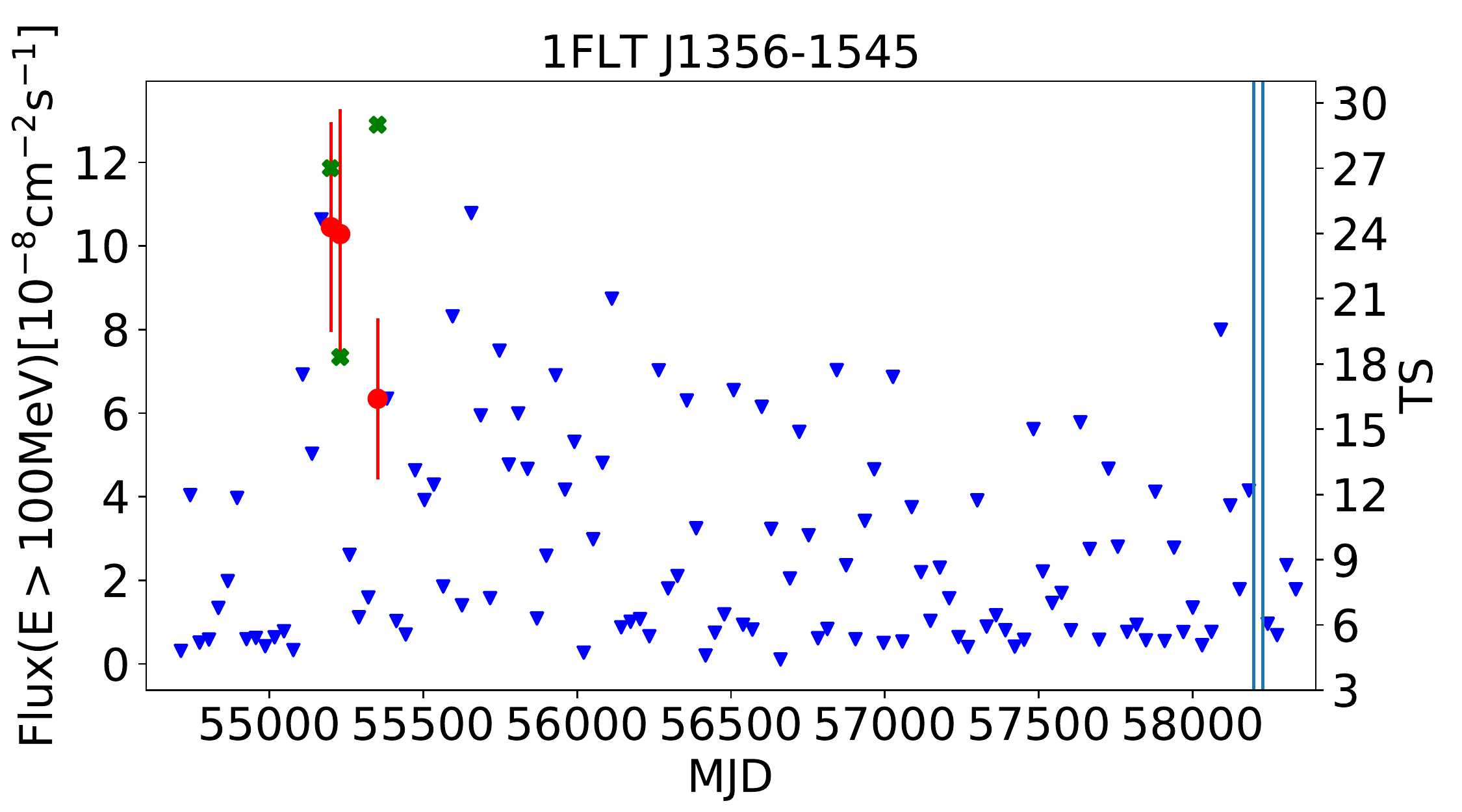}\label{fig:1FLTJ1356-1545}&
  \includegraphics[width=0.35\textwidth]{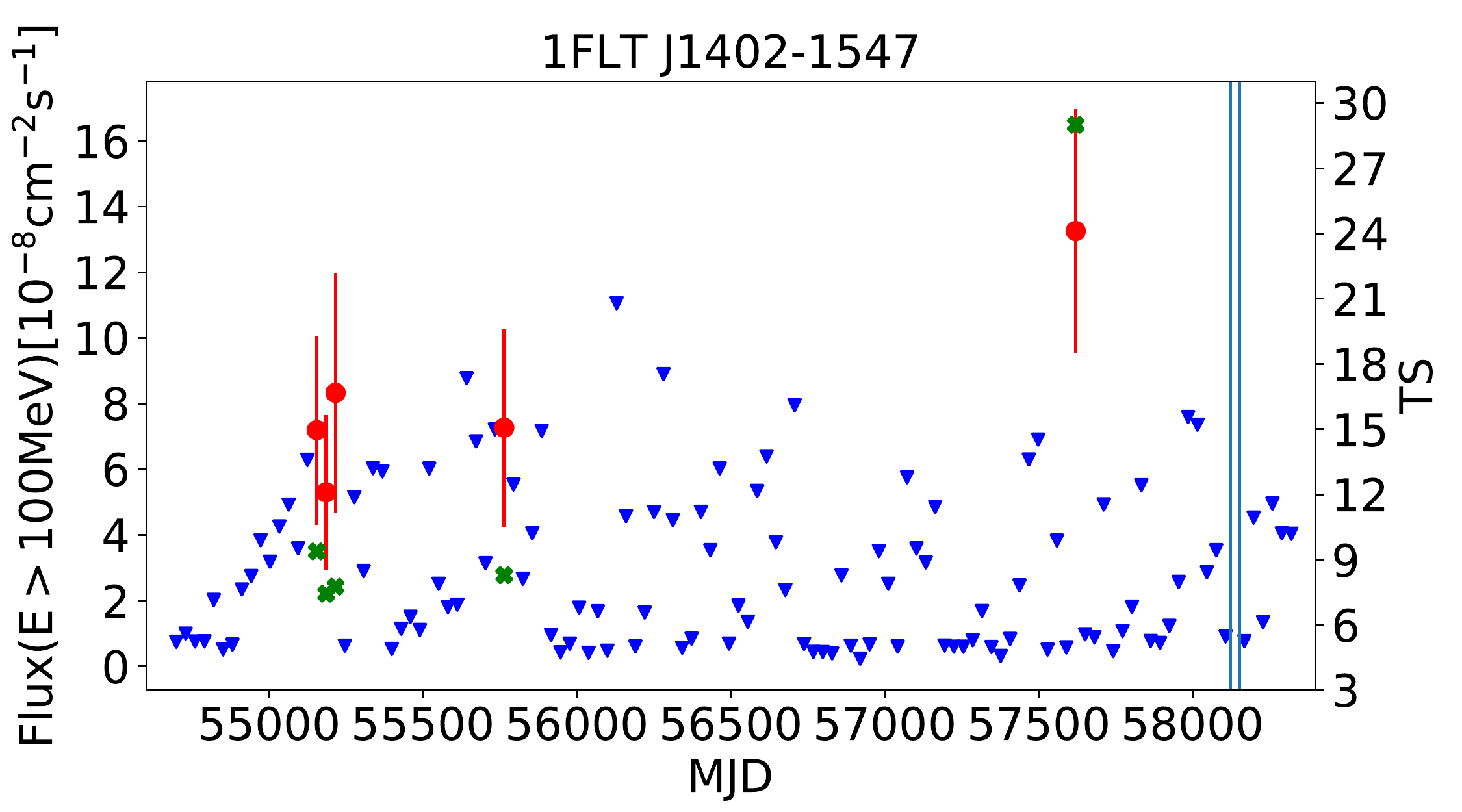}\label{fig:1FLTJ1402-1547}\\
  %[1FLTJ1331+1346]&%[1FLTJ1356-1545]&%[1FLTJ1402-1547]\\
  \includegraphics[width=0.35\textwidth]{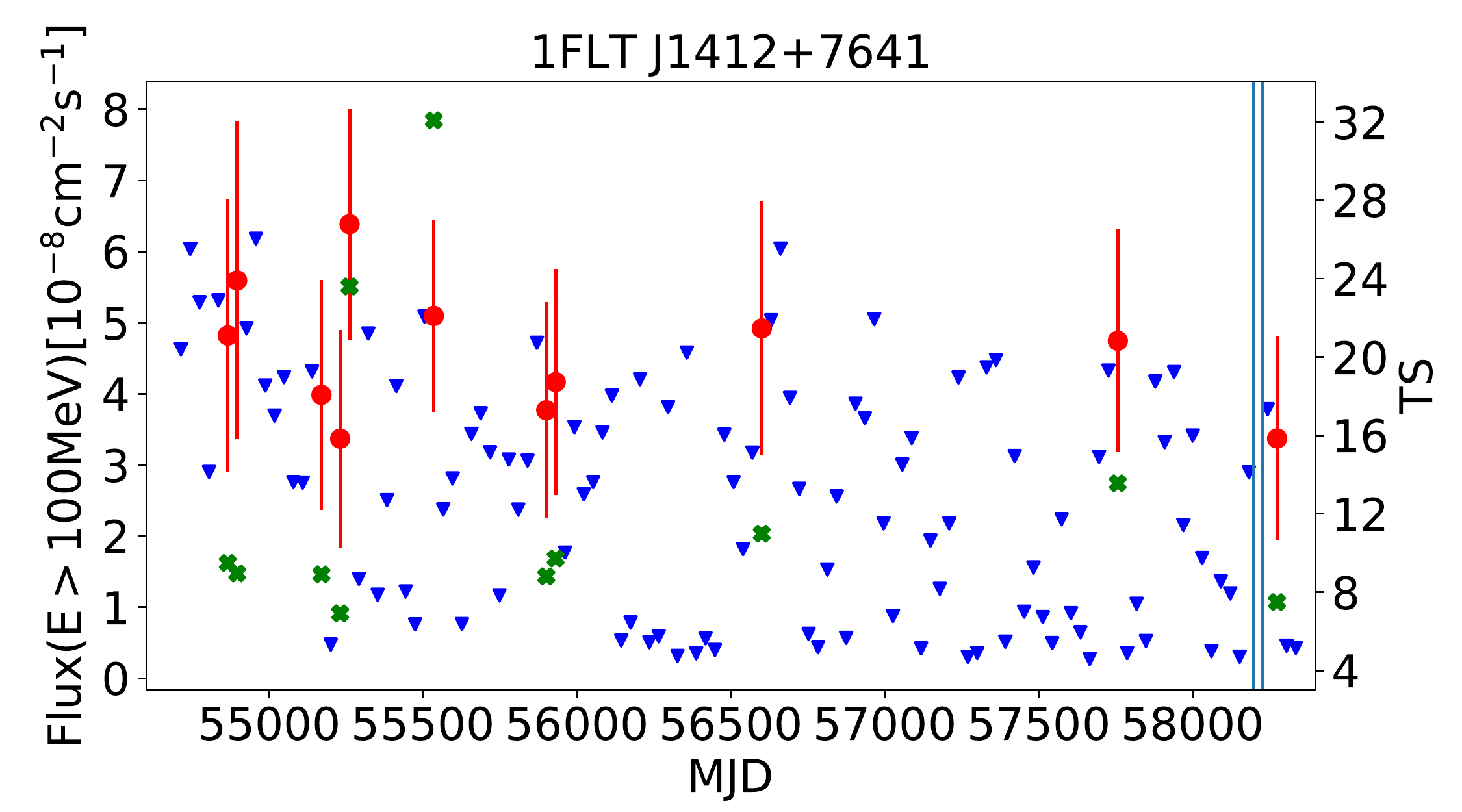}\label{fig:1FLTJ1412+7641}&
  \includegraphics[width=0.35\textwidth]{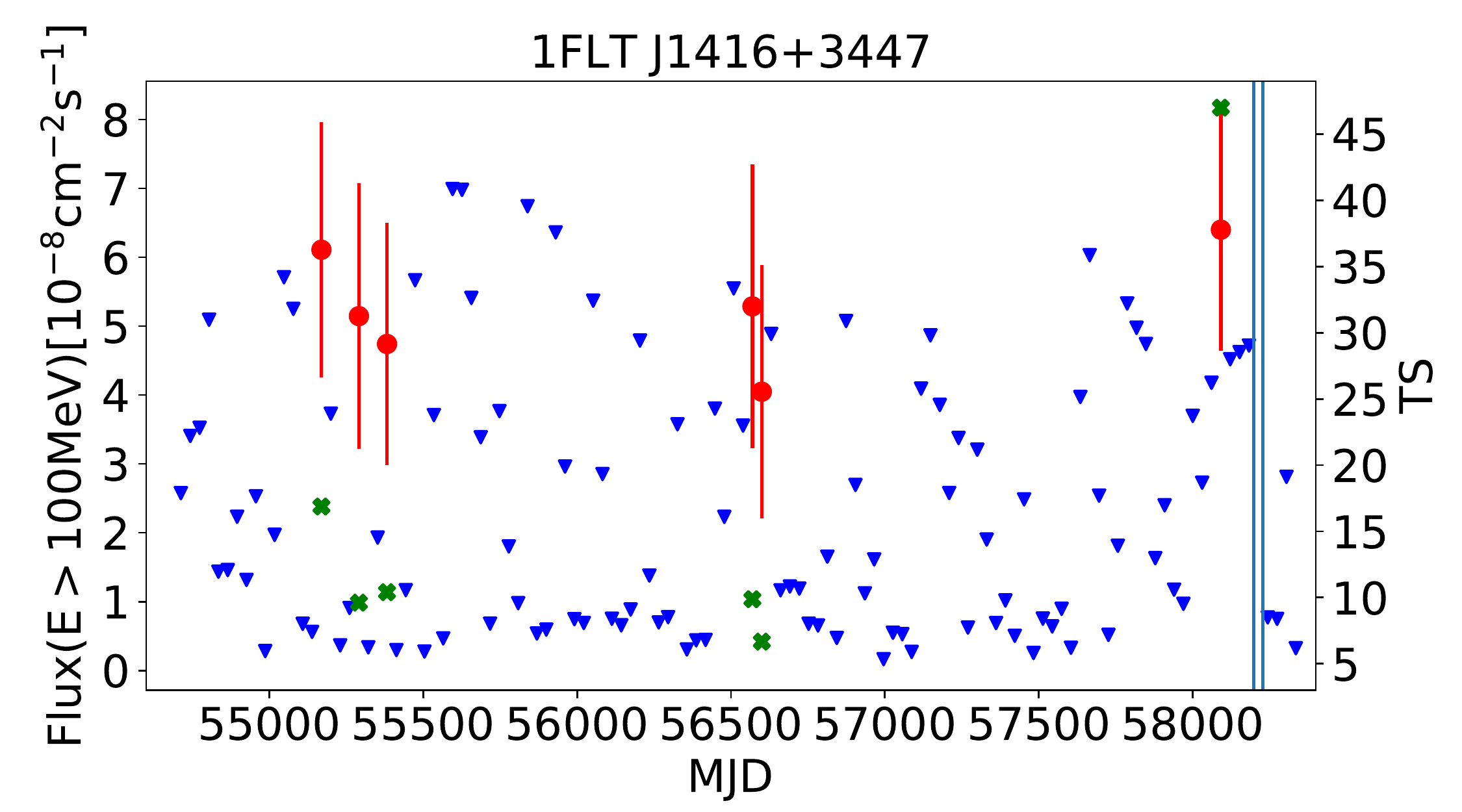}\label{fig:1FLTJ1416+3447}&
  \includegraphics[width=0.35\textwidth]{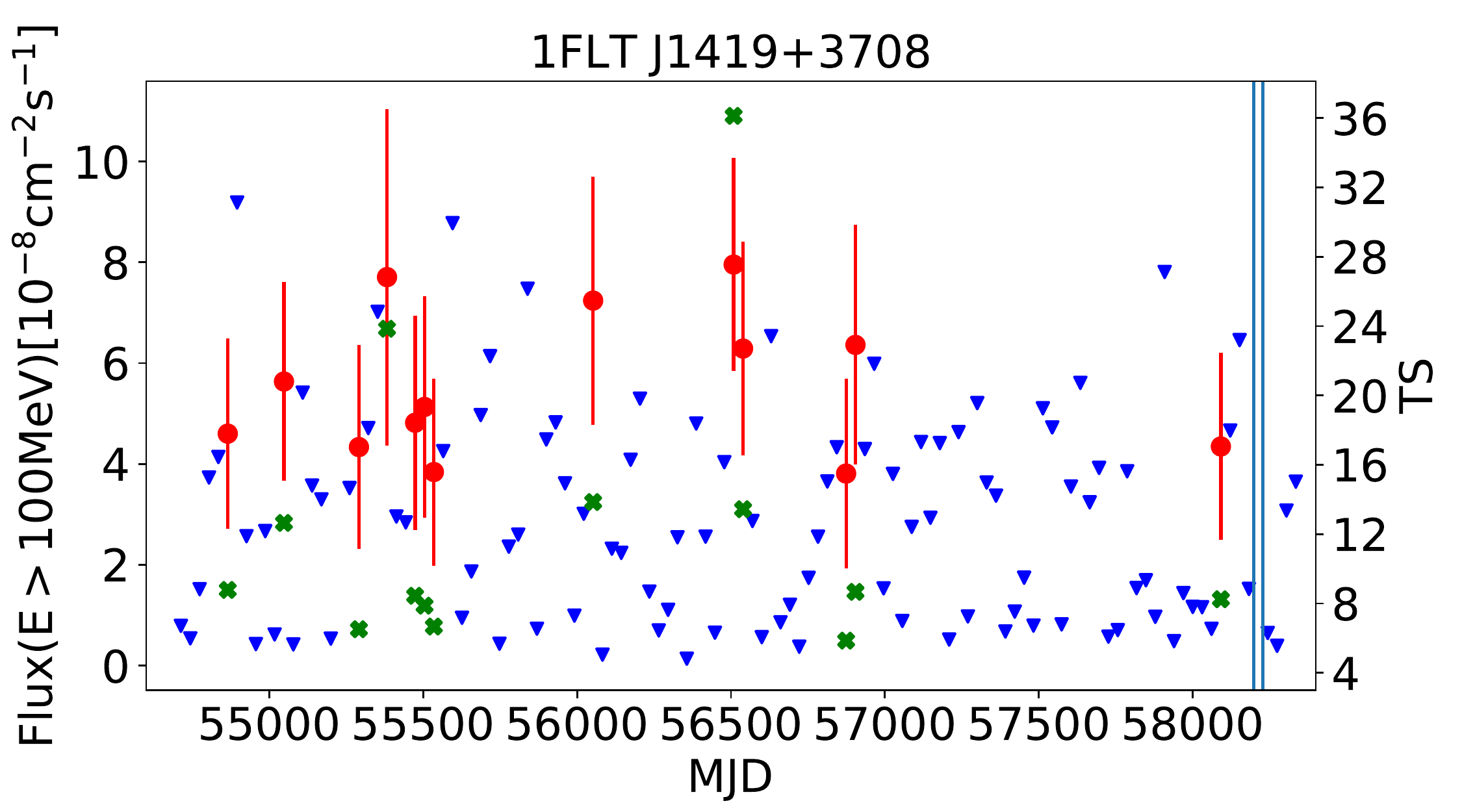}\label{fig:1FLTJ1419+3708}\\
  %[1FLTJ1412+7641]&%[1FLTJ1416+3447]&%[1FLTJ1419+3708]\\
\end{tabular}
\end{figure}
\begin{figure}[!t]
	\centering            
	\ContinuedFloat 
\setlength\tabcolsep{0.0pt}
\begin{tabular}{ccc} 
  \includegraphics[width=0.35\textwidth]{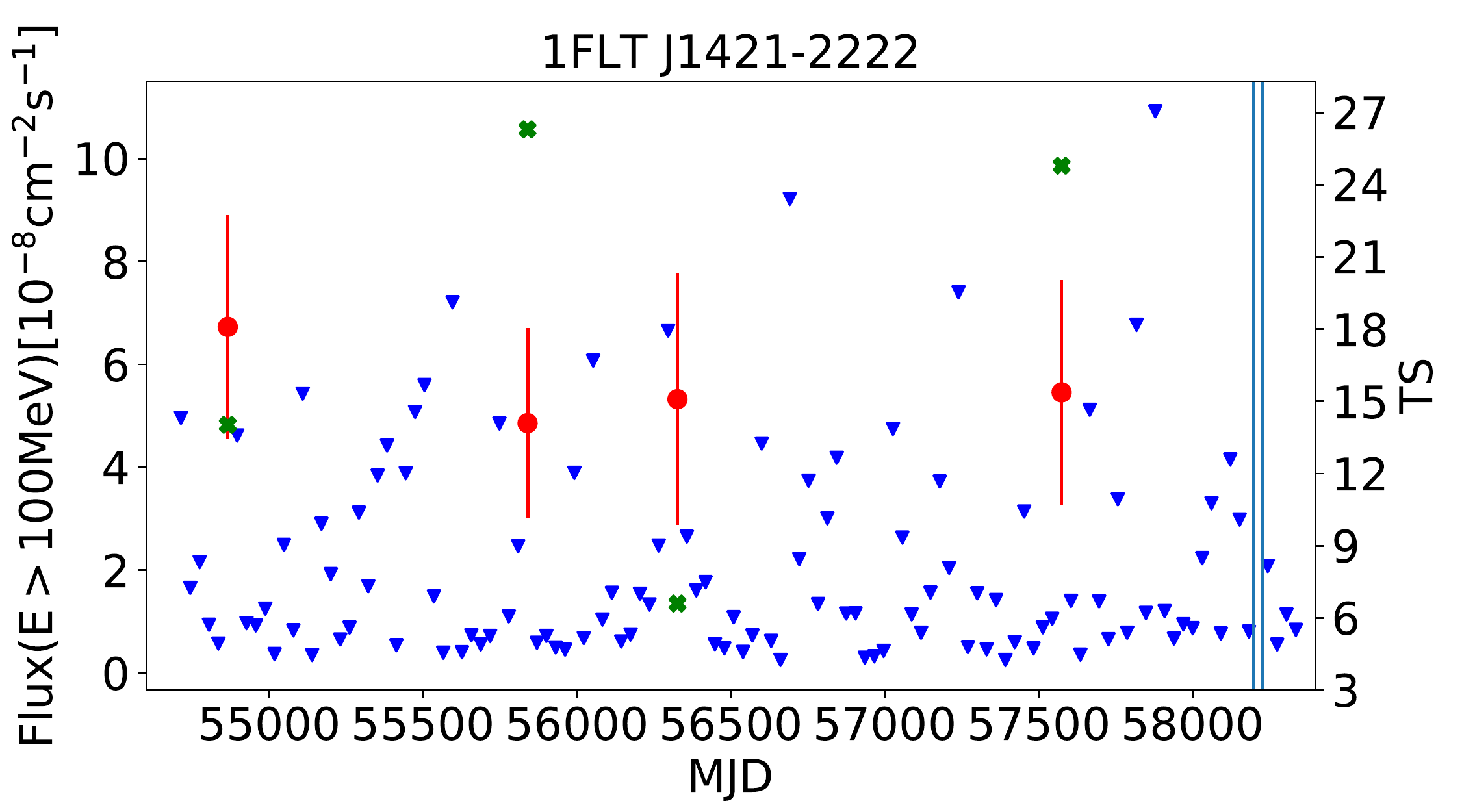}\label{fig:1FLTJ1421-2222}&
  \includegraphics[width=0.35\textwidth]{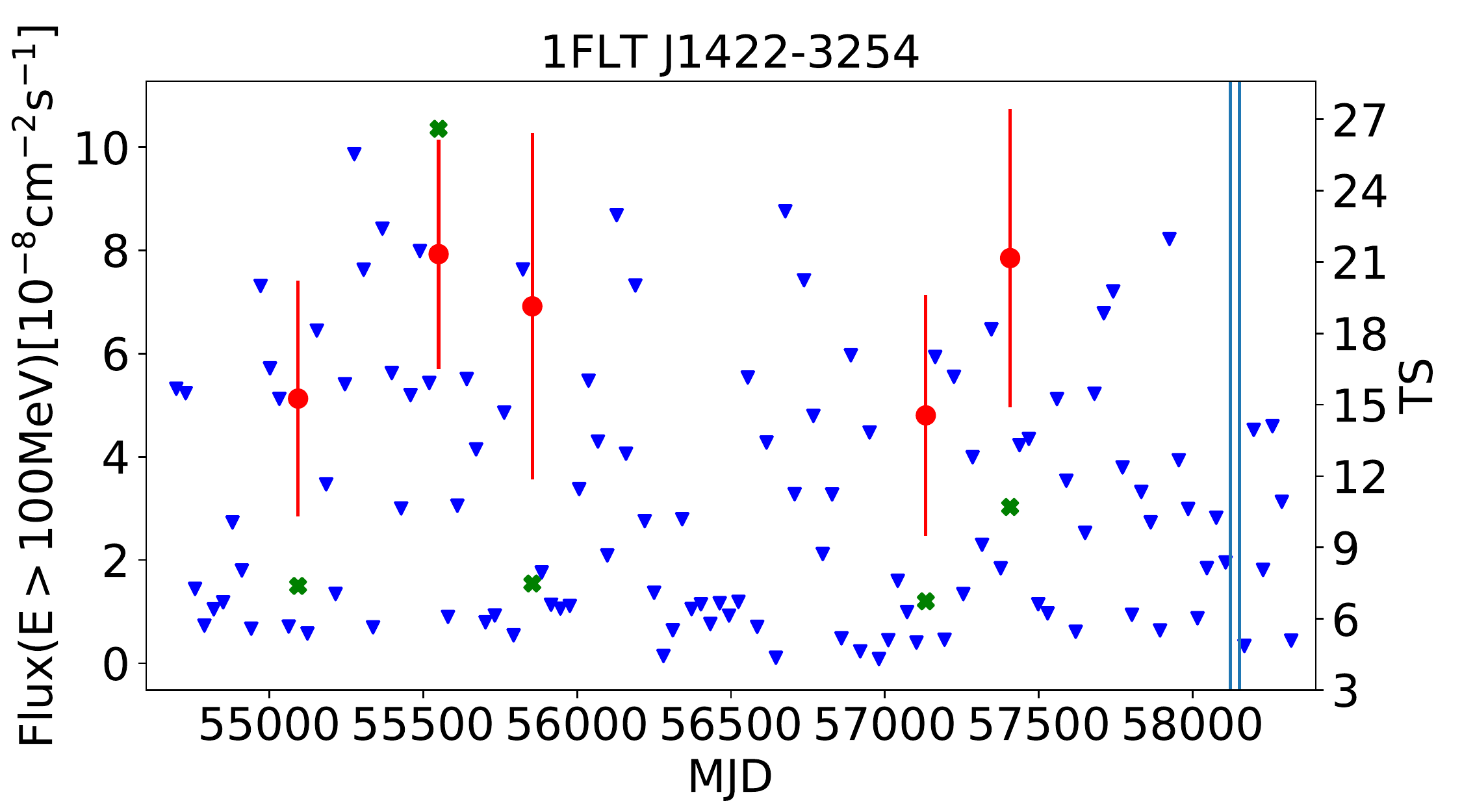}\label{fig:1FLTJ1422-3254}&
  \includegraphics[width=0.35\textwidth]{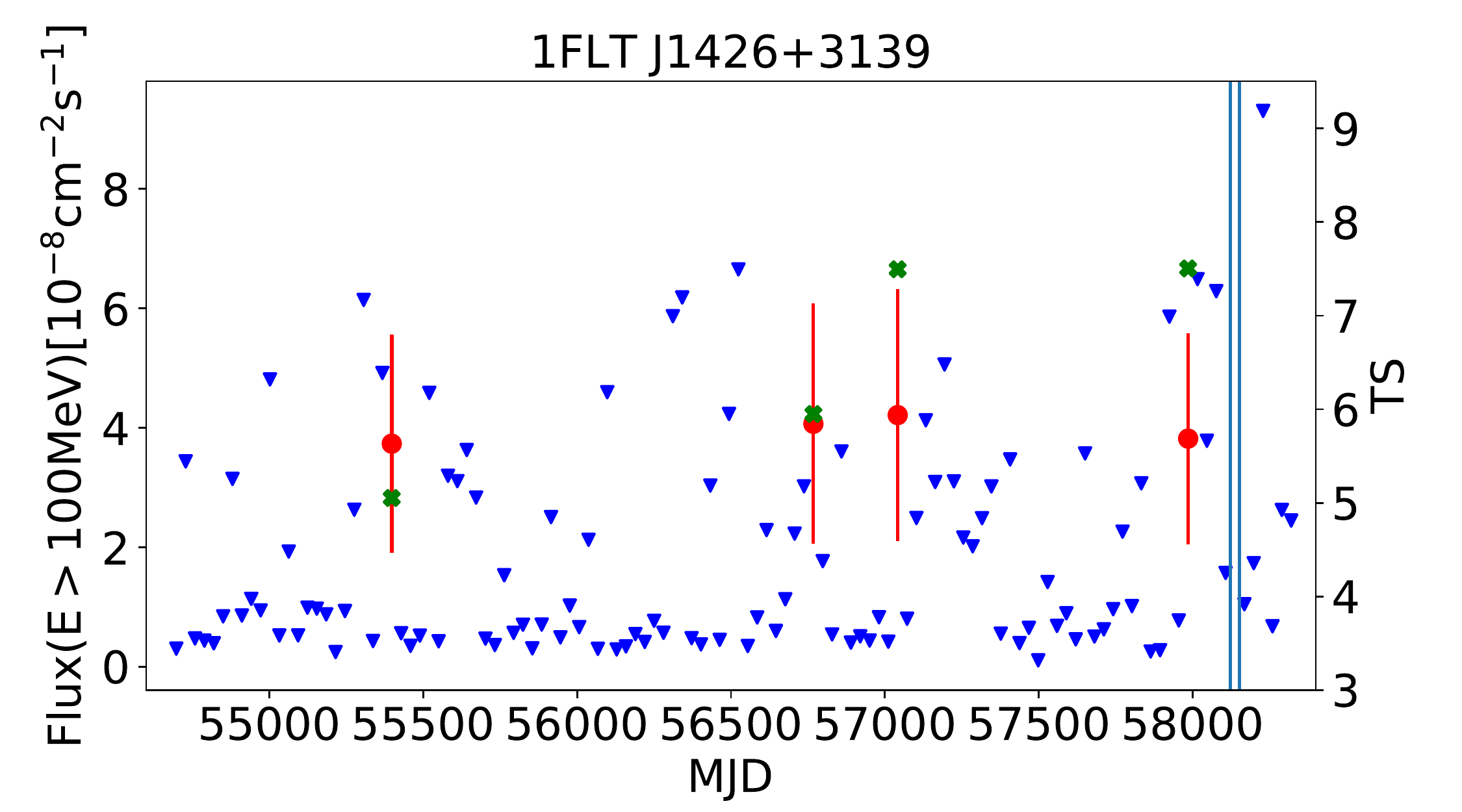}\label{fig:1FLTJ1426+3139}\\
  %[1FLTJ1422-3254]&%[1FLTJ1426+3139]&%[1FLTJ1502-2420]
  \includegraphics[width=0.35\textwidth]{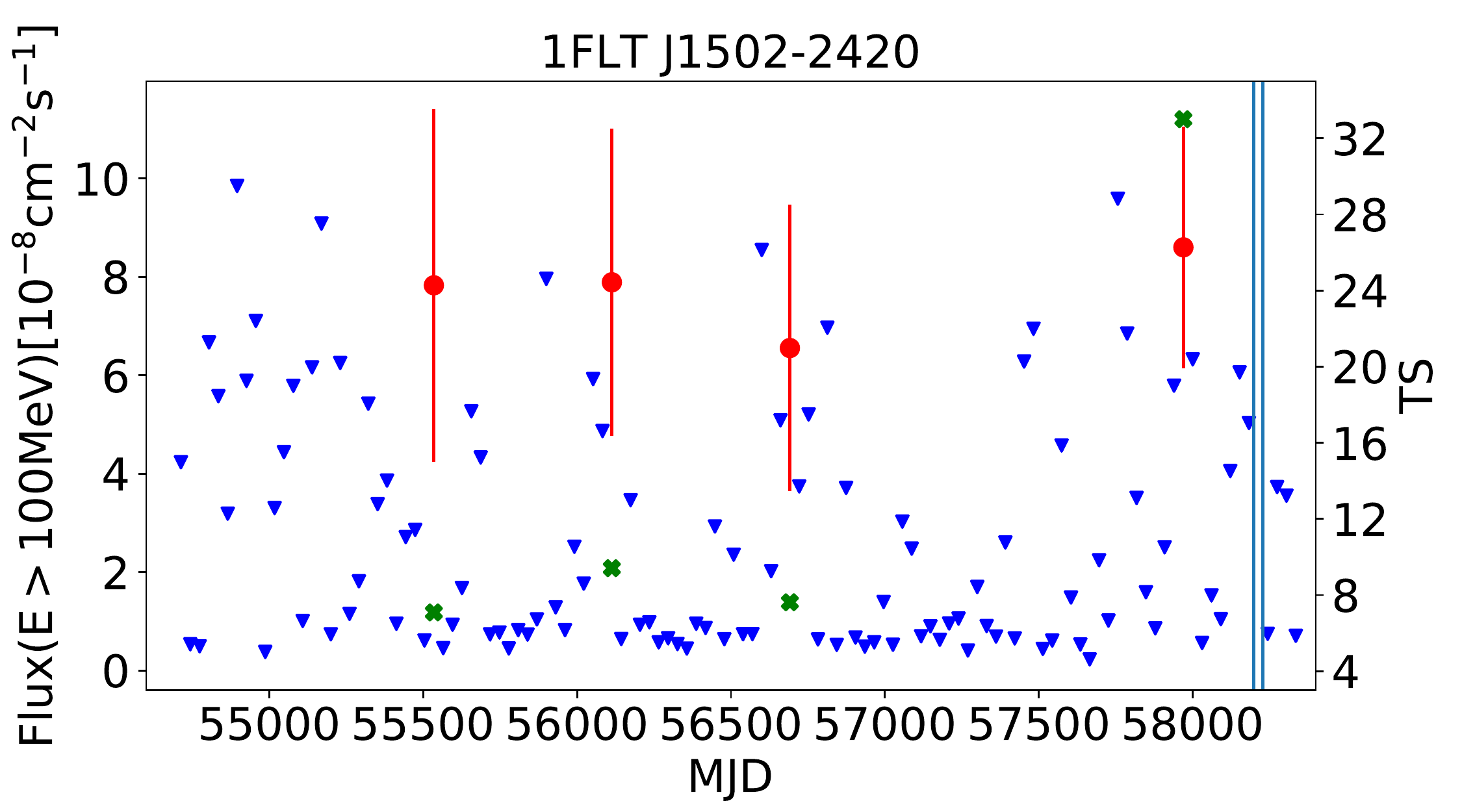}\label{fig:1FLTJ1502-2420}&
  \includegraphics[width=0.35\textwidth]{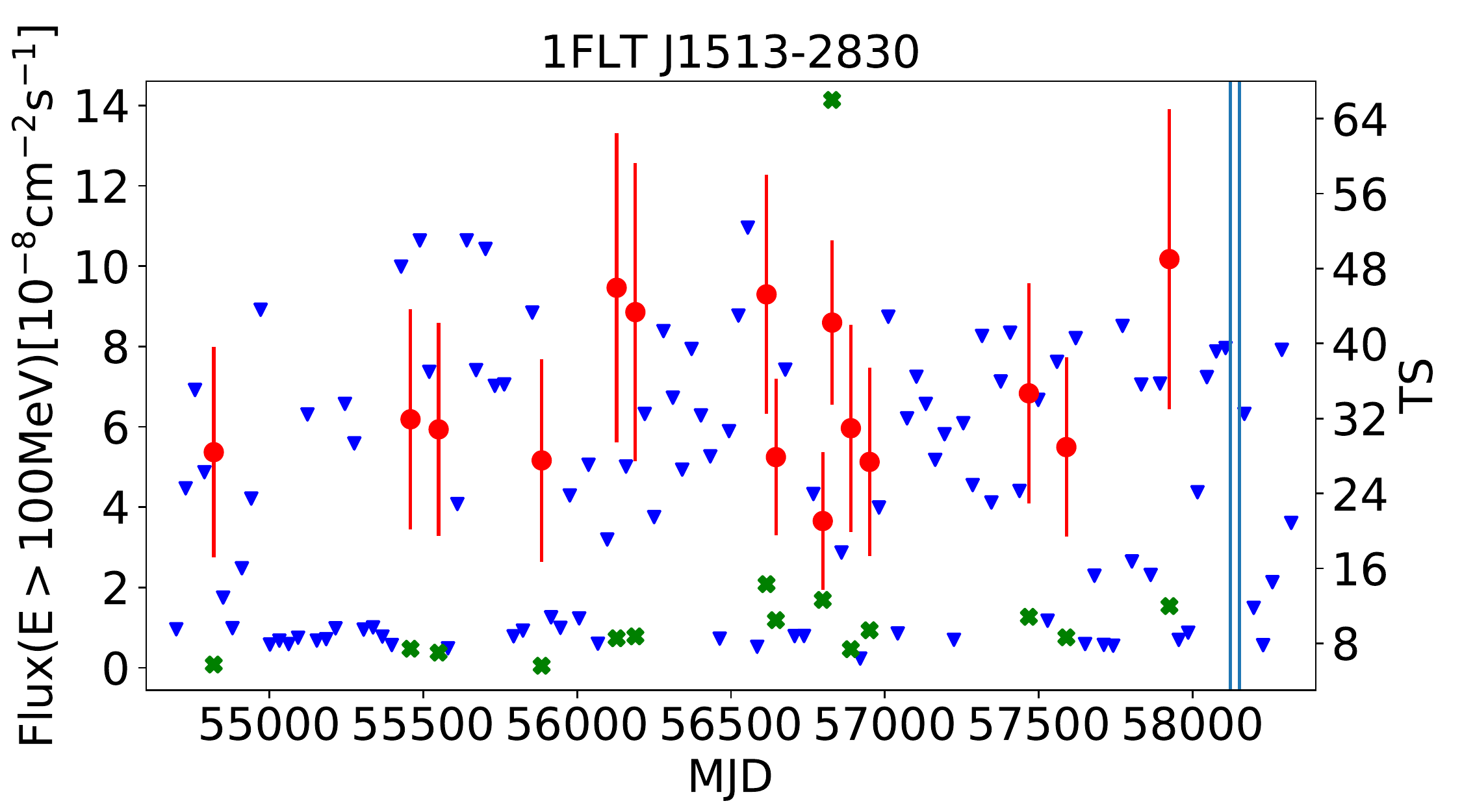}\label{fig:1FLTJ1513-2830}&
  \includegraphics[width=0.35\textwidth]{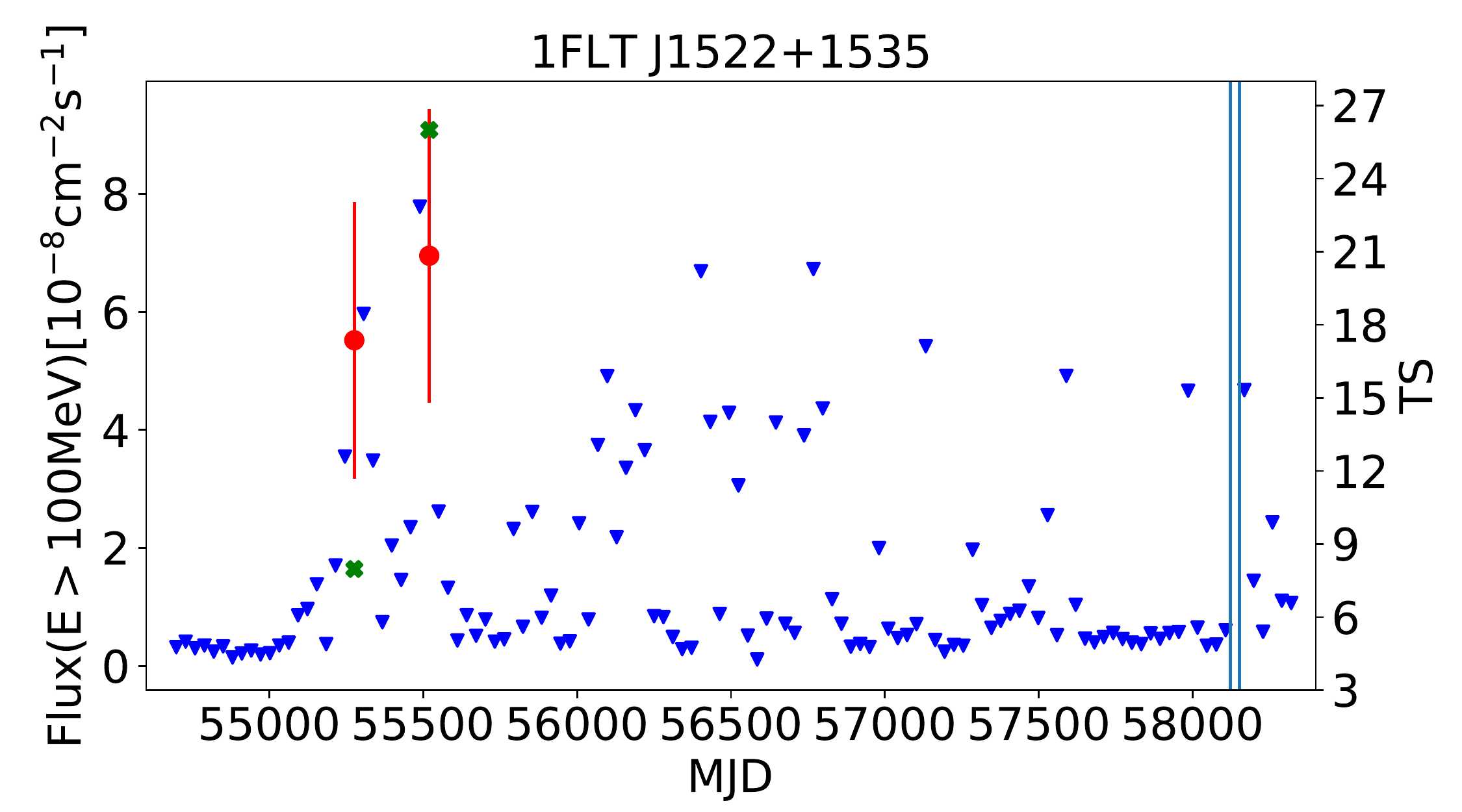}\label{fig:1FLTJ1522+1535}\\
  %[1FLTJ1513-2830]&%[1FLTJ1522+1535]&%[1FLTJ1528-1348]\\
  \includegraphics[width=0.35\textwidth]{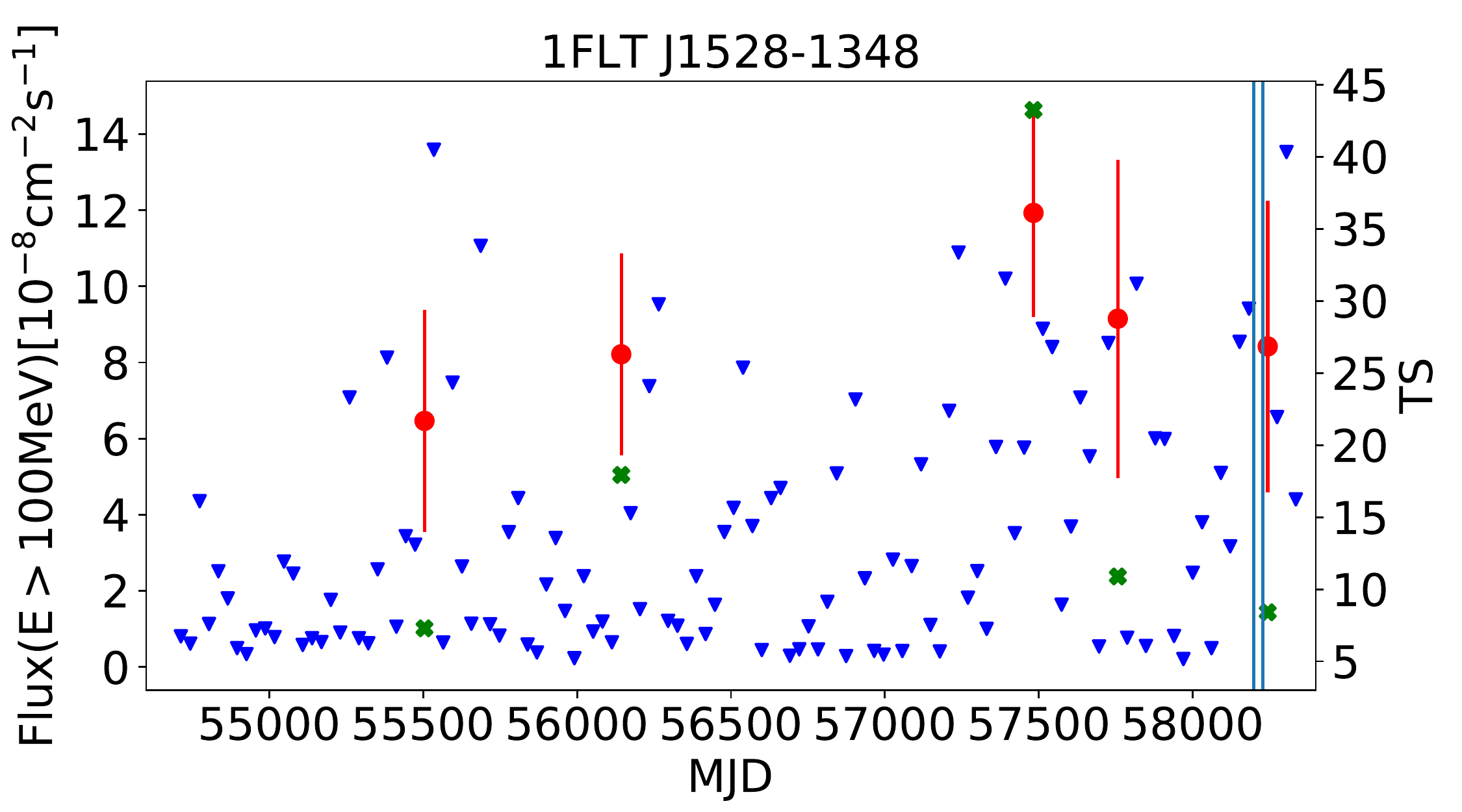}\label{fig:1FLTJ1528-1348}&
  \includegraphics[width=0.35\textwidth]{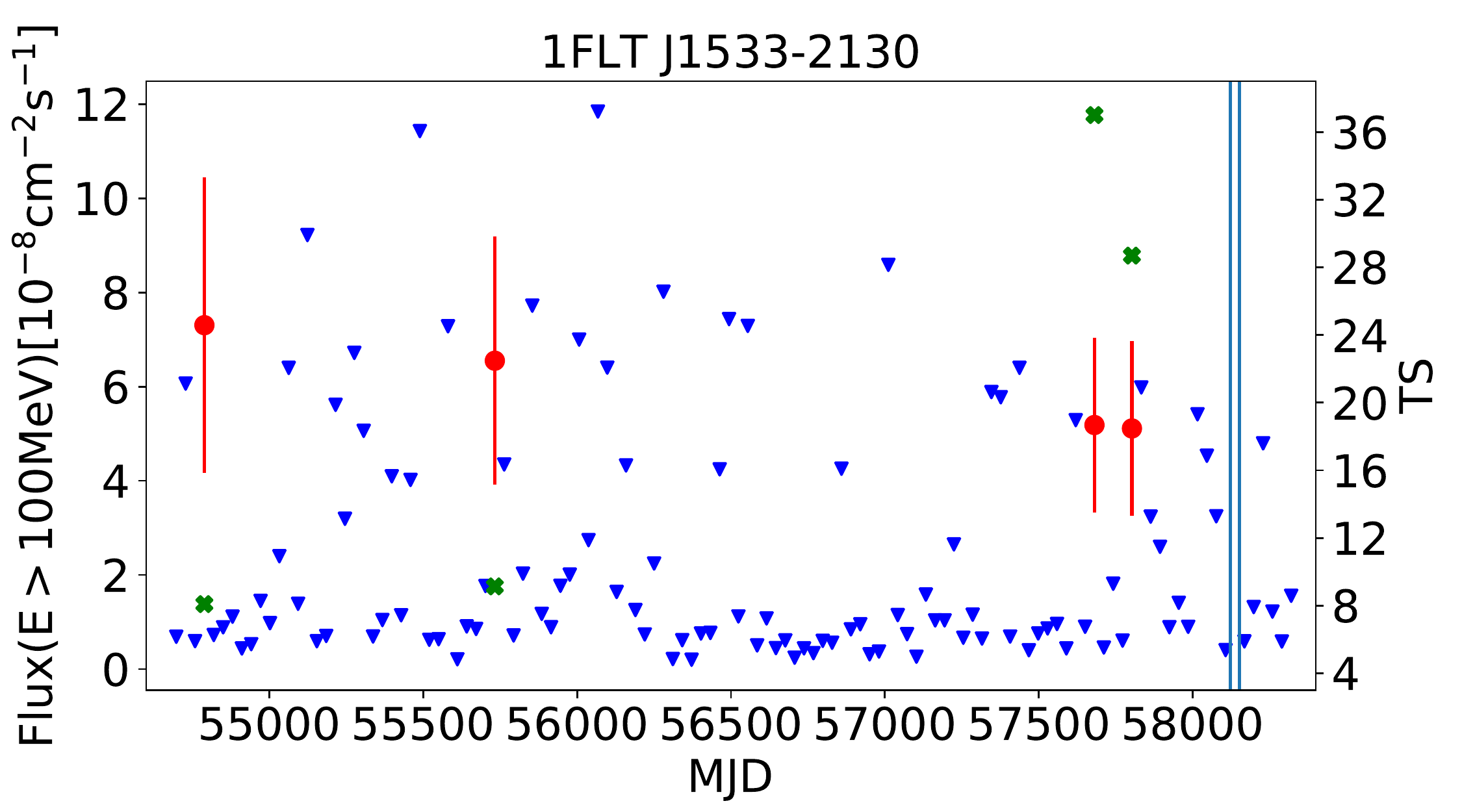}\label{fig:1FLTJ1533-2130}&
  \includegraphics[width=0.35\textwidth]{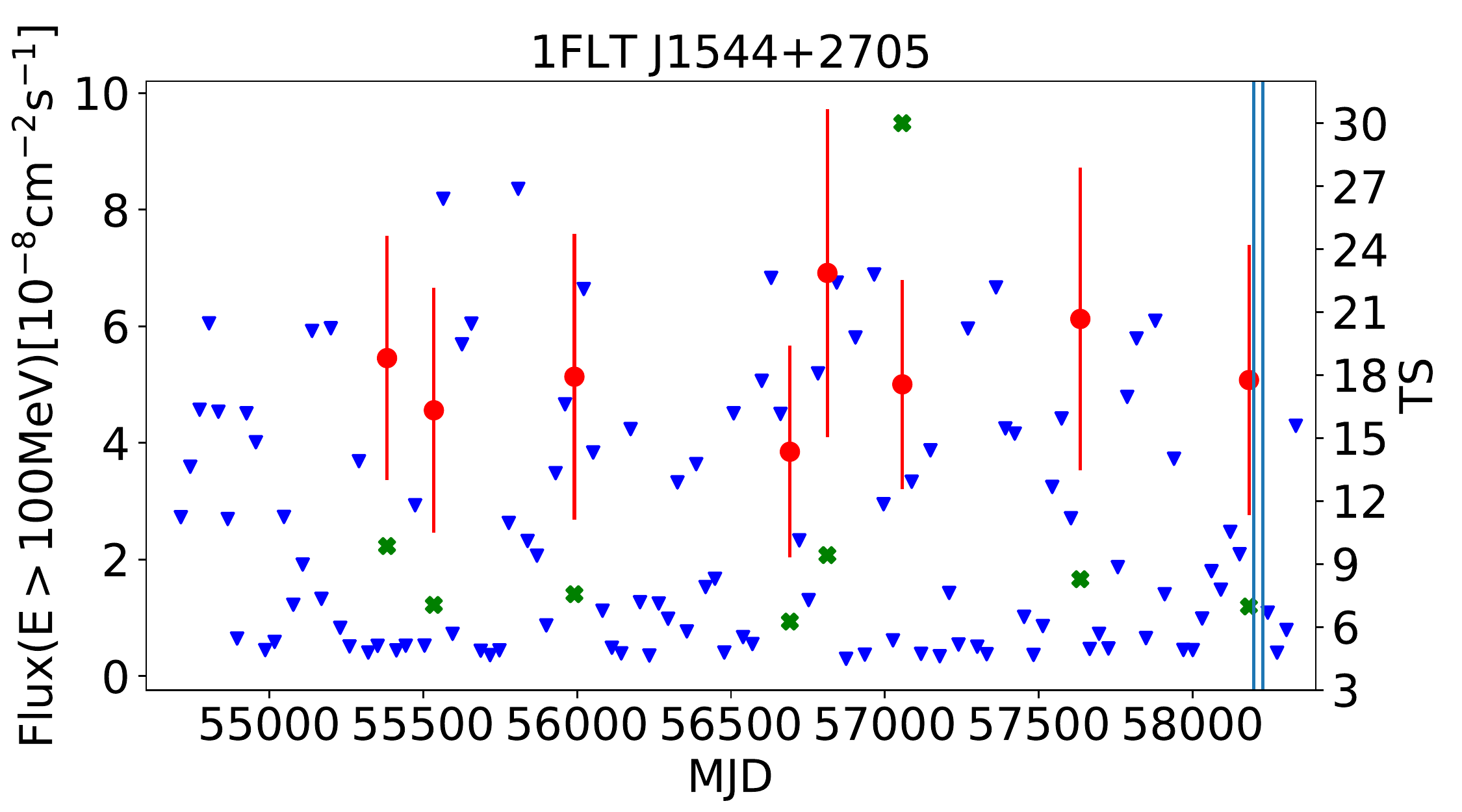}\label{fig:1FLTJ1544+2705}\\
  \includegraphics[width=0.35\textwidth]{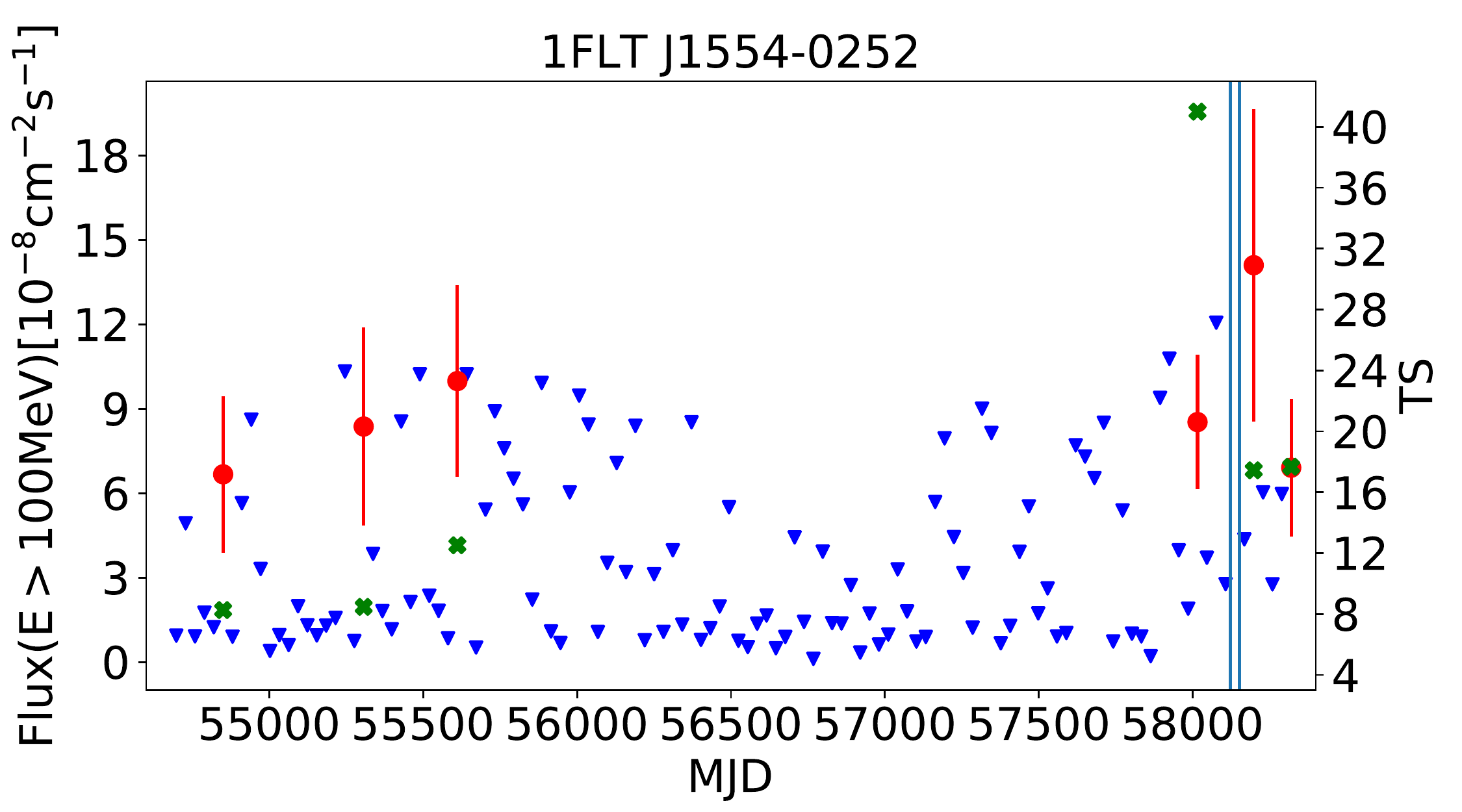}\label{fig:1FLTJ1554-0252}&
  %[1FLTJ1533-2130]&%[1FLTJ1544+2705]&%[1FLTJ1554-0252]\\
  \includegraphics[width=0.35\textwidth]{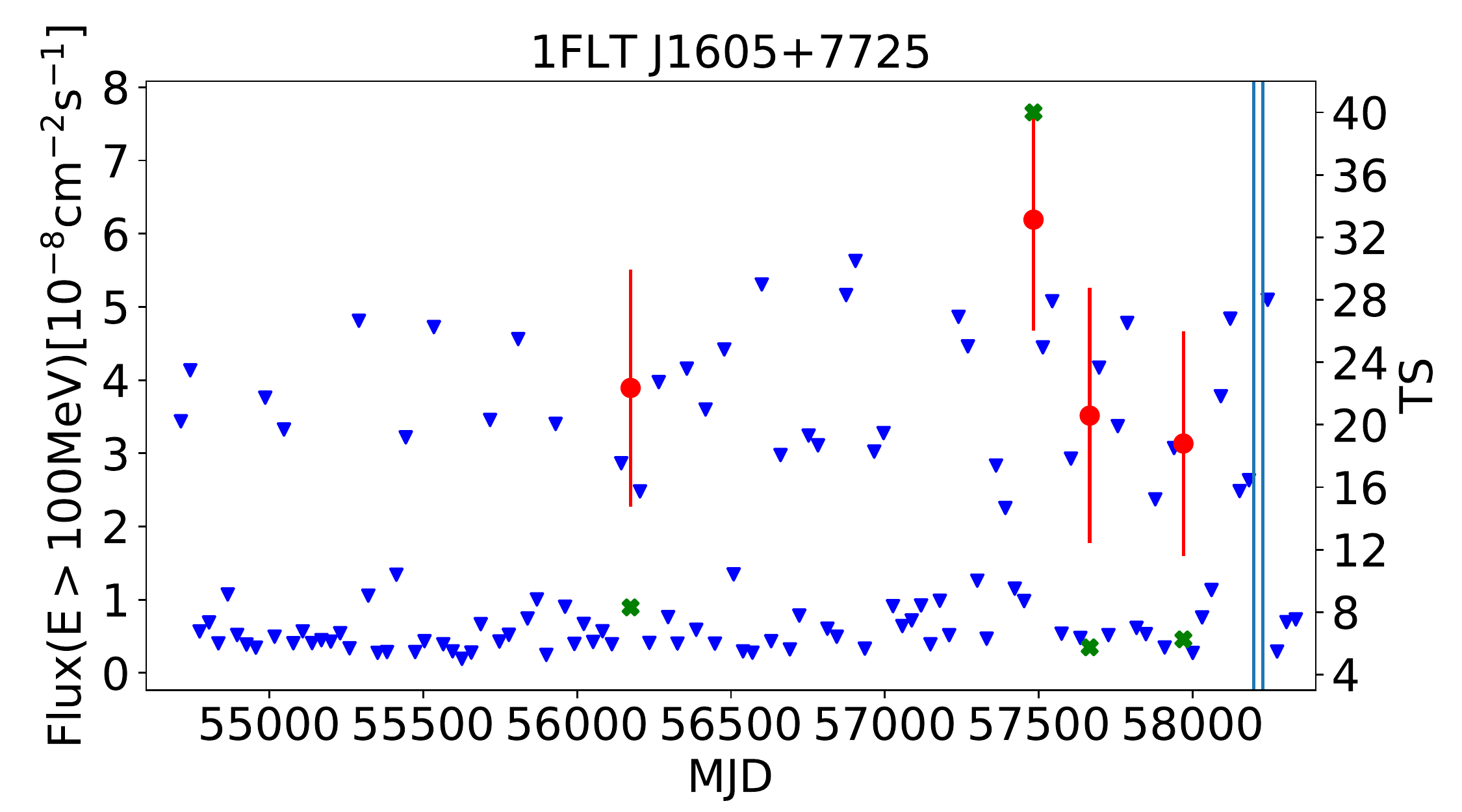}\label{fig:1FLTJ1605+7725}&
  \includegraphics[width=0.35\textwidth]{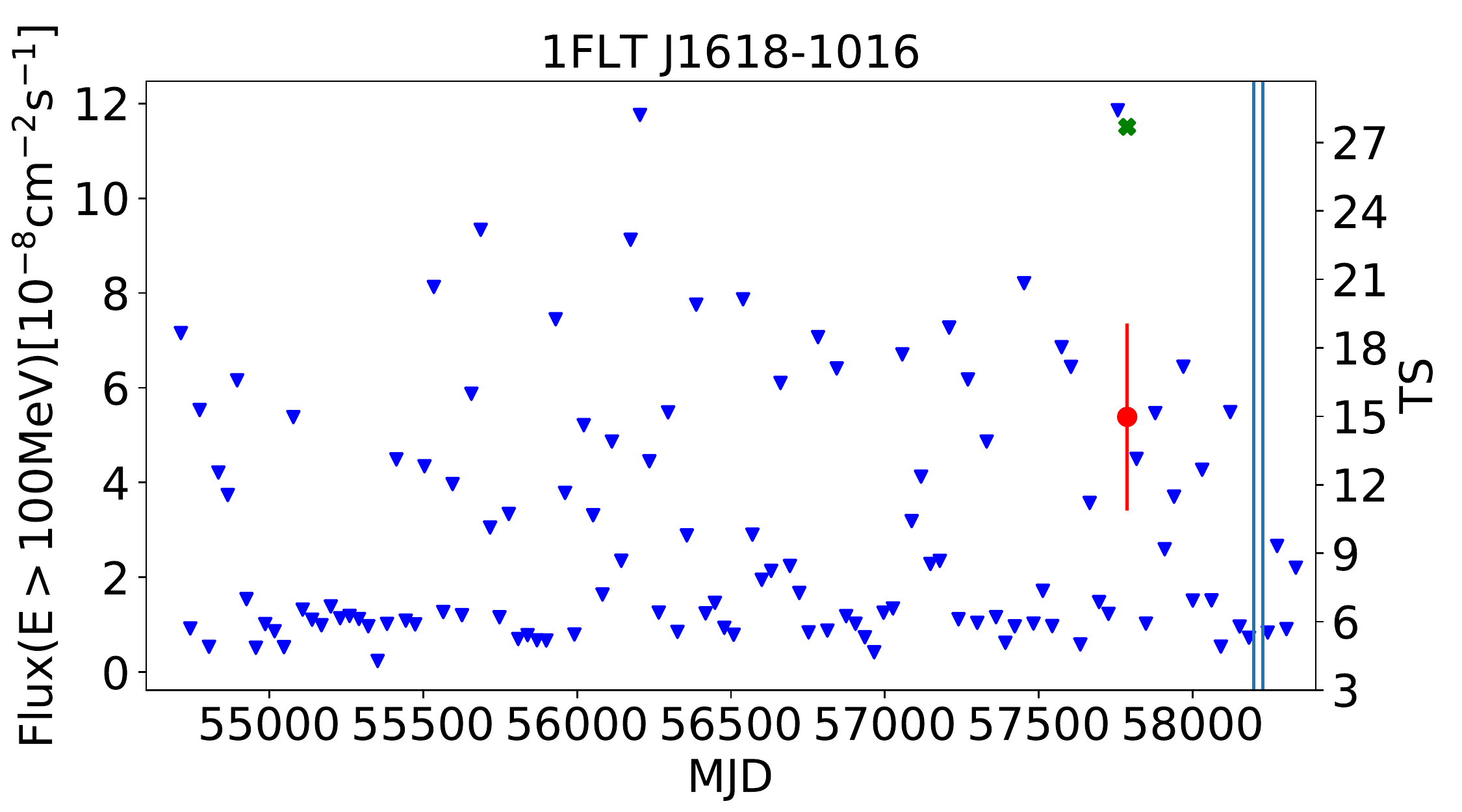}\label{fig:1FLTJ1618-1016}\\
  \includegraphics[width=0.35\textwidth]{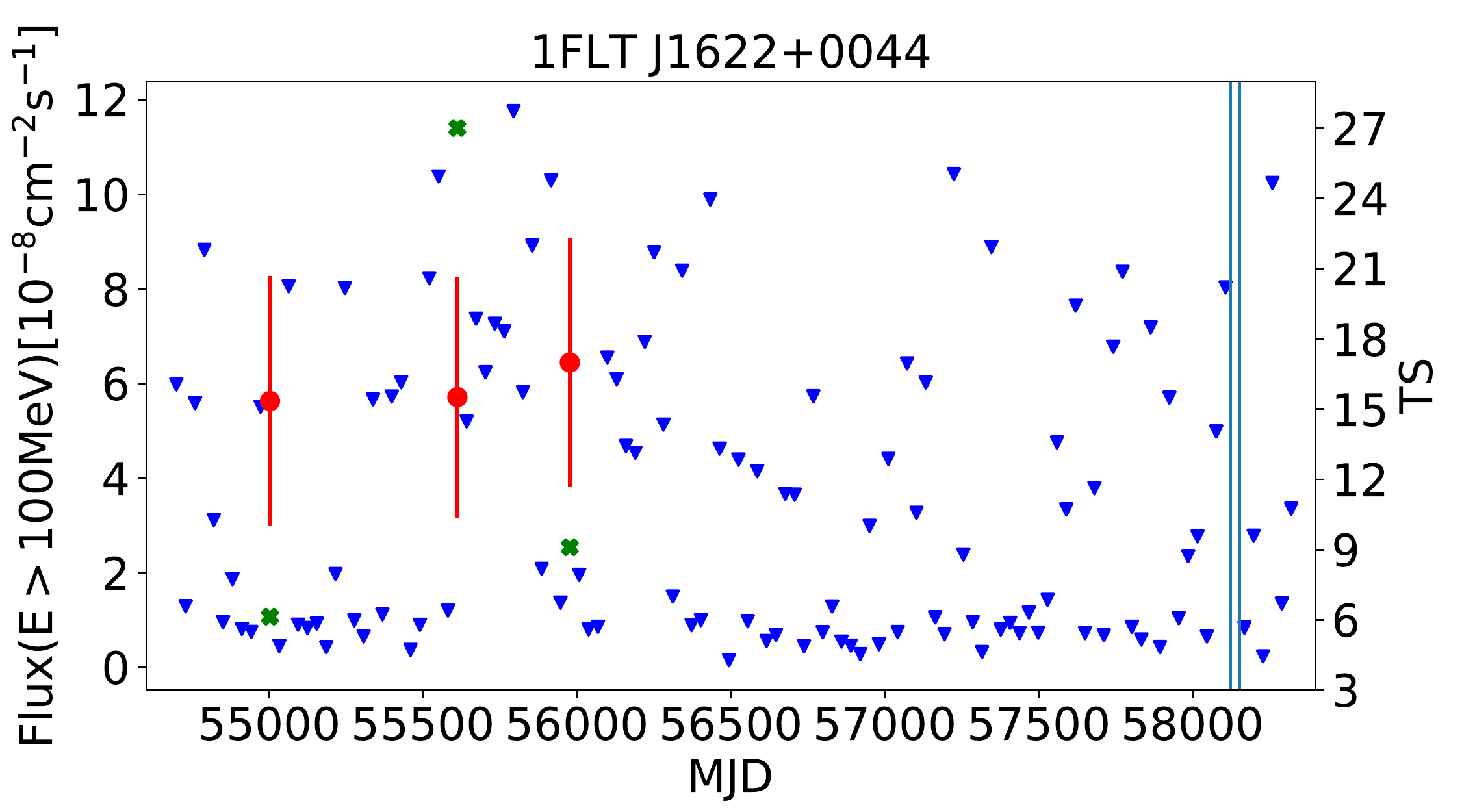}\label{fig:1FLTJ1622+0044}&
  %[1FLTJ1605+7725]&%[1FLTJ1618-1016]&%[1FLTJ1622+0044]\\
  \includegraphics[width=0.35\textwidth]{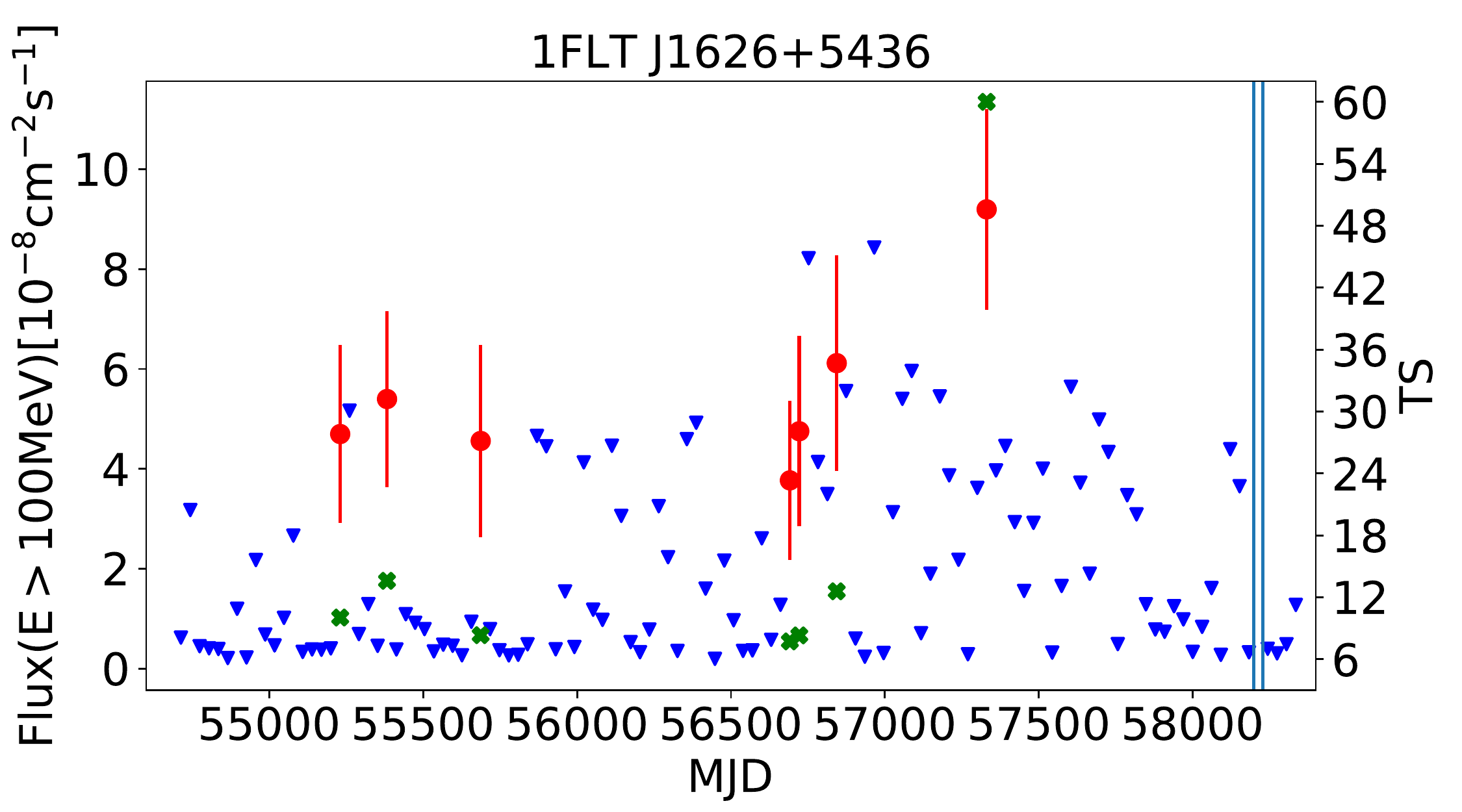}\label{fig:1FLTJ1626+5436}&
  \includegraphics[width=0.35\textwidth]{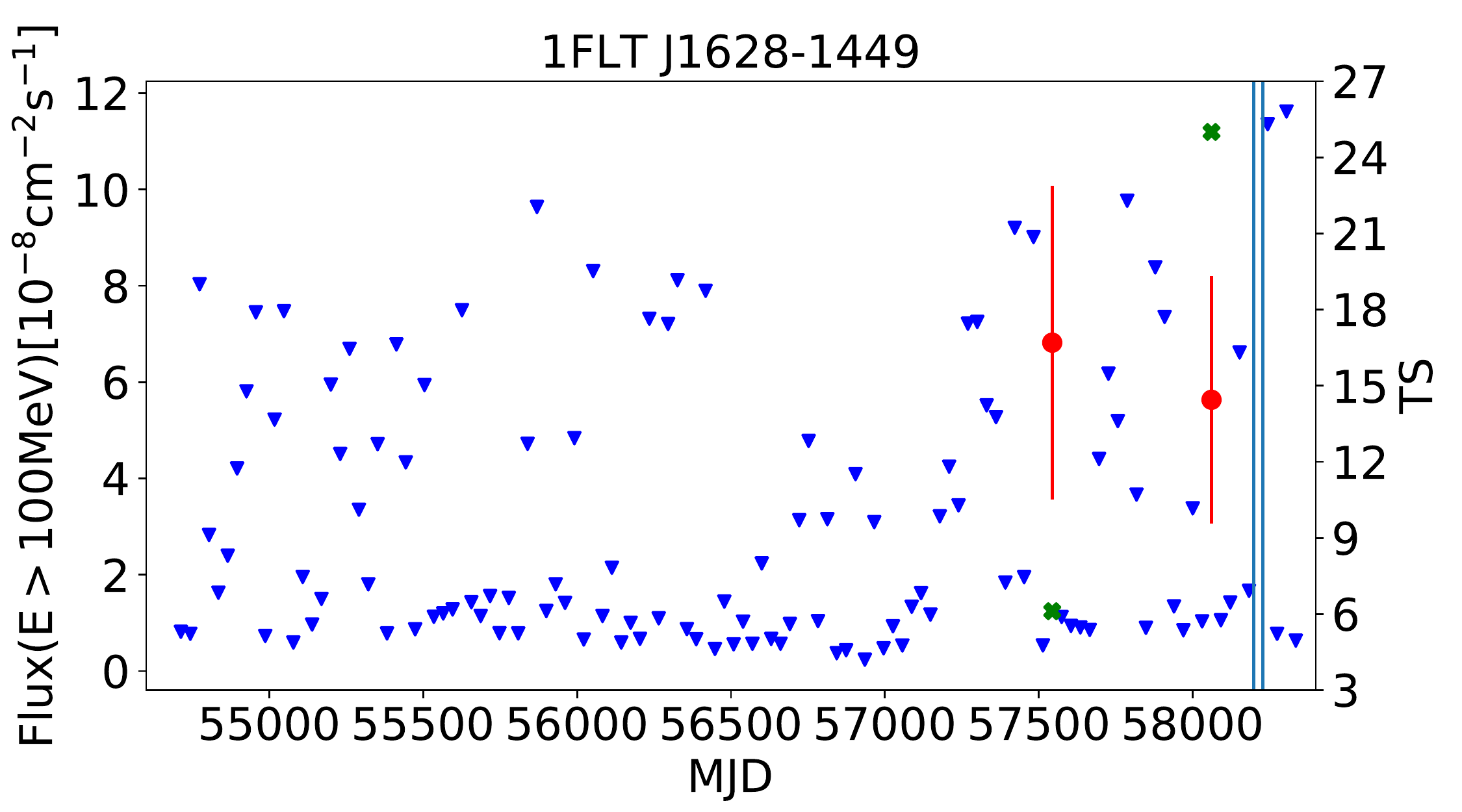}\label{fig:1FLTJ1628-1449}\\
  \includegraphics[width=0.35\textwidth]{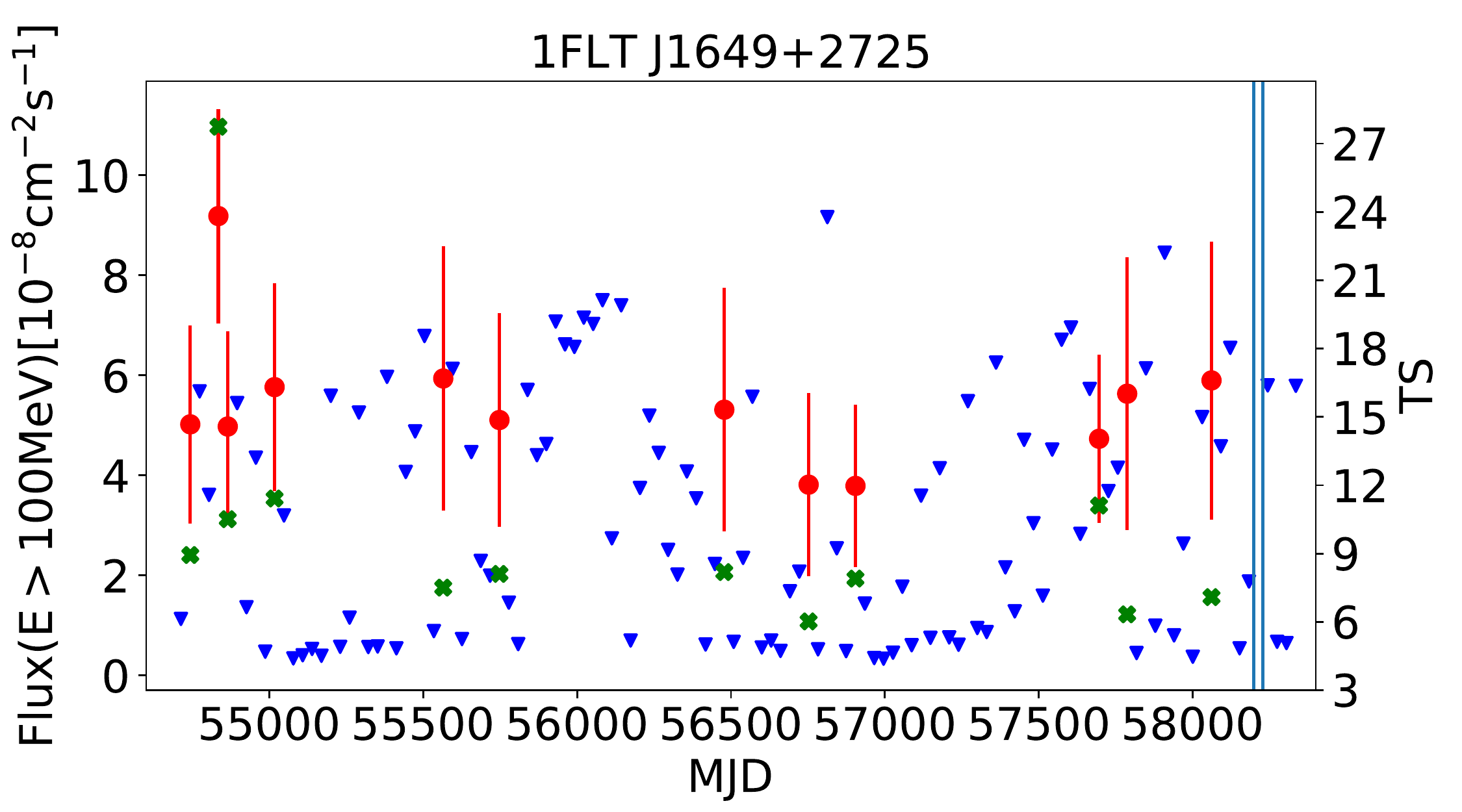}\label{fig:1FLTJ1649+2725}&
  %[1FLTJ1626+5436]&%[1FLTJ1628-1449]&%[1FLTJ1649+2725]\\
  \includegraphics[width=0.35\textwidth]{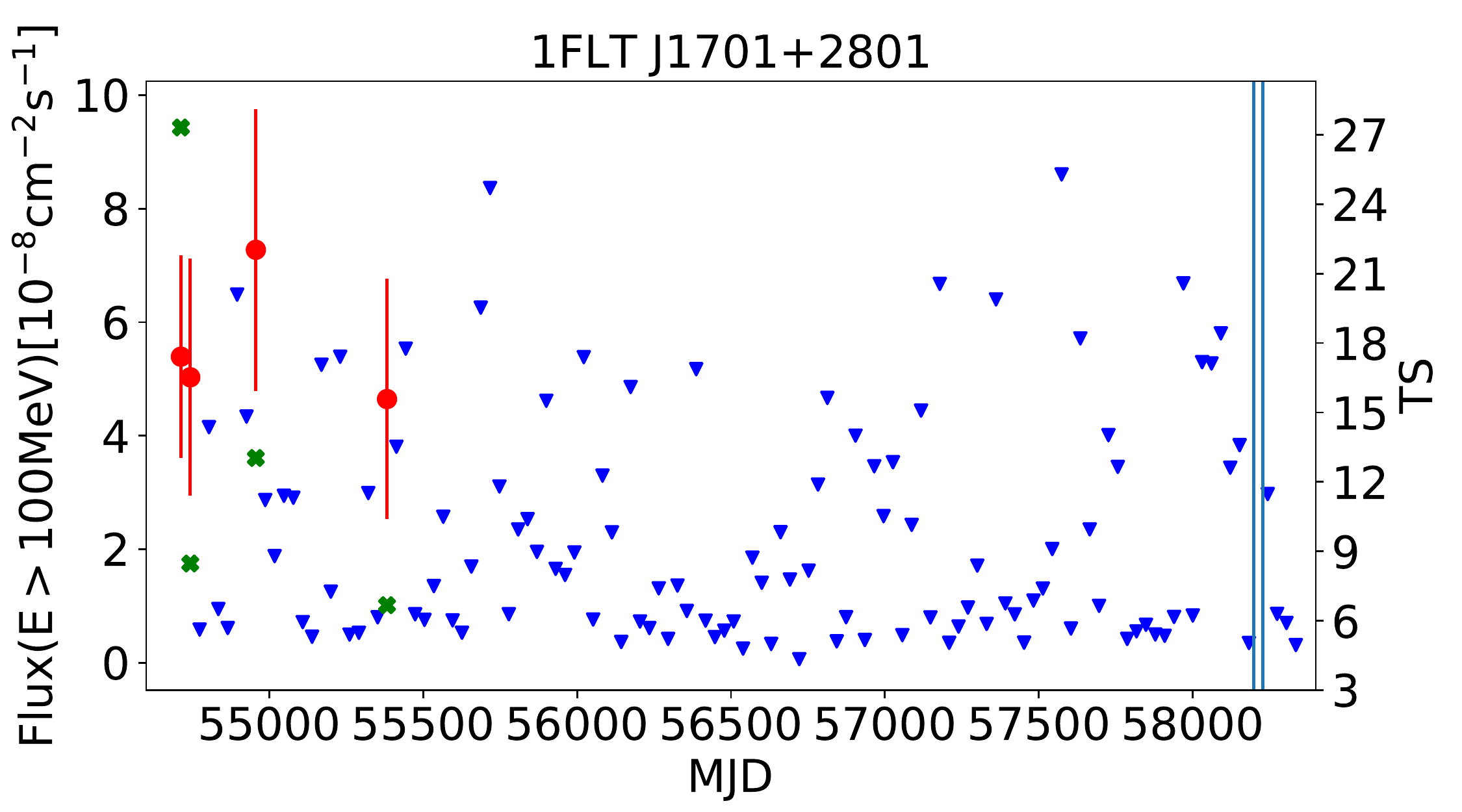}\label{fig:1FLTJ1701+2801}&
  \includegraphics[width=0.35\textwidth]{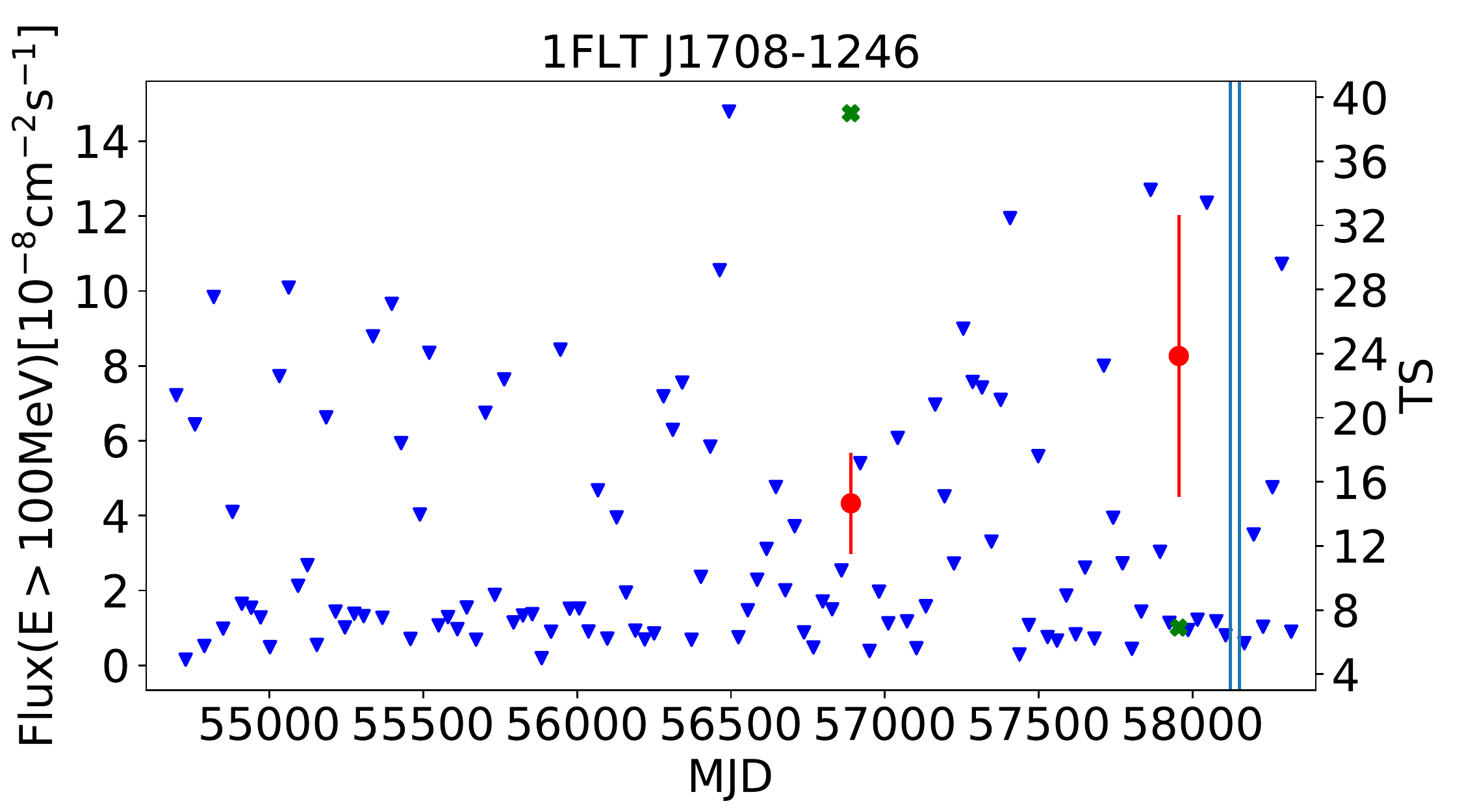}\label{fig:1FLTJ1708-1246}\\
\end{tabular}
\end{figure}
\begin{figure}[!t]
	\centering            
	\ContinuedFloat 
\setlength\tabcolsep{0.0pt}
\begin{tabular}{ccc}
  \includegraphics[width=0.35\textwidth]{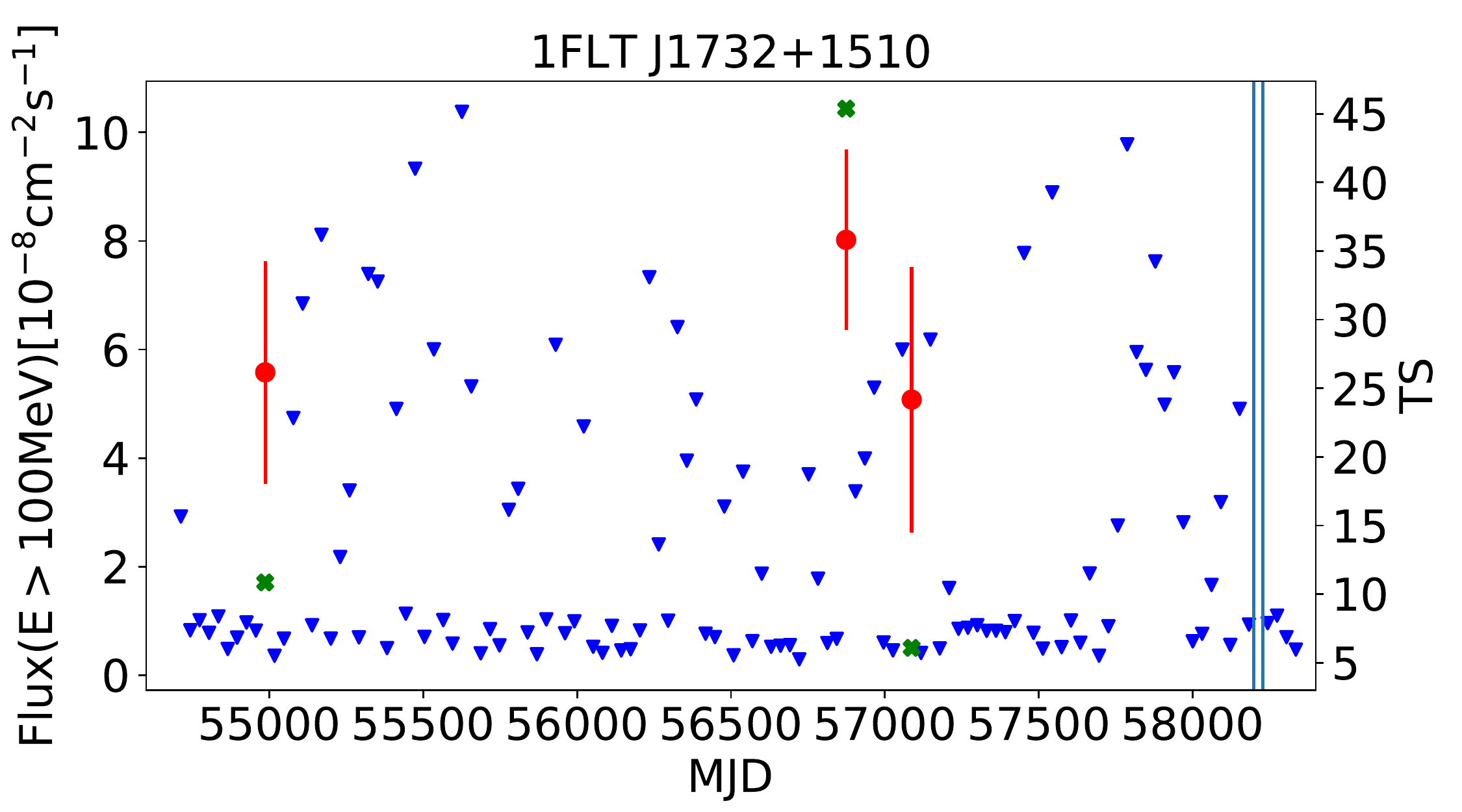}\label{fig:1FLTJ1732+1510}&
  %[1FLTJ1701+2801]&%[1FLTJ1708-1246]&%[1FLTJ1732+1510]\\
  \includegraphics[width=0.35\textwidth]{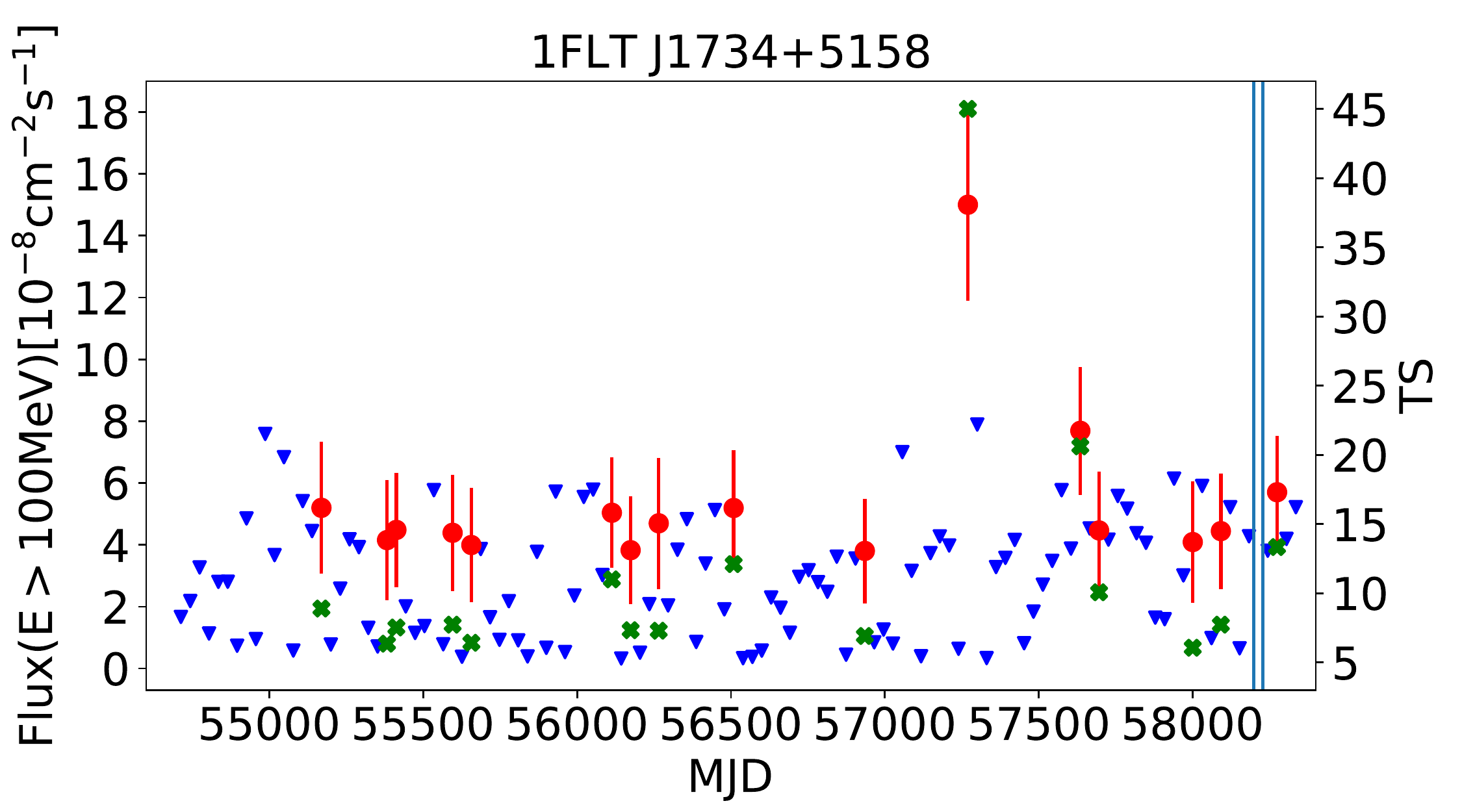}\label{fig:1FLTJ1734+5158}&
  \includegraphics[width=0.35\textwidth]{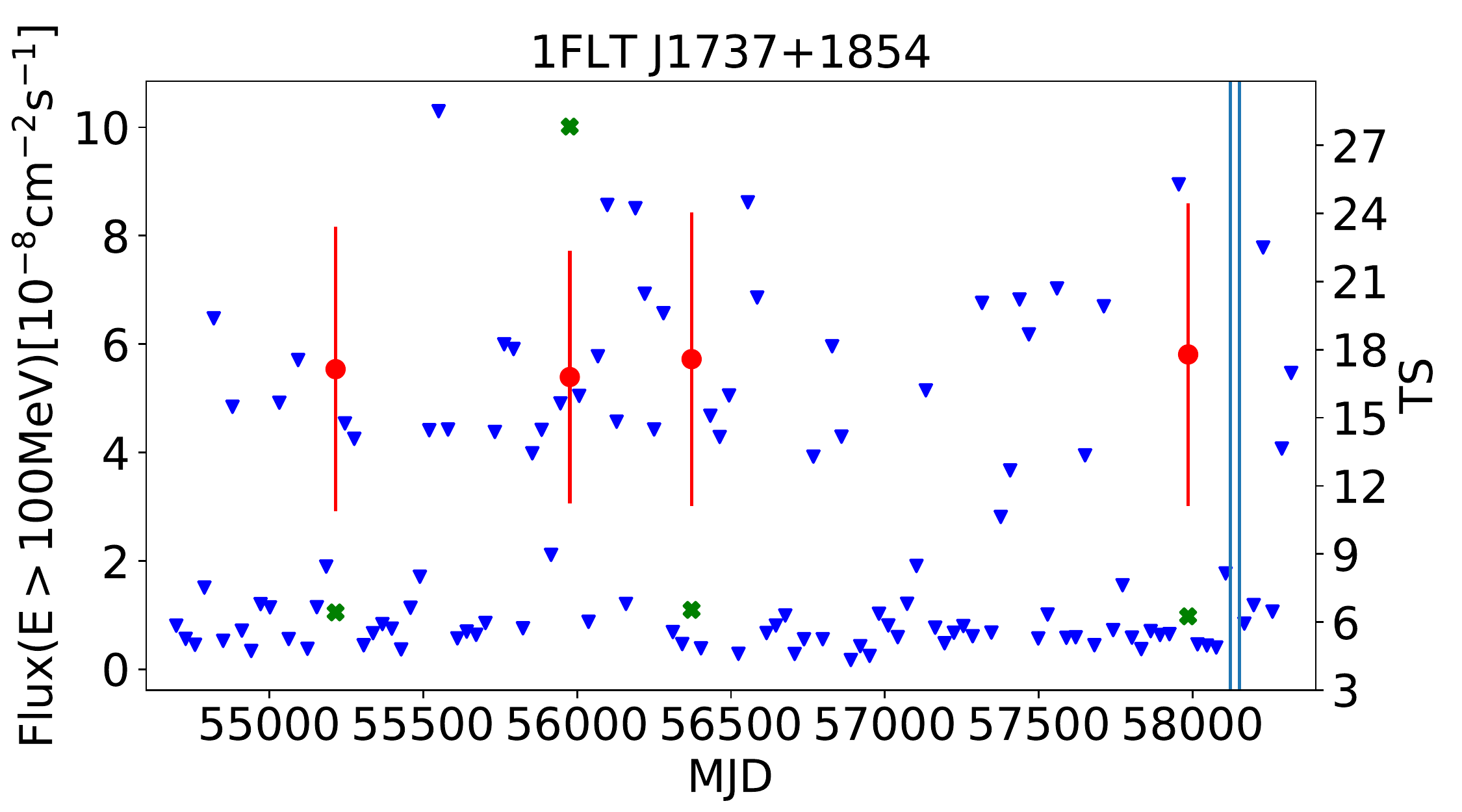}\label{fig:1FLTJ1737+1854}\\
  %[1FLTJ1734+5158]&%[1FLTJ1737+1854]&%[1FLTJ1752+4355]
  \includegraphics[width=0.35\textwidth]{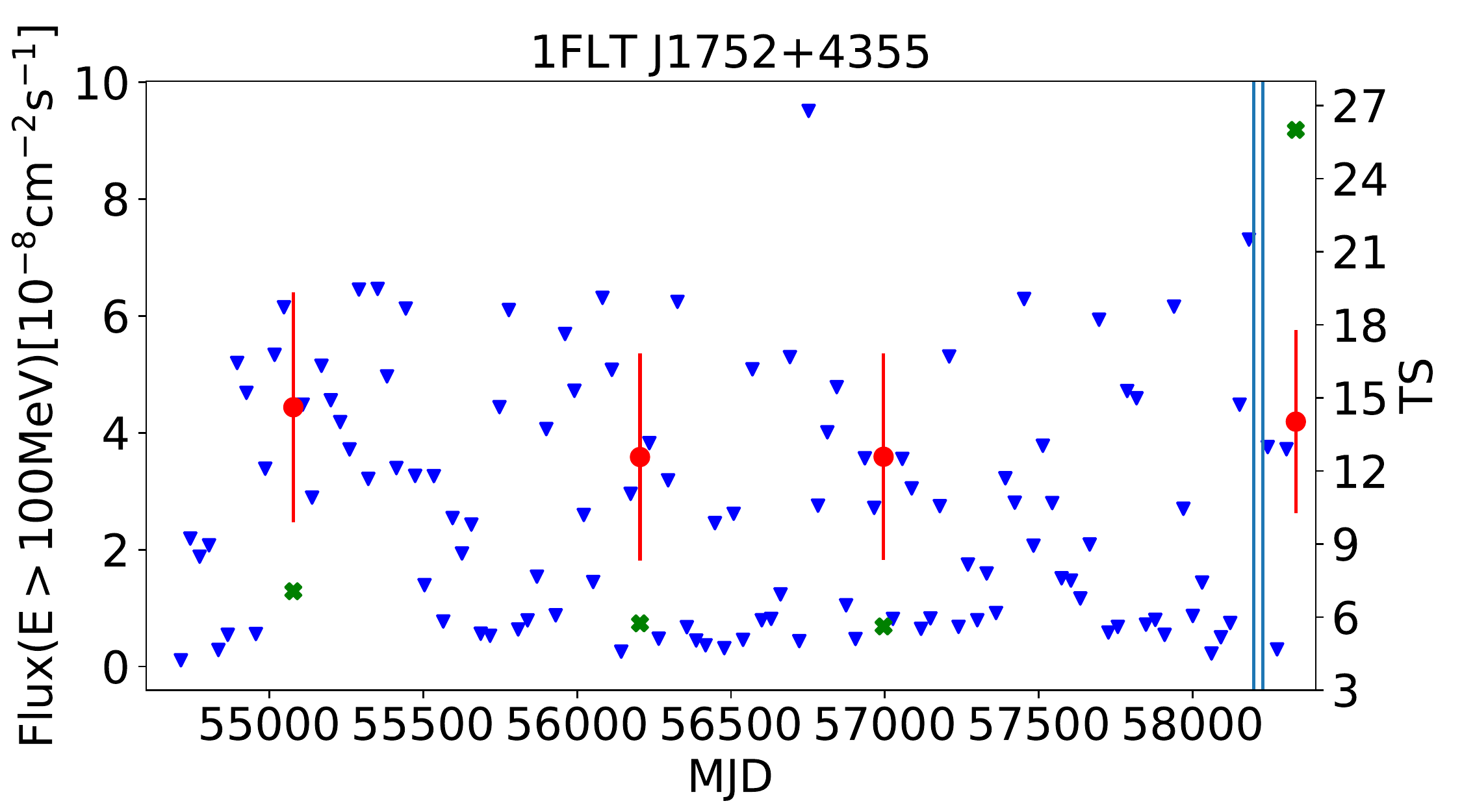}\label{fig:1FLTJ1752+4355}& 
  \includegraphics[width=0.35\textwidth]{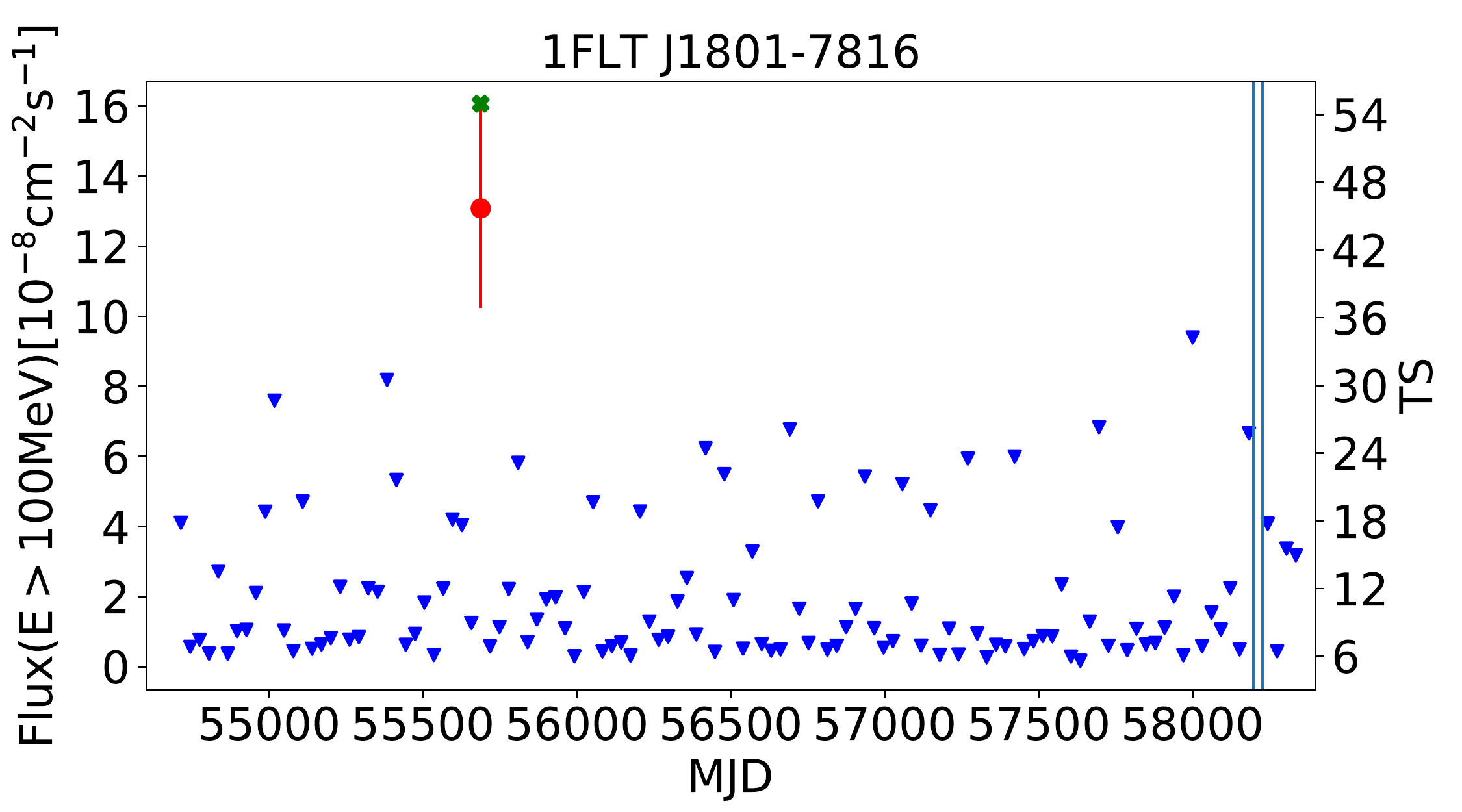}\label{fig:1FLTJ1801-7816}&
  \includegraphics[width=0.35\textwidth]{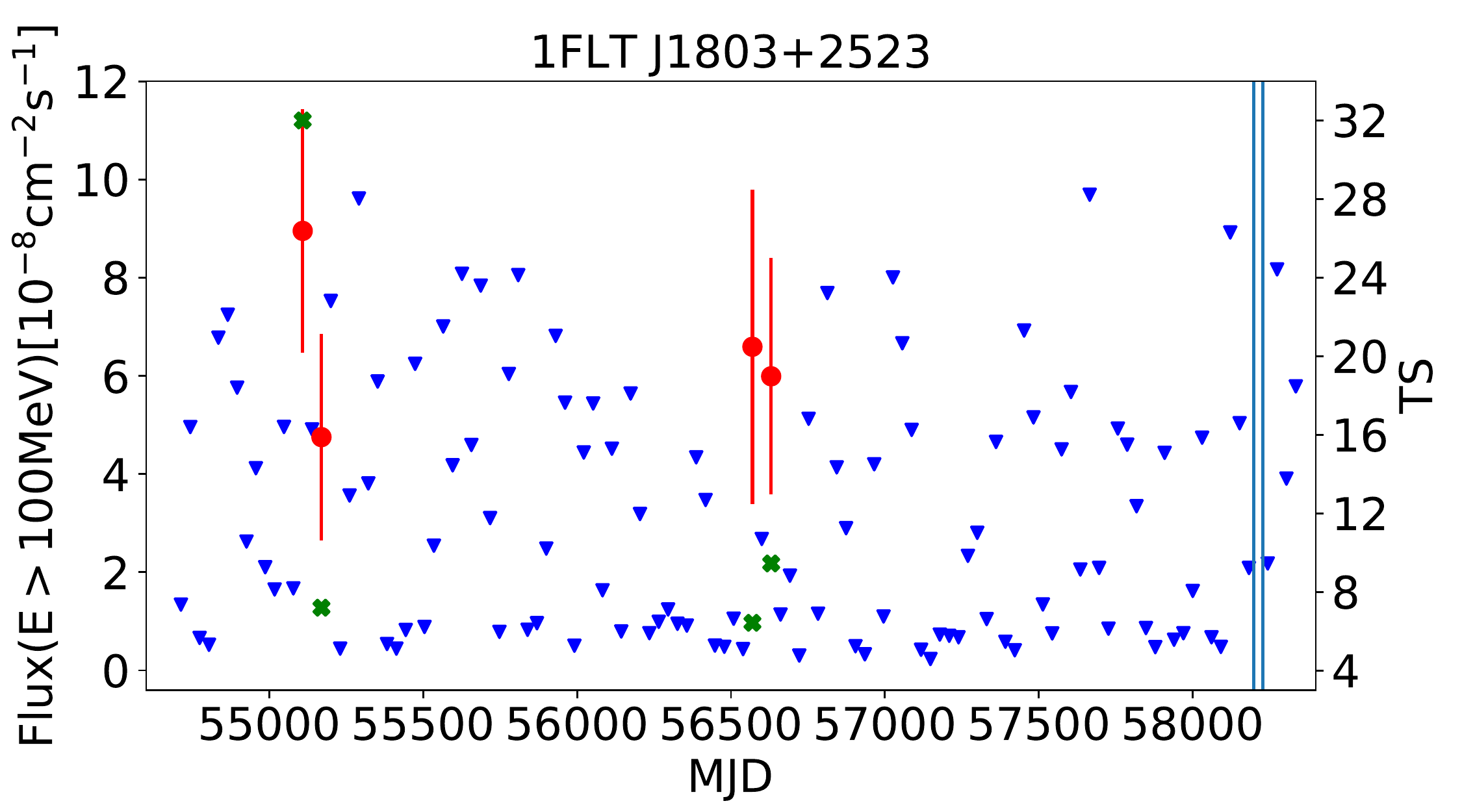}\label{fig:1FLTJ1803+2523}\\
  \includegraphics[width=0.35\textwidth]{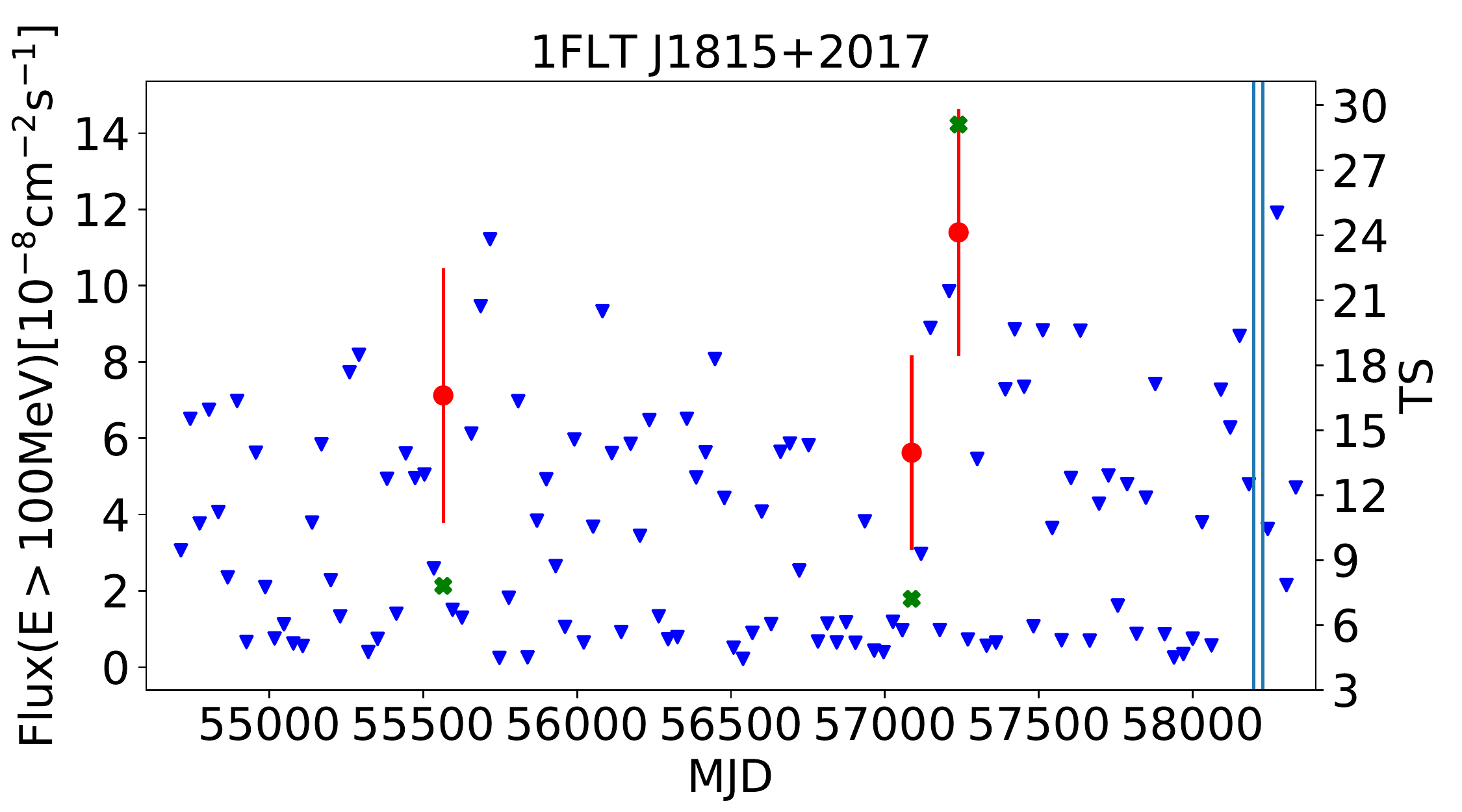}\label{fig:1FLTJ1815+2017}&
  %[1FLTJ1801-7816]&%[1FLTJ1803+2523]&%[1FLTJ1815+2017]\\
  \includegraphics[width=0.35\textwidth]{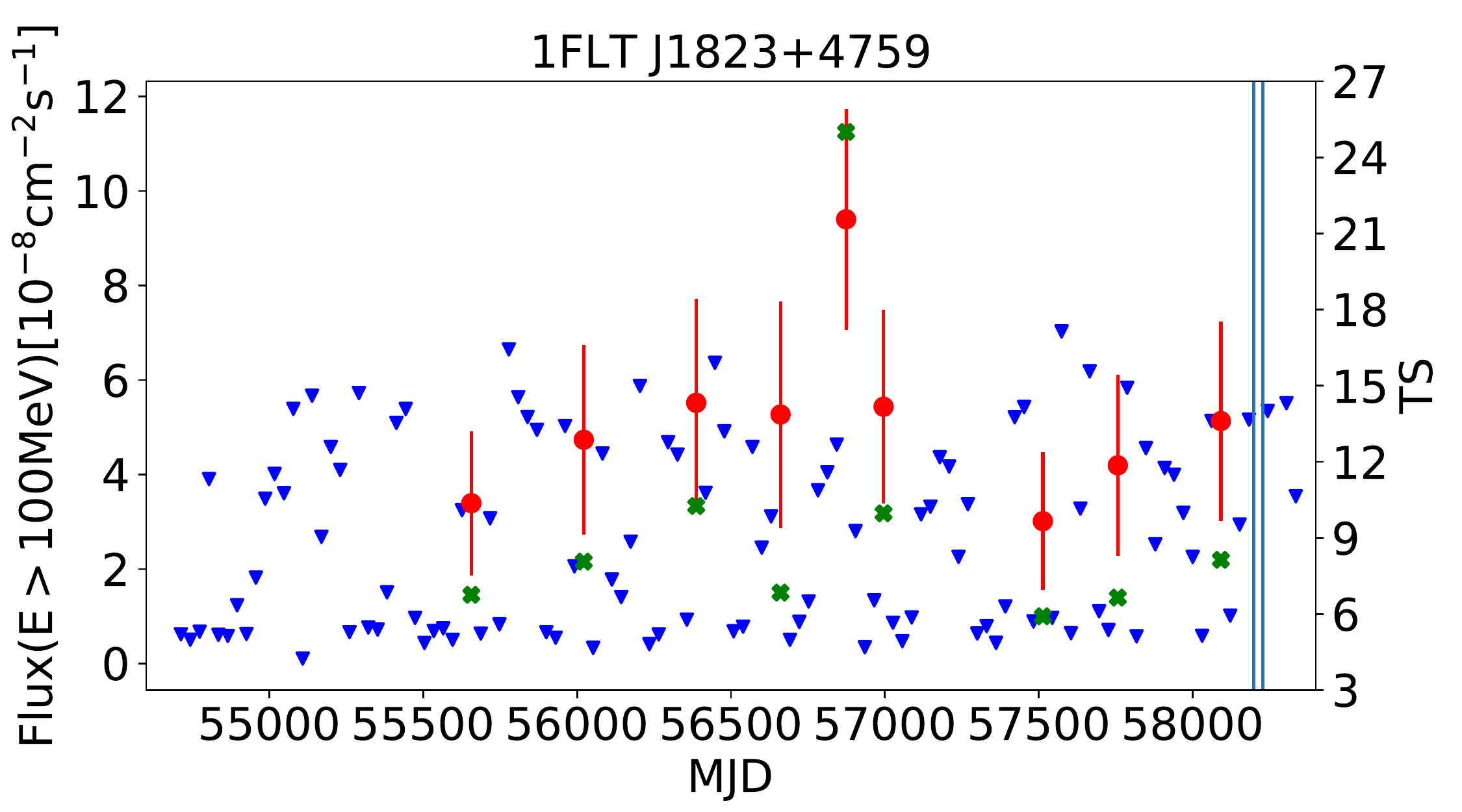}\label{fig:1FLTJ1823+4759}&
  \includegraphics[width=0.35\textwidth]{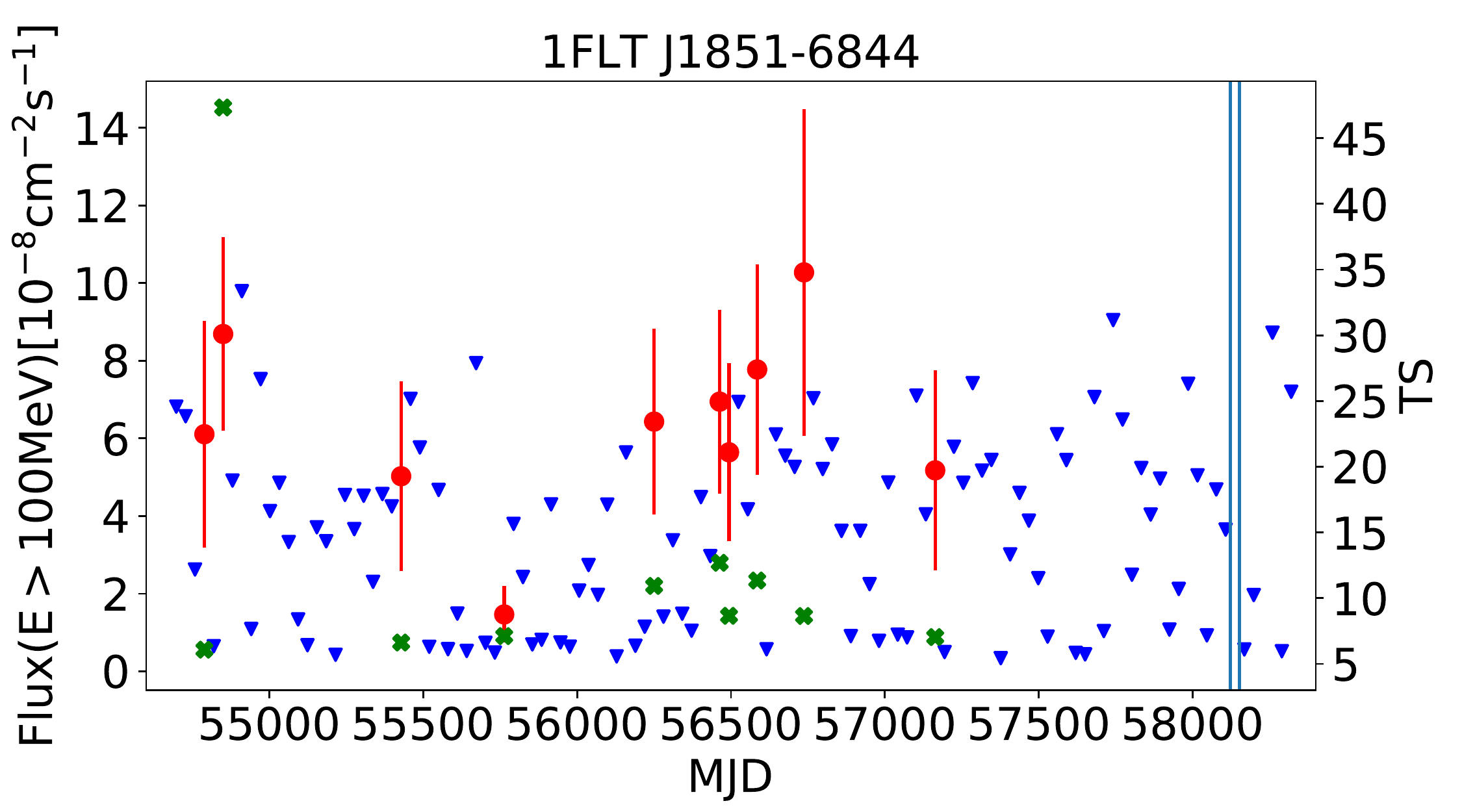}\label{fig:1FLTJ1851-6844}\\
  \includegraphics[width=0.35\textwidth]{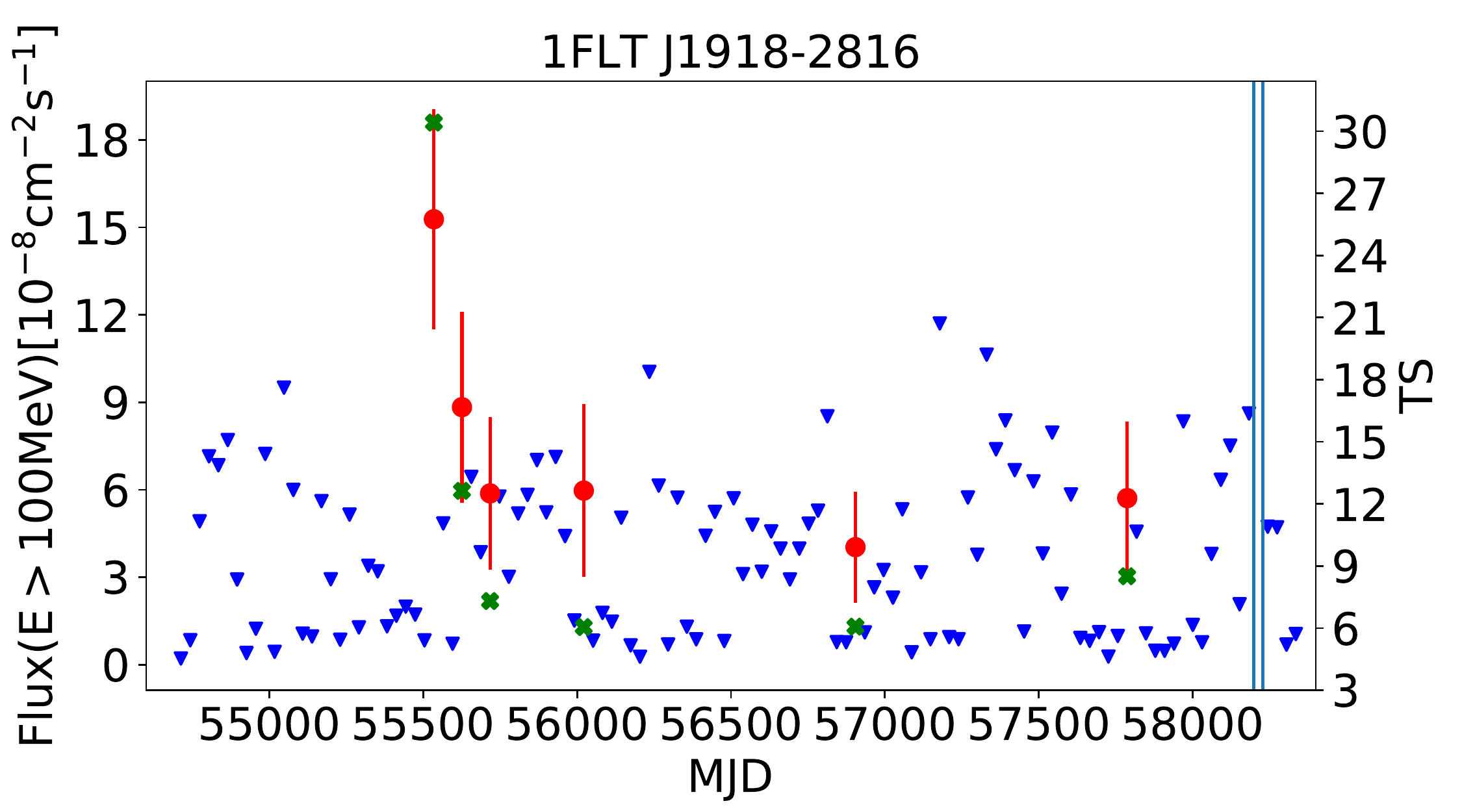}\label{fig:1FLTJ1918-2816}&
  %[1FLTJ1823+4759]&%[1FLTJ1851-6844]&%[1FLTJ1918-2816]\\
  \includegraphics[width=0.35\textwidth]{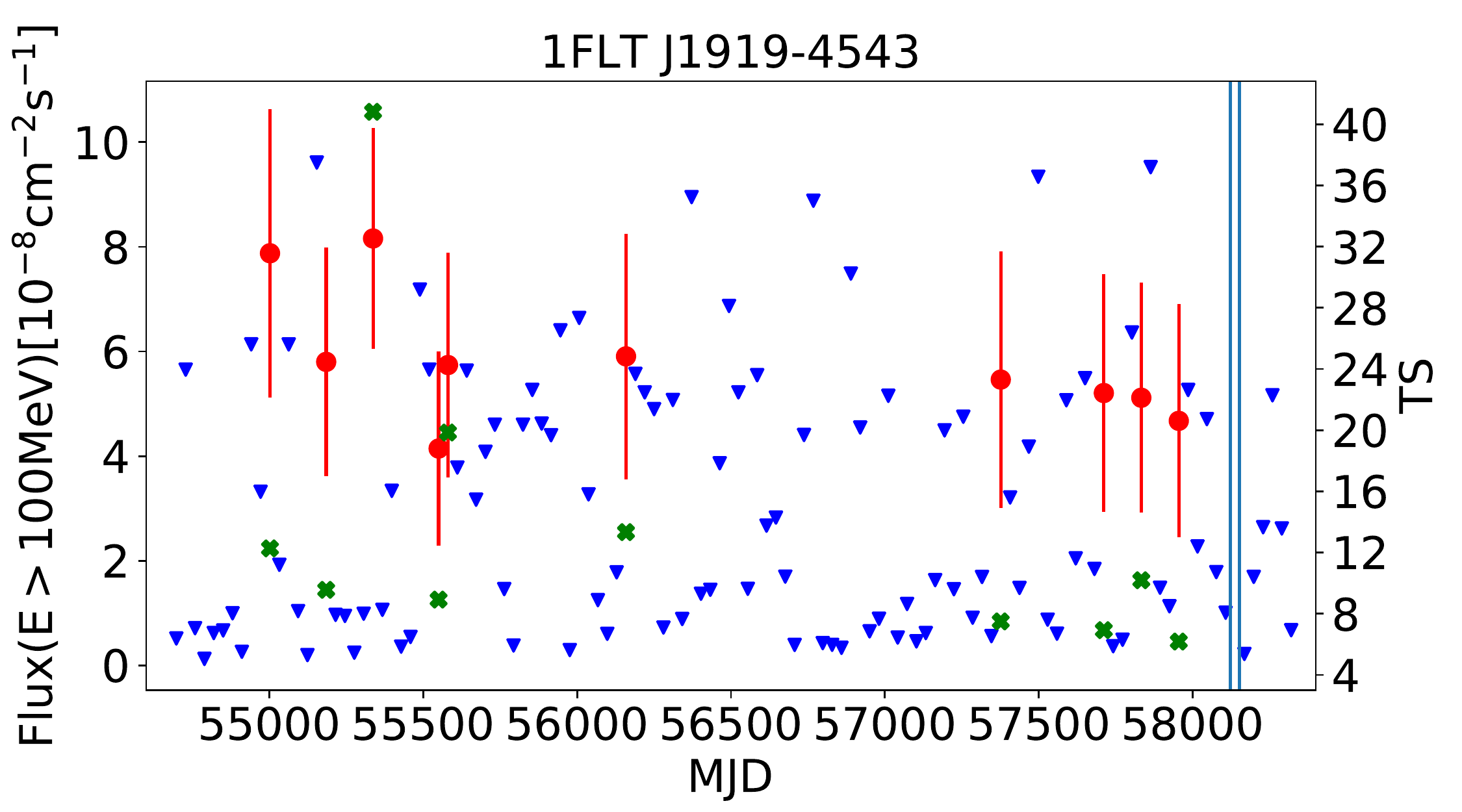}\label{fig:1FLTJ1919-4543}&
  \includegraphics[width=0.35\textwidth]{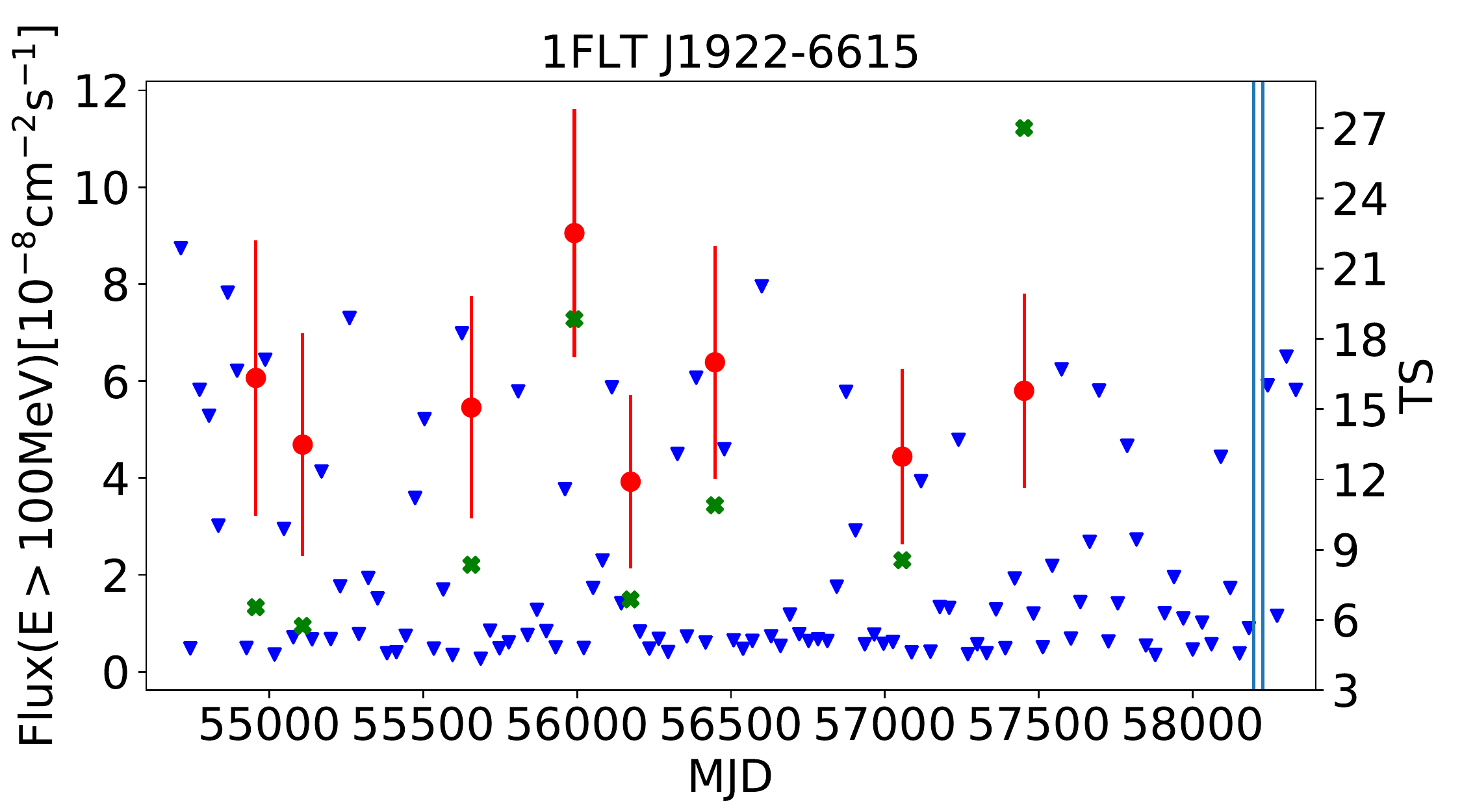}\label{fig:1FLTJ1922-6615}\\
  \includegraphics[width=0.35\textwidth]{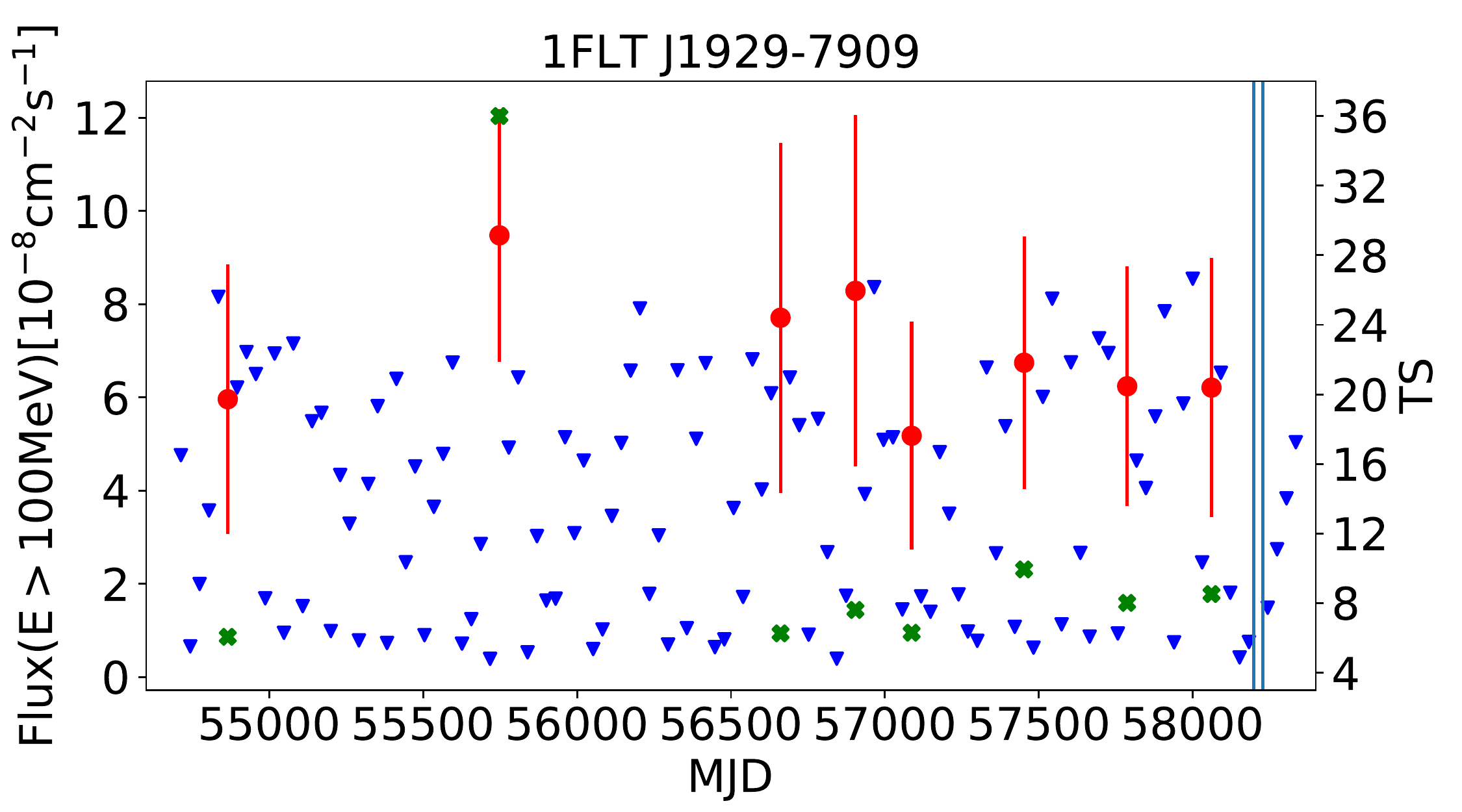}\label{fig:1FLTJ1929-7909}&
  %[1FLTJ1919-4543]&%[1FLTJ1922-6615]&%[1FLTJ1929-7909]\\
  \includegraphics[width=0.35\textwidth]{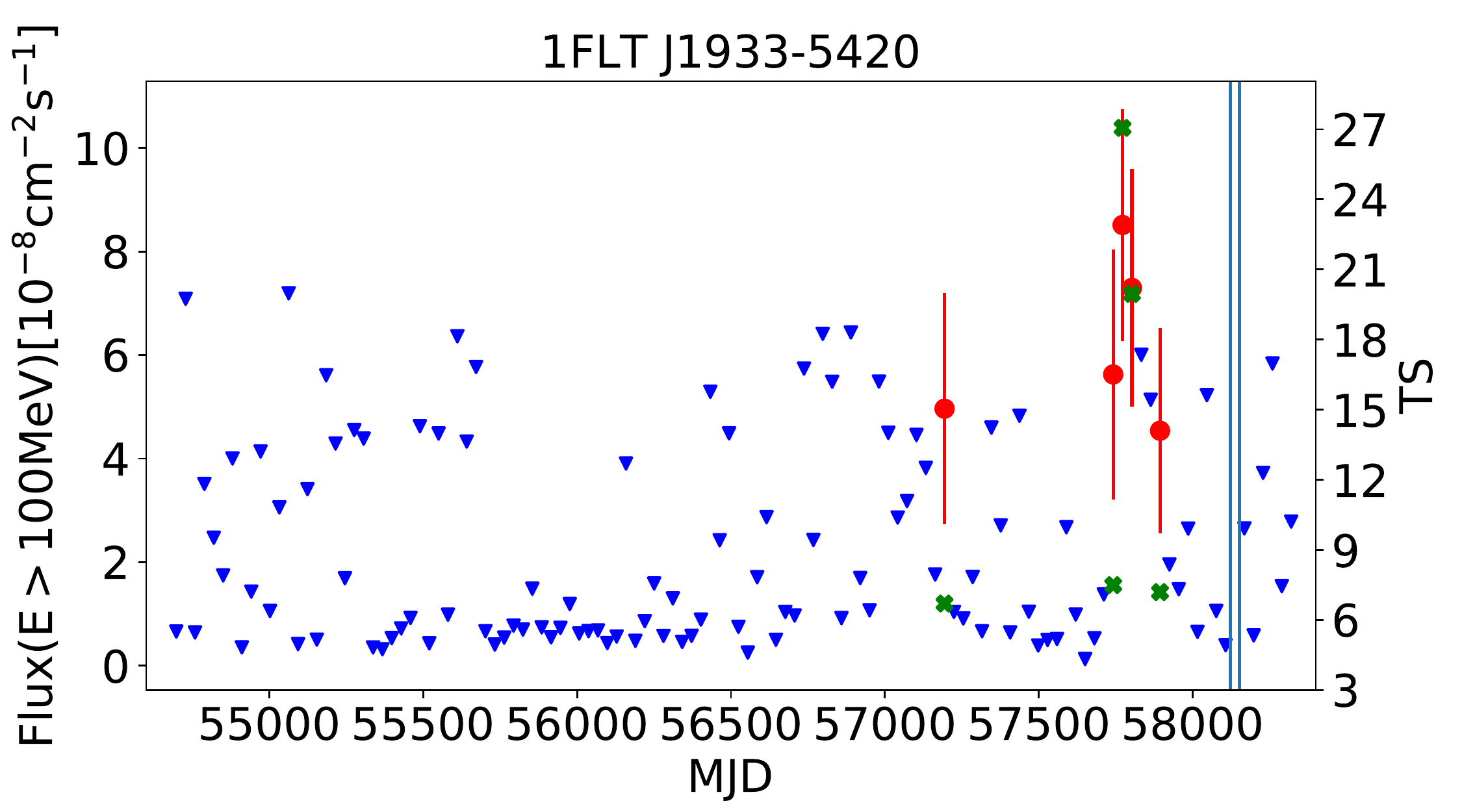}\label{fig:1FLTJ1933-5420}&
  \includegraphics[width=0.35\textwidth]{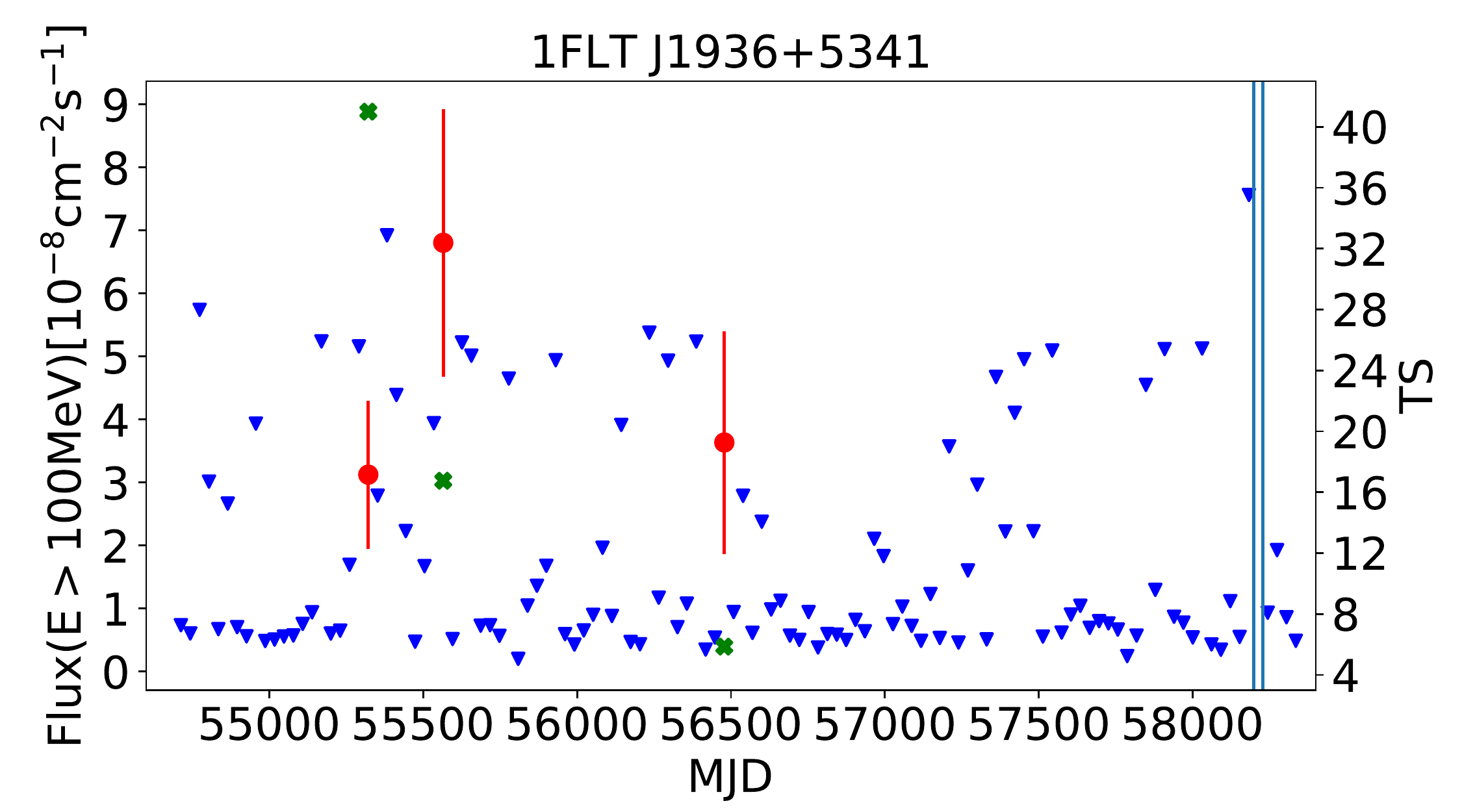}\label{fig:1FLTJ1936+5341}\\
  \includegraphics[width=0.35\textwidth]{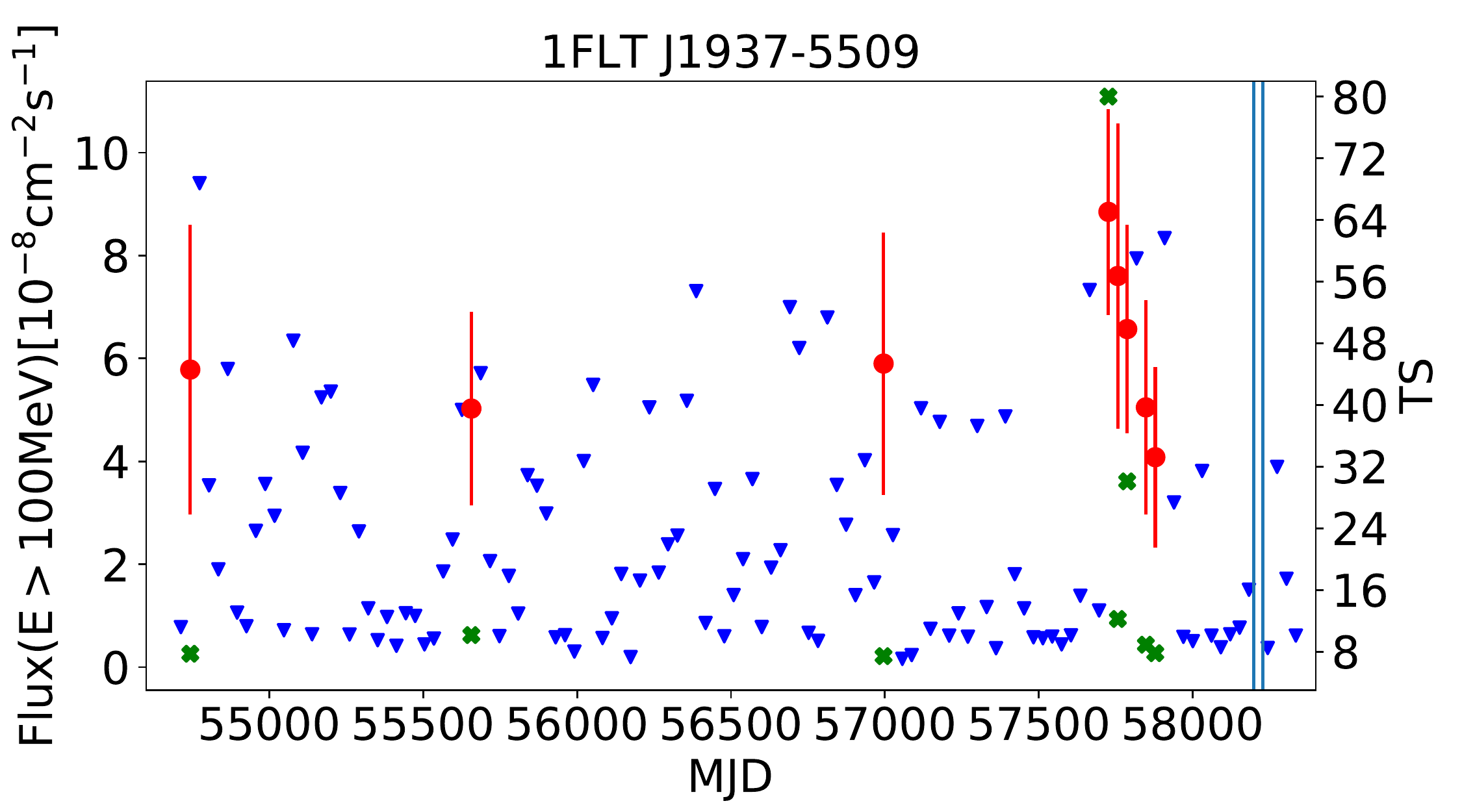}\label{fig:1FLTJ1937-5509}&
  %[1FLTJ1933-5420]&%[1FLTJ1936+5341]&%[1FLTJ1937-5509]\\
  \includegraphics[width=0.35\textwidth]{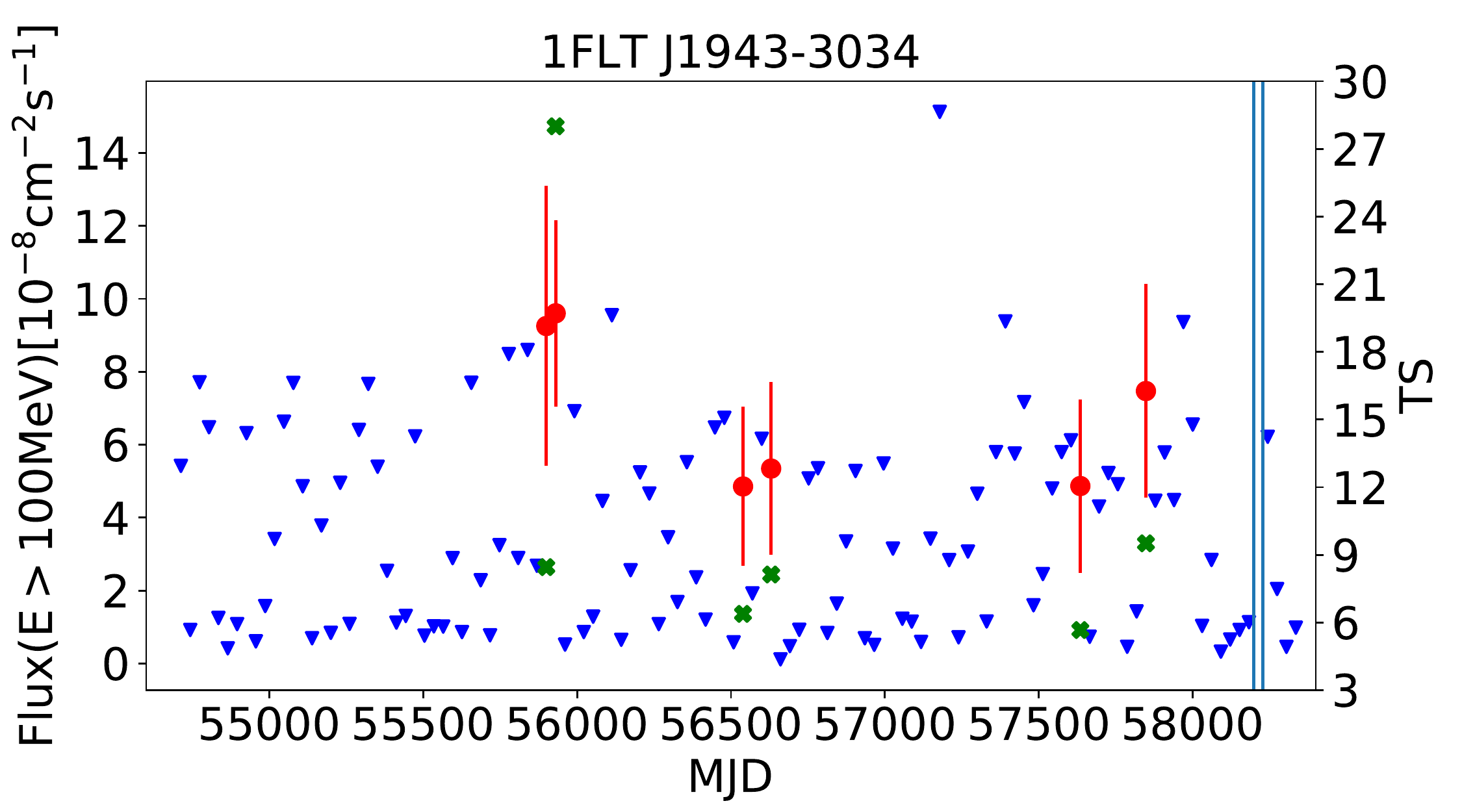}\label{fig:1FLTJ1943-3034}&
  \includegraphics[width=0.35\textwidth]{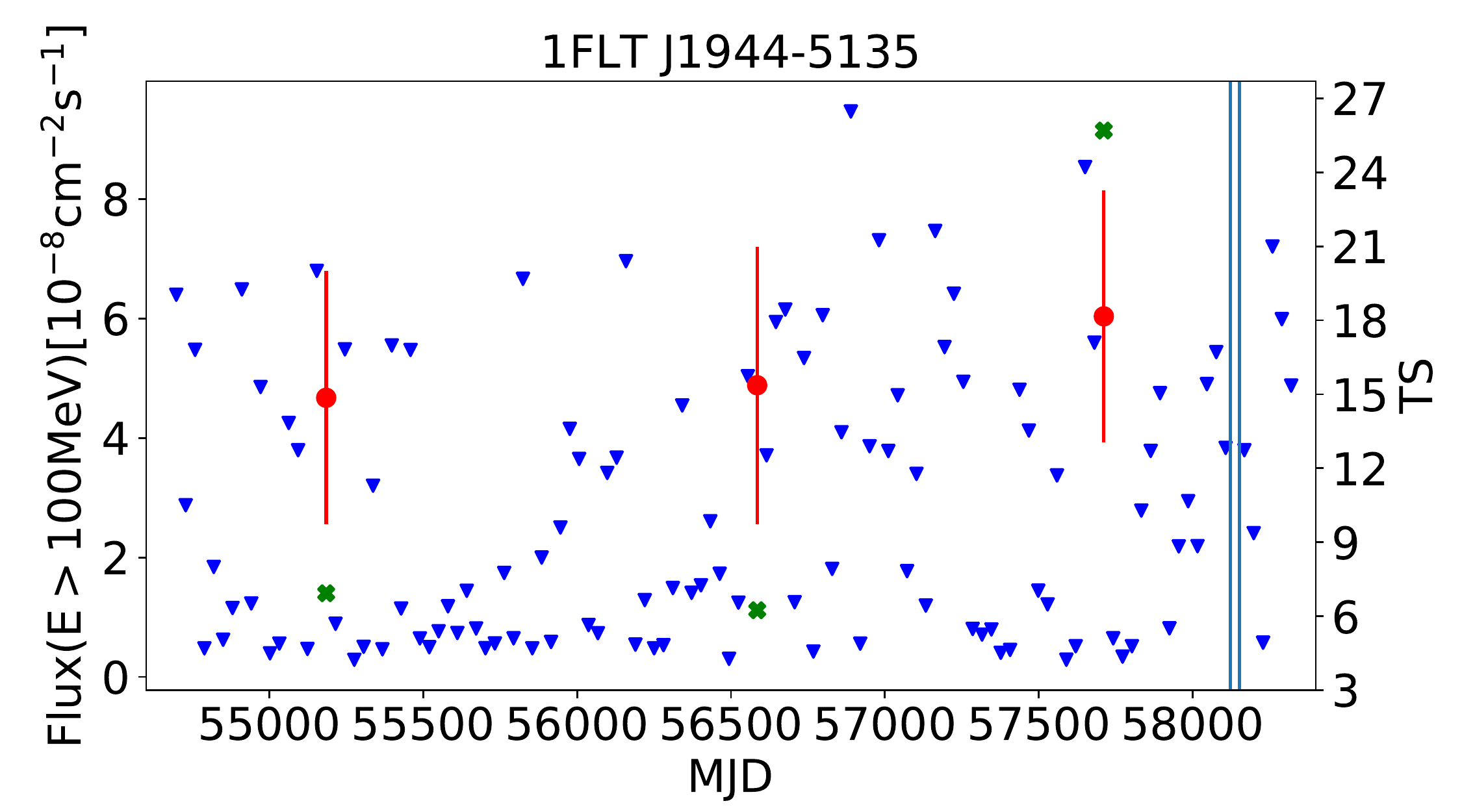}\label{fig:1FLTJ1944-5135}\\
\end{tabular}
\end{figure}
\begin{figure}[!t]
	\centering            
	\ContinuedFloat 
\setlength\tabcolsep{0.0pt}
\begin{tabular}{ccc}
  \includegraphics[width=0.35\textwidth]{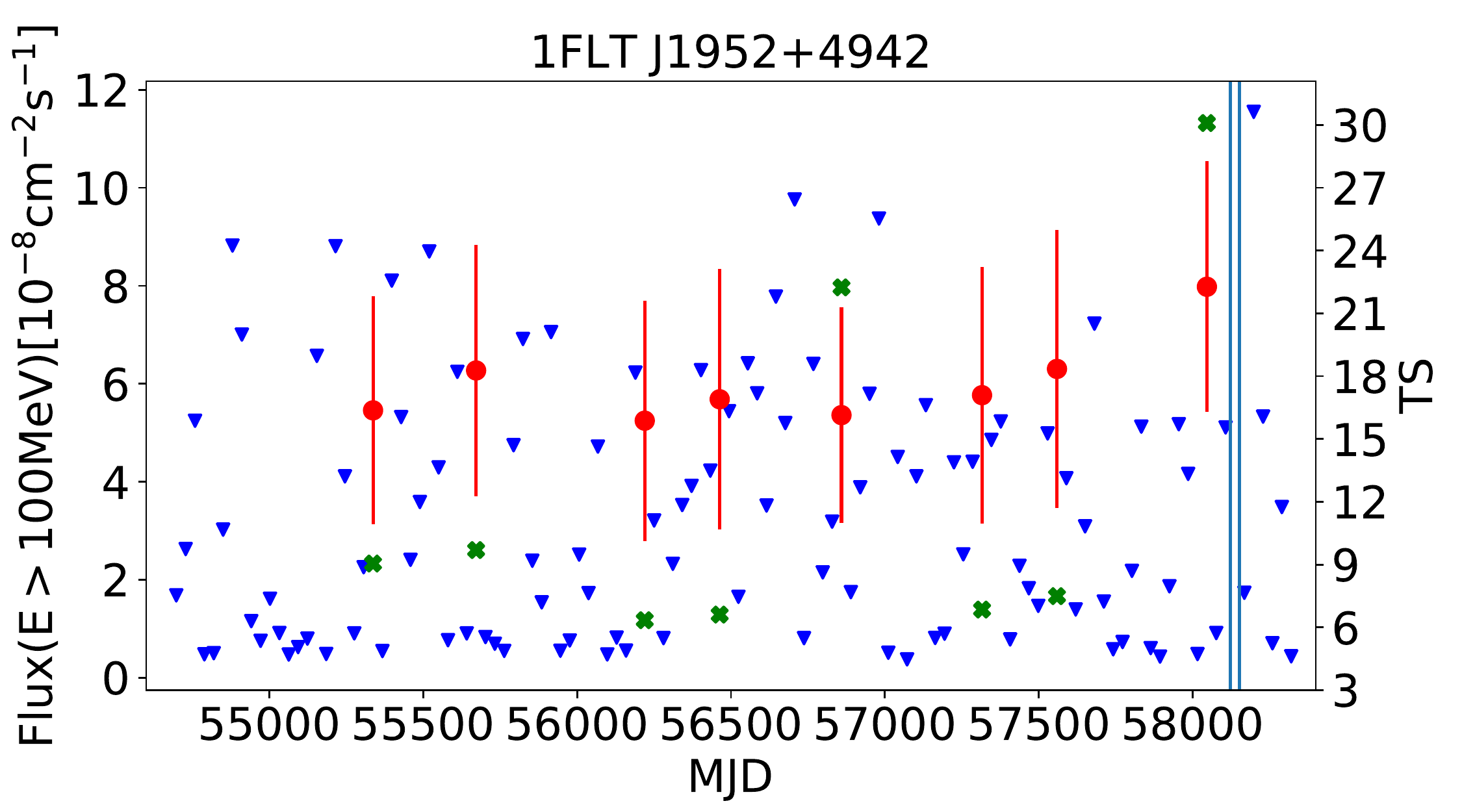}\label{fig:1FLTJ1952+4942}&
  %[1FLTJ1943-3034]&%[1FLTJ1944-5135]&%[1FLTJ1952+4942]\\
  \includegraphics[width=0.35\textwidth]{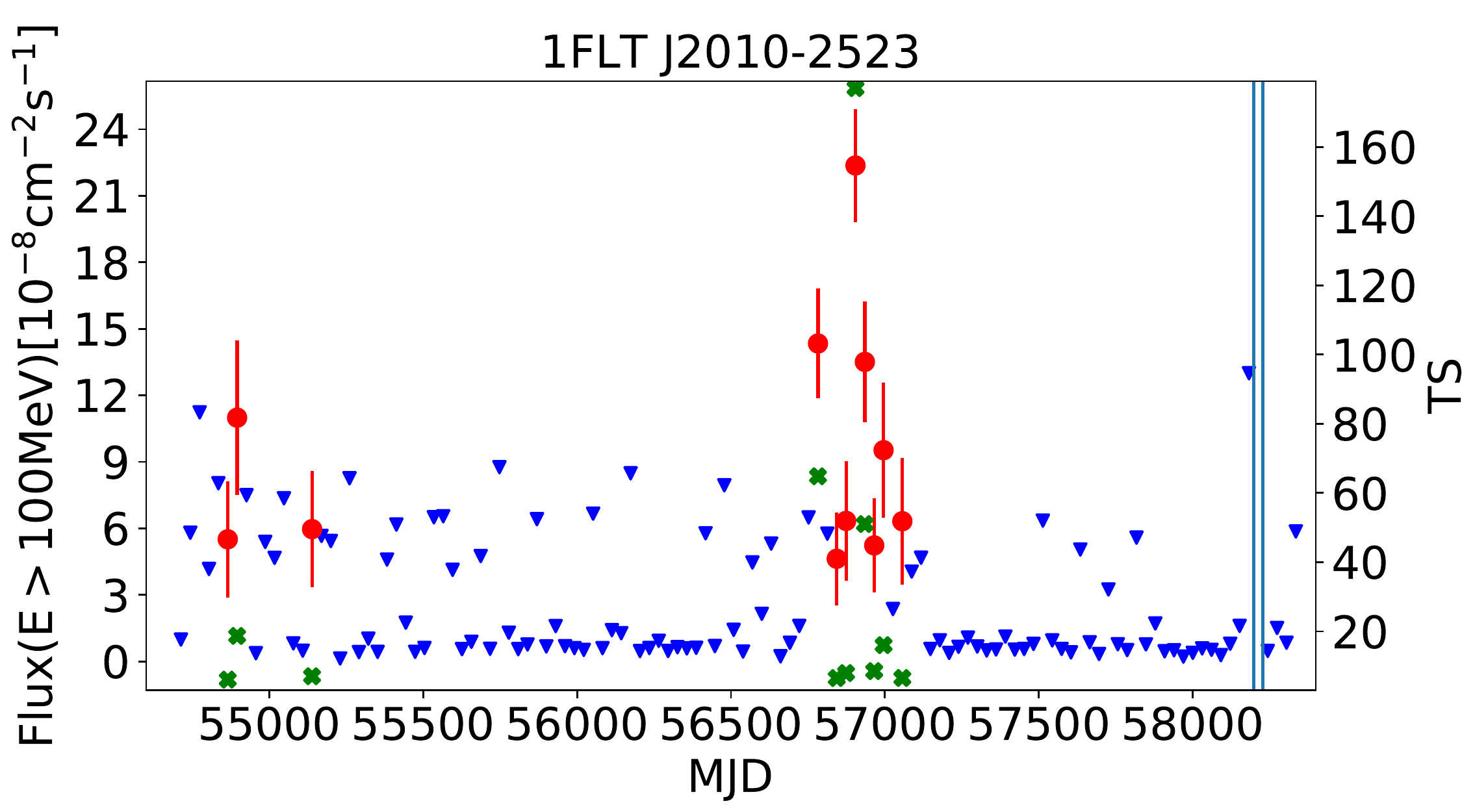}\label{fig:1FLTJ2010-2523}&
  \includegraphics[width=0.35\textwidth]{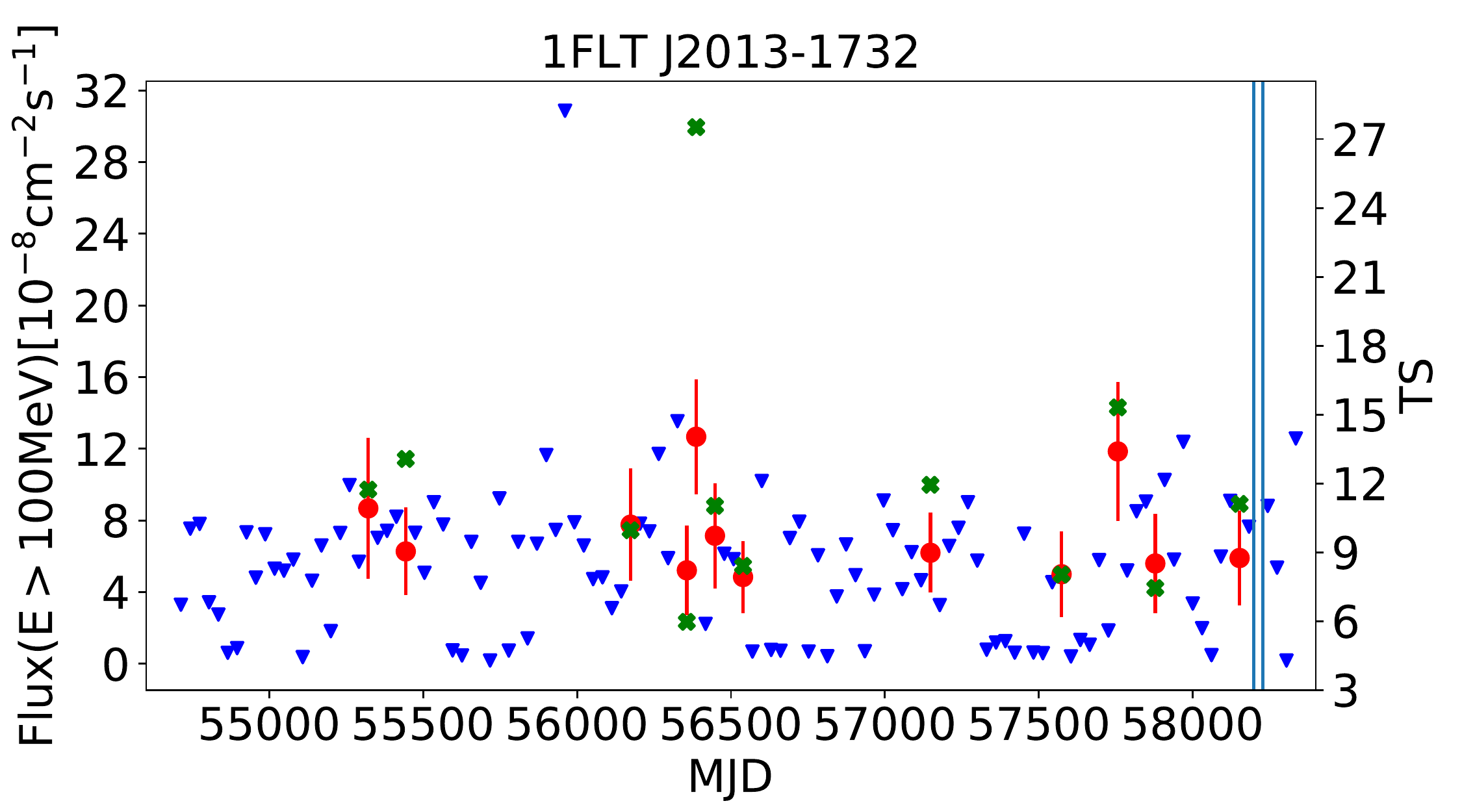}\label{fig:1FLTJ2013-1732}\\
  %[1FLTJ2010-2523]&%[1FLTJ2013-1732]&%[1FLTJ2035-2708]
  \includegraphics[width=0.35\textwidth]{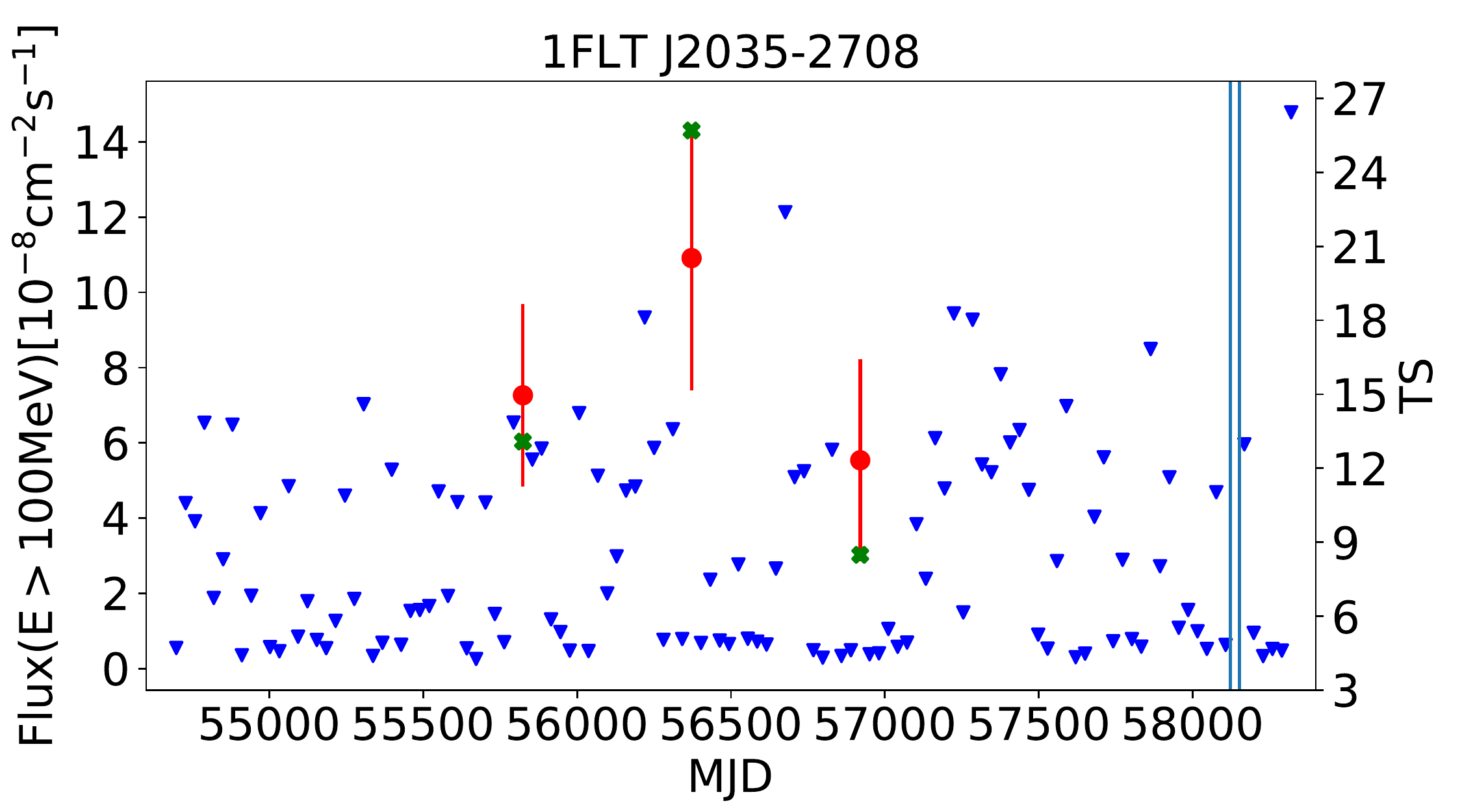}\label{fig:1FLTJ2035-2708}&
  \includegraphics[width=0.35\textwidth]{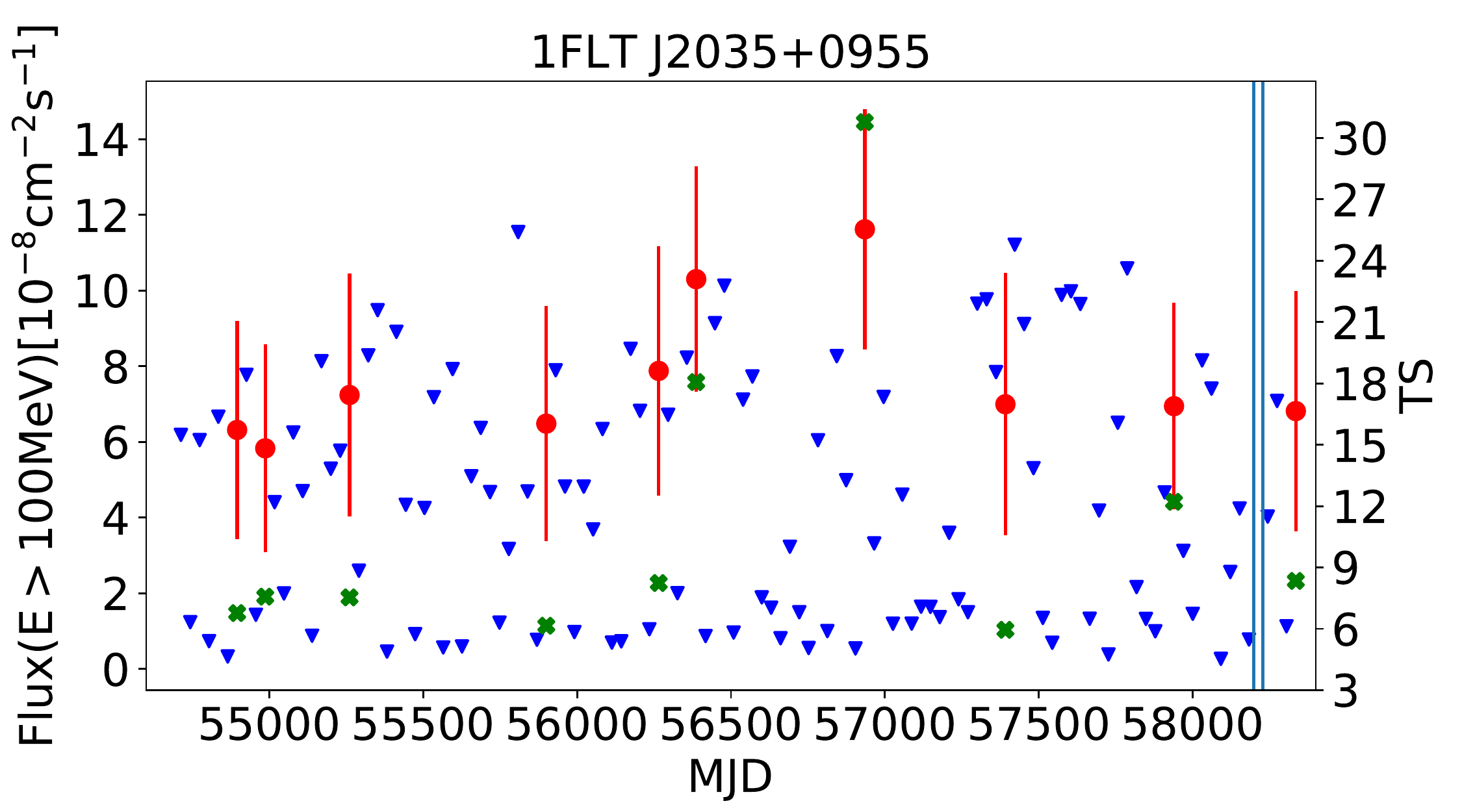}\label{fig:1FLTJ2035+0955}&
  \includegraphics[width=0.35\textwidth]{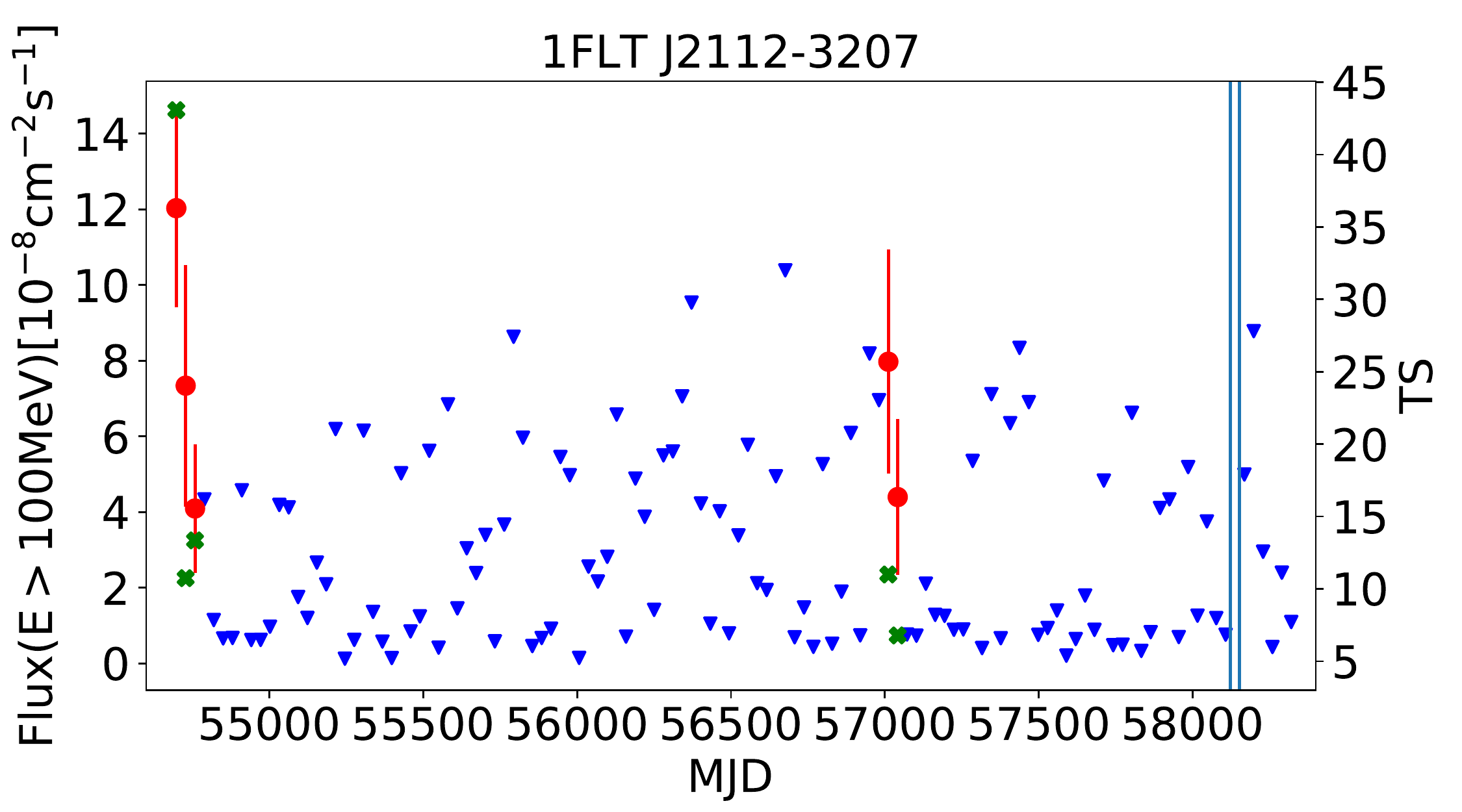}\label{fig:1FLTJ2112-3207}\\
  \includegraphics[width=0.35\textwidth]{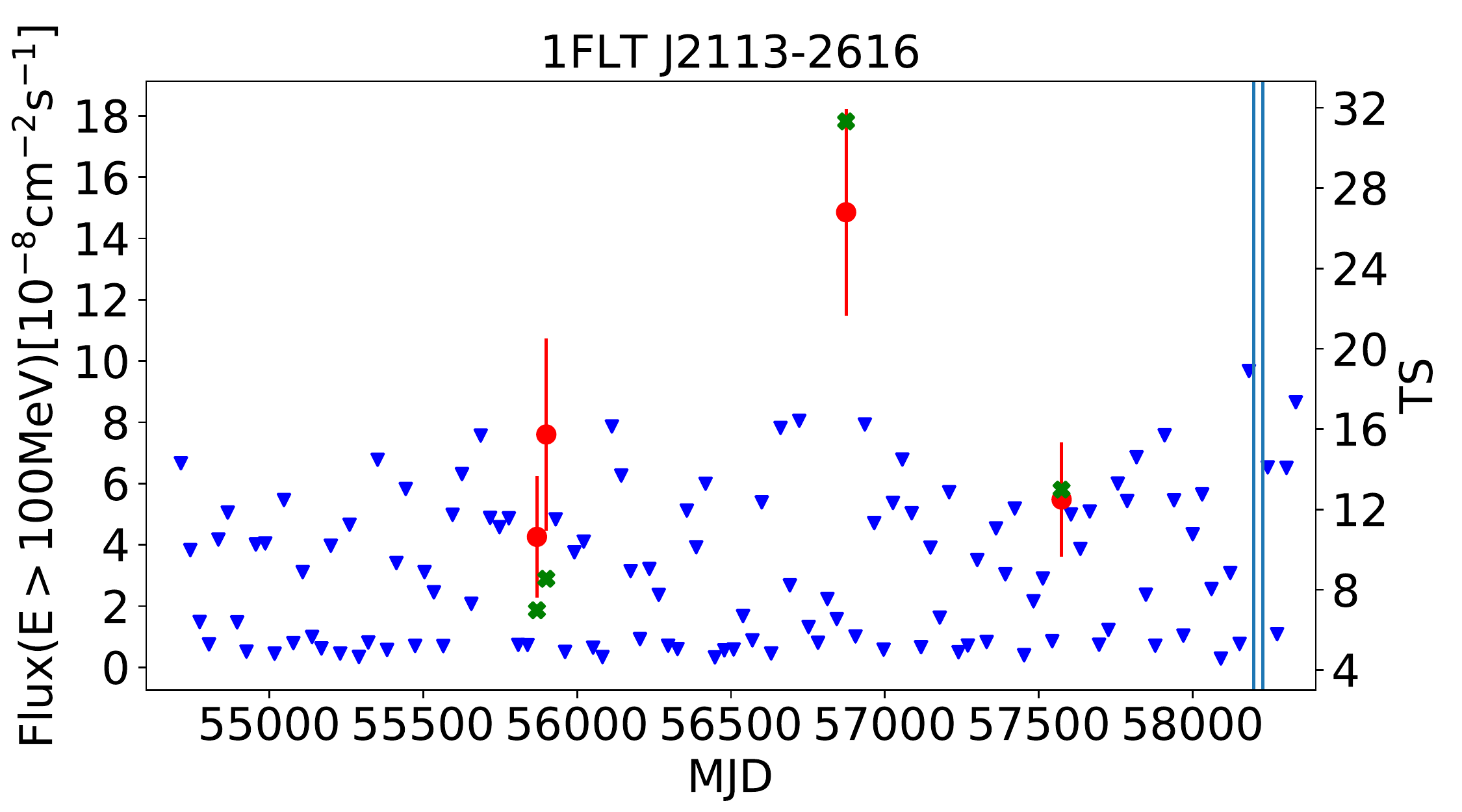}\label{fig:1FLTJ2113-2616}&
  %[1FLTJ2035+0955]&%[1FLTJ2112-3207]&%[1FLTJ2113-2616]\\
  \includegraphics[width=0.35\textwidth]{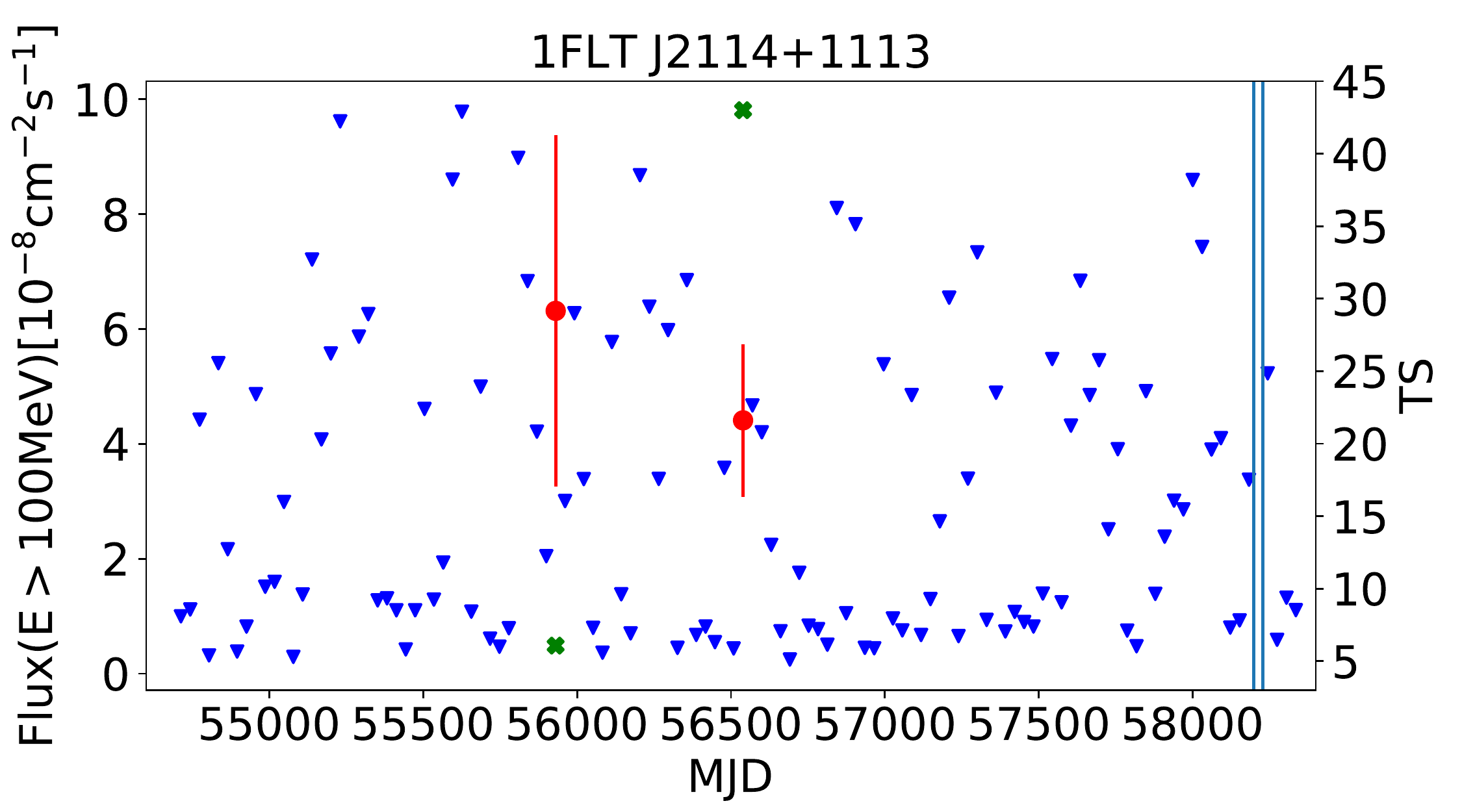}\label{fig:1FLTJ2114+1113}&
  \includegraphics[width=0.35\textwidth]{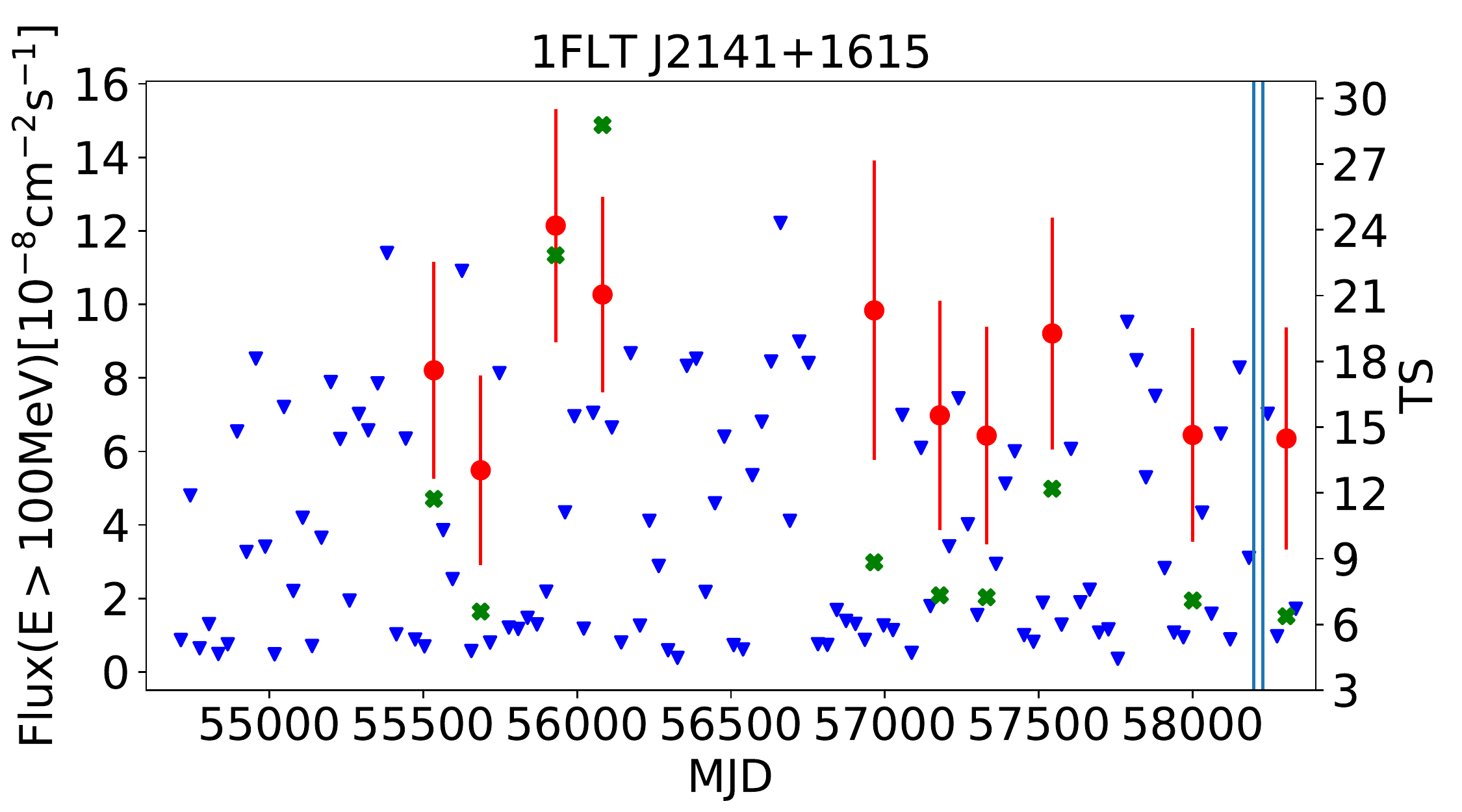}\label{fig:1FLTJ2141+1615}\\
  \includegraphics[width=0.35\textwidth]{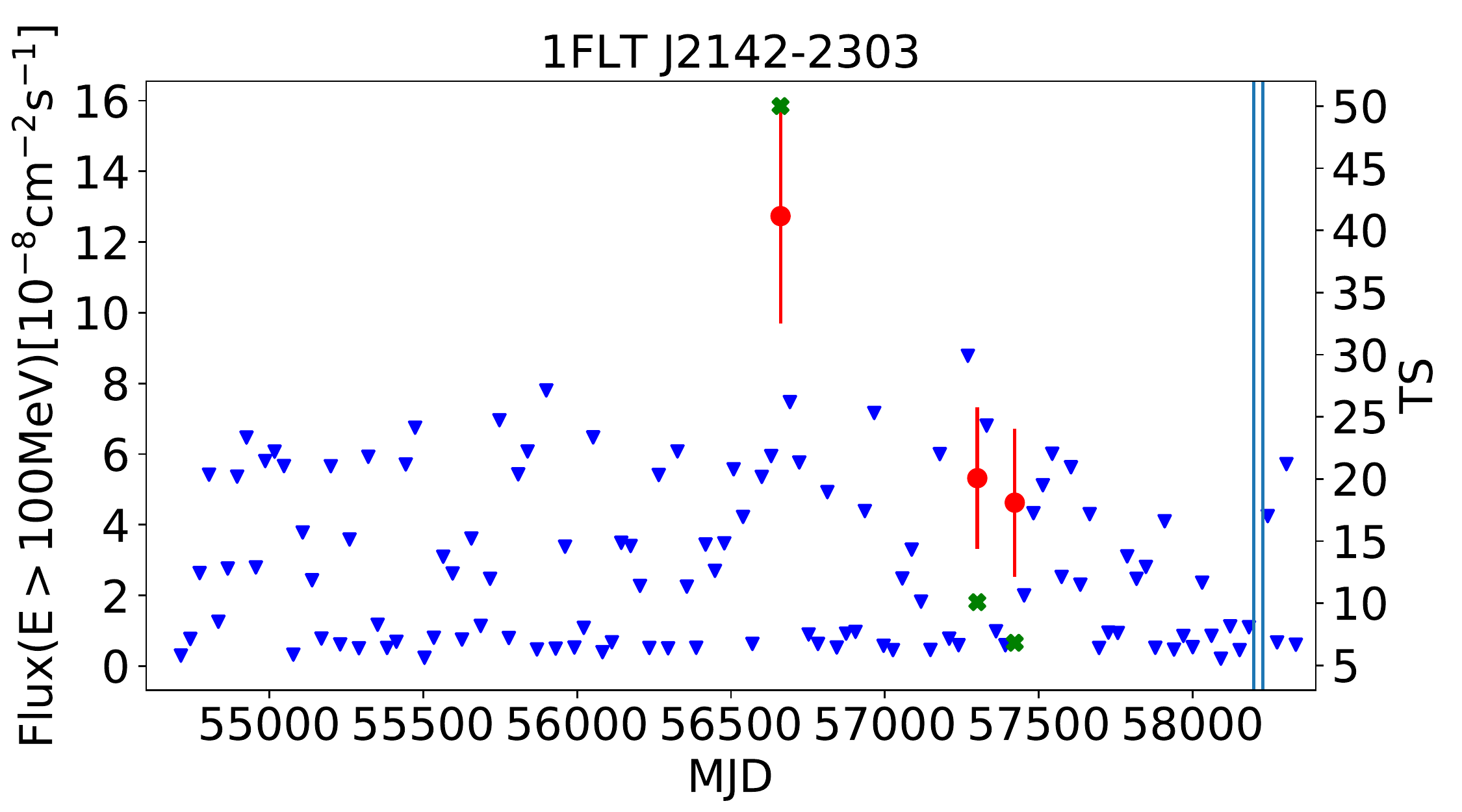}\label{fig:1FLTJ2142-2303}&
  \includegraphics[width=0.35\textwidth]{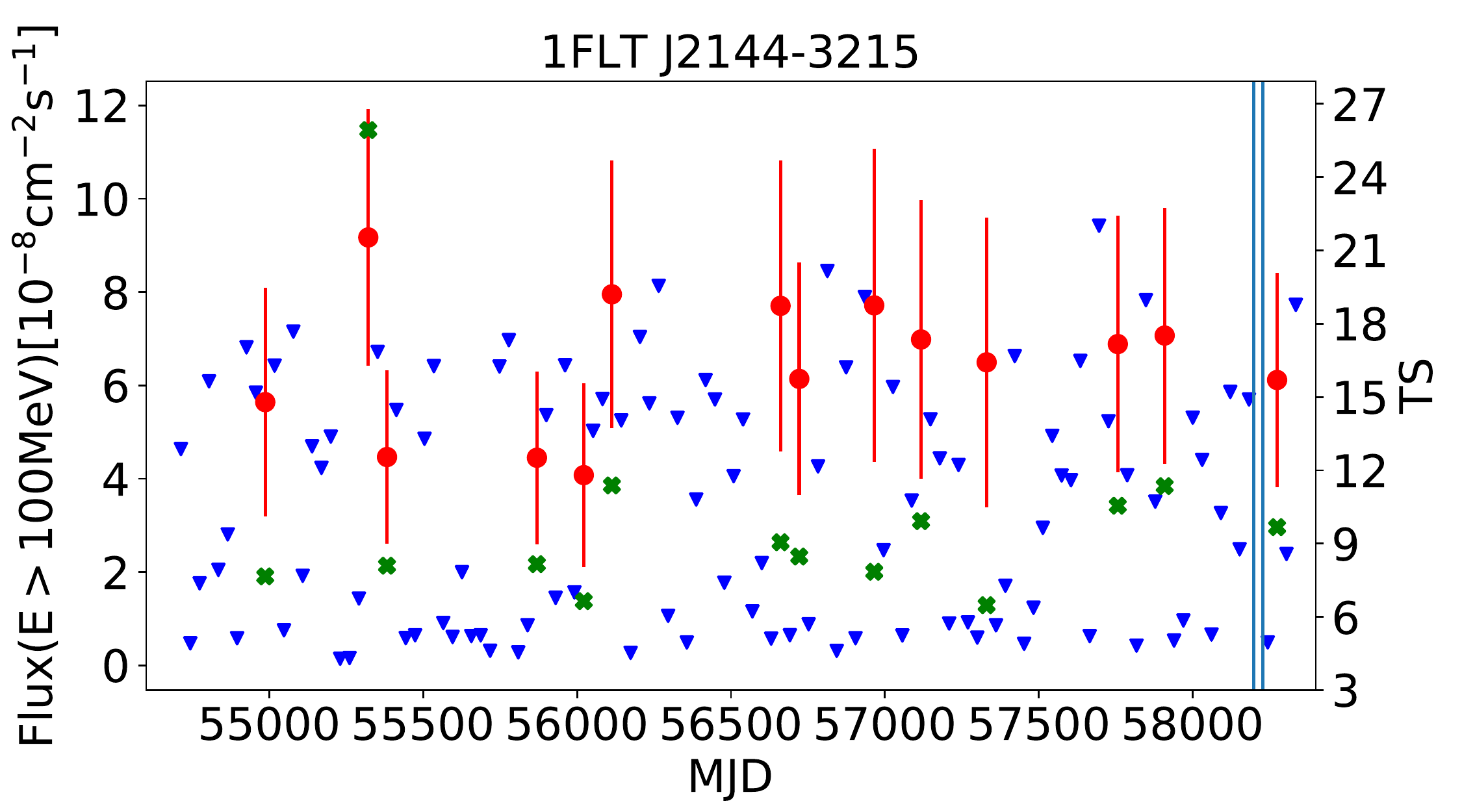}\label{fig:1FLTJ2144-3215}&
  %[1FLTJ2114+1113]&%[1FLTJ2141+1615]&%[1FLTJ2142-2303]\\
  \includegraphics[width=0.35\textwidth]{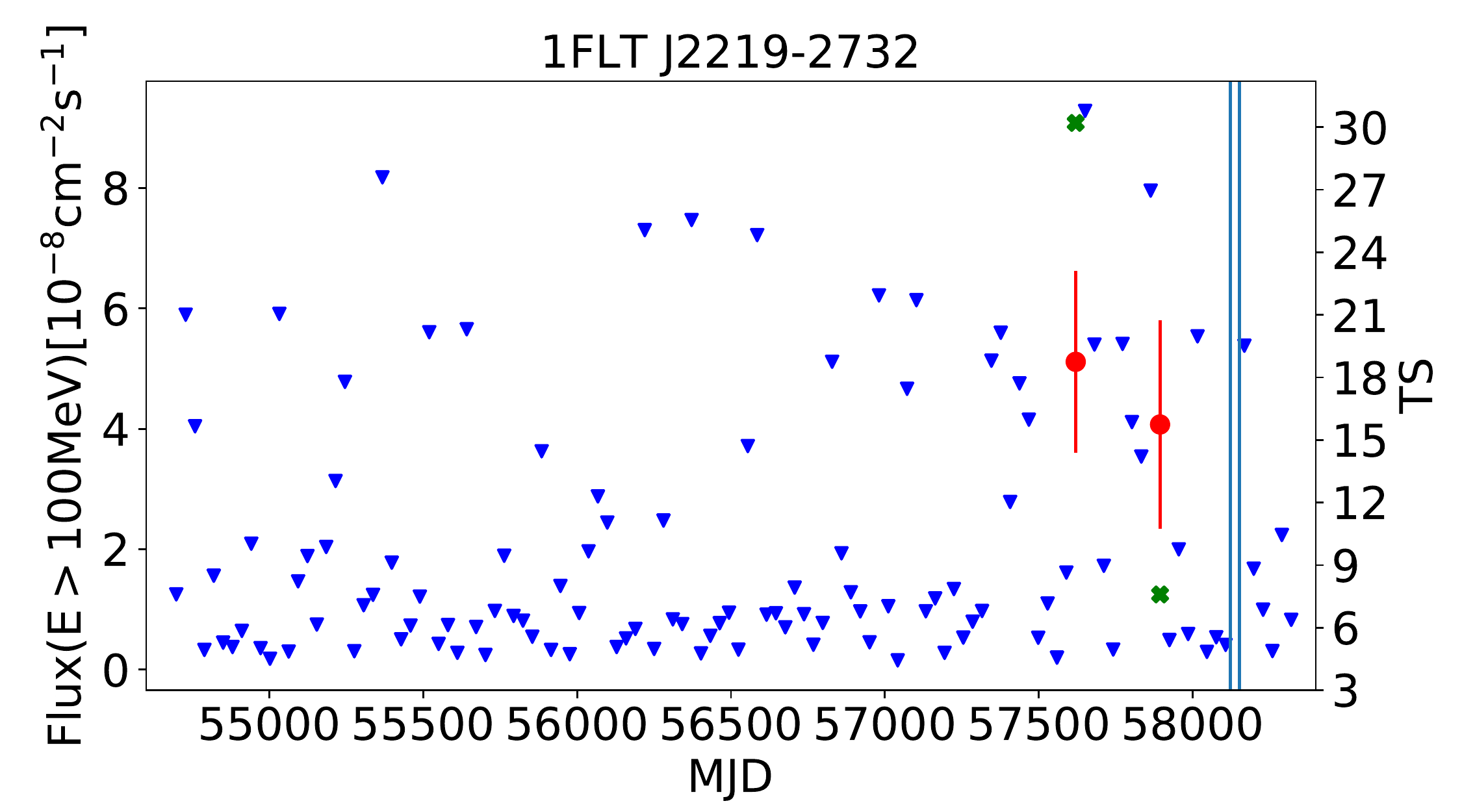}\label{fig:1FLTJ2219-2732}\\
  \includegraphics[width=0.35\textwidth]{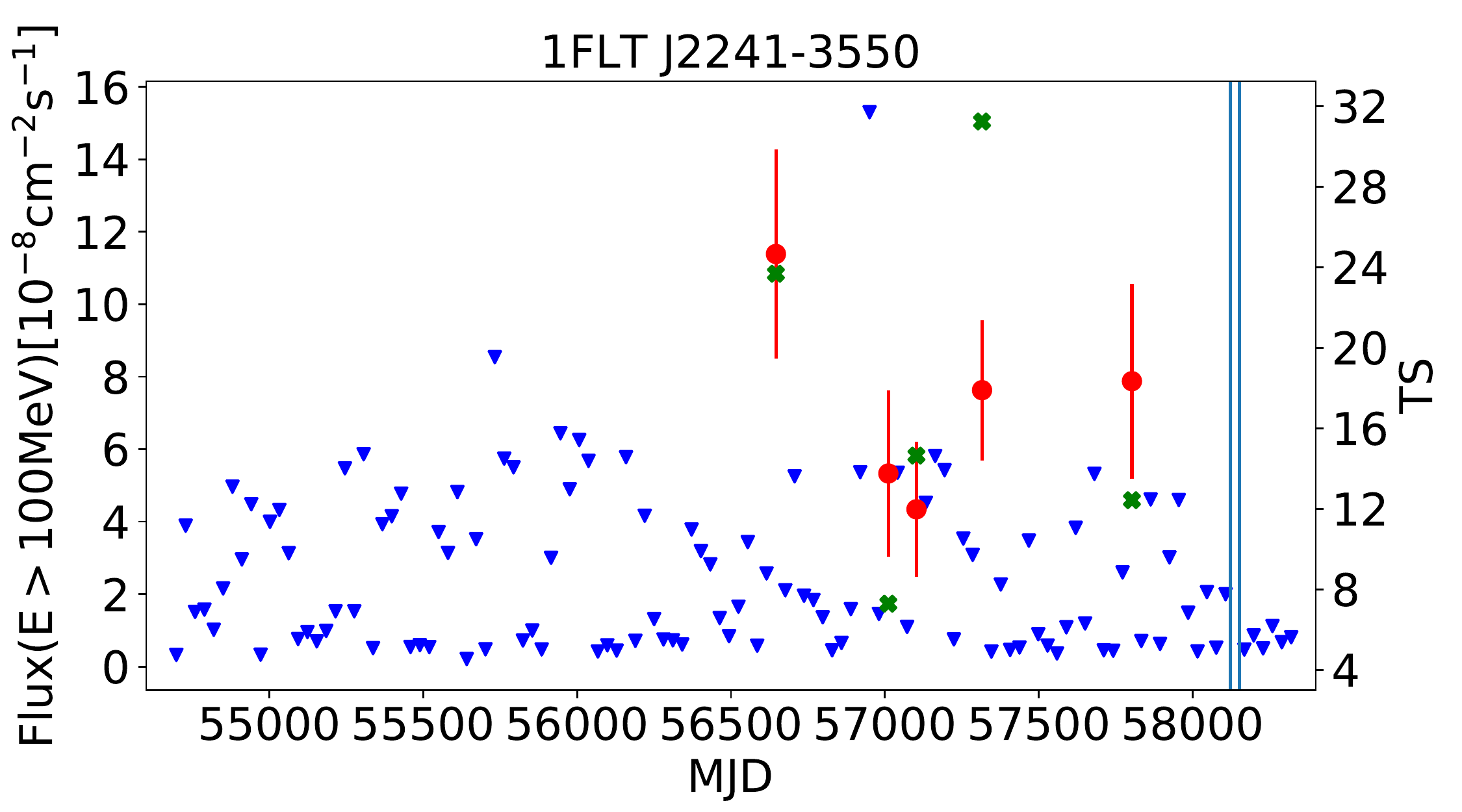}\label{fig:1FLTJ2241-3550}&
  \includegraphics[width=0.35\textwidth]{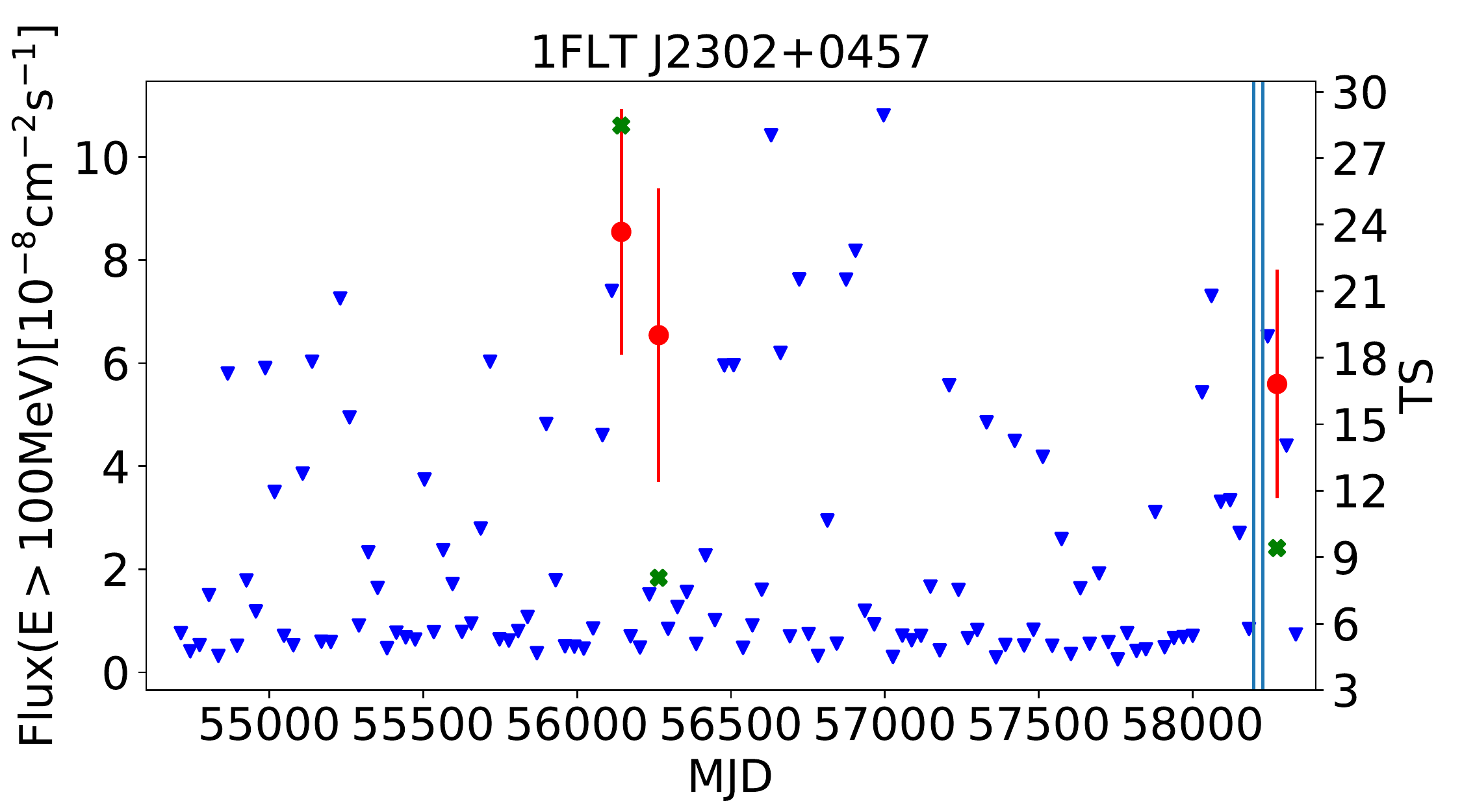}\label{fig:1FLTJ2302+0457}&
  %[1FLTJ2219-2732]&%[1FLTJ2241-3550]&%[1FLTJ2302+0457]\\
  \includegraphics[width=0.35\textwidth]{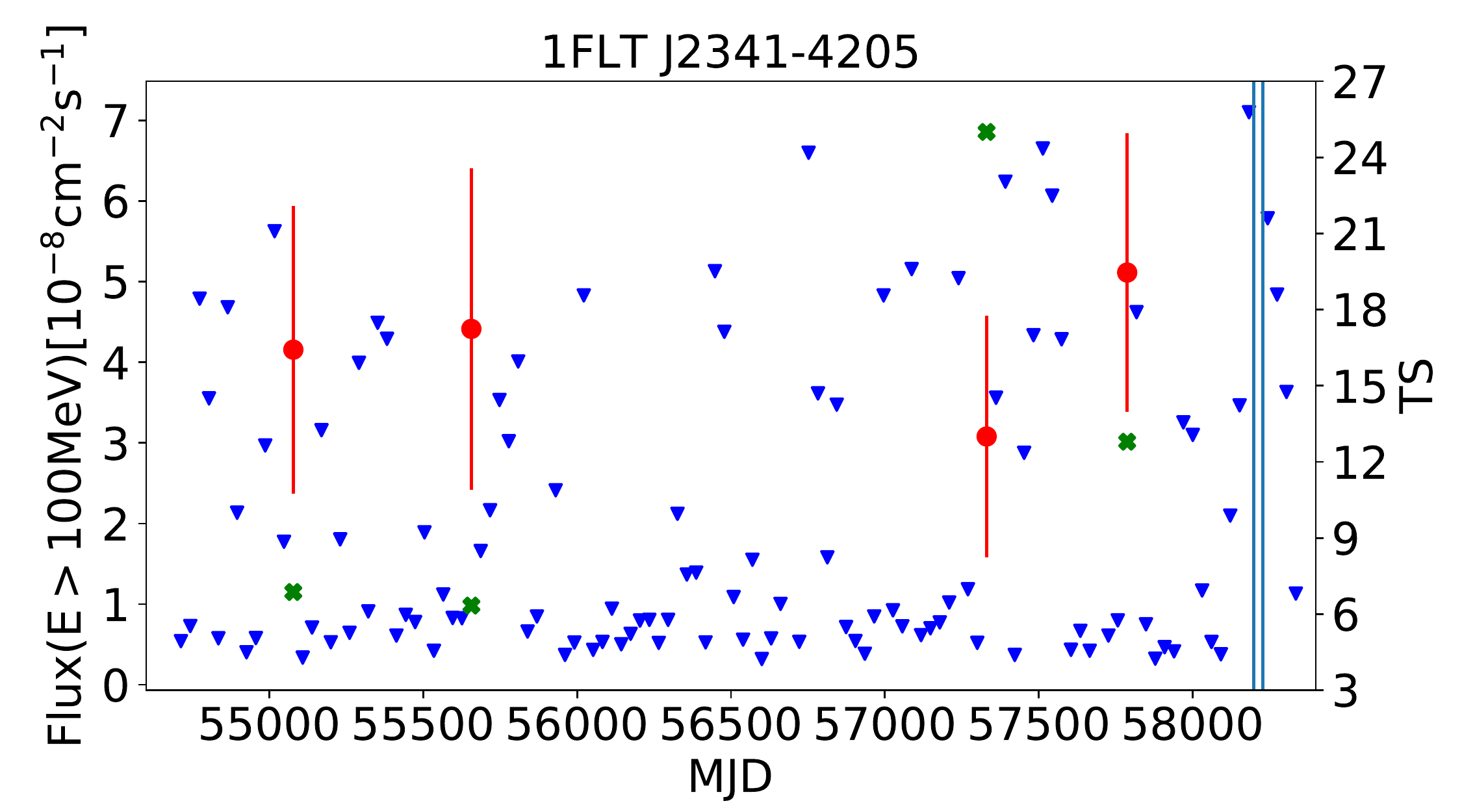}\label{fig:1FLTJ2341-420}\\
  %[1FLTJ2341-4205]& 
\end{tabular}
\end{figure}

%\end{linenumbers}

\end{document}